\documentclass[sunil1]{sunil-rpb}

\usepackage{amssymb}
\usepackage{amsmath}
\usepackage{fancybox}
\usepackage{color}
\usepackage{graphicx}
\usepackage{nicefrac}
\usepackage{dcolumn}
\usepackage{eurosym}
\usepackage{url}
\usepackage{dsfont}
\usepackage{arydshln}
\usepackage{multirow}
\usepackage{pict2e}
\usepackage{multicol}

\newlength{\figurewidth}
\newlength{\figureheight}
\newlength{\minipagewidth}
\newlength{\questionwidth}
\def\tableskip{\vskip 5pt plus 2pt minus 2pt\relax}

\newtheorem{remark}{Remark}

\def\limfunc#1{\mathop{\rm #1}}%
\def\func#1{\mathop{\rm #1}\nolimits}%
\def\slantfrac#1#2{\kern.1em^{#1}\kern-.3em/\kern-.1em_{#2}}

\newcommand{\bp}{\mathrel{\;}}
\newcommand{\bn}{\mathrel{\;\,}}

\newcommand{\bP}{\mathrel{\;\;}}
\newcommand{\bPp}{\mathrel{\;\;\,}}

\newcommand{\bPP}{\mathrel{\;\;\;\;}}

\newcommand{\bN}{\mathrel{\;\;\,}}
\newcommand{\bNP}{\mathrel{\;\;\,\;\;}}
\newcommand{\bNPP}{\mathrel{\;\;\,\;\;\;\;}}
\newcommand{\bnP}{\mathrel{\;\,\;\;}}

\definecolor{lyxor_light_blue}{RGB}{112,153,189}
\definecolor{lyxor_dark_blue}{RGB}{0,28,73}
\definecolor{lyxor_tan}{RGB}{195,187,175}
\definecolor{lyxor_cyan}{RGB}{0,164,167}
\definecolor{lyxor_pink}{RGB}{196,0,102}
\definecolor{lyxor_magenta}{RGB}{238,115,44}

\begin{document}

%\graphicspath{{Figures/exercises/pdf/}}

\title{Introduction to Risk Parity and Budgeting}

\author{Thierry Roncalli}

\maketitle

\pagestyle{empty}

\frontmatter

%\chapter*{Introduction}

\vspace*{1cm}
This book contains solutions of the tutorial exercises which are
provided in Appendix B of TR-RPB:
\begin{center}
\begin{minipage}{9cm}
\begin{itemize}
\item[(TR-RPB)] \textsc{Roncalli} T. (2013),
\textit{Introduction to Risk Parity and Budgeting},
Chapman \& Hall/CRC Financial Mathematics Series, 410 pages.
\end{itemize}
\end{minipage}
\end{center}
\begin{figure}[h]
\centering
\includegraphics[scale = 0.35]{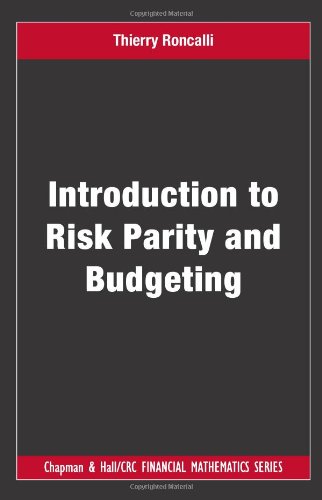}
\end{figure}
Description and materials of \emph{Introduction to Risk Parity and Budgeting} are available
on the author's website:
\begin{center}
\url{http://www.thierry-roncalli.com/riskparitybook.html}
\end{center}
or on the Chapman \& Hall website:
\begin{center}
\url{http://www.crcpress.com/product/isbn/9781482207156}
\end{center}
\vspace*{1cm}

I am grateful to Pierre Grison, Pierre Hereil and Zhengwei Wu for their careful reading of this
solution book.\vspace*{4cm}

\tableofcontents

\mainmatter

\setcounter{page}{1}

\chapter{Exercises related to modern portfolio theory}

\section{Markowitz optimized portfolios}

\begin{enumerate}
\item The weights of the minimum variance portfolio are:
$x_{1}^{\star }=3.05\%$, $x_{2}^{\star }=3.05\%$ and $x_{3}^{\star }=93.89\%$.
We have $\sigma \left( x^{\star} \right) = 4.94\% $.

\item We have to solve a $\sigma $-problem (TR-RPB, page 5). The optimal
value of $\phi $ is $49.99$ and the optimized portfolio is: $x_{1}^{\star }=6.11\%$, $x_{2}^{\star }=6.11\%$ and $x_{3}^{\star }=87.79\%$.

\item If the ex-ante volatility is equal to $10\%$, the optimal value of $%
\phi $ becomes $4.49$ and the optimized portfolio is: $x_{1}^{\star }=37.%
03\%$, $x_{2}^{\star }=37.03\%$ and $x_{3}^{\star }=25.94\%$.

\item We notice that $x_{1}^{\star }=x_{2}^{\star }$. This is normal
because the first and second assets present the same characteristics in terms
of expected return, volatility and correlation with the third asset.

\item
\begin{enumerate}
\item We obtain the following results:%
\begin{equation*}
\begin{tabular}{|c|ccc|}
\hline
$i$    & MV              & $\sigma \left( x\right) =5\%$ & $\sigma\left( x\right) =10\%$ \\ \hline
1      & ${\bP}8.00\%$ & ${\bP}8.00\%$ & $37.03\%$ \\
2      & ${\bP}0.64\%$ & ${\bP}3.66\%$ & $37.03\%$ \\
3      & $91.36\%$     & $88.34\%$     & $25.94\%$ \\ \hdashline
$\phi$ & $+\infty$     & $75.19$       & $4.49$    \\ \hline
\end{tabular}%
\end{equation*}
For the MV portfolio, we have $\sigma \left( x^{\star} \right) = 4.98\% $.

\item We consider the $\gamma $-formulation (TR-RPB, page 7). The
corresponding dual program is (TR-RPB, page 302):%
\begin{eqnarray*}
\lambda ^{\star } &=&\arg \min \frac{1}{2}\lambda ^{\top }\bar{Q}\lambda
-\lambda ^{\top }\bar{R} \\
&\text{u.c.}&%
\begin{array}{l}
\lambda \geq 0%
\end{array}%
\end{eqnarray*}%
with\footnote{%
We recall that $\mu $ and $\Sigma $ are the vector of expected returns and
the covariance matrix of asset returns.} $\bar{Q}=S\Sigma
^{-1}S^{\top }$, $\bar{R}=\gamma S\Sigma ^{-1}\mu -T$, $\gamma
=\phi ^{-1}$,%
\begin{equation*}
S=\left(
\begin{array}{rrr}
-1 & 0 & 0 \\
1 & 1 & 1 \\
-1 & -1 & -1 \\
-1 & 0 & 0 \\
0 & -1 & 0 \\
0 & 0 & -1%
\end{array}%
\right) \text{\quad and \quad }T=\left(
\begin{array}{r}
-8\% \\
1 \\
-1 \\
0 \\
0 \\
0%
\end{array}%
\right)
\end{equation*}%
$\lambda _{1}^{\star }$ is the Lagrange coefficient associated to the $8\%$
minimum exposure for the first asset ($x_{1}\geq 8\%$ in the primal program
and first row of the $S$ matrix in the dual program). $\max \left( \lambda
_{2}^{\star },\lambda _{3}^{\star }\right) $ is the Lagrange coefficient
associated to the fully invested portfolio constraint ($\sum_{i=1}^{3}x_{i}=100\%$
in the primal program and second and third rows of the $S$ matrix in the
dual program). Finally, the Lagrange coefficients $\lambda _{4}^{\star }$, $%
\lambda _{5}^{\star }$ and $\lambda _{6}^{\star }$ are associated to the
positivity constraints of the weights $x_{1}$, $x_{2}$ and $x_{3}$.

\item We have to solve the previous quadratic programming problem by
considering the value of $\phi $ corresponding to the results of Question
5(a). We obtain $\lambda _{1}^{\star }=0.0828\%$ for the minimum variance
portfolio, $\lambda _{1}^{\star }=0.0488\%$ for the optimized portfolio
with a $5\%$ ex-ante volatility and $\lambda _{1}^{\star }=0$ for the
optimized portfolio with a $10\%$ ex-ante volatility.

\item We verify that the Lagrange coefficient is zero for the optimized
portfolio with a 10\% ex-ante volatility, because the constraint $x_{1}\geq
8\%$ is not reached. The cost of this constraint is larger for the
minimum variance portfolio. Indeed, a relaxation $\varepsilon $ of this
constraint permits to reduce the variance by a factor equal to $2 \cdot
0.0828\%\cdot \varepsilon $.
\end{enumerate}

\item If we solve the minimum variance problem with $x_{1}\geq 20\%$, we
obtain a portfolio which has an ex-ante volatility equal to $5.46\%$.
There isn't a portfolio whose volatility is smaller than this lower bound.
We know that the constraints $x_{i}\geq 0$ are not reached for the minimum
variance problem regardless of the constraint $x_{1}\geq 20\%$. Let $%
\xi $ be the lower bound of $x_{1}$. Because of the previous results,
we have $0\%\leq \xi \leq 20\%$. We would like to find the minimum
variance portfolio $x^{\star }$ such that the constraint $x_{1}\geq \xi $ is reached
and $\sigma \left( x^{\star }\right) =\sigma ^{\star }=5\%$. In this case,
the optimization problem with three variables reduces to a minimum variance
problem with two variables with the constraint $x_{2}+x_{3}=1-\xi $
because $x_{1}^{\star }=\xi $. We then have:%
\begin{eqnarray*}
x^{\top }\Sigma x &=&x_{2}^{2}\sigma _{2}^{2}+2x_{2}x_{3}\rho _{2,3}\sigma
_{2}\sigma _{3}+x_{3}^{2}\sigma _{3}^{2}+ \\
&&\xi ^{2}\sigma _{1}^{2}+2\xi x_{2}\rho _{1,2}\sigma _{1}\sigma
_{2}+2\xi x_{3}\rho _{1,3}\sigma _{1}\sigma _{3}
\end{eqnarray*}%
The objective function becomes:%
\begin{eqnarray*}
x^{\top }\Sigma x &=&\left( 1-\xi -x_{3}\right) ^{2}\sigma
_{2}^{2}+2\left( 1-\xi -x_{3}\right) x_{3}\rho _{2,3}\sigma _{2}\sigma
_{3}+x_{3}^{2}\sigma _{3}^{2}+ \\
&&\xi ^{2}\sigma _{1}^{2}+2\xi \left( 1-\xi -x_{3}\right)
\rho _{1,2}\sigma _{1}\sigma _{2}+2\xi x_{3}\rho _{1,3}\sigma
_{1}\sigma _{3} \\
&=&x_{3}^{2}\left( \sigma _{2}^{2}-2\rho _{2,3}\sigma _{2}\sigma _{3}+\sigma
_{3}^{2}\right) + \\
&&2x_{3}\left( \left( 1-\xi \right) \left( \rho _{2,3}\sigma
_{2}\sigma _{3}-\sigma _{2}^{2}\right) -\xi \rho _{1,2}\sigma
_{1}\sigma _{2}+\xi \rho _{1,3}\sigma _{1}\sigma _{3}\right) + \\
&&\left( 1-\xi \right) ^{2}\sigma _{2}^{2}+\xi ^{2}\sigma
_{1}^{2}+2\xi \left( 1-\xi \right) \rho _{1,2}\sigma _{1}\sigma
_{2}
\end{eqnarray*}%
We deduce that:%
\begin{equation*}
\frac{\partial \,x^{\top }\Sigma x}{\partial \,x_{3}}=0\Leftrightarrow
x_{3}^{\star }=\frac{\left( 1-\xi \right) \left( \sigma
_{2}^{2}-\rho _{2,3}\sigma _{2}\sigma _{3}\right) +\xi \sigma
_{1}\left( \rho _{1,2}\sigma _{2}-\rho _{1,3}\sigma _{3}\right) }{\sigma
_{2}^{2}-2\rho _{2,3}\sigma _{2}\sigma _{3}+\sigma _{3}^{2}}
\end{equation*}%
The minimum variance portfolio is then:%
\begin{equation*}
\left\{
\begin{array}{l}
x_{1}^{\star }=\xi  \\
x_{2}^{\star }=a-\left( a+c\right) \xi  \\
x_{3}^{\star }=b-\left( b-c\right) \xi
\end{array}%
\right.
\end{equation*}%
with $a=\left( \sigma _{3}^{2}-\rho _{2,3}\sigma _{2}\sigma _{3}\right) /d$,
$b=\left( \sigma _{2}^{2}-\rho _{2,3}\sigma _{2}\sigma _{3}\right) /d$, $%
c=\sigma _{1}\left( \rho _{1,2}\sigma _{2}-\rho _{1,3}\sigma _{3}\right) /d$
and $d=\sigma _{2}^{2}-2\rho _{2,3}\sigma _{2}\sigma _{3}+\sigma _{3}^{2}$.
We also have:%
\begin{eqnarray*}
\sigma ^{2}\left( x\right)  &=&x_{1}^{2}\sigma _{1}^{2}+x_{2}^{2}\sigma
_{2}^{2}+x_{3}^{2}\sigma _{3}^{2}+2x_{1}x_{2}\rho _{1,2}\sigma _{1}\sigma
_{2}+2x_{1}x_{3}\rho _{1,3}\sigma _{1}\sigma _{3}+ \\
  & & 2x_{2}x_{3}\rho_{2,3}\sigma _{2}\sigma _{3} \\
&=&\xi ^{2}\sigma _{1}^{2}+\left( a-\left( a+c\right) \xi
\right) ^{2}\sigma _{2}^{2}+\left( b-\left( b-c\right) \xi \right)
^{2}\sigma _{3}^{2}+ \\
&&2\xi \left( a-\left( a+c\right) \xi \right) \rho _{1,2}\sigma
_{1}\sigma _{2}+ \\
&&2\xi \left( b-\left( b-c\right) \xi \right) \rho _{1,3}\sigma
_{1}\sigma _{3}+ \\
&&2\left( a-\left( a+c\right) \xi \right) \left( b-\left( b-c\right)
\xi \right) \rho _{2,3}\sigma _{2}\sigma _{3}
\end{eqnarray*}%
We deduce that the optimal value $\xi ^{\star }$ such that $\sigma
\left( x^{\star }\right) =\sigma ^{\star }$ satisfies the polynomial
equation of the second degree:%
\begin{equation*}
\alpha \xi ^{2}+2\beta \xi +\left( \gamma -\sigma ^{\star
^{2}}\right) =0
\end{equation*}%
with:
\begin{equation*}
\left\{
\begin{array}{lll}
\alpha  & = & \sigma _{1}^{2}+\left( a+c\right) ^{2}\sigma _{2}^{2}+\left(
b-c\right) ^{2}\sigma _{3}^{2}-2\left( a+c\right) \rho _{1,2}\sigma
_{1}\sigma _{2}- \\
& & 2\left( b-c\right) \rho _{1,3}\sigma _{1}\sigma _{3}+ 2\left( a+c\right) \left( b-c\right) \rho _{2,3}\sigma _{2}\sigma _{3}
\\
\beta  & = & -a\left( a+c\right) \sigma _{2}^{2}-b\left( b-c\right) \sigma
_{3}^{2}+a\rho _{1,2}\sigma _{1}\sigma _{2}+b\rho _{1,3}\sigma _{1}\sigma
_{3}-\\
& & \left( a\left( b-c\right) +b\left( a+c\right) \right) \rho _{2,3}\sigma
_{2}\sigma _{3} \\
\gamma  & = & a^{2}\sigma _{2}^{2}+b^{2}\sigma _{3}^{2}+2ab\rho _{2,3}\sigma
_{2}\sigma _{3}%
\end{array}%
\right.
\end{equation*}%
By using the numerical values, the solutions of the quadratic equation are
$\xi _{1}=9.09207\%$ and $\xi _{2}=-2.98520\%$. The optimal
solution is then $\xi ^{\star }=9.09207\%$. In order to check this result, we report in Figure \ref{fig:app2-1-1-1}
the volatility of the minimum variance portfolio when we impose the constraint $x_1 \geq x_1^{-}$.
We verify that the volatility is larger than $5\%$ when $x_1 \geq \xi ^{\star }$.
\end{enumerate}

\begin{figure}[tbph]
\centering
\includegraphics[width = \figurewidth, height = \figureheight]{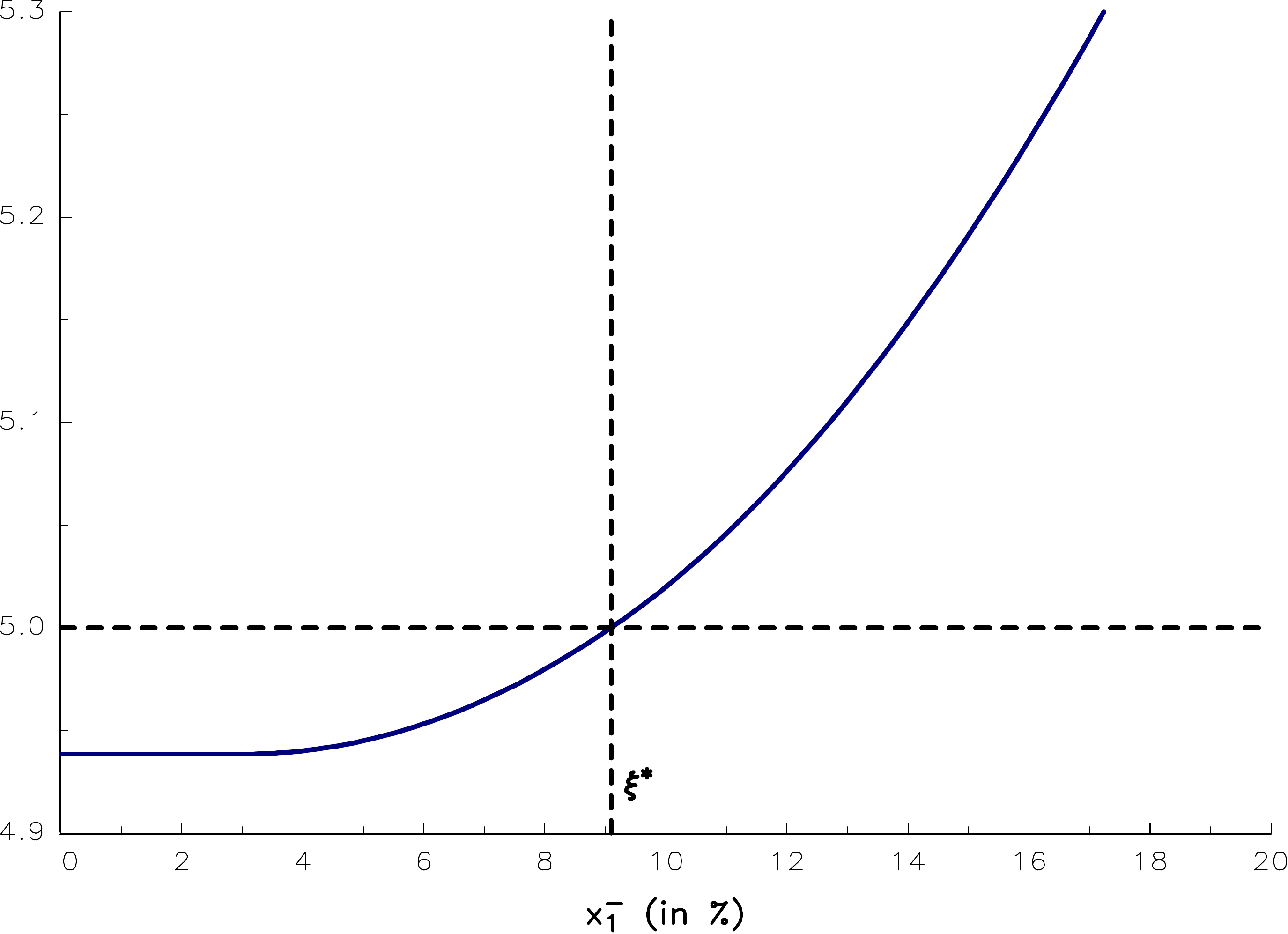}
\caption{Volatility of the minimum variance portfolio (in \%)}
\label{fig:app2-1-1-1}
\end{figure}

\section{Variations on the efficient frontier}
\label{section:exercise1-2}

\begin{enumerate}
\item We deduce that the covariance matrix is:
\begin{equation*}
\Sigma =\left(
\begin{array}{cccc}
2.250 & 0.300 & 1.500 & 2.250 \\
0.300 & 4.000 & 3.500 & 2.400 \\
1.500 & 3.500 & 6.250 & 6.000 \\
2.250 & 2.400 & 6.000 & 9.000%
\end{array}%
\right) \times 10^{-2}
\end{equation*}%
We then have to solve the $\gamma $-formulation of the Markowitz problem
(TR-RPB, page 7). We obtain the results\footnote{%
The weights, expected returns and volatilities are expressed in $\%$.} given in Table
\ref{tab:app2-1-2-1}. We represent the efficient frontier in Figure \ref{fig:app2-1-2-1}.

\begin{table}[tbh]
\centering
\caption{Solution of Question 1}
\label{tab:app2-1-2-1}
\tableskip
\begin{footnotesize}
\begin{tabular}{|c|rrrrrrrr|}
\hline
$\gamma$                       & $  -1.00$        & $ -0.50$        & $ -0.25$        & $  0.00$        & $  0.25$        & $  0.50$        & $  1.00$        & $  2.00$       \\
\hline
$x_1^\star$                    & $  94.04$        & $  83.39$        & $ 78.07$        & $ 72.74$        & $ 67.42$        & $ 62.09$        & $ 51.44$        & $  30.15$       \\
$x_2^\star$                    & $ 120.05$        & $  84.76$        & $ 67.11$        & $ 49.46$        & $ 31.82$        & $ 14.17$        & $-21.13$        & $ -91.72$       \\
$x_3^\star$                    & $-185.79$        & $-103.12$        & $-61.79$        & $-20.45$        & $ 20.88$        & $ 62.21$        & $144.88$        & $ 310.22$       \\
$x_4^\star$                    & $  71.69$        & $  34.97$        & $ 16.61$        & $ -1.75$        & $-20.12$        & $-38.48$        & $-75.20$        & $-148.65$       \\
\hdashline
$\mu \left( x^\star\right)$    & $   1.34$        & $   3.10$        & $  3.98$        & $  4.86$        & $  5.74$        & $  6.62$        & $  8.38$        & $  11.90$       \\
$\sigma \left( x^\star\right)$ & $  22.27$        & $  15.23$        & $ 12.88$        & $ 12.00$        & $ 12.88$        & $ 15.23$        & $ 22.27$        & $  39.39$       \\
\hline
\end{tabular}
\end{footnotesize}
\end{table}

\begin{figure}[tbph]
\centering
\includegraphics[width = \figurewidth, height = \figureheight]{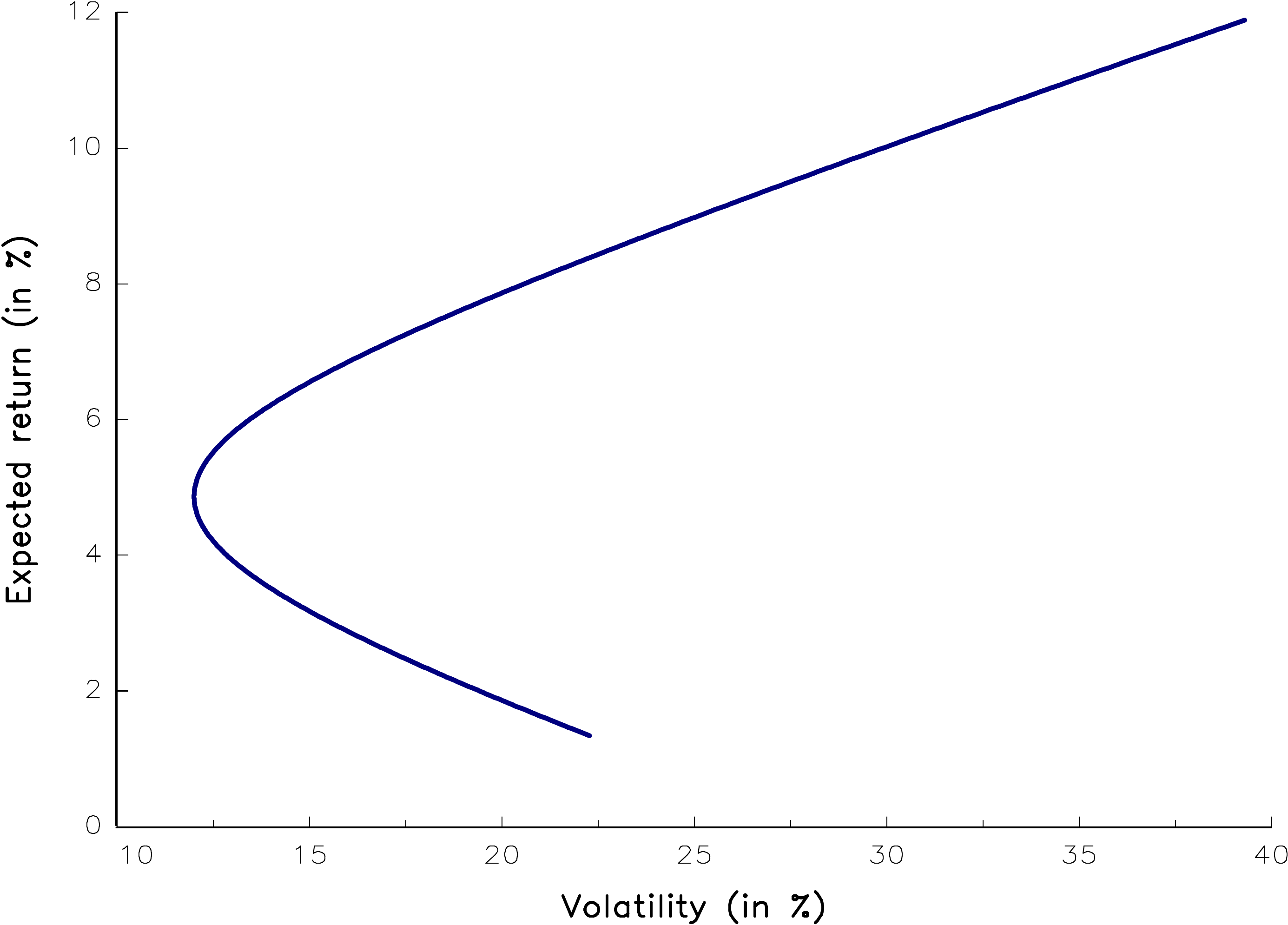}
\caption{Markowitz efficient frontier}
\label{fig:app2-1-2-1}
\end{figure}

\item We solve the $\gamma $-problem with $\gamma =0$. The minimum variance
portfolio is then $x_{1}^{\star }=72.74\%$, $x_{2}^{\star }=49.46\%$, $%
x_{3}^{\star }=-20.45\%$ and $x_{4}^{\star }=-1.75\%$. We deduce that $\mu
\left( x^{\star }\right) =4.86\%$ and $\sigma \left( x^{\star }\right)
=12.00\%$.

\item There is no solution when the target volatility $\sigma ^{\star }
$ is equal to $10\%$ because the minimum variance portfolio has a
volatility larger than $10\%$. Finding the optimized portfolio for $\sigma
^{\star }=15\%$ or $\sigma ^{\star }=20\%$ is equivalent to solving a $%
\sigma $-problem (TR-RPB, page 5). If $\sigma ^{\star }=15\%$ (resp. $%
\sigma ^{\star }=20\%$), we obtain an implied value of $\gamma $ equal to
$0.48$ (resp. $0.85$). Results are given in the following Table:
\begin{equation*}
\begin{tabular}{|c|rr|}
\hline
$\sigma^{\star}$               & $ 15.00$ & $ 20.00$ \\
\hline
$x_1^\star$                    & $ 62.52$ & $ 54.57$ \\
$x_2^\star$                    & $ 15.58$ & $-10.75$ \\
$x_3^\star$                    & $ 58.92$ & $120.58$ \\
$x_4^\star$                    & $-37.01$ & $-64.41$ \\
\hdashline
$\mu \left( x^\star\right)$    & $  6.55$ & $  7.87$ \\
$\gamma$                       & $  0.48$ & $  0.85$ \\
\hline
\end{tabular}
\end{equation*}

\item Let $x^{\left( \alpha \right) }$ be the portfolio defined by the
relationship $x^{\left( \alpha \right) }=\left( 1-\alpha \right) x^{\left(
1\right) }+\alpha x^{\left( 2\right) }$ where $x^{\left( 1\right) }$ is the
minium variance portfolio and $x^{\left( 2\right) }$ is the optimized
portfolio with a $20\%$ ex-ante volatility. We obtain the following results:
\begin{equation*}
\begin{tabular}{|r|cc|}
\hline
\multicolumn{1}{|c|}{$\alpha$} &
\multicolumn{1}{c}{$\sigma \left(x^{\left( \alpha \right) }\right)$} &
\multicolumn{1}{c|}{$\mu \left( x^{\left( \alpha \right) }\right)$} \\
\hline
$-0.50$ & $14.42$ & $3.36$ \\
$-0.25$ & $12.64$ & $4.11$ \\
$ 0.00$ & $12.00$ & $4.86$ \\
$ 0.10$ & $12.10$ & $5.16$ \\
$ 0.20$ & $12.41$ & $5.46$ \\
$ 0.50$ & $14.42$ & $6.36$ \\
$ 0.70$ & $16.41$ & $6.97$ \\
$ 1.00$ & $20.00$ & $7.87$ \\
\hline
\end{tabular}
\end{equation*}
We have reported these portfolios in Figure \ref{fig:app2-1-2-2}. We notice
that they are located on the efficient frontier. This is perfectly normal
because we know that a combination of two optimal portfolios corresponds to
another optimal portfolio.

\begin{figure}[tbph]
\centering
\includegraphics[width = \figurewidth, height = \figureheight]{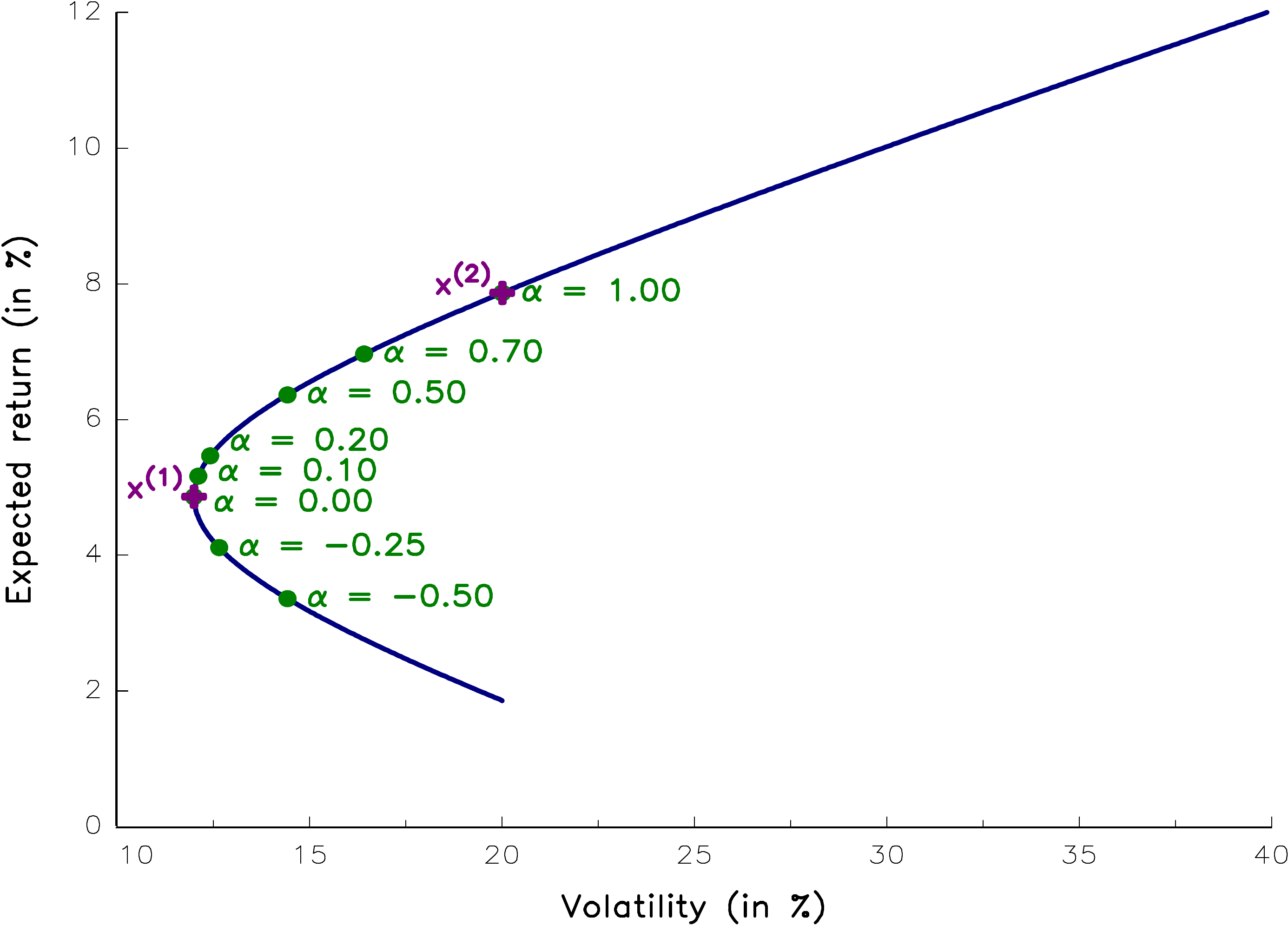}
\caption{Mean-variance diagram of portfolios $x^{\left( \alpha \right) }$}
\label{fig:app2-1-2-2}
\end{figure}

\item If we consider the constraint $0\leq x_{i}\leq 1$, we obtain the
following results:
\begin{equation*}
\begin{tabular}{|c|rcrr|}
\hline
$\sigma^{\star}$               & MV       & $ 12.00$ & $ 15.00$ & $ 20.00$ \\
\hline
$x_1^\star$                    & $65.49$  &  $\checkmark$  &  $45.59$  &  $24.88$  \\
$x_2^\star$                    & $34.51$  &  $\checkmark$  &  $24.74$  &  $ 4.96$  \\
$x_3^\star$                    & $ 0.00$  &  $\checkmark$  &  $29.67$  &  $70.15$  \\
$x_4^\star$                    & $ 0.00$  &  $\checkmark$  &  $ 0.00$  &  $ 0.00$  \\
\hdashline
$\mu \left( x^\star\right)$    & $ 5.35$  &  $\checkmark$  &  $ 6.14$  &  $ 7.15$  \\
$\sigma \left( x^\star\right)$ & $12.56$  &  $\checkmark$  &  $15.00$  &  $20.00$  \\
$\gamma$                       & $ 0.00$  &  $\checkmark$  &  $ 0.62$  &  $ 1.10$  \\
\hline
\end{tabular}
\end{equation*}

\item
\begin{enumerate}
\item We have:
\begin{equation*}
\mu =\left(
\begin{array}{c}
5.0 \\
6.0 \\
8.0 \\
6.0 \\
3.0%
\end{array}%
\right) \times 10^{-2}
\end{equation*}%
and:
\begin{equation*}
\Sigma =\left(
\begin{array}{ccccc}
2.250 & 0.300 & 1.500 & 2.250 & 0.000 \\
0.300 & 4.000 & 3.500 & 2.400 & 0.000 \\
1.500 & 3.500 & 6.250 & 6.000 & 0.000 \\
2.250 & 2.400 & 6.000 & 9.000 & 0.000 \\
0.000 & 0.000 & 0.000 & 0.000 & 0.000%
\end{array}%
\right) \times 10^{-2}
\end{equation*}

\item We solve the $\gamma $-problem and obtain the efficient frontier
given in Figure \ref{fig:app2-1-2-3}.

\begin{figure}[tbph]
\centering
\includegraphics[width = \figurewidth, height = \figureheight]{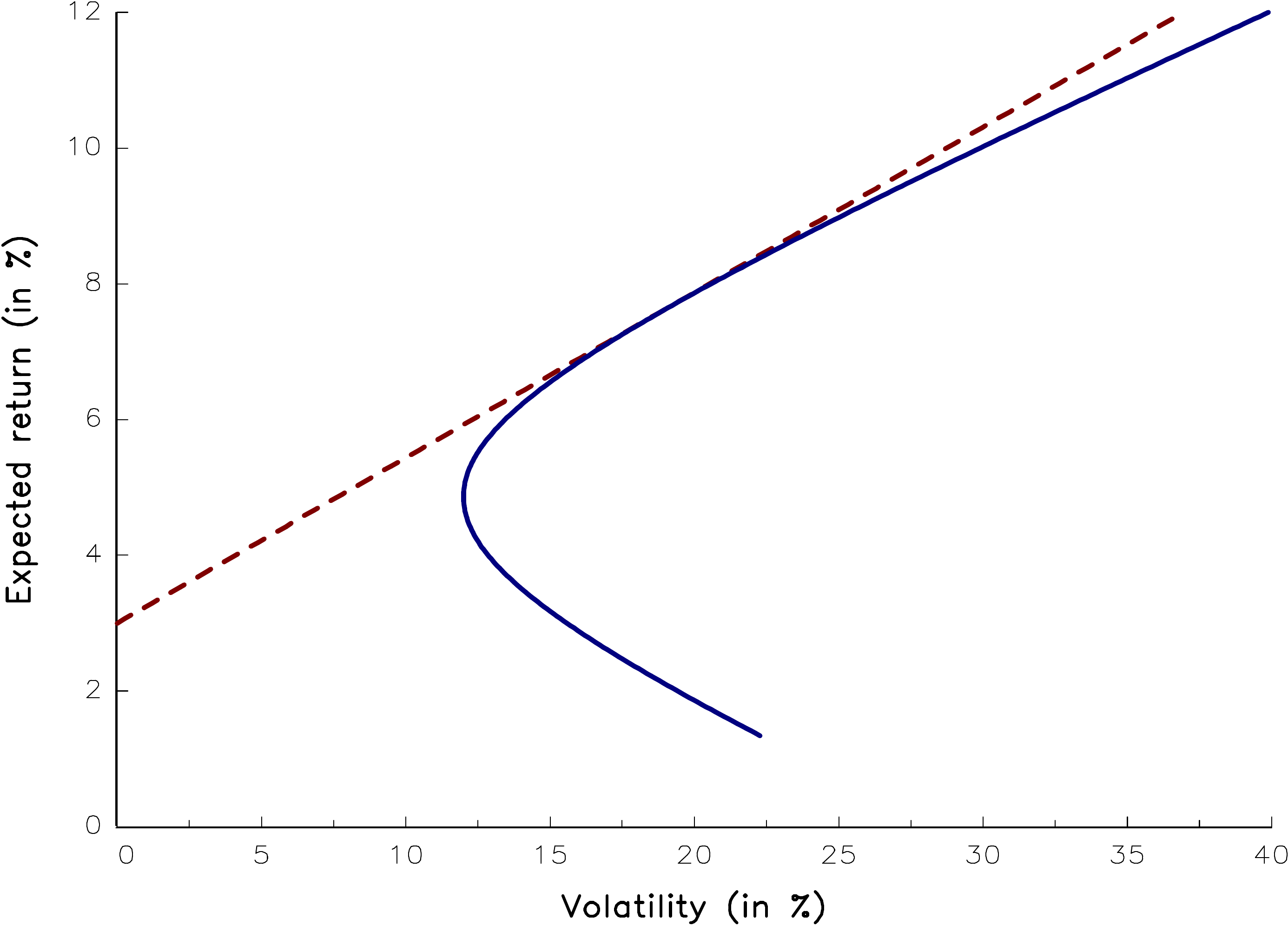}
\caption{Efficient frontier when the risk-free asset is introduced}
\label{fig:app2-1-2-3}
\end{figure}

\item This efficient frontier is a straight line. This line passes through
the risk-free asset and is tangent to the efficient frontier of Figure \ref%
{fig:app2-1-2-1}. This exercise is a direct application of the \textit{Separation
Theorem} of Tobin.

\item We consider two optimized portfolios of this efficient frontier. They
corresponds to $\gamma =0.25$ and $\gamma = 0.50$. We obtain the following
results:%
\begin{equation*}
\begin{tabular}{|c|rr|} \hline
$\gamma$                       & $  0.25$        & $  0.50$       \\ \hline
$x_1^\star$                    & $ 18.23$        & $ 36.46$       \\
$x_2^\star$                    & $ -1.63$        & $ -3.26$       \\
$x_3^\star$                    & $ 34.71$        & $ 69.42$       \\
$x_4^\star$                    & $-18.93$        & $-37.86$       \\
$x_5^\star$                    & $ 67.62$        & $ 35.24$       \\ \hdashline
$\mu \left( x^\star\right)$    & $  4.48$        & $  5.97$       \\
$\sigma \left( x^\star\right)$ & $  6.09$        & $ 12.18$       \\
\hline
\end{tabular}
\end{equation*}
The first portfolio has an expected return equal to $4.48\%$ and a
volatility equal to $6.09\%$. The weight of the risk-free asset is $67.62\%$.
This explains the low volatility of this portfolio. For the second
portfolio, the weight of the risk-free asset is lower and equal to $35.24\%$.
The expected return and the volatility are then equal to $5.97\%$ and
$12.18\%$. We note $x^{\left( 1\right) }$ and $x^{\left( 2\right) }$ these two
portfolios. By definition, the Sharpe ratio of the market portfolio $%
x^{\star }$ is the tangency of the line. We deduce that:%
\begin{eqnarray*}
\func{SR}\left( x^{\star }\mid r\right)  &=&\frac{\mu \left( x^{\left(
2\right) }\right) -\mu \left( x^{\left( 1\right) }\right) }{\sigma \left(
x^{\left( 2\right) }\right) -\sigma \left( x^{\left( 1\right) }\right) } \\
&=&\frac{5.97-4.48}{12.18-6.09} \\
&=&0.2436
\end{eqnarray*}%
The Sharpe ratio of the market portfolio $x^{\star }$ is then equal to $0.2436$.

\item By construction, every portfolio $x^{\left( \alpha \right) }$ which
belongs to the tangency line is a linear combination of two portfolios $%
x^{\left( 1\right) }$ and $x^{\left( 2\right) }$ of this efficient frontier:
\begin{equation*}
x^{\left( \alpha \right) }=\left( 1-\alpha \right) x^{\left( 1\right)
}+\alpha x^{\left( 2\right) }
\end{equation*}%
The market portfolio $x^{\star }$ is the portfolio $x^{\left( \alpha \right)
}$ which has a zero weight in the risk-free asset. We deduce that the value $%
\alpha ^{\star }$ which corresponds to the market portfolio satisfies the
following relationship:
\begin{equation*}
\left( 1-\alpha ^{\star }\right) x_{5}^{\left( 1\right) }+\alpha ^{\star
}x_{5}^{\left( 2\right) }=0
\end{equation*}%
because the risk-free asset is the fifth asset of the portfolio. It follows
that:
\begin{eqnarray*}
\alpha ^{\star } &=&\frac{x_{5}^{\left( 1\right) }}{x_{5}^{\left( 1\right)
}-x_{5}^{\left( 2\right) }} \\
&=&\frac{67.62}{67.62-35.24} \\
&=&2.09
\end{eqnarray*}%
We deduce that the market portfolio is:
\begin{equation*}
x^{\star }=\left( 1-2.09\right) \cdot \left(
\begin{array}{r}
 18.23 \\
 -1.63 \\
 34.71 \\
-18.93 \\
 67.62
\end{array}%
\right) + 2.09\cdot \left(
\begin{array}{r}
 36.46 \\
 -3.26 \\
 69.42 \\
-37.86 \\
 35.24
\end{array}%
\right)  \\
=\left(
\begin{array}{r}
56.30 \\
-5.04 \\
107.21 \\
-58.46 \\
0.00%
\end{array}%
\right)
\end{equation*}
We check that the Sharpe ratio of this portfolio is $0.2436$.
\end{enumerate}

\begin{enumerate}
\item We have:
\begin{equation*}
\tilde{\mu}=\left(
\begin{array}{c}
\mu  \\
r%
\end{array}%
\right)
\end{equation*}%
and:
\begin{equation*}
\tilde{\Sigma}=\left(
\begin{array}{cc}
\Sigma  & \mathbf{0} \\
\mathbf{0} & 0%
\end{array}%
\right)
\end{equation*}

\item This problem is entirely solved in TR-RPB on page 13.
\end{enumerate}
\end{enumerate}

\section{Sharpe ratio}

\begin{enumerate}
\item
\begin{enumerate}
\item We have (TR-RPB, page 12):
\begin{equation*}
\limfunc{SR}\nolimits_{i}=\frac{\mu _{i}-r}{\sigma _{i}}
\end{equation*}

\item We have:%
\begin{equation*}
\limfunc{SR}\left( x\mid r\right) =\frac{x_{1}\mu _{1}+x_{2}\mu _{2}-r}{%
\sqrt{x_{1}^{2}\sigma _{1}^{2}+2x_{1}x_{2}\rho \sigma _{1}\sigma
_{2}+x_{2}^{2}\sigma _{2}^{2}}}
\end{equation*}

\item If the second asset corresponds to the risk-free asset, its volatility
$\sigma _{2}$ and its correlation $\rho $ with the first asset are equal to
zero. We deduce that:%
\begin{eqnarray*}
\limfunc{SR}\left( x\mid r\right)  &=&\frac{x_{1}\mu _{1}+\left(
1-x_{1}\right) r-r}{\sqrt{x_{1}^{2}\sigma _{1}^{2}}} \\
&=&\frac{x_{1}\left( \mu _{1}-r\right) }{\left\vert x_{1}\right\vert \sigma
_{1}} \\
&=&\limfunc{sgn}\left( x_{1}\right) \cdot \limfunc{SR}\nolimits_{1}
\end{eqnarray*}%
We finally obtain that:%
\begin{equation*}
\limfunc{SR}\left( x\mid r\right) =\left\{
\begin{array}{ll}
-\limfunc{SR}_{1} & \text{if }x_{1}<0 \\
+\limfunc{SR}_{1} & \text{if }x_{1}>0%
\end{array}%
\right.
\end{equation*}
\end{enumerate}

\item

\begin{enumerate}
\item Let $R\left( x\right) $ be the return of the portfolio $x$. We have:%
\begin{equation*}
\mathbb{E}\left[ R\left( x\right) \right] =\sum_{i=1}^{n}n^{-1}\mu
_{i}=n^{-1}\sum_{i=1}^{n}\mu _{i}
\end{equation*}%
and:%
\begin{equation*}
\sigma \left( R\left( x\right) \right) =\sqrt{\sum_{i=1}^{n}\left(
n^{-1}\sigma _{i}\right) ^{2}}=n^{-1}\sqrt{\sum_{i=1}^{n}\sigma _{i}^{2}}
\end{equation*}%
We deduce that the Sharpe ratio of the portfolio $x$ is:%
\begin{eqnarray*}
\limfunc{SR}\left( x\mid r\right)  &=&\frac{n^{-1}\sum_{i=1}^{n}\mu _{i}-r}{%
n^{-1}\sqrt{\sum_{i=1}^{n}\sigma _{i}^{2}}} \\
&=&\frac{\sum_{i=1}^{n}\left( \mu _{i}-r\right) }{\sqrt{\sum_{i=1}^{n}\sigma
_{i}^{2}}}
\end{eqnarray*}%
because $r=n^{-1}\sum_{i=1}^{n}r$.

\item Another expression of the Sharpe ratio is:%
\begin{eqnarray*}
\limfunc{SR}\left( x\mid r\right)  &=&\sum_{i=1}^{n}\frac{\sigma _{i}}{\sqrt{%
\sum_{j=1}^{n}\sigma _{i}^{2}}}\cdot \frac{\left( \mu _{i}-r\right) }{\sigma
_{i}} \\
&=&\sum_{i=1}^{n}w_{i}\limfunc{SR}\nolimits_{i}
\end{eqnarray*}%
with:%
\begin{equation*}
w_{i}=\frac{\sigma _{i}}{\sqrt{\sum_{j=1}^{n}\sigma _{i}^{2}}}
\end{equation*}

\item Because $0<\sigma _{i}<\sqrt{\sum_{j=1}^{n}\sigma _{i}^{2}}$, we
deduce that:%
\begin{equation*}
0<w_{i}<1
\end{equation*}

\item We obtain the following results:
\begin{equation*}
\begin{small}
\begin{tabular}{|l|cccccc|c|}
\hline
& $w_{1}$ & $w_{2}$ & $w_{3}$ & $w_{4}$ & $w{_{5}}$ & $\sum_{i=1}^{n}w_{i}$ & $\limfunc{SR}\left( x\mid r\right) $ \\ \hline
$\mathcal{A}_{1}$ & $38.5\%$ & $38.5\%$ & $57.7\%$ & $19.2\%$ & $57.7\%$ & $211.7\%$ & $0.828$ \\
$\mathcal{A}_{2}$ & $25.5\%$ & $25.5\%$ & $34.1\%$ & $17.0\%$ & $85.1\%$ & $187.3\%$ & $0.856$ \\
\hline
\end{tabular}%
\end{small}
\end{equation*}%
It may be surprising that the portfolio based on the set $\mathcal{A}_{2}$ has a larger
Sharpe ratio than the portfolio based on the set $\mathcal{A}_{1}$, because four assets of $%
\mathcal{A}_{2}$ are all dominated by the assets of $\mathcal{A}_{1}$. Only
the fifth asset of $\mathcal{A}_{2}$ has a higher Sharpe ratio. However, we
easily understand this result if we consider the previous decomposition.
Indeed, this fifth asset has a higher volatility than the other assets. It
follows that its contribution $w_{5}$ to the Sharpe ratio is then much
greater.
\end{enumerate}

\item

\begin{enumerate}
\item We have:%
\begin{eqnarray*}
\sigma \left( R\left( x\right) \right)  &=&\sqrt{\sum_{i=1}^{n}\left(
n^{-1}\sigma \right) ^{2}+2\sum_{i>j}^{n}\rho \left( n^{-1}\sigma \right)
^{2}} \\
&=&\sigma \sqrt{\rho +n^{-1}\left( 1-\rho \right) }
\end{eqnarray*}%
We deduce that the Sharpe ratio is:%
\begin{equation*}
\limfunc{SR}\left( x\mid r\right) =\frac{n^{-1}\sum_{i=1}^{n}\mu _{i}-r}{%
\sigma \sqrt{\rho +n^{-1}\left( 1-\rho \right) }}
\end{equation*}

\item It follows that:%
\begin{eqnarray*}
\limfunc{SR}\left( x\mid r\right)  &=&\frac{1}{\sqrt{\rho +n^{-1}\left(
1-\rho \right) }}n^{-1}\sum_{i=1}^{n}\frac{\left( \mu _{i}-r\right) }{\sigma
} \\
&=&w\cdot \left( \frac{1}{n}\sum_{i=1}^{n}\limfunc{SR}\nolimits_{i}\right)
\end{eqnarray*}%
with:%
\begin{equation*}
w=\frac{1}{\sqrt{\rho +n^{-1}\left( 1-\rho \right) }}
\end{equation*}

\item One seeks $n$ such that:%
\begin{equation*}
\frac{1}{\sqrt{\rho +n^{-1}\left( 1-\rho \right) }}=w
\end{equation*}%
We deduce that:%
\begin{equation*}
n^{\star }=w^{2}\frac{1-\rho }{1-\rho w^{2}}
\end{equation*}%
If $\rho =50\%$ and $w=1.25$, we obtain:%
\begin{eqnarray*}
n^{\star } &=&1.25^{2}\frac{1-0.5}{1-0.5\cdot 1.25^{2}} \\
&=&3.57
\end{eqnarray*}%
Four assets are sufficient to improve the Sharpe ratio by a factor of $25\%$.

\item We notice that:%
\begin{equation*}
w=\frac{1}{\sqrt{\rho +n^{-1}\left( 1-\rho \right) }}<\frac{1}{\sqrt{\rho }}
\end{equation*}%
If $\rho =80\%$, then $w<1.12$. We cannot improve the Sharpe ratio by $%
25\%$ when the correlation is equal to $80\%$.

\item The most important parameter is the correlation $\rho $. The lower this correlation,
the larger the increase of the Sharpe ratio. If the correlation is high, the
gain in terms of Sharpe ratio is negligible. For instance, if $\rho \geq 80\%
$, the gain cannot exceed $12\%$.
\end{enumerate}

\item

\begin{enumerate}
\item Let $R^{g}\left( x\right) $ be the gross performance of the portfolio.
We note $m$ and $p$ the management and performance fees. The net performance $R^{n}\left(
x\right) $ is equal to:%
\begin{equation*}
R^{n}\left( x\right) =\left( R^{g}\left( x\right) -m\right) -p\left(
R^{g}\left( x\right) -m-\func{Libor}\right) _{+}
\end{equation*}%
If we assume that $R^{g}\left( x\right) -m-\func{Libor}>0$, we obtain:%
\begin{eqnarray*}
R^{n}\left( x\right)  &=&\left( R^{g}\left( x\right) -m\right) -p\left(
R^{g}\left( x\right) -m-\func{Libor}\right)  \\
&=&\left( 1-p\right) \left( R^{g}\left( x\right) -m\right) +p\func{Libor}
\end{eqnarray*}%
We deduce that:%
\begin{equation*}
R^{g}\left( x\right) =m+\frac{\left( R^{n}\left( x\right) -p\func{Libor}%
\right) }{1-p}
\end{equation*}%
Using the numerical values, we obtain:%
\begin{eqnarray*}
R^{g}\left( x\right)  &=&1\%+\frac{\left( \func{Libor}+4\%-10\%\cdot \func{%
Libor}\right) }{\left( 1-10\%\right) } \\
&=&\func{Libor}+544\text{ bps}
\end{eqnarray*}%
Moreover, if we assume that the performance fees have little influence on the
volatility of the portfolio%
\footnote{This is not true in practice.}, the Sharpe ratio of the hedge funds portfolio
is equal to:%
\begin{eqnarray*}
\limfunc{SR}\left( x\mid r\right)  &=&\frac{\func{Libor}+544\text{ bps}-%
\func{Libor}}{4\%} \\
&=&1.36
\end{eqnarray*}

\item We obtain the following results:
\begin{equation*}
\begin{tabular}{|c|ccccccc|}
\hline
         & \multicolumn{7}{c|}{$\rho$}                  \\
           &    $0.00$ & $0.10$ & $0.20$ & $0.30$ & $0.50$ & $0.75$ & $0.90$ \\ \hline
$n=10$     &    $3.16$ & $2.29$ & $1.89$ & $1.64$ & $1.35$ & $1.14$ & $1.05$ \\
$n=20$     &    $4.47$ & $2.63$ & $2.04$ & $1.73$ & $1.38$ & $1.15$ & $1.05$ \\
$n=30$     &    $5.48$ & $2.77$ & $2.10$ & $1.76$ & $1.39$ & $1.15$ & $1.05$ \\
$n=50$     &    $7.07$ & $2.91$ & $2.15$ & $1.78$ & $1.40$ & $1.15$ & $1.05$ \\
$+\infty$  & $+\infty$ & $3.16$ & $2.24$ & $1.83$ & $1.41$ & $1.15$ & $1.05$ \\

\hline
\end{tabular}
\end{equation*}
This means for instance that if the correlation among the hedge funds
is equal to $20\%$, the Sharpe ratio of a portfolio of 30 hedge funds is
multiplied by a factor of $2.10$ with respect to the average Sharpe ratio.

\item If we assume that the average Sharpe ratio of single hedge funds is $%
0.5$ and if we target a Sharpe ratio equal to $1.36$ gross of fees, the
multiplication factor $w$ must satisfy the following inequality:%
\begin{eqnarray*}
w &\geq &\frac{\limfunc{SR}\left( x\mid r\right) }{n^{-1}\sum_{i=1}^{n}%
\limfunc{SR}\nolimits_{i}} \\
&=&\frac{1.36}{0.50} \\
&=&2.72
\end{eqnarray*}%
It is then not possible to achieve a net performance of Libor + 400 bps with a
volatility of $4\%$ if the correlation between these hedge funds is larger
than $20\%$.
\end{enumerate}
\end{enumerate}

\section{Beta coefficient}

\begin{enumerate}
\item
\begin{enumerate}
\item The beta of an asset is the ratio between its covariance with the
market portfolio return and the variance of the market portfolio return
(TR-RPB, page 16). In the CAPM theory, we have:%
\begin{equation*}
\mathbb{E}\left[ R_{i}\right] =r+\beta _{i}\left( \mathbb{E}\left[ R\left(
b\right) \right] -r\right)
\end{equation*}%
where $R_{i}$ is the return of asset $i$, $R\left( b\right) $ is the return of the
market portfolio and $r$ is the risk-free rate. The beta $\beta _{i}$ of asset $%
i$ is:%
\begin{equation*}
\beta _{i}=\frac{\limfunc{cov}\left( R_{i},R\left( b\right) \right) }{%
\limfunc{var}\left( R\left( b\right) \right) }
\end{equation*}%
Let $\Sigma $ be the covariance matrix of asset returns. We have $\limfunc{%
cov}\left( R,R\left( b\right) \right) =\Sigma b$ and $\limfunc{var}\left(
R\left( b\right) \right) =b^{\top }\Sigma b$. We deduce that:%
\begin{equation*}
\beta _{i}=\frac{\left( \Sigma b\right) _{i}}{b^{\top }\Sigma b}
\end{equation*}

\item We recall that the mathematical operator $\mathbb{E}$ is bilinear.
Let $c$ be the covariance $\func{cov}\left( c_{1}X_{1}+c_{2}X_{2},X_{3}\right)$.
We then have:%
\begin{eqnarray*}
c &=&\mathbb{E}\left[
\left( c_{1}X_{1}+c_{2}X_{2}-\mathbb{E}\left[ c_{1}X_{1}+c_{2}X_{2}\right]
\right) \left( X_{3}-\mathbb{E}\left[ X_{3}\right] \right) \right]  \\
&=&\mathbb{E}\left[ \left( c_{1}\left( X_{1}-\mathbb{E}\left[ X_{1}\right]
\right) +c_{2}\left( X_{2}-\mathbb{E}\left[ X_{2}\right] \right) \right)
\left( X_{3}-\mathbb{E}\left[ X_{3}\right] \right) \right]  \\
&=&c_{1}\mathbb{E}\left[ \left( X_{1}-\mathbb{E}\left[ X_{1}\right] \right)
\left( X_{3}-\mathbb{E}\left[ X_{3}\right] \right) \right] + \\
&&c_{2}\mathbb{E}\left[ \left( X_{2}-\mathbb{E}\left[ X_{2}\right] \right)
\left( X_{3}-\mathbb{E}\left[ X_{3}\right] \right) \right]  \\
&=&c_{1}\func{cov}\left( X_{1},X_{3}\right) +c_{2}\func{cov}\left(
X_{2},X_{3}\right)
\end{eqnarray*}

\item We have:%
\begin{eqnarray*}
\beta \left( x\mid b\right)  &=&\frac{\limfunc{cov}\left( R\left( x\right)
,R\left( b\right) \right) }{\limfunc{var}\left( R\left( b\right) \right) } \\
&=&\frac{\limfunc{cov}\left( x^{\top }R,b^{\top }R\right) }{\limfunc{var}%
\left( b^{\top }R\right) } \\
&=&\frac{x^{\top }\mathbb{E}\left[ \left( R-\mu \right) \left( R-\mu \right)
^{\top }\right] b}{b^{\top }\mathbb{E}\left[ \left( R-\mu \right) \left(
R-\mu \right) ^{\top }\right] b} \\
&=&\frac{x^{\top }\Sigma b}{b^{\top }\Sigma b} \\
&=&x^{\top }\frac{\Sigma b}{b^{\top }\Sigma b} \\
&=&x^{\top }\beta  \\
&=&\sum_{i=1}^{n}x_{i}\beta _{i}
\end{eqnarray*}%
with $\beta =\left( \beta _{1},\ldots ,\beta _{n}\right) $. The beta of
portfolio $x$ is then the weighted mean of asset betas. Another way to
show this result is to exploit the result of Question 1(b). We have:%
\begin{eqnarray*}
\beta \left( x\mid b\right)  &=&\frac{\limfunc{cov}\left(
\sum_{i=1}^{n}x_{i}R_{i},R\left( b\right) \right) }{\limfunc{var}\left(
R\left( b\right) \right) } \\
&=&\sum_{i=1}^{n}x_{i}\frac{\limfunc{cov}\left( R_{i},R\left( b\right)
\right) }{\limfunc{var}\left( R\left( b\right) \right) } \\
&=&\sum_{i=1}^{n}x_{i}\beta _{i}
\end{eqnarray*}

\item We obtain $\beta \left( x^{\left( 1\right) }\mid b\right) =0.80$ and $%
\beta \left( x^{\left( 2\right) }\mid b\right) =0.85$.
\end{enumerate}

\item The weights of the market portfolio are then $b=n^{-1}\mathbf{1}$.

\begin{enumerate}
\item We have:%
\begin{eqnarray*}
\beta  &=&\frac{\limfunc{cov}\left( R,R\left( b\right) \right) }{\limfunc{var%
}\left( R\left( b\right) \right) } \\
&=&\frac{\Sigma b}{b^{\top }\Sigma b} \\
&=&\frac{n^{-1}\Sigma \mathbf{1}}{n^{-2}\left( \mathbf{1}^{\top }\Sigma
\mathbf{1}\right) } \\
&=&n\frac{\Sigma \mathbf{1}}{\left( \mathbf{1}^{\top }\Sigma \mathbf{1}%
\right) }
\end{eqnarray*}%
We deduce that:%
\begin{eqnarray*}
\sum_{i=1}^{n}\beta _{i} &=&\mathbf{1}^{\top }\beta  \\
&=&\mathbf{1}^{\top }n\frac{\Sigma \mathbf{1}}{\left( \mathbf{1}^{\top
}\Sigma \mathbf{1}\right) } \\
&=&n\frac{\mathbf{1}^{\top }\Sigma \mathbf{1}}{\left( \mathbf{1}^{\top
}\Sigma \mathbf{1}\right) } \\
&=&n
\end{eqnarray*}

\item If $\rho _{i,j}=0$, we have:%
\begin{equation*}
\beta _{i}=n\frac{\sigma _{i}^{2}}{\sum_{j=1}^{n}\sigma _{j}^{2}}
\end{equation*}%
We deduce that:%
\begin{eqnarray*}
\beta _{1}\geq \beta _{2}\geq \beta _{3} &\Rightarrow &n\frac{\sigma _{1}^{2}%
}{\sum_{j=1}^{3}\sigma _{j}^{2}}\geq n\frac{\sigma _{2}^{2}}{%
\sum_{j=1}^{3}\sigma _{j}^{2}}\geq n\frac{\sigma _{3}^{2}}{%
\sum_{j=1}^{3}\sigma _{j}^{2}} \\
&\Rightarrow &\sigma _{1}^{2}\geq \sigma _{2}^{2}\geq \sigma _{3}^{2} \\
&\Rightarrow &\sigma _{1}\geq \sigma _{2}\geq \sigma _{3}
\end{eqnarray*}

\item If $\rho _{i,j}=\rho $, it follows that:%
\begin{eqnarray*}
\beta _{i} &\propto &\sigma _{i}^{2}+\sum_{j\neq i}\rho \sigma _{i}\sigma
_{j} \\
&=&\sigma _{i}^{2}+\rho \sigma _{i}\sum_{j\neq i}\sigma _{j}+\rho \sigma
_{i}^{2}-\rho \sigma _{i}^{2} \\
&=&\left( 1-\rho \right) \sigma _{i}^{2}+\rho \sigma
_{i}\sum_{j=1}^{n}\sigma _{j} \\
&=&f\left( \sigma _{i}\right)
\end{eqnarray*}%
with:%
\begin{equation*}
f\left( z\right) =\left( 1-\rho \right) z^{2}+\rho z\sum_{j=1}^{n}\sigma _{j}
\end{equation*}%
The first derivative of $f\left( z\right) $ is:%
\begin{equation*}
f^{\prime }\left( z\right) =2\left( 1-\rho \right) z+\rho
\sum_{j=1}^{n}\sigma _{j}
\end{equation*}%
If $\rho \geq 0$, then $f\left( z\right) $ is an increasing function for $%
z\geq 0$ because $\left( 1-\rho \right) \geq 0$ and $\rho
\sum_{j=1}^{n}\sigma _{j}\geq 0$. This explains why the previous result remains
valid:%
\begin{equation*}
\beta _{1}\geq \beta _{2}\geq \beta _{3}\Rightarrow \sigma _{1}\geq \sigma
_{2}\geq \sigma _{3}\quad \text{if} \quad \rho _{i,j}=\rho \geq 0
\end{equation*}%
If $-\left( n-1\right) ^{-1}\leq \rho <0$, then $f^{\prime }$ is decreasing
if $z<-2^{-1}\rho \left( 1-\rho \right) ^{-1}\sum_{j=1}^{n}\sigma _{j}$ and
increasing otherwise. We then have:%
\begin{equation*}
\beta _{1}\geq \beta _{2}\geq \beta _{3}\nRightarrow \sigma _{1}\geq \sigma
_{2}\geq \sigma _{3}\quad \text{if} \quad \rho _{i,j}=\rho <0
\end{equation*}%
In fact, the result remains valid in most cases. To obtain a
counter-example, we must have large differences between the volatilities and
a correlation close to $-\left( n-1\right) ^{-1}$. For example, if
$\sigma _{1}=5\%$, $\sigma _{2}=6\%$, $\sigma
_{3}=80\%$ and $\rho=-49\%$, we have $\beta _{1}=-0.100$, $\beta _{2}=-0.115$ and $\beta _{3}=3.215$.

\item We assume that $\sigma _{1}=15\%$, $\sigma _{2}=20\%$, $\sigma
_{3}=22\%$, $\rho _{1,2}=70\%$, $\rho _{1,3}=20\%$ and $\rho _{2,3}=-50\%$.
It follows that $\beta _{1}=1.231$, $\beta _{2}=0.958$ and $\beta _{3}=0.811$.
We thus have found an example such that $\beta _{1}>\beta _{2}>\beta _{3}$
and $\sigma _{1}<\sigma _{2}<\sigma _{3}$.

\item There is no reason that we have either $\sum_{i=1}^{n}\beta _{i}<n$ or
$\sum_{i=1}^{n}\beta _{i}>n$. Let us consider the previous numerical
example. If $b=\left( 5\%,25\%,70\%\right) $, we obtain $\sum_{i=1}^{3}\beta
_{i}=1.808$ whereas if $b=\left( 20\%,40\%,40\%\right) $, we have $%
\sum_{i=1}^{3}\beta _{i}=3.126$.
\end{enumerate}

\item

\begin{enumerate}
\item We have:%
\begin{eqnarray*}
\sum_{i=1}^{n}b_{i}\beta _{i} &=&\sum_{i=1}^{n}b_{i}\frac{\left( \Sigma
b\right) _{i}}{b^{\top }\Sigma b} \\
&=&b^{\top }\frac{\Sigma b}{b^{\top }\Sigma b} \\
&=&1
\end{eqnarray*}%
If $\beta _{i}=\beta _{j}=\beta $, then $\beta =1$ is an obvious solution
because the previous relationship is satisfied:%
\begin{equation*}
\sum_{i=1}^{n}b_{i}\beta _{i}=\sum_{i=1}^{n}b_{i}=1
\end{equation*}

\item If $\beta _{i}=\beta _{j}=\beta $, then we have:%
\begin{equation*}
\sum_{i=1}^{n}b_{i}\beta =1\Leftrightarrow \beta =\frac{1}{%
\sum_{i=1}^{n}b_{i}}=1
\end{equation*}%
$\beta $ can only take one value, the solution is then unique. We know that
the marginal volatilities are the same in the case of the minimum variance
portfolio $x$ (TR-RPB, page 173):%
\begin{equation*}
\frac{\partial \,\sigma \left( x\right) }{\partial \,x_{i}}=\frac{\partial
\,\sigma \left( x\right) }{\partial \,x_{j}}
\end{equation*}%
with $\sigma \left( x\right) =\sqrt{x^{\top }\Sigma x}$ the volatility of
the portfolio $x$. It follows that:%
\begin{equation*}
\frac{\left( \Sigma x\right) _{i}}{\sqrt{x^{\top }\Sigma x}}=\frac{\left(
\Sigma x\right) _{j}}{\sqrt{x^{\top }\Sigma x}}
\end{equation*}%
By dividing the two terms by $\sqrt{x^{\top }\Sigma x}$, we obtain:%
\begin{equation*}
\frac{\left( \Sigma x\right) _{i}}{x^{\top }\Sigma x}=\frac{\left( \Sigma
x\right) _{j}}{x^{\top }\Sigma x}
\end{equation*}%
The asset betas are then the same in the minimum variance portfolio.
Because we have:%
\begin{equation*}
\left\{
\begin{array}{l}
\beta _{i}=\beta _{j} \\
\sum_{i=1}^{n}x_{i}\beta _{i}=1%
\end{array}%
\right.
\end{equation*}%
we deduce that:%
\begin{equation*}
\beta _{i}=1
\end{equation*}
\end{enumerate}

\item

\begin{enumerate}
\item We have:%
\begin{eqnarray*}
&&\sum_{i=1}^{n}b_{i}\beta _{i}=1 \\
&\Leftrightarrow &\sum_{i=1}^{n}b_{i}\beta _{i}=\sum_{i=1}^{n}b_{i} \\
&\Leftrightarrow &\sum_{i=1}^{n}b_{i}\beta _{i}-\sum_{i=1}^{n}b_{i}=0 \\
&\Leftrightarrow &\sum_{i=1}^{n}b_{i}\left( \beta _{i}-1\right) =0
\end{eqnarray*}%
We obtain the following system of equations:%
\begin{equation*}
\left\{
\begin{array}{l}
\sum_{i=1}^{n}b_{i}\left( \beta _{i}-1\right) =0 \\
b_{i}\geq 0%
\end{array}%
\right.
\end{equation*}%
Let us assume that the asset $j$ has a beta greater than 1. We then have:%
\begin{equation*}
\left\{
\begin{array}{l}
b_{j}\left( \beta _{j}-1\right) +\sum_{i\neq j}b_{i}\left( \beta
_{i}-1\right) =0 \\
b_{i}\geq 0%
\end{array}%
\right.
\end{equation*}%
It follows that $b_{j}\left( \beta _{j}-1\right) >0$ because $b_{j}>0$
(otherwise the beta is zero). We must therefore have $\sum_{i\neq j}x_{i}\left(
\beta _{i}-1\right) <0$. Because $b_{i}\geq 0$, it is necessary that at
least one asset has a beta smaller than 1.

\item We use standard notations to represent $\Sigma $. We seek a portfolio
such that $\beta _{1}>0$, $\beta _{2}>0$ and $\beta _{3}<0$. To simplify
this problem, we assume that the three assets have the same volatility. We also
obtain the following system of inequalities:%
\begin{equation*}
\left\{
\begin{array}{l}
b_{1}+b_{2}\rho _{1,2}+b_{3}\rho _{1,3}>0 \\
b_{1}\rho _{1,2}+b_{2}+b_{3}\rho _{2,3}>0 \\
b_{1}\rho _{1,3}+b_{2}\rho _{2,3}+b_{3}<0%
\end{array}%
\right.
\end{equation*}%
It is sufficient that $b_{1}\rho _{1,3}+b_{2}\rho _{2,3}$ is negative and $%
b_{3}$ is small. For example, we may consider $b_{1}=50\%$, $b_{2}=45\%$, $%
b_{3}=5\%$, $\rho _{1,2}=50\%$, $\rho _{1,3}=0\%$ and $\rho _{2,3}=-50\%$.
We obtain $\beta _{1}=1.10$, $\beta _{2}=1.03$ and $\beta _{3}=-0.27$.
\end{enumerate}

\item

\begin{enumerate}
\item We perform the linear regression $R_{i,t}=\beta _{i}R_{t}\left(
b\right) +\varepsilon _{i,t}$ (TR-RPB, page 16) and we obtain $\hat{\beta}_{i}=1.06$.

\item We deduce that the contribution $c_{i}$ of the market factor is (TR-RPB, page
16):%
\begin{equation*}
c_{i}=\frac{\beta _{i}^{2}\limfunc{var}\left( R\left( b\right) \right) }{%
\limfunc{var}\left( R_{i}\right) }=90.62\%
\end{equation*}
\end{enumerate}
\end{enumerate}

\section{Tangency portfolio}

\begin{enumerate}
\item To find the tangency portfolio, we can maximize the Sharpe ratio or
determine the efficient frontier by including the risk-free asset in the
asset universe (see Exercise \ref{section:exercise1-2} on page \pageref{section:exercise1-2}). We obtain the following result:
\begin{equation*}
\begin{tabular}{|c|ccc|}
\hline
$r$                               & $2\%$    & $3\%$    & $4\%$     \\
\hline
$x_1$                             & ${\bP}10.72\%$    & ${\bP}13.25\%$    &     $17.43\%$     \\
$x_2$                             & ${\bP}12.06\%$    & ${\bP}12.34\%$    &     $12.80\%$     \\
$x_3$                             & ${\bP}28.92\%$    & ${\bP}29.23\%$    &     $29.73\%$     \\
$x_4$                             & ${\bP}48.30\%$    & ${\bP}45.19\%$    &     $40.04\%$     \\
\hdashline
$\mu \left( x\right)$             & ${\bPP}8.03\%$    & ${\bPP}8.27\%$    & ${\bP}8.68\%$     \\
$\sigma \left( x\right)$          & ${\bPP}4.26\%$    & ${\bPP}4.45\%$    & ${\bP}4.84\%$     \\
$\func{SR}\left( x\mid r\right) $ &     $141.29\%$    & $118.30\%$        & $96.65\%$         \\
\hline
\end{tabular}
\end{equation*}

\begin{enumerate}
\item The tangency portfolio is $x=\left( 10.72\%,12.06\%,28.92\%,48.%
30\%\right) $ if the return of the risk-free asset is equal to $2\%$. Its
Sharpe ratio is $1.41$.

\item The tangency portfolio becomes:
\begin{equation*}
x=\left( 13.25\%,12.34\%,29.23\%,45.19\%\right)
\end{equation*}
and $\func{SR}\left( x\mid r\right)$ is equal to $1.18$.

\item The tangency portfolio becomes
\begin{equation*}
x=\left( 17.43\%,12.80\%,29.73\%,40.04\%\right)
\end{equation*}
and $\func{SR}\left( x\mid r\right)$ is equal to $0.97$.

\item When $r$ rises, the weight of the first asset increases whereas the
weight of the fourth asset decreases. This is because the tangency portfolio
must have a higher expected return, that is a higher volatility when $r$
increases. The tangency portfolio will then be more exposed to high
volatility assets (typically, the first asset) and less exposed to low
volatility assets (typically, the fourth asset).
\end{enumerate}

\item We recall that the optimization problem is (TR-RPB, page 19):
\begin{equation*}
x^{\star }=\arg \max x^{\top }\left( \mu +\phi \Sigma b\right) -\frac{\phi }{%
2}x^{\top }\Sigma x-\left( \frac{\phi }{2}b^{\top }\Sigma b+b^{\top }\mu
\right)
\end{equation*}%
We write it as a QP program:
\begin{equation*}
x^{\star }=\arg \min \frac{1}{2}x^{\top }\Sigma x-x^{\top }\left( \gamma \mu
+\Sigma b\right)
\end{equation*}
with $\gamma =\phi ^{-1}$. With the long-only constraint, we obtain the results given in
Table \ref{tab:app2-1-5-1}.

\begin{table}[tbph]
\centering
\caption{Solution of Question 2}
\label{tab:app2-1-5-1}
\tableskip
\begin{tabular}{|c|ccccc|}
\hline
                                         & $b$           & $\min \sigma \left(e\right)$
                                                               & $\max \sigma \left(e\right)$
                                                               & $\sigma \left(e\right) = 3\%$
                                                               & $\max \func{IR}\left( x\mid b\right)$ \\
\hline
$x_1$                               &         $60.00\%$ &     $60.00\%$ &   ${\bNP}0.00\%$ &     $83.01\%$ &     $60.33\%$   \\
$x_2$                               &         $30.00\%$ &     $30.00\%$ &   ${\bNP}0.00\%$ &     $16.99\%$ &     $29.92\%$   \\
$x_3$                               &         $10.00\%$ &     $10.00\%$ &   ${\bNP}0.00\%$ & ${\bP}0.00\%$ & ${\bP}9.75\%$   \\
$x_4$                               &     ${\bP}0.00\%$ & ${\bP}0.00\%$ &       $100.00\%$ & ${\bP}0.00\%$ & ${\bP}0.00\%$   \\
\hline
$\mu \left( x\right)$               &         $12.80\%$ &     $12.80\%$ &   ${\bNP}6.00\%$ &     $14.15\%$ &     $12.82\%$   \\
$\sigma \left( x\right)$            &         $10.99\%$ &     $10.99\%$ &   ${\bNP}5.00\%$ &     $13.38\%$ &     $11.03\%$   \\
$\func{SR}\left( x \mid 3\%\right)$ &         $89.15\%$ &     $89.15\%$ &         $60.00\%$ &     $83.32\%$ &     $89.04\%$   \\
\hline
$\mu \left( x\mid b\right)$         &     ${\bP}0.00\%$ & ${\bP}0.00\%$ & ${\bP}$$-6.80\%$ & ${\bP}1.35\%$ & ${\bP}0.02\%$   \\
$\sigma \left( x\mid b\right)$      &     ${\bP}0.00\%$ & ${\bP}0.00\%$ &   ${\bN}12.08\%$ & ${\bP}3.00\%$ & ${\bP}0.05\%$   \\
$\func{IR}\left( x\mid b\right)$    &     ${\bP}0.00\%$ & ${\bP}0.00\%$ &       $-56.31\%$ &     $45.01\%$ &     $46.54\%$   \\
\hline
\end{tabular}
\end{table}

\begin{figure}[tbph]
\centering
\includegraphics[width = \figurewidth, height = \figureheight]{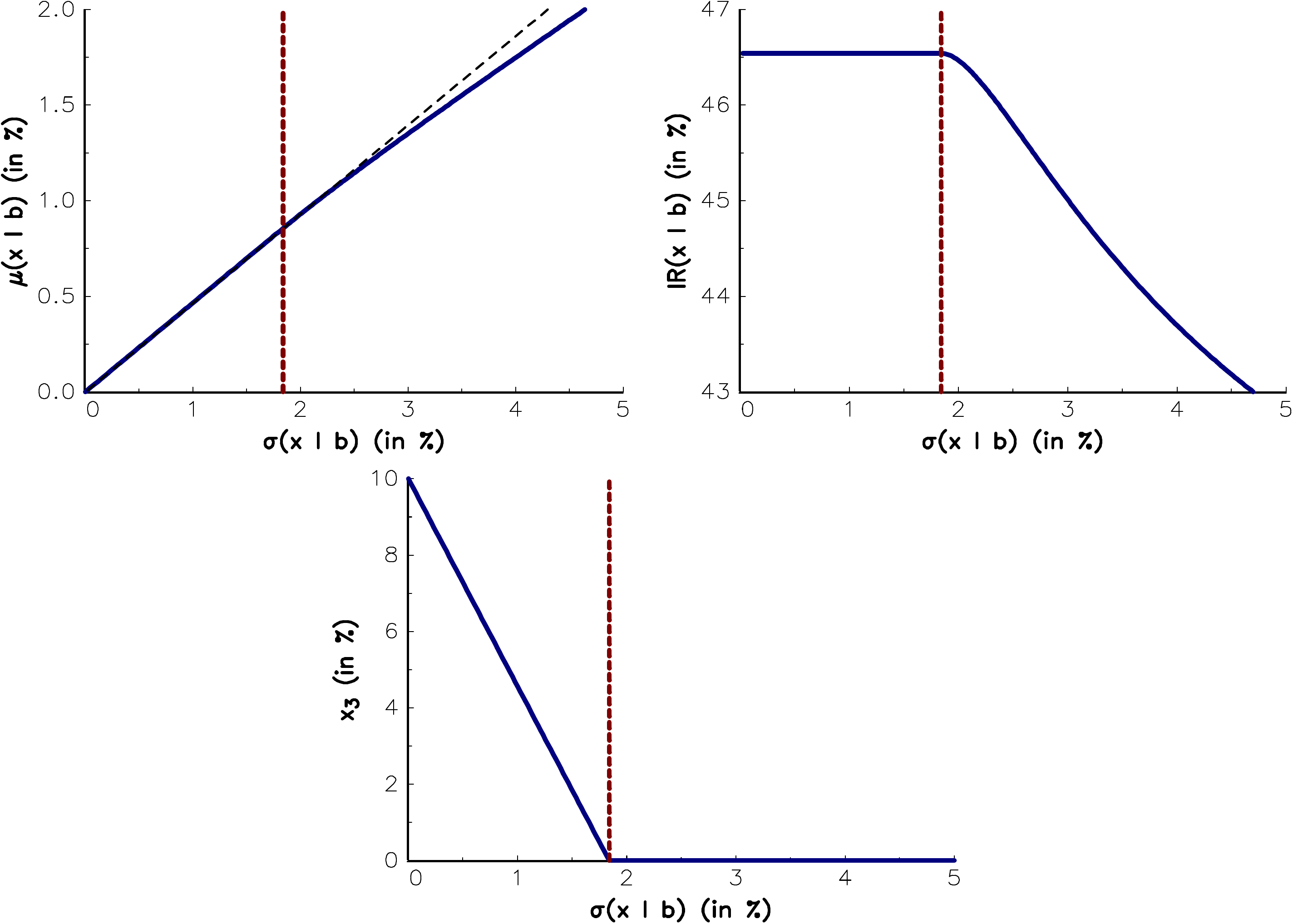}
\caption{Maximizing the information ratio}
\label{fig:app2-1-5-2}
\end{figure}

\begin{enumerate}
\item The portfolio which minimizes the tracking error volatility is the
benchmark. The portfolio which maximizes the tracking error volatility is
the solution of the optimization problem:%
\begin{eqnarray*}
x^{\star } &=&\arg \max \left( x-b\right) ^{\top }\Sigma \left( x-b\right)
\\
&=&\arg \min -\frac{1}{2}x^{\top }\Sigma x+x^{\top }\Sigma b
\end{eqnarray*}%
We obtain $x=\left( 0\%,0\%,0\%,100\%\right) $.

\item There are an infinite number of solutions. In Figure \ref{fig:app2-1-5-2},
we report the relationship between the excess performance $\mu\left(x \mid b\right)$
and the tracking error volatility $\sigma\left(x \mid b\right)$. We notice that the first part
of this relationship is a straight line. In the second panel, we verify that the information
ratio is constant and is equal to $46.5419\%$. In fact, the solutions which maximize
the information ratio correspond to optimized portfolios such that the weight
of the third asset remains positive (third panel). This implies that $\sigma\left(x \mid b\right) \leq 1.8384\%$.
For instance, one possible solution is $x=\left( 60.33\%,29.92\%,9.75\%,0.00\%\right) $.
Another solution is $x=\left( 66.47\%,28.46\%,5.06\%,0.00\%\right) $.

\item With the constraint $x_{i}\in \left[ 10\%,50\%\right] $, the portfolio
with the lowest tracking error volatility is $x =\left( 50\%,30\%,10\%,10\%\right) $.
Its information ratio is negative and is equal to $-0.57$. This means that the portfolio has a
negative excess return. The portfolio with the highest tracking error volatility
is $x =\left( 10\%,10\%,30\%,50\%\right) $ and $\sigma\left(e\right)$
is equal to $8.84\%$. In fact, there is no portfolio which satisfies the constraint
$x_{i}\in \left[ 10\%,50\%\right] $ and has a positive information ratio.

\item When $r=3\%$, the tangency portfolio is:
$$x=\left( 13.25\%,12.%
34\%,29.23\%,45.19\%\right) $$ and has an information ratio equal to
$-0.55$. This implies that there is no equivalence between the
Sharpe ratio ordering and the information ratio ordering.
\end{enumerate}
\end{enumerate}

\section{Information ratio}

\begin{enumerate}
\item

\begin{enumerate}
\item We have $R\left( b\right) =b^{\top }R$ and $R\left( x\right) =x^{\top
}R$. The tracking error is then:%
\begin{equation*}
e=R\left( x\right) -R\left( b\right) =\left( x-b\right) ^{\top }R
\end{equation*}%
It follows that the volatility of the tracking error is:%
\begin{equation*}
\sigma \left( x\mid b\right) =\sigma \left( r\right) =\sqrt{\left(
x-b\right) ^{\top }\Sigma \left( x-b\right) }
\end{equation*}

\item The definition of $\rho \left( x,b\right) $ is:%
\begin{equation*}
\rho \left( x,b\right) = \frac{\mathbb{E}\left[ \left( R\left( x\right)
-\mu \left( x\right) \right) \left( R\left( b\right) -\mu \left( b\right)
\right) \right] }{\sigma \left( x\right) \sigma \left( b\right) }
\end{equation*}
We obtain:
\begin{eqnarray*}
\rho \left( x,b\right)&=&\frac{\mathbb{E}\left[ \left( x^{\top }R-x^{\top }\mu \right) \left(
b^{\top }R-b^{\top }\mu \right) \right] }{\sigma \left( x\right) \sigma
\left( b\right) } \\
&=&\frac{\mathbb{E}\left[ \left( x^{\top }R-x^{\top }\mu \right) \left(
R^{\top }b-\mu ^{\top }b\right) \right] }{\sigma \left( x\right) \sigma
\left( b\right) } \\
&=&\frac{x^{\top }\mathbb{E}\left[ \left( R-\mu \right) \left( R-\mu \right)
^{\top }\right] b}{\sigma \left( x\right) \sigma \left( b\right) } \\
&=&\frac{x^{\top }\Sigma b}{\sqrt{x^{\top }\Sigma x}\sqrt{b^{\top }\Sigma b}}
\end{eqnarray*}

\item We have:%
\begin{eqnarray}
\sigma ^{2}\left( x\mid b\right)  &=&\left( x-b\right) ^{\top }\Sigma \left(
x-b\right)   \notag \\
&=&x^{\top }\Sigma x+b^{\top }\Sigma b-2x^{\top }\Sigma b  \notag \\
&=&\sigma ^{2}\left( x\right) +\sigma ^{2}\left( b\right) -2\rho \left(
x,b\right) \sigma \left( x\right) \sigma \left( b\right)
\label{eq:app2-te1}
\end{eqnarray}%
We deduce that the correlation between portfolio $x$ and benchmark $b$ is:%
\begin{equation}
\rho \left( x,b\right) =\frac{\sigma ^{2}\left( x\right) +\sigma ^{2}\left(
b\right) -\sigma ^{2}\left( x\mid b\right) }{2\sigma \left( x\right) \sigma
\left( b\right) }  \label{eq:app2-te2}
\end{equation}

\item Using Equation (\ref{eq:app2-te1}), we deduce that:%
\begin{equation*}
\sigma ^{2}\left( x\mid b\right) \leq \sigma ^{2}\left( x\right) +\sigma
^{2}\left( b\right) +2\sigma \left( x\right) \sigma \left( b\right)
\end{equation*}%
because $\rho \left( x,b\right) \geq -1$. We then have:%
\begin{eqnarray*}
\sigma \left( x\mid b\right)  &\leq &\sqrt{\sigma ^{2}\left( x\right)
+\sigma ^{2}\left( b\right) +2\sigma \left( x\right) \sigma \left( b\right) }
\\
&\leq &\sigma \left( x\right) +\sigma \left( b\right)
\end{eqnarray*}%
Using Equation (\ref{eq:app2-te2}), we obtain:%
\begin{equation*}
\frac{\sigma ^{2}\left( x\right) +\sigma ^{2}\left( b\right) -\sigma
^{2}\left( x\mid b\right) }{2\sigma \left( x\right) \sigma \left( b\right) }%
\leq 1
\end{equation*}%
It follows that:%
\begin{equation*}
\sigma ^{2}\left( x\right) +\sigma ^{2}\left( b\right) -2\sigma \left(
x\right) \sigma \left( b\right) \leq \sigma ^{2}\left( x\mid b\right)
\end{equation*}%
and:%
\begin{eqnarray*}
\sigma \left( x\mid b\right)  &\geq &\sqrt{\left( \sigma \left( x\right)
-\sigma \left( b\right) \right) ^{2}} \\
&\geq &\left\vert \sigma \left( x\right) -\sigma \left( b\right) \right\vert
\end{eqnarray*}

\item The lower bound is $\left\vert \sigma \left( x\right) -\sigma \left(
b\right) \right\vert $. Even if the correlation is close to one, the
volatility of the tracking error may be high because portfolio $x$ and
benchmark $b$ don't have the same level of volatility. This happens
when the portfolio is leveraged with respect to the benchmark.
\end{enumerate}

\item

\begin{enumerate}
\item If $\sigma \left( x\mid b\right) =\sigma \left( y\mid b\right) $, then:%
\begin{equation*}
\limfunc{IR}\left( x\mid b\right) \geq \limfunc{IR}\left( y\mid b\right)
\Leftrightarrow \mu \left( x\mid b\right) \geq \mu \left( y\mid b\right)
\end{equation*}%
The two portfolios have the same tracking error volatility, but one
portfolio has a greater excess return. In this case, it is obvious that $x$
is preferred to $y$.

\item If $\sigma \left( x\mid b\right) \neq \sigma \left( y\mid b\right) $
and $\limfunc{IR}\left( x\mid b\right) \geq \limfunc{IR}\left( y\mid
b\right) $, we consider a combination of benchmark $b$ and portfolio
$x$:%
\begin{equation*}
z=\left( 1-\alpha \right) b+\alpha x
\end{equation*}%
with $\alpha \geq 0$.  It follows that:%
\begin{equation*}
z-b=\alpha \left( x-b\right)
\end{equation*}%
We deduce that:%
\begin{equation*}
\mu \left( z\mid b\right) =\left( z-b\right) ^{\top }\mu =\alpha \mu \left(
x\mid b\right)
\end{equation*}%
and:%
\begin{equation*}
\sigma ^{2}\left( z\mid b\right) =\left( z-b\right) ^{\top }\Sigma \left(
z-b\right) =\alpha ^{2}\sigma ^{2}\left( x\mid b\right)
\end{equation*}%
We finally obtain that:%
\begin{equation*}
\mu \left( z\mid b\right) =\limfunc{IR}\left( x\mid b\right) \cdot \sigma
\left( z\mid b\right)
\end{equation*}%
Every combination of benchmark $b$ and portfolio $x$ has then the
same information ratio than portfolio $x$. In particular, we can take:%
\begin{equation*}
\alpha =\frac{\sigma \left( y\mid b\right) }{\sigma \left( x\mid b\right) }
\end{equation*}%
In this case, portfolio $z$ has the same tracking error volatility than
portfolio $y$:%
\begin{eqnarray*}
\sigma \left( z\mid b\right)  &=&\alpha \sigma \left( x\mid b\right)  \\
&=&\sigma \left( y\mid b\right)
\end{eqnarray*}%
but a higher excess return:%
\begin{eqnarray*}
\mu \left( z\mid b\right)  &=&\limfunc{IR}\left( x\mid b\right) \cdot \sigma
\left( z\mid b\right)  \\
&=&\limfunc{IR}\left( x\mid b\right) \cdot \sigma \left( y\mid b\right)  \\
&\geq &\limfunc{IR}\left( y\mid b\right) \cdot \sigma \left( y\mid b\right)
\\
&\geq &\mu \left( y\mid b\right)
\end{eqnarray*}%
So, we prefer portfolio $x$ to portfolio $y$.

\item We have:%
\begin{equation*}
\alpha =\frac{3\%}{5\%}=60\%
\end{equation*}%
Portfolio $z$ which is defined by:%
\begin{equation*}
z=0.4\cdot b+0.6\cdot x
\end{equation*}%
has then the same tracking error volatility than portfolio $y$, but a
higher excess return:%
\begin{eqnarray*}
\mu \left( z\mid b\right)  &=&0.6\cdot 5\% \\
&=&3\%
\end{eqnarray*}%
In Figure \ref{fig:app2-1-6-1}, we have represented portfolios $x$, $y$
and $z$. We verify that $z\succ y$ implying that $x\succ y$.

\begin{figure}[tbph]
\centering
\includegraphics[width = \figurewidth, height = \figureheight]{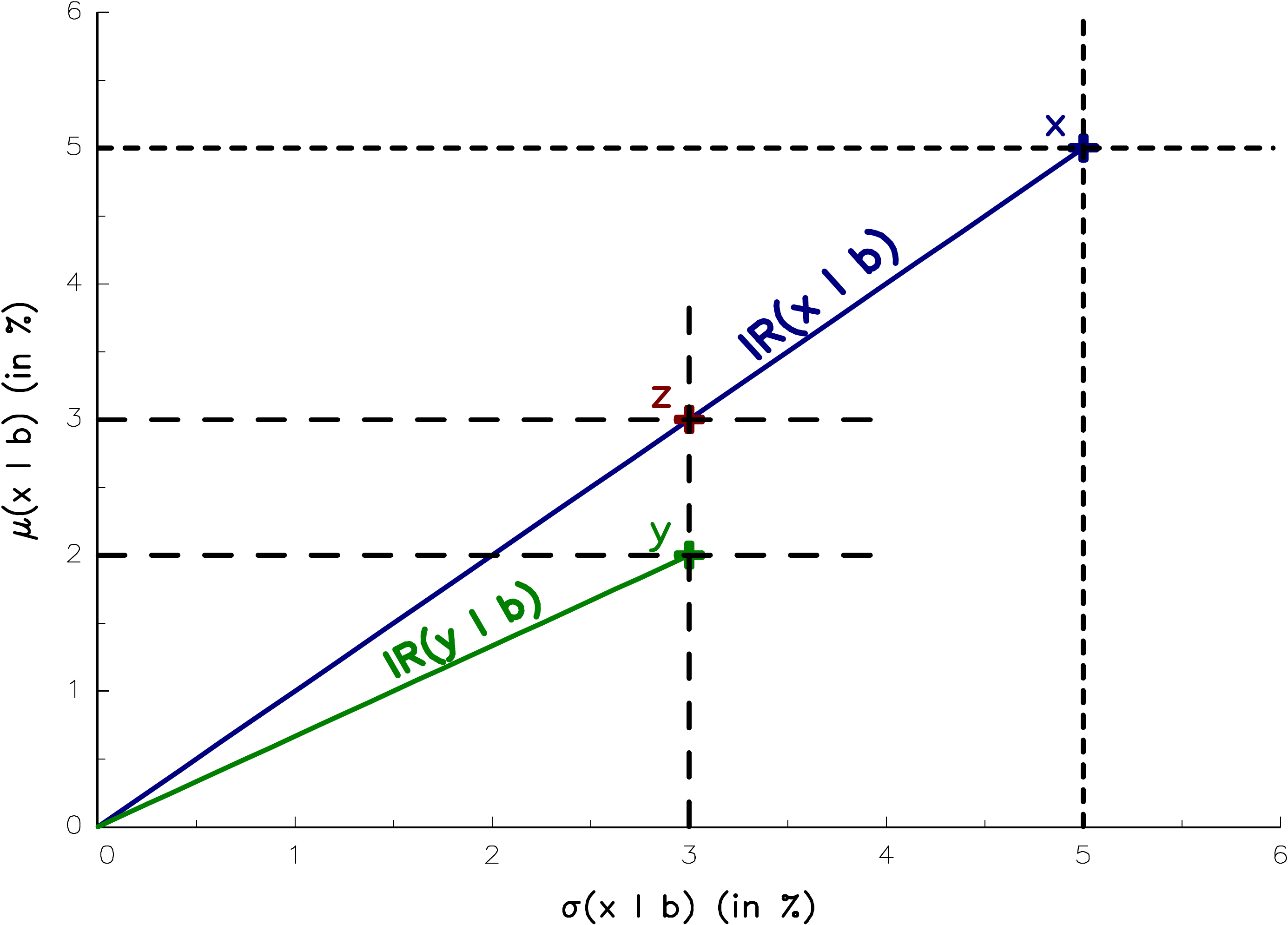}
\caption{Information ratio of portfolio $z$}
\label{fig:app2-1-6-1}
\end{figure}

\end{enumerate}

\item

\begin{enumerate}
\item Let $z\left( x_{0}\right) $ be the combination of the tracker $x_{0}$
and the portfolio $x$. We have:%
\begin{equation*}
z\left( x_{0}\right) =\left( 1-\alpha \right) x_{0}+\alpha x
\end{equation*}%
and:%
\begin{equation*}
z\left( x_{0}\right) -b=\left( 1-\alpha \right) \left( x_{0}-b\right)
+\alpha \left( x-b\right)
\end{equation*}%
It follows that:%
\begin{equation*}
\mu \left( z\left( x_{0}\right) \mid b\right) =\left( 1-\alpha \right) \mu
\left( x_{0}\mid b\right) +\alpha \mu \left( x\mid b\right)
\end{equation*}%
and:%
\begin{eqnarray*}
\sigma ^{2}\left( z\left( x_{0}\right) \mid b\right)  &=&\left( z\left(
x_{0}\right) -b\right) ^{\top }\Sigma \left( z\left( x_{0}\right) -b\right)
\\
&=&\left( 1-\alpha \right) ^{2}\left( x_{0}-b\right) ^{\top }\Sigma \left(
x_{0}-b\right) + \\
&&\alpha ^{2}\left( x-b\right) ^{\top }\Sigma \left( x-b\right) + \\
&&2\alpha \left( 1-\alpha \right) \left( x_{0}-b\right) ^{\top }\Sigma
\left( x-b\right)  \\
&=&\left( 1-\alpha \right) ^{2}\sigma ^{2}\left( x_{0}\mid b\right) +\alpha
^{2}\sigma ^{2}\left( x\mid b\right) + \\
&&\alpha \left( 1-\alpha \right) \left( \sigma ^{2}\left( x_{0}\mid b\right)
+\sigma ^{2}\left( x\mid b\right) -\sigma ^{2}\left( x\mid x_{0}\right)
\right)  \\
&=&\left( 1-\alpha \right) \sigma ^{2}\left( x_{0}\mid b\right) +\alpha
\sigma ^{2}\left( x\mid b\right) + \\
& & \left( \alpha ^{2}-\alpha \right) \sigma
^{2}\left( x\mid x_{0}\right)
\end{eqnarray*}%
We deduce that:%
\begin{eqnarray*}
\limfunc{IR}\left( z\left( x_{0}\right) \mid b\right)  &=&\frac{\mu \left(
z\left( x_{0}\right) \mid b\right) }{\sigma \left( z\left( x_{0}\right) \mid
b\right) } \\
&=&\frac{\left( 1-\alpha \right) \mu \left( x_{0}\mid b\right) +\alpha \mu
\left( x\mid b\right) }{\sqrt{%
\begin{array}{c}
\left( 1-\alpha \right) \sigma ^{2}\left( x_{0}\mid b\right) +\alpha \sigma
^{2}\left( x\mid b\right) + \\
\left( \alpha ^{2}-\alpha \right) \sigma ^{2}\left( x\mid x_{0}\right)
\end{array}%
}}
\end{eqnarray*}

\item We have to find $\alpha $ such that $\sigma \left( z\left(
x_{0}\right) \mid b\right) =\sigma \left( y\mid b\right) $. The equation is:
\begin{equation*}
\left( 1-\alpha \right) \sigma ^{2}\left( x_{0}\mid b\right) +\alpha \sigma
^{2}\left( x\mid b\right) +\left( \alpha ^{2}-\alpha \right) \sigma
^{2}\left( x\mid x_{0}\right) =\sigma ^{2}\left( y\mid b\right)
\end{equation*}%
It is a second-order polynomial equation:%
\begin{equation*}
A\alpha ^{2}+B\alpha +C=0
\end{equation*}%
with $A=\sigma ^{2}\left( x\mid x_{0}\right) $, $B=\sigma ^{2}\left( x\mid
b\right) -\sigma ^{2}\left( x\mid x_{0}\right) -\sigma ^{2}\left( x_{0}\mid
b\right) $ and $C=\sigma ^{2}\left( x_{0}\mid b\right) -\sigma ^{2}\left(
y\mid b\right) $. Using the numerical values, we obtain $\alpha =42.4\%$. We
deduce that $\mu \left( z\left( x_{0}\right) \mid b\right) =97$ bps and $%
\limfunc{IR}\left( z\left( x_{0}\right) \mid b\right) =0.32$.

\item In Figure \ref{fig:app2-1-6-2}, we have represented portfolios $%
x_{0}$, $x$, $y$, $z$ and $z\left( x_{0}\right) $. In this case, we have $%
y\succ z\left( x_{0}\right) $. We conclude that the preference ordering
based on the information ratio is not valid when it is difficult to
replicate the benchmark $b$.

\begin{figure}[tbph]
\centering
\includegraphics[width = \figurewidth, height = \figureheight]{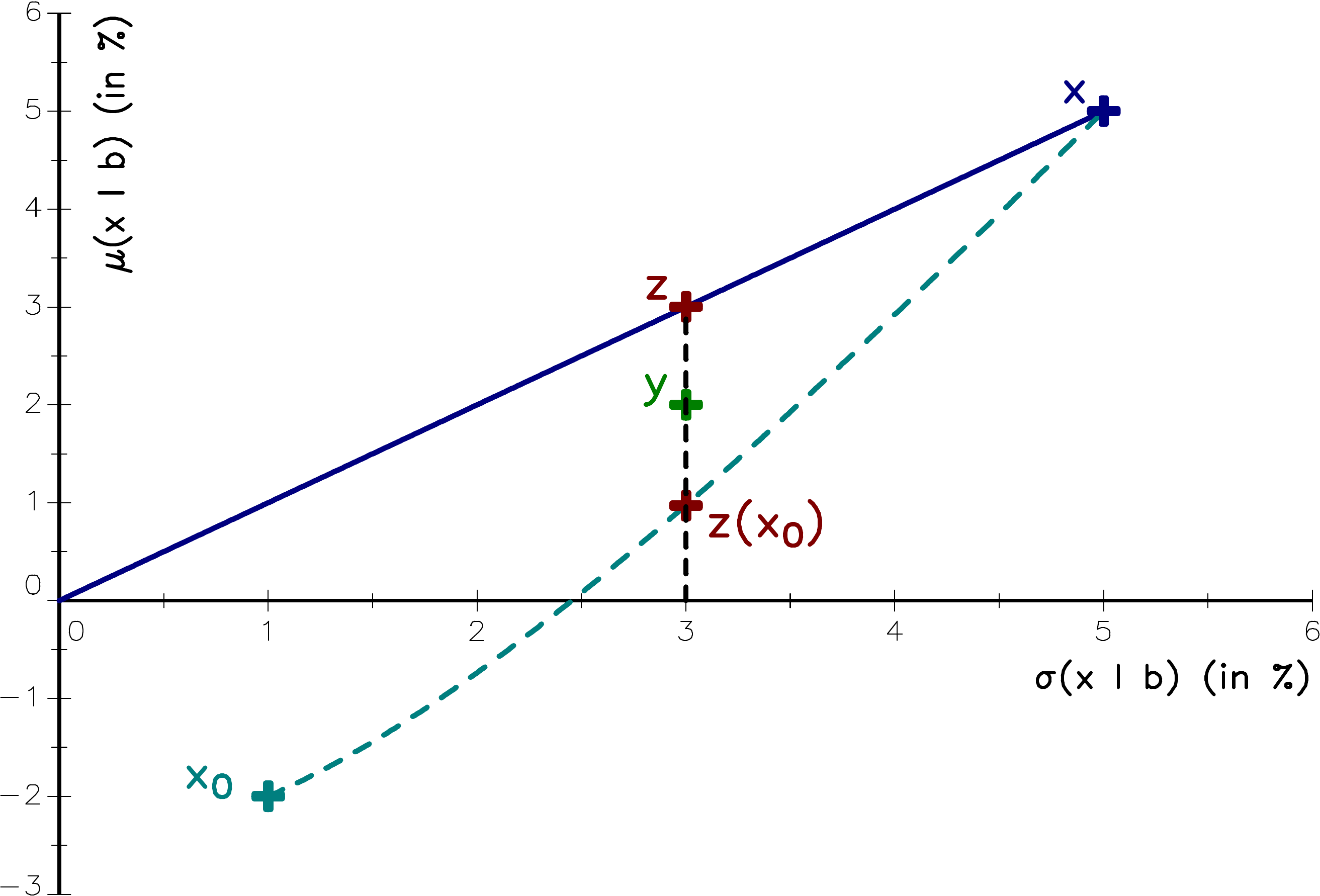}
\caption{Information ratio of portfolio $z\left(x_0\right)$}
\label{fig:app2-1-6-2}
\end{figure}

\end{enumerate}
\end{enumerate}

\section{Building a tilted portfolio}

\begin{enumerate}
\item The ERC portfolio is defined in TR-RPB page 119. We obtain the
following results:
\begin{equation*}
\begin{tabular}{|c|cccc|}
\hline
Asset & $x_{i}$ & $\mathcal{MR}_{i}$ & $\mathcal{RC}_{i}$ & $\mathcal{RC}_{i}^{\star }$ \\ \hline
1 & $32.47\%$ & $10.83\%$ & $3.52\%$ & $25.00\%$ \\
2 & $25.41\%$ & $13.84\%$ & $3.52\%$ & $25.00\%$ \\
3 & $21.09\%$ & $16.67\%$ & $3.52\%$ & $25.00\%$ \\
4 & $21.04\%$ & $16.71\%$ & $3.52\%$ & $25.00\%$ \\ \hline
\end{tabular}%
\end{equation*}

\item The benchmark $b$ is the ERC portfolio. Using the tracking-error
optimization problem (TR-RPB, page 19), we obtain the optimized
portfolios given in Table \ref{tab:app2-1-7-1}.

\begin{table}[tbh]
\centering
\caption{Solution of Question 2}
\label{tab:app2-1-7-1}
\tableskip
\begin{tabular}{|c|ccccc|} \hline
$\sigma \left( e\right) $ & $0\%$ & $1\%$ & $5\%$ & $10\%$ & $\max$ \\
\hline
$x_1$                             &  $32.47\%$ &     $38.50\%$    &     $63.48\%$    &     $96.26\%$    & ${\bPP}0.00\%$      \\
$x_2$                             &  $25.41\%$ &     $20.16\%$    & ${\bP}0.00\%$    & ${\bP}0.00\%$    & ${\bPP}0.00\%$      \\
$x_3$                             &  $21.09\%$ &     $20.18\%$    &     $15.15\%$    & ${\bP}0.00\%$    & ${\bPP}0.00\%$      \\
$x_4$                             &  $21.04\%$ &     $21.16\%$    &     $21.37\%$    & ${\bP}3.74\%$    &     $100.00\%$      \\ \hdashline
$\mu \left( x\mid b\right)$       &              & ${\bP}1.13\%$    & ${\bP}5.66\%$    & ${\bP}8.05\%$    & ${\bPP}3.24\%$    \\
$\sigma \left( x\mid b\right)$    &              & ${\bP}1.00\%$    & ${\bP}5.00\%$    &     $10.00\%$    & ${\bP}25.05\%$    \\
$\func{IR}\left( x \mid b\right)$ &              & ${\bP}1.13{\bP}$ & ${\bP}1.13{\bP}$ & ${\bP}0.81{\bP}$ & ${\bPP}0.13{\bP}$ \\ \hdashline
$\sigma \left( x\right)$          &  $14.06\%$ &     $13.89\%$    &     $13.86\%$    &     $14.59\%$    & ${\bP}30.00\%$      \\
$\rho\left( x \mid b\right)$      &              &     $99.75\%$    &     $93.60\%$    &     $75.70\%$    & ${\bP}55.71\%$    \\
\hline
\end{tabular}
\end{table}

\begin{enumerate}
\item If the tracking error volatility is set to $1\%$, the optimal
portfolio is $\left( 38.50\%,20.16\%,20.18\%,21.16\%\right) $. The
excess return is equal to $1.13\%$, which implies an information ratio
equal to $1.13$.

\item If the tracking error is
equal to $10\%$, the information ratio of the optimal portfolio decreases
to $0.81$.

\item We have\footnote{%
We recall that the correlation between portfolio $x$ and benchmark $b
$ is equal to:%
\begin{equation*}
\rho \left( x\mid b\right) =\frac{x^{\top }\Sigma b}{\sqrt{x^{\top }\Sigma x}%
\sqrt{b^{\top }\Sigma b}}
\end{equation*}%
}:%
\begin{equation*}
\sigma \left( x\mid b\right) =\sqrt{\sigma ^{2}\left( x\right) -2\rho \left(
x\mid b\right) \sigma \left( x\right) \sigma \left( b\right) +\sigma
^{2}\left( b\right) }
\end{equation*}%
We suppose that $\rho \left( x\mid b\right) \in \left[ \rho _{\min },\rho
_{\max }\right] $. Because $x$ may be equal to $b$, $\rho _{\max }$ is equal
to $1$. We deduce that:%
\begin{equation*}
0\leq \sigma \left( x\mid b\right) \leq \sqrt{\sigma ^{2}\left( x\right)
-2\rho _{\min }\sigma \left( x\right) \sigma \left( b\right) +\sigma
^{2}\left( b\right) }
\end{equation*}%
If $\rho _{\min }=-1$, the upper bound of the tracking error volatility is:%
\begin{equation*}
\sigma \left( x\mid b\right) \leq \sigma \left( x\right) +\sigma \left(
b\right)
\end{equation*}%
If $\rho _{\min }=0$, the upper bound becomes:%
\begin{equation*}
\sigma \left( x\mid b\right) \leq \sqrt{\sigma ^{2}\left( x\right) +\sigma
^{2}\left( b\right) }
\end{equation*}%
If $\rho _{\min }=50\%$, we use the Cauchy-Schwarz inequality and we obtain:%
\begin{eqnarray*}
\sigma \left( x\mid b\right)  &\leq &\sqrt{\sigma ^{2}\left( x\right)
-\sigma \left( x\right) \sigma \left( b\right) +\sigma ^{2}\left( b\right) }
\\
&\leq &\sqrt{\left( \sigma \left( x\right) -\sigma \left( b\right) \right)
^{2}+\sigma \left( x\right) \sigma \left( b\right) } \\
&\leq &\left\vert \sigma \left( x\right) -\sigma \left( b\right) \right\vert
+\sqrt{\sigma \left( x\right) \sigma \left( b\right) }
\end{eqnarray*}%
Because we have imposed a long-only constraint, it is difficult to find a
portfolio which has a negative correlation. For instance, if we consider the
previous results, we observe that the correlation is larger than $50\%$. In
this case, $\sigma \left( x\right) \simeq \sigma \left( b\right) $ and the
order of magnitude of $\sigma \left( x\mid b\right) $ is $\sigma \left(
b\right) $. Because $\sigma \left( b\right) $ is equal to $14.06\%$, it is
not possible to find a portfolio which has a tracking error volatility equal
to $35\%$. Even if we consider that $\rho \left( x\mid b\right) =0$, the
order of magnitude of $\sigma \left( x\mid b\right) $ is $\sqrt{2}\sigma
\left( b\right) $, that is $28\%$. We are far from the target value which is equal
to $35\%$. In fact, the portfolio which maximizes the tracking error
volatility is $\left( 0\%,0\%,0\%,100\%\right) $ and the maximum tracking
error volatility is $25.05\%$. We conclude that there is no solution to
this question.
\end{enumerate}

\item We obtain the results given in Table \ref{tab:app2-1-7-2}.
\begin{table}[tb]
\centering
\caption{Solution of Question 3}
\label{tab:app2-1-7-2}
\tableskip
\begin{tabular}{|c|ccccc|}
\hline
$\sigma \left( e\right) $ & $0\%$ & $1\%$ & $5\%$ & $10\%$ & $35\%$ \\
\hline
$x_1$                             &  $32.47\%$ &     $38.50\%$    &  ${\bn}62.65\%$    &  ${\bN}$$92.82\%$    &  ${\bN}243.72\%$      \\
$x_2$                             &  $25.41\%$ &     $20.16\%$    &       $-0.83\%$    &        $-27.07\%$    &      $-158.28\%$      \\
$x_3$                             &  $21.09\%$ &     $20.18\%$    &  ${\bn}16.54\%$    &  ${\bN}$$11.99\%$    &  ${\bP}$$-10.77\%$    \\
$x_4$                             &  $21.04\%$ &     $21.16\%$    &  ${\bn}21.65\%$    &  ${\bN}$$22.27\%$    &  ${\bNP}25.34\%$      \\ \hdashline
$\mu \left( x\mid b\right)$       &              & ${\bP}1.13\%$    &  ${\bnP}5.67\%$    &  ${\bN}$$11.34\%$    &  ${\bNP}39.71\%$    \\
$\sigma \left( x\mid b\right)$    &              & ${\bP}1.00\%$    &  ${\bnP}5.00\%$    &  ${\bN}$$10.00\%$    &  ${\bNP}35.00\%$    \\
$\func{IR}\left( x \mid b\right)$ &              & ${\bP}1.13{\bP}$ &  ${\bnP}1.13{\bP}$ &  ${\bNP}$$1.13{\bP}$ &  ${\bNPP}1.13{\bP}$ \\ \hdashline
$\sigma \left( x\right)$          &  $14.06\%$ &     $13.89\%$    &  ${\bn}13.93\%$    &  ${\bN}$$15.50\%$    &  ${\bNP}34.96\%$      \\
$\rho\left( x \mid b\right)$      &              &     $99.75\%$    &  ${\bn}93.62\%$    &  ${\bN}$$77.55\%$    &  ${\bNP}19.81\%$    \\
\hline
\end{tabular}
\end{table}
The deletion of the long-only constraint permits now to find a portfolio
with a tracking error volatility which is equal to $35\%$. We notice that
optimal portfolios have the same information ratio. This is perfectly
normal because the efficient frontier $\left\{ \sigma \left( x^{\star }\mid
b\right) ,\mu \left( x^{\star }\mid b\right) \right\} $ is a straight line
when there is no constraint%
\footnote{For instance, we have reported
the constrained and unconstrained efficient frontiers in Figure \ref{fig:app2-1-7-1}.} (TR-RPB, page 21).
\begin{figure}[tbph]
\centering
\includegraphics[width = \figurewidth, height = \figureheight]{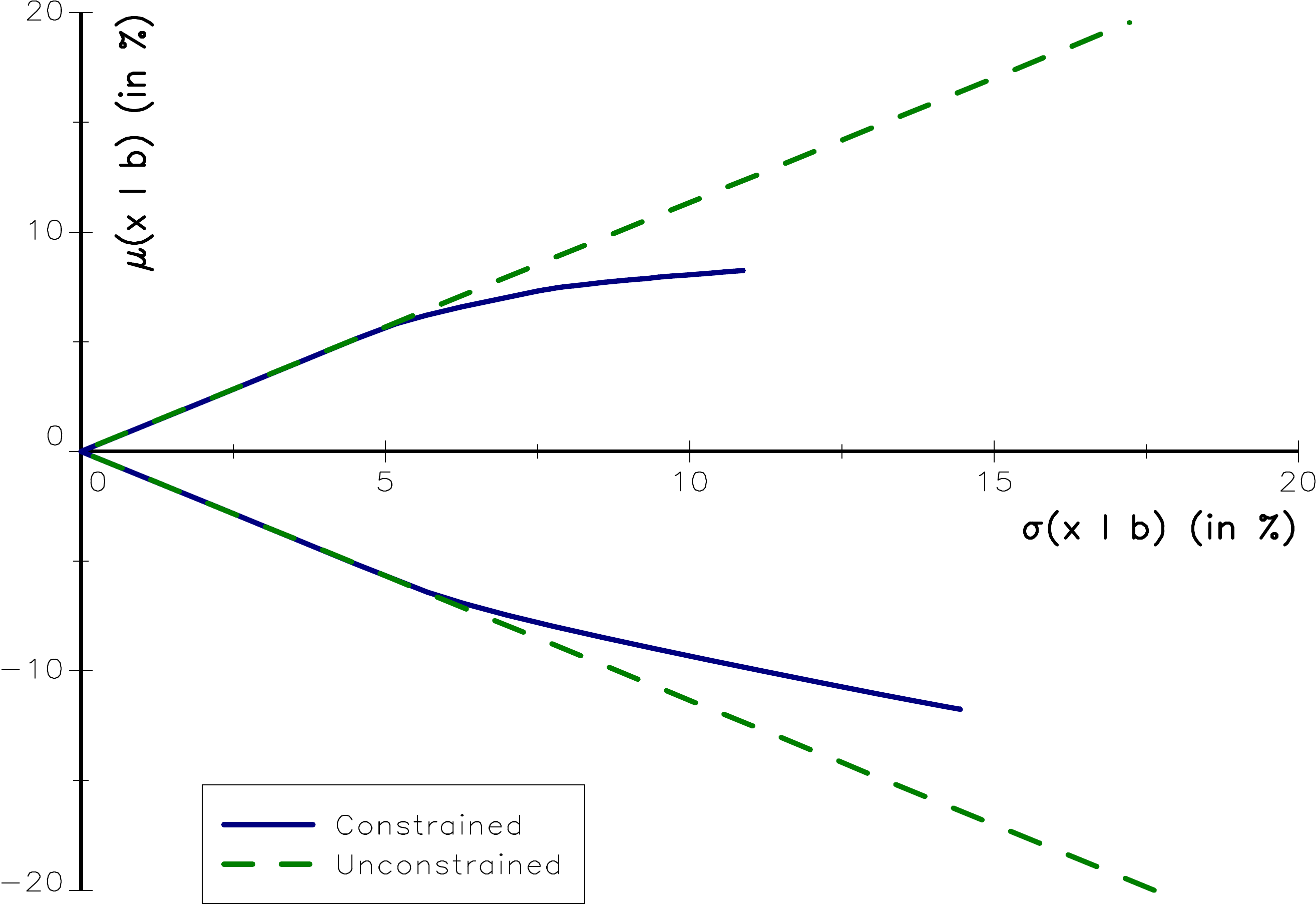}
\caption{Constrained and unconstrained efficient frontier}
\label{fig:app2-1-7-1}
\end{figure}
It follows that:%
\begin{equation*}
\func{IR}\left( x^{\star }\mid b\right) =\frac{\mu \left( x^{\star }\mid
b\right) }{\sigma \left( x^{\star }\mid b\right) }=\text{constant}
\end{equation*}%
Let $x_{0}$ be one optimized portfolio corresponding to a given tracking
error volatility. Without any constraints, the optimized portfolios
may be written as:%
\begin{equation*}
x^{\star }=b+\ell \cdot \left( x_{0}-b\right)
\end{equation*}%
We then decompose the optimized portfolio $x^{\star }$ as the sum of the
benchmark $b$\ and a leveraged long-short portfolio $x_{0}-b$. Let us
consider the previous results with $x_{0}$ corresponding to the optimal
portfolio for a 1\% tracking error volatility. We verify that the optimal
portfolio which has a tracking error volatility equal to $5\%$ (resp. $10\%$
and $35\%$) is a portfolio leveraged $5$ times (resp. $10$ and $35$ times)
with respect to $x_{0}$. Indeed, we have:%
\begin{eqnarray*}
\sigma \left( x^{\star }\mid b\right)  &=&\sigma \left( b+\ell \cdot \left(
x_{0}-b\right) \mid b\right)  \\
&=&\ell \cdot \sigma \left( x_{0}-b\mid b\right)  \\
&=&\ell \cdot \sigma \left( x_{0}\mid b\right)
\end{eqnarray*}%
We deduce that the leverage is the ratio of tracking error volatilities:%
\begin{equation*}
\ell =\frac{\sigma \left( x^{\star }\mid b\right) }{\sigma \left( x_{0}\mid
b\right) }
\end{equation*}%
In this case, we verify that:%
\begin{eqnarray*}
\func{IR}\left( x^{\star }\mid b\right)  &=&\frac{\mu \left( b+\ell \cdot
\left( x_{0}-b\right) \mid b\right) }{\ell \cdot \sigma \left( x_{0}\mid
b\right) } \\
&=&\frac{\ell \cdot \mu \left( x_{0}\mid b\right) }{\ell \cdot \sigma \left(
x_{0}\mid b\right) } \\
&=&\frac{\mu \left( x_{0}\mid b\right) }{\sigma \left( x_{0}\mid b\right) }
\end{eqnarray*}
\end{enumerate}

\section{Implied risk premium}

\begin{enumerate}
\item

\begin{enumerate}
\item The optimal portfolio is the solution of the following optimization
problem:%
\begin{equation*}
x^{\star }=\arg \max \mathcal{U}\left( x\right)
\end{equation*}%
The first-order condition $\partial _{x}\,\mathcal{U}\left( x\right) =0$ is:
\begin{equation*}
\left( \mu -r\right) -\phi \Sigma x^{\star }=0
\end{equation*}%
We deduce that:%
\begin{eqnarray*}
x^{\star } &=&\frac{1}{\phi }\Sigma ^{-1}\left( \mu -r \mathbf{1}\right)  \\
&=&\frac{1}{\phi }\Sigma ^{-1}\pi
\end{eqnarray*}%
We verify that the optimal portfolio is a linear function of the risk
premium $\pi =\mu -r \mathbf{1}$.

\item If the investor holds the portfolio $x_{0}$, he thinks that it is an
optimal investment. We then have:%
\begin{equation*}
\pi -\phi \Sigma x_{0}=0
\end{equation*}%
We deduce that the implied risk premium is:%
\begin{equation*}
\pi =\phi \Sigma x_{0}
\end{equation*}%
The risk premium is related to three parameters which depend on the investor (the risk aversion $%
\phi $ and the composition of the portfolio $x_{0}$) and a market parameter
(the covariance matrix $\Sigma $).

\item Because $\pi =\phi \Sigma x_{0}$, we have:%
\begin{equation*}
x_{0}^{\top }\pi =\phi x_{0}^{\top }\Sigma x_{0}
\end{equation*}%
We deduce that:%
\begin{eqnarray*}
\phi  &=&\frac{x_{0}^{\top }\pi }{x_{0}^{\top }\Sigma x_{0}} \\
&=&\frac{1}{\sqrt{x_{0}^{\top }\Sigma x_{0}}}\cdot \frac{x_{0}^{\top }\pi }{%
\sqrt{x_{0}^{\top }\Sigma x_{0}}} \\
&=&\frac{\limfunc{SR}\left( x_{0}\mid r\right) }{\sqrt{x_{0}^{\top }\Sigma
x_{0}}}
\end{eqnarray*}

\item It follows that:%
\begin{eqnarray*}
\pi  &=&\phi \Sigma x_{0} \\
&=&\frac{\limfunc{SR}\left( x_{0}\mid r\right) }{\sqrt{x_{0}^{\top }\Sigma
x_{0}}}\Sigma x_{0} \\
&=&\limfunc{SR}\left( x_{0}\mid r\right) \frac{\Sigma x_{0}}{\sqrt{%
x_{0}^{\top }\Sigma x_{0}}}
\end{eqnarray*}%
We know that:%
\begin{equation*}
\frac{\partial \,\sigma \left( x_{0}\right) }{\partial \,x}=\frac{\Sigma
x_{0}}{\sqrt{x_{0}^{\top }\Sigma x_{0}}}
\end{equation*}%
We deduce that:%
\begin{equation*}
\pi _{i}=\limfunc{SR}\left( x_{0}\mid r\right) \cdot \mathcal{MR}_{i}
\end{equation*}%
The implied risk premium of asset $i$ is then a linear function of its
marginal volatility and the proportionality factor is the Sharpe ratio of
the portfolio.

\item In microeconomics, the price of a good is equal to its marginal cost
at the equilibrium. We retrieve this marginalism principle in the
relationship between the asset price $\pi_{i}$ and the asset risk,
which is equal to the product of the Sharpe ratio and the marginal
volatility of the asset.

\item We have:%
\begin{equation*}
\sum_{i=1}^{n}\pi _{i}=\sum_{i=1}^{n}\limfunc{SR}\left( x_{0}\mid r\right)
\cdot \mathcal{MR}_{i}
\end{equation*}%
Another expression of the Sharpe ratio is then:%
\begin{equation*}
\limfunc{SR}\left( x_{0}\mid r\right) =\frac{\sum_{i=1}^{n}\pi _{i}}{%
\sum_{i=1}^{n}\mathcal{MR}_{i}}
\end{equation*}%
It is the ratio of the sum of implied risk premia divided by the sum of
marginal volatilities. We also notice that:%
\begin{equation*}
x_{i}\pi _{i}=\limfunc{SR}\left( x_{0}\mid r\right) \cdot \left( x_{i}\cdot
\mathcal{MR}_{i}\right)
\end{equation*}%
We deduce that:%
\begin{equation*}
\limfunc{SR}\left( x_{0}\mid r\right) =\frac{\sum_{i=1}^{n}x_{i}\pi _{i}}{%
\sum_{i=1}^{n}\mathcal{RC}_{i}}
\end{equation*}%
In this case, the Sharpe ratio is the weighted sum of implied risk premia
divided by the sum of risk contributions. In fact, it is the definition of
the Sharpe ratio:%
\begin{equation*}
\limfunc{SR}\left( x_{0}\mid r\right) =\frac{\sum_{i=1}^{n}x_{i}\pi _{i}}{%
\mathcal{R}\left( x_{0}\right) }
\end{equation*}%
with $\mathcal{R}\left( x_{0}\right) =\sum_{i=1}^{n}\mathcal{RC}_{i}=\sqrt{%
x_{0}^{\top }\Sigma x_{0}}$.
\end{enumerate}

\item

\begin{enumerate}
\item Let $x^{\star }$ be the market portfolio. The implied risk premium is:%
\begin{equation*}
\pi =\limfunc{SR}\left( x^{\star }\mid r\right) \frac{\Sigma x^{\star }}{%
\sigma \left( x^{\star }\right) }
\end{equation*}%
The vector of asset betas is:%
\begin{eqnarray*}
\beta  &=&\frac{\limfunc{cov}\left( R,R\left( x^{\star }\right) \right) }{%
\limfunc{var}\left( R\left( x^{\star }\right) \right) } \\
&=&\frac{\Sigma x^{\star }}{\sigma ^{2}\left( x^{\star }\right) }
\end{eqnarray*}%
We deduce that:%
\begin{equation*}
\mu -r=\left( \frac{\mu \left( x^{\star }\right) -r}{\sigma \left( x^{\star
}\right) }\right) \frac{\beta \sigma ^{2}\left( x^{\star }\right) }{\sigma
\left( x^{\star }\right) }
\end{equation*}
or:%
\begin{equation*}
\mu -r=\beta \left( \mu \left( x^{\star }\right) -r\right)
\end{equation*}%
For asset $i$, we obtain:%
\begin{equation*}
\mu _{i}-r=\beta _{i}\left( \mu \left( x^{\star }\right) -r\right)
\end{equation*}%
or equivalently:%
\begin{equation*}
\mathbb{E}\left[ R_{i}\right] -r=\beta _{i}\left( \mathbb{E}\left[ R\left(
x^{\star }\right) \right] -r\right)
\end{equation*}%
We retrieve the CAPM relationship.

\item The beta is generally defined in terms of risk:%
\begin{equation*}
\beta _{i}=\frac{\limfunc{cov}\left( R_{i},R\left( x^{\star }\right) \right)
}{\limfunc{var}\left( R\left( x^{\star }\right) \right) }
\end{equation*}%
We sometimes forget that it is also equal to:%
\begin{equation*}
\beta _{i}=\frac{\mathbb{E}\left[ R_{i}\right] -r}{\mathbb{E}\left[ R\left(
x^{\star }\right) \right] -r}
\end{equation*}%
It is the ratio between the risk premium of the asset and the excess return
of the market portfolio.
\end{enumerate}

\item

\begin{enumerate}
\item As the volatility of the portfolio $\sigma \left( x\right) $ is a
convex risk measure, we have (TR-RPB, page 78):%
\begin{equation*}
\mathcal{RC}_{i}\leq x_{i}\sigma _{i}
\end{equation*}%
We deduce that $\mathcal{MR}_{i}\leq \sigma _{i}$. Moreover, we have $%
\mathcal{MR}_{i}\geq 0$ because $\rho _{i,j}\geq 0$. The marginal volatility
is then bounded:%
\begin{equation*}
0\leq \mathcal{MR}_{i}\leq \sigma _{i}
\end{equation*}%
Using the fact that $\pi _{i}=\limfunc{SR}\left( x\mid r\right) \cdot
\mathcal{MR}_{i}$, we deduce that:%
\begin{equation*}
0\leq \pi _{i}\leq \limfunc{SR}\left( x\mid r\right) \cdot \sigma _{i}
\end{equation*}

\item $\pi _{i}$ is equal to the upper bound when $\mathcal{MR}_{i}=\sigma
_{i}$, that is when the portfolio is fully invested in the $i^{\mathsf{th}}$
asset:%
\begin{equation*}
x_{j}=\left\{
\begin{array}{lll}
1 & \text{if} & j=i \\
0 & \text{if} & j\neq i%
\end{array}%
\right.
\end{equation*}

\item We have (TR-RPB, page 101):%
\begin{equation*}
\mathcal{MR}_{i}=\frac{x_{i}\sigma _{i}^{2}+\sigma _{i}\sum_{j\neq
i}x_{j}\rho _{i,j}\sigma _{j}}{\sigma \left( x\right) }
\end{equation*}%
If $\rho _{i,j}=0$ and $x_{i}=0$, then $\mathcal{MR}_{i}=0$ and $\pi _{i}=0$%
. The risk premium of the asset reaches then the lower bound when this asset
is not correlated to the other assets and when it is not invested.

\item Negative correlations do not change the upper bound, but the lower
bound may be negative because the marginal volatility may be negative.
\end{enumerate}

\item

\begin{enumerate}
\item Results are given in the following table:%
\begin{equation*}
\begin{tabular}{|c|ccc:cc|}
\hline
$i$ & $x_{i}$ & $\mathcal{MR}_{i}$ & $\pi _{i}$ & $\beta _{i}$ & $\pi_{i}/\pi \left( x\right) $ \\
\hline
                  $1$  & $25.00\%$ & $20.08\%$ &     $10.04\%$ & $1.52$ & $1.52$ \\
                  $2$  & $25.00\%$ & $12.28\%$ & ${\bP}6.14\%$ & $0.93$ & $0.93$ \\
                  $3$  & $50.00\%$ & $10.28\%$ & ${\bP}5.14\%$ & $0.78$ & $0.78$ \\ \hdashline
$\pi \left( x\right) $ &           &           & ${\bP}6.61\%$ &        &        \\
\hline
\end{tabular}%
\end{equation*}

\item Results are given in the following table:%
\begin{equation*}
\begin{tabular}{|c|ccc:cc|}
\hline
$i$ & $x_{i}$ & $\mathcal{MR}_{i}$ & $\pi _{i}$ & $\beta _{i}$ & $\pi_{i}/\pi \left( x\right) $ \\
\hline
                  $1$  & ${\bP}5.00\%$ & ${\bP}9.19\%$ & $4.59\%$ & $0.66$ & $0.66$ \\
                  $2$  & ${\bP}5.00\%$ & ${\bP}2.33\%$ & $1.17\%$ & $0.17$ & $0.17$ \\
                  $3$  &     $90.00\%$ &     $14.86\%$ & $7.43\%$ & $1.07$ & $1.07$ \\ \hdashline
$\pi \left( x\right) $ &               &               & $6.97\%$ &        &        \\
\hline
\end{tabular}%
\end{equation*}

\item Results are given in the following table:%
\begin{equation*}
\begin{tabular}{|c|ccc:cc|}
\hline
$i$ & $x_{i}$ & $\mathcal{MR}_{i}$ & $\pi _{i}$ & $\beta _{i}$ & $\pi_{i}/\pi \left( x\right) $ \\
\hline
                  $1$  &     $100.00\%$ &     $25.00\%$ &     $12.50\%$ & $1.00$ & $1.00$ \\
                  $2$  & ${\bPP}0.00\%$ &     $10.00\%$ & ${\bP}5.00\%$ & $0.40$ & $0.40$ \\
                  $3$  & ${\bPP}0.00\%$ & ${\bP}3.75\%$ & ${\bP}1.88\%$ & $0.15$ & $0.15$ \\ \hdashline
$\pi \left( x\right) $ &                &               &     $12.50\%$ &        &        \\
\hline
\end{tabular}%
\end{equation*}

\item If we compare the results of the second portfolio with respect to the results of the first
portfolio, we notice that the risk premium of the third asset increases whereas the risk premium
of the first and second assets decreases. The second investor is then overweighted in the third asset,
because he implicitly considers that the third asset is very well rewarded. If we consider the results of the
third portfolio, we verify that the risk premium may be strictly positive even if the weight of the asset
is equal to zero.
\end{enumerate}

\end{enumerate}

\section{Black-Litterman model}

\begin{enumerate}
\item
\begin{enumerate}
\item We consider the portfolio optimization problem in the presence of a
benchmark (TR-RPB, page 17). We obtain the following results (expressed in \%):

\begin{equation*}
\begin{tabular}{|c|rrrrr|}
\hline
$\sigma \left( x^\star \mid b\right)$ & $1.00$        & $2.00$        & $3.00$        & $4.00$        & ${\bN}5.00$ \\ \hline
$x_1^\star$                           & $35.15$        & $36.97$        & $38.78$        & $40.60$        & $42.42$ \\
$x_2^\star$                           & $26.32$        & $19.30$        & $12.28$        & $ 5.26$        & $-1.76$ \\
$x_3^\star$                           & $38.53$        & $43.74$        & $48.94$        & $54.14$        & $59.34$ \\ \hdashline
$\mu \left( x^\star \mid b\right)$    & $1.31$        & $2.63$        & $3.94$        & $5.25$        & ${\bN}6.56$ \\
\hline
\end{tabular}
\end{equation*}

\end{enumerate}

\item

\begin{enumerate}
\item Let $b$ be the benchmark (that is the equally weighted portfolio). We
recall that the implied risk aversion parameter is:%
\begin{equation*}
\phi =\frac{\limfunc{SR}\left( b\mid r\right) }{\sqrt{b^{\top }\Sigma b}}
\end{equation*}%
and the implied risk premium is:%
\begin{equation*}
\tilde{\mu}=r+\limfunc{SR}\left( b\mid r\right) \frac{\Sigma b}{\sqrt{%
b^{\top }\Sigma b}}
\end{equation*}%
We obtain $\phi =3.4367$, $\tilde{\mu}_{1}=7.56\%$, $\tilde{\mu}_{2}=8.94\%$
and $\tilde{\mu}_{3}=5.33\%$.

\item In this case, the views of the portfolio manager corresponds to the
trends observed in the market. We then have $P=I_{n}$, $Q=\hat{\mu}$ and%
\footnote{%
If we suppose that the trends are not correlated.} $\Omega =\limfunc{diag}%
\left( \sigma ^{2}\left( \hat{\mu}_{1}\right) ,\ldots ,\sigma ^{2}\left(
\hat{\mu}_{n}\right) \right) $. The views $P\mu =Q+\varepsilon $ become:%
\begin{equation*}
\mu =\hat{\mu}+\varepsilon
\end{equation*}%
with $\varepsilon \sim \mathcal{N}\left( \mathbf{0},\Omega \right) $.

\item We have (TR-RPB, page 25):%
\begin{eqnarray*}
\bar{\mu} &=&E\left[ \mu \mid P\mu =Q+\varepsilon \right]  \\
&=&\tilde{\mu}+\Gamma P^{\top }\left( P\Gamma P^{\top }+\Omega \right)
^{-1}\left( Q-P\tilde{\mu}\right)  \\
&=&\tilde{\mu}+\tau \Sigma \left( \tau \Sigma +\Omega \right) ^{-1}\left(
\hat{\mu}-\tilde{\mu}\right)
\end{eqnarray*}%
We obtain $\bar{\mu}_{1}=5.16\%$, $\bar{\mu}_{2}=2.38\%$ and $\bar{\mu}%
_{3}=2.47\%$.

\item We optimize the quadratic utility function with $\phi =3.4367$. The
Black-Litterman portfolio is then $x_{1}=56.81\%$, $x_{2}=-23.61\%$ and $%
x_{3}=66.80\%$. Its volatility tracking error is $\sigma \left( x\mid
b\right) =8.02\%$ and its alpha is $\mu \left( x\mid b\right) =10.21\%$.
\end{enumerate}

\item

\begin{enumerate}
\item If $\tau =0$, $\bar{\mu}=\tilde{\mu}$. The BL portfolio $x$ is then
equal to the neutral portfolio $b$. We also have:%
\begin{eqnarray*}
\lim_{\tau \rightarrow \infty }\bar{\mu} &=&\tilde{\mu}+\lim_{\tau
\rightarrow \infty }\tau \Sigma ^{\top }\left( \tau \Sigma +\Omega \right)
^{-1}\left( \hat{\mu}-\tilde{\mu}\right)  \\
&=&\tilde{\mu}+\left( \hat{\mu}-\tilde{\mu}\right)  \\
&=&\hat{\mu}
\end{eqnarray*}%
In this case, $\bar{\mu}$ is independent from the implied risk premium $\hat{%
\mu}$ and is exactly equal to the estimated trends $\hat{\mu}$. The BL
portfolio $x$ is then the Markowitz optimized portfolio with the given value
of $\phi $.

\begin{figure}[tbp]
\centering
\includegraphics[width = \figurewidth, height = \figureheight]{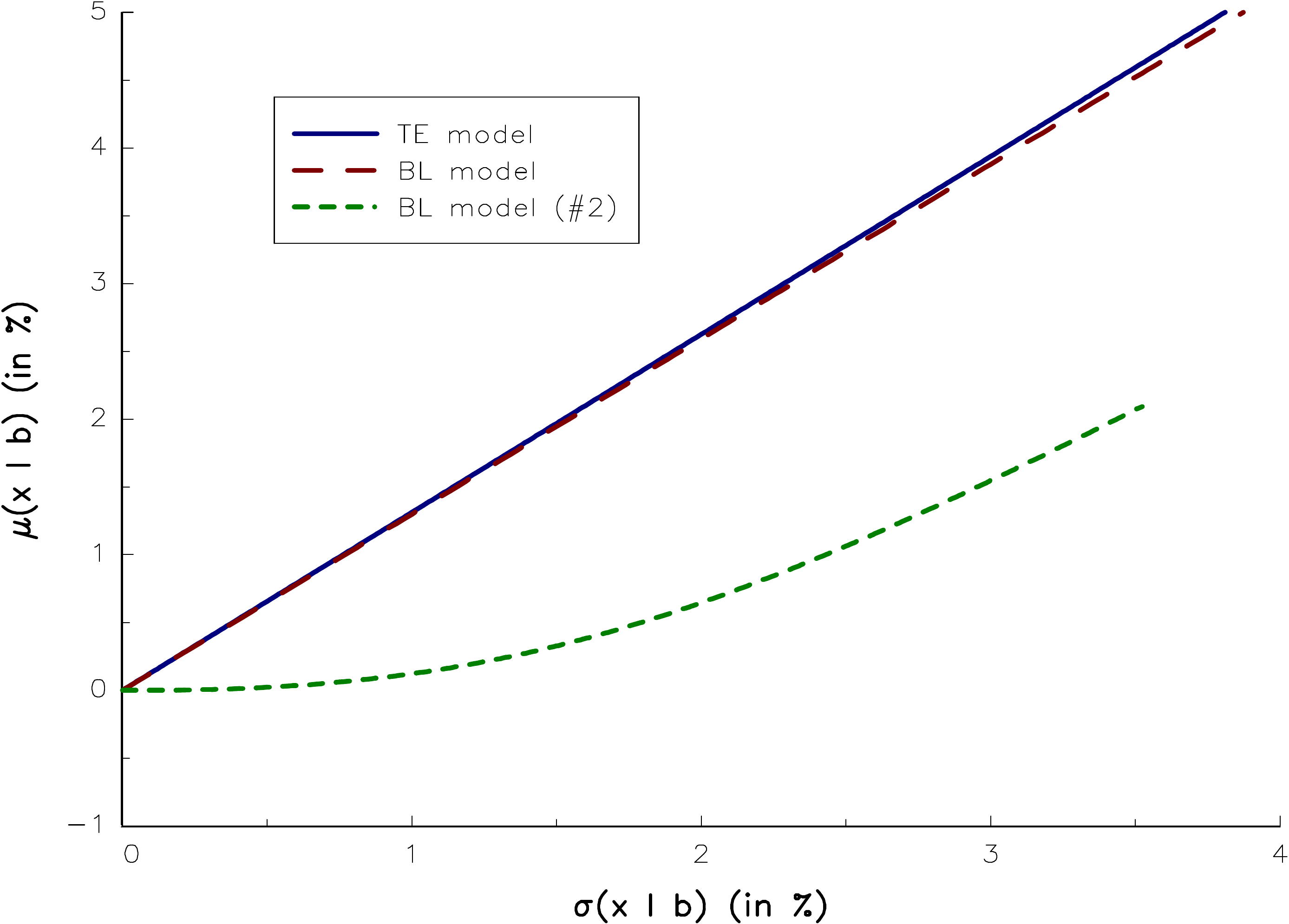}
\caption{Efficient frontier of TE and BL portfolios}
\label{fig:app2-1-9-1}
\end{figure}

\item We would like to find the BL portfolio such that $\sigma \left( x\mid
b\right) =3\%$. We know that $\sigma \left( x\mid b\right) =0$ if $\tau =0$.
Thanks to Question 2(d), we also know that $\sigma \left( x\mid b\right)
=8.02\%$ if $\tau =1\%$. It implies that the optimal portfolio corresponds
to a specific value of $\tau $ which is between $0$ and $1\%$. If we apply
the bi-section algorithm, we find that $\tau ^{\star }=0.242\%$. The
composition of the optimal portfolio is then $x_{1}^{\star }=41.18\%$, $%
x_{2}^{\star }=11.96\%$ and $x_{3}^{\star }=46.85\%$. We obtain an alpha
equal to $3.88\%$, which is a little bit smaller than the alpha of $3.94\%$
obtained for the TE portfolio.

\item We have reported the relationship between $\sigma \left( x\mid
b\right) $ and $\mu \left( x\mid b\right) $ in Figure \ref{fig:app2-1-9-1}.
We notice that the information ratio of BL portfolios is very close to
the information ratio of TE portfolios. We may explain that because of the
homogeneity of the estimated trends $\hat{\mu}_{i}$ and the
volatilities $\sigma \left( \hat{\mu}_{i}\right) $. If we suppose that $%
\sigma \left( \hat{\mu}_{1}\right) =1\%$, $\sigma \left( \hat{\mu}%
_{2}\right) =5\%$ and $\sigma \left( \hat{\mu}_{3}\right) =15\%$, we obtain
the relationship \#2. In this case, the BL model produces a smaller information
ratio than the TE model. We explain this because $\bar{\mu}$ is the right
measure of expected return for the BL model whereas it is $\hat{\mu}$ for
the TE model. We deduce that the ratios $\bar{\mu}_{i}/\hat{\mu}_{i}$\ are
more volatile for the parameter set \#2, in particular when $\tau $ is small.
\end{enumerate}

\end{enumerate}

\section{Portfolio optimization with transaction costs}

\begin{enumerate}
\item

\begin{enumerate}
\item The turnover is defined in TR-RPB on page 58. Results are given
in Table \ref{tab:app2-1-10-1}.

\begin{table}[b]
\centering
\caption{Solution of Question 1(a)}
\label{tab:app2-1-10-1}
\tableskip
\begin{tabular}{|c|rrrrrr|}
\hline
$\sigma ^\star$                                 & EW & $ 4.00$ & $ 4.50$ & $ 5.00$ & $ 5.50$ & $ 6.00$ \\
\hline
$x_1^\star$                                     & $16.67$ & $28.00$ & $14.44$ & $ 4.60$ & $ 0.00$ & $ 0.00$ \\
$x_2^\star$                                     & $16.67$ & $41.44$ & $40.11$ & $39.14$ & $34.34$ & $26.18$ \\
$x_3^\star$                                     & $16.67$ & $11.99$ & $14.86$ & $16.94$ & $17.99$ & $18.13$ \\
$x_4^\star$                                     & $16.67$ & $17.24$ & $24.89$ & $30.44$ & $35.38$ & $39.79$ \\
$x_5^\star$                                     & $16.67$ & $ 0.00$ & $ 0.00$ & $ 0.00$ & $ 0.00$ & $ 0.00$ \\
$x_6^\star$                                     & $16.67$ & $ 1.33$ & $ 5.70$ & $ 8.88$ & $12.29$ & $15.91$ \\ \hdashline
$\mu \left( x^\star\right)$                     & $ 6.33$ & $ 6.26$ & $ 6.84$ & $ 7.26$ & $ 7.62$ & $ 7.93$ \\
$\sigma \left( x^\star\right)$                  & $ 5.63$ & $ 4.00$ & $ 4.50$ & $ 5.00$ & $ 5.50$ & $ 6.00$ \\
$\tau \left( x\mid x^{\left( 0\right) }\right) $ & $ 0.00$ & $73.36$ & $63.32$ & $73.04$ & $75.42$ & $68.17$ \\
\hline
\end{tabular}
\end{table}

\item The relationship is reported in Figure \ref{fig:app2-1-10-1}. We notice that the turnover is
not an increasing function of the tracking error volatility. Controlling the
last one does not then permit to control the turnover.

\begin{figure}[tbph]
\centering
\includegraphics[width = \figurewidth, height = \figureheight]{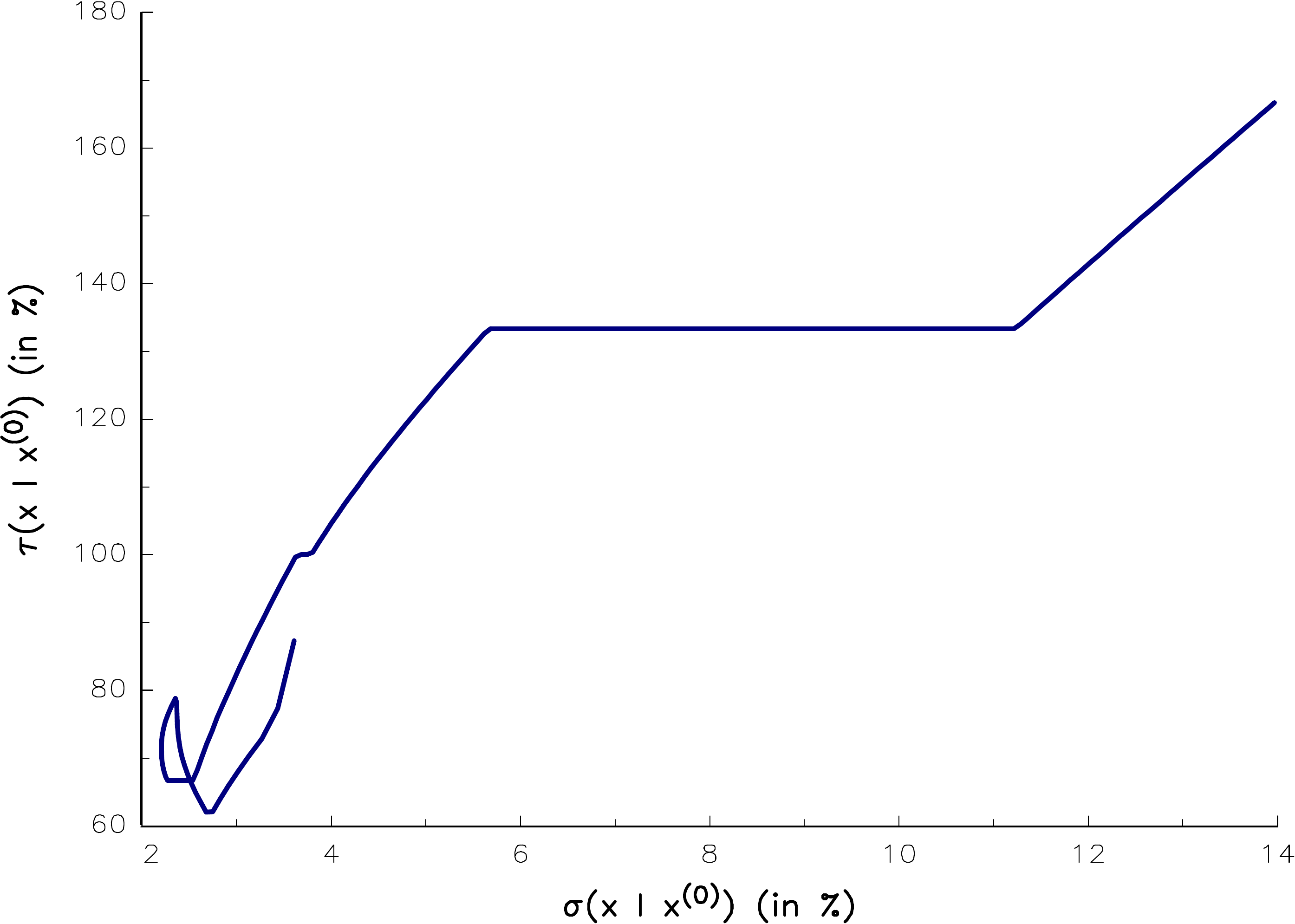}
\caption{Relationship between tracking error volatility and turnover}
\label{fig:app2-1-10-1}
\end{figure}

\item We consider the optimization program given in TR-RPB on page 59.
Results are reported in Figure \ref{fig:app2-1-10-2}. We note that the turnover
constraint reduces the risk/return tradeoff of MVO portfolios.

\begin{figure}[tbph]
\centering
\includegraphics[width = \figurewidth, height = \figureheight]{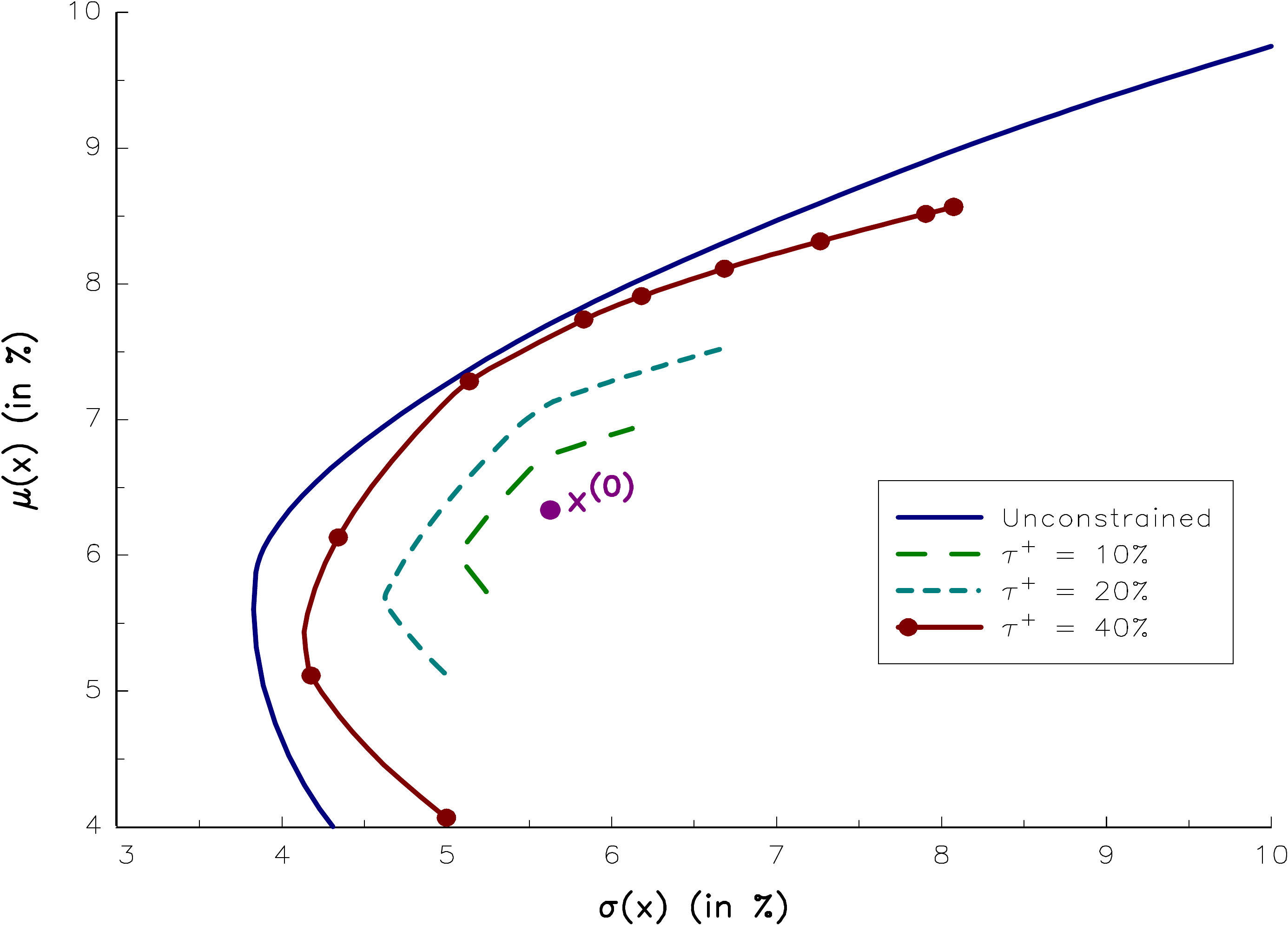}
\caption{Efficient frontier with turnover constraints}
\label{fig:app2-1-10-2}
\end{figure}

\item We obtain the results reported in Table \ref{tab:app2-1-10-2}.
\begin{table}[tbph]
\centering
\caption{Solution of Question 1(d)}
\label{tab:app2-1-10-2}
\tableskip
\begin{tabular}{|c|rrrrr|}
\hline
$\tau^{+}$                                       & EW   & $10.00$ & $20.00$ & $40.00$ & $80.00$ \\
\hline
$x_1^\star$                                      & $16.67$ &      & $16.67$ & $16.67$ & $ 4.60$ \\
$x_2^\star$                                      & $16.67$ &      & $26.67$ & $34.82$ & $39.14$ \\
$x_3^\star$                                      & $16.67$ &      & $16.67$ & $16.67$ & $16.94$ \\
$x_4^\star$                                      & $16.67$ &      & $16.67$ & $18.51$ & $30.44$ \\
$x_5^\star$                                      & $16.67$ &      & $11.15$ & $ 0.26$ & $ 0.00$ \\
$x_6^\star$                                      & $16.67$ &      & $12.18$ & $13.07$ & $ 8.88$ \\ \hdashline
$\mu \left( x^\star\right)$                      & $ 6.33$ &      & $ 6.39$ & $ 7.14$ & $ 7.26$ \\
$\sigma \left( x^\star\right)$                   & $ 5.63$ &      & $ 5.00$ & $ 5.00$ & $ 5.00$ \\
$\tau \left( x\mid x^{\left( 0\right) }\right) $ & $ 0.00$ &      & $20.00$ & $40.00$ & $73.04$ \\
\hline
\end{tabular}
\end{table}
We notice that there is no solution
if $\tau ^{+}=10\%$.  if $\tau ^{+}=80\%$, we retrieve the unconstrained
optimized portfolio.

\item Results are reported in Figure \ref{fig:app2-1-10-3}. After having rebalanced the allocation
seven times, we
obtain a portfolio which is located on the efficient frontier.

\begin{figure}[tbph]
\centering
\includegraphics[width = \figurewidth, height = \figureheight]{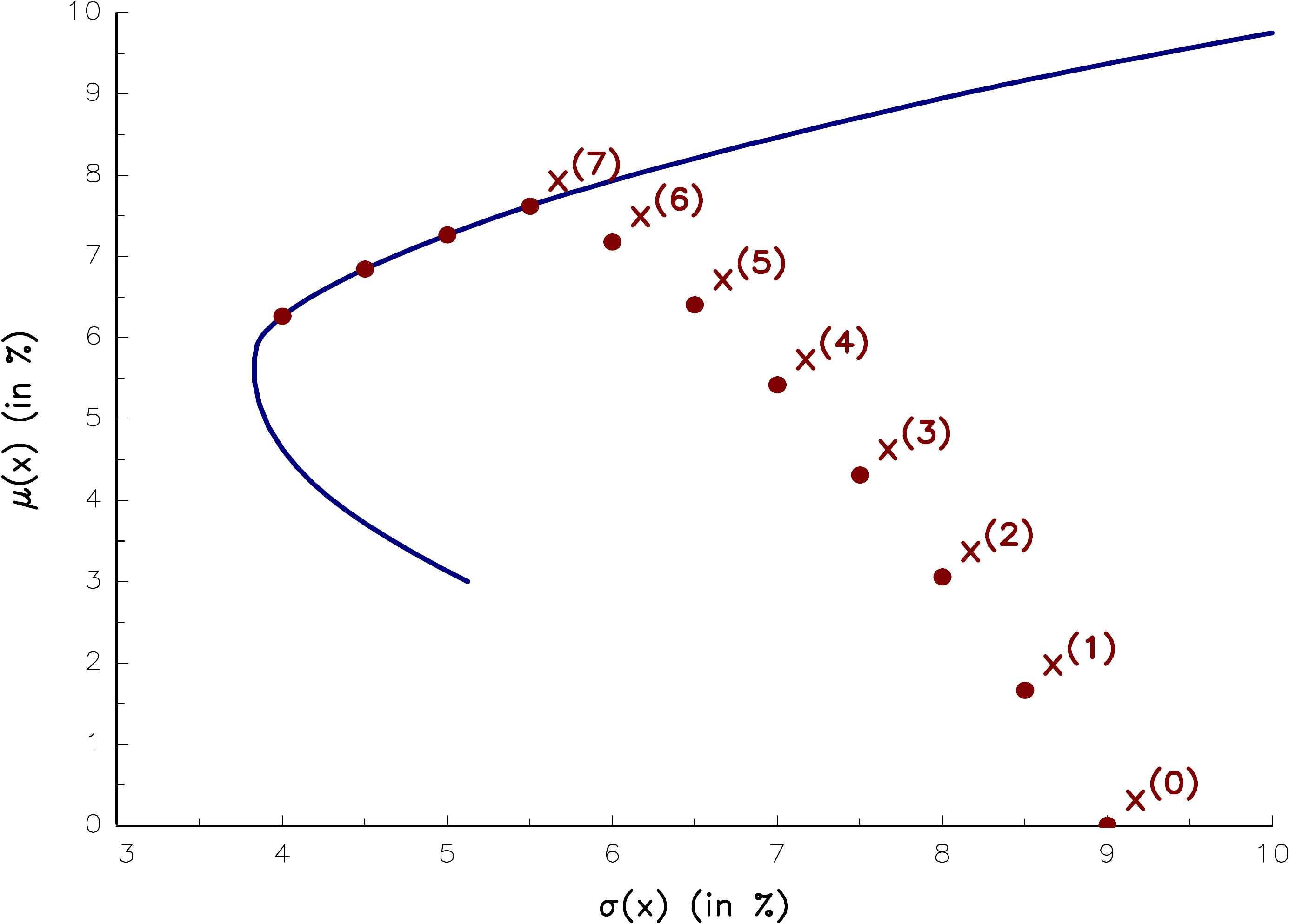}
\caption{Path of rebalanced portfolios}
\label{fig:app2-1-10-3}
\end{figure}

\end{enumerate}

\item

\begin{enumerate}
\item The weight $x_{i}$ of asset $i$ is equal to the actual weight $%
x_{i}^{\left( 0\right) }$ plus the positive change $x_{i}^{+}$ minus the
negative change $x_{i}^{-}$:%
\begin{equation*}
x_{i}=x_{i}^{0}+x_{i}^{+}-x_{i}^{-}
\end{equation*}%
The transactions costs are equal to:%
\begin{equation*}
\mathcal{C}=\sum_{i=1}^{n}x_{i}^{-}c_{i}^{-}+\sum_{i=1}^{n}x_{i}^{+}c_{i}^{+}
\end{equation*}%
Their financing are done by considering a part of the actual wealth:%
\begin{equation*}
\sum_{i=1}^{n}x_{i}+\mathcal{C}=1
\end{equation*}%
Moreover, the expected return of the portfolio is equal to $\mu \left(
x\right) -\mathcal{C}$. We deduce that the $\gamma $-problem of Markowitz
becomes:%
\begin{eqnarray*}
x^{\star } &=&\arg \min \frac{1}{2}x^{\top }\Sigma x-\gamma \left( x^{\top
}\mu -\mathcal{C}\right)  \\
&\text{u.c.}&\left\{
\begin{array}{l}
x=x^{0}+x^{+}-x^{-} \\
\mathbf{1}^{\top }x+\mathcal{C}=1 \\
\mathbf{0}\leq x\leq \mathbf{1} \\
\mathbf{0}\leq x^{-}\leq \mathbf{1} \\
\mathbf{0}\leq x^{+}\leq \mathbf{1}%
\end{array}%
\right.
\end{eqnarray*}%
The associated QP problem is:%
\begin{eqnarray*}
x^{\star } &=&\arg \min \frac{1}{2}X^{\top }QX-x^{\top }R \\
&&\text{u.c.}\left\{
\begin{array}{l}
AX=B \\
\mathbf{0}\leq X\leq \mathbf{1}%
\end{array}%
\right.
\end{eqnarray*}%
with:
\begin{eqnarray*}
Q &=&\left(
\begin{array}{ccc}
\Sigma  & \mathbf{0} & \mathbf{0} \\
\mathbf{0} & \mathbf{0} & \mathbf{0} \\
\mathbf{0} & \mathbf{0} & \mathbf{0}%
\end{array}%
\right) \text{,}\quad R=\gamma \left(
\begin{array}{c}
\mu  \\
-c^{-} \\
-c^{+}%
\end{array}%
\right) \text{, } \\
A &=&\left(
\begin{array}{ccc}
\mathbf{1}^{\top } & \left( c^{-}\right) ^{\top } & \left( c^{+}\right)
^{\top } \\
I_{n} & I_{n} & -I_{n}%
\end{array}%
\right) \quad \text{and}\quad B=\left(
\begin{array}{c}
1 \\
x^{0}%
\end{array}%
\right)
\end{eqnarray*}

\item We obtain the following results:%
\begin{equation*}
\begin{tabular}{|c|rrr|}
\hline
                                                 & \multicolumn{1}{c}{EW}      & \multicolumn{1}{c}{\#1}      & \multicolumn{1}{c|}{\#2} \\
\hline
$x_1^\star$                                      & $16.67$ & $ 4.60$ & $16.67$ \\
$x_2^\star$                                      & $16.67$ & $39.14$ & $30.81$ \\
$x_3^\star$                                      & $16.67$ & $16.94$ & $16.67$ \\
$x_4^\star$                                      & $16.67$ & $30.44$ & $22.77$ \\
$x_5^\star$                                      & $16.67$ & $ 0.00$ & $ 0.00$ \\
$x_6^\star$                                      & $16.67$ & $ 8.88$ & $12.46$ \\ \hdashline
$\mu \left( x^\star\right)$                      & $ 6.33$ & $ 7.26$ & $ 7.17$ \\
$\sigma \left( x^\star\right)$                   & $ 5.63$ & $ 5.00$ & $ 5.00$ \\
$\mathcal{C}$                                    &         & $ 1.10$ & $ 0.62$ \\
$\mu \left( x^\star\right) - \mathcal{C}$        &         & $ 6.17$ & $ 6.55$ \\
\hline
\end{tabular}
\end{equation*}
The portfolio \#1 is optimized without taking into account the transaction costs.
We obtain an expected return equal to $7.26\%$. However, the trading costs $\mathcal{C}$ are
equal to $1.10\%$ and reduce the net expected return to $6.17\%$. By taking into account
the transaction costs, it is possible to find an optimized portfolio $\#2$ which has
a net expected return equal to $6.55\%$.

\item In the case of a long-only portfolio, the financing of transaction
costs is done by the long positions:%
\begin{equation*}
\sum_{i=1}^{n}x_{i}+\mathcal{C}=1
\end{equation*}%
In a long-short portfolio, the cost $\mathcal{C}$ may be financed by both
the long and short positions. We then have to choose how to finance it. For
instance, if we suppose that 50\% (resp. 50\%) of the cost is financed by
the short (resp. long) positions, we obtain:%
\begin{eqnarray*}
x^{\star } &=&\arg \min \frac{1}{2}x^{\top }\Sigma x-\gamma \left( x^{\top
}\mu -\mathcal{C}\right)  \\
&\text{u.c.}&\left\{
\begin{array}{l}
x=x^{0}+x^{+}-x^{-} \\
\mathbf{1}^{\top }x+\mathcal{C}=1 \\
\sum_{i=1}^{n}\left( x_{i}+\frac{1}{2}\mathcal{C}\right)
\mathds{1}_{i}=\sum_{i=1}^{n}\left( x_{i}-\frac{1}{2}\mathcal{C}\right) \left(
1-\mathds{1}_{i}\right)  \\
-\left( 1-\mathds{1}_{i}\right) x^{S}\leq x_{i}\leq \mathds{1}_{i}x^{L} \\
0\leq x_{i}^{-}\leq 1 \\
0\leq x_{i}^{+}\leq 1%
\end{array}%
\right.
\end{eqnarray*}%
with $\mathds{1}_{i}$ an indicator function which takes the value $1$ if we want to
be long in the asset $i$ or $0$ if we want to be short. $x^{S}$ and $x^{L}$
indicate the maximum short and long exposures by asset. As previously, it is
then easy to write the corresponding QP program.
\end{enumerate}
\end{enumerate}

\section{Impact of constraints on the CAPM theory}

\begin{enumerate}
\item
\begin{enumerate}
\item At the equilibrium, we have:%
\begin{equation*}
\mathbb{E}\left[ R_{i}\right] =r+\beta _{i}\left( \mathbb{E}\left[ R\left(
x^{\star }\right) \right] -r\right)
\end{equation*}%
We introduce the notation $\beta \left( \mathbf{e}_{i}\mid x\right) $ to design the
beta of asset $i$ with respect to portfolio $x$. The previous
relationship can be written as follows:%
\begin{equation*}
\mu _{i}-r=\beta \left( \mathbf{e}_{i}\mid x^{\star }\right) \left( \mu \left(
x^{\star }\right) -r\right)
\end{equation*}

\item We have:%
\begin{eqnarray*}
\mu _{i}-r &=&\beta \left( \mathbf{e}_{i}\mid x^{\star }\right) \left( \mu \left(
x^{\star }\right) -r\right) + \\
&&\beta \left( \mathbf{e}_{i}\mid x\right) \left( \mu \left( x\right) -r\right)
-\beta \left( \mathbf{e}_{i}\mid x\right) \left( \mu \left( x\right) -r\right) \\
&=&\pi \left( \mathbf{e}_{i}\mid x\right) +\beta \left( \mathbf{e}_{i}\mid x^{\star }\right)
\left( \mu \left( x^{\star }\right) -r\right) -\beta \left( \mathbf{e}_{i}\mid
x\right) \left( \mu \left( x\right) -r\right)
\end{eqnarray*}%
We recall that:%
\begin{equation*}
\beta \left( \mathbf{e}_{i}\mid x^{\star }\right) \left( \mu \left( x^{\star }\right)
-r\right) =\mathcal{MR}_{i}\left( x^{\star }\right) \limfunc{SR}\left(
x^{\star }\mid r\right)
\end{equation*}%
We deduce that:%
\begin{equation*}
\mu _{i}-r=\pi \left( \mathbf{e}_{i}\mid x\right) +\delta _{i}\left( x^{\star
},x\right)
\end{equation*}%
with:%
\begin{equation*}
\delta _{i}\left( x^{\star },x\right) =\mathcal{MR}_{i}\left( x^{\star
}\right) \limfunc{SR}\left( x^{\star }\mid r\right) -\mathcal{MR}_{i}\left(
x\right) \limfunc{SR}\left( x\mid r\right)
\end{equation*}%
The risk premium of asset $i$ can be decomposed as the sum of the
beta return $\pi \left( \mathbf{e}_{i}\mid x\right) $ and the deviation $\delta
_{i}\left( x^{\star },x\right) $, which depends on the marginal volatilities
and the Sharpe ratios.

\item The beta return overestimates the risk premium of asset $i$ if $%
\pi \left( \mathbf{e}_{i}\mid x\right) >\mu _{i}-r$, that is when $\delta _{i}\left(
x^{\star },x\right) <0$. We then have:%
\begin{equation*}
\mathcal{MR}_{i}\left( x\right) \limfunc{SR}\left( x\mid r\right) >\mathcal{%
MR}_{i}\left( x^{\star }\right) \limfunc{SR}\left( x^{\star }\mid r\right)
\end{equation*}%
or:%
\begin{equation*}
\mathcal{MR}_{i}\left( x\right) >\mathcal{MR}_{i}\left( x^{\star }\right)
\frac{\limfunc{SR}\left( x^{\star }\mid r\right) }{\limfunc{SR}\left( x\mid
r\right) }
\end{equation*}%
Because $\limfunc{SR}\left( x^{\star }\mid r\right) >\limfunc{SR}\left(
x\mid r\right) $, it follows that $\mathcal{MR}_{i}\left( x\right) \gg
\mathcal{MR}_{i}\left( x^{\star }\right) $. We conclude that the beta return
overestimates the risk premium of asset $i$ if its marginal volatility $%
\mathcal{MR}_{i}\left( x\right) $ in the portfolio $x$ is large enough
compared to its marginal volatility $\mathcal{MR}_{i}\left( x^{\star
}\right) $ in the market portfolio $x^{\star }$.

\item We have:%
\begin{eqnarray*}
\limfunc{SR}\left( x\mid r\right)  &=&\frac{\mu \left( x\right) -r}{\sigma
\left( x\right) } \\
&=&\frac{\mu \left( x\right) -x^{\top }r}{\sigma \left( x\right) }
\end{eqnarray*}%
because $\mathbf{1}^{\top }x=1$. The optimization program is then:%
\begin{equation*}
x^{\star }=\arg \max \frac{\mu \left( x\right) -x^{\top }r}{\sigma \left(
x\right) }
\end{equation*}%
We deduce that the first-order condition is:
\begin{equation*}
\frac{\left( \partial _{x}\,\mu \left( x^{\star }\right) -r\mathbf{1}\right)
\sigma \left( x^{\star }\right) -\left( \mu \left( x^{\star }\right)
-r\right) \partial _{x}\,\sigma \left( x^{\star }\right) }{\sigma ^{2}\left(
x^{\star }\right) }=\mathbf{0}
\end{equation*}%
or:%
\begin{equation*}
\frac{\partial _{x}\,\mu \left( x^{\star }\right) -r\mathbf{1}}{\mu \left(
x^{\star }\right) -r}=\frac{\partial _{x}\,\sigma \left( x^{\star }\right) }{%
\sigma \left( x^{\star }\right) }
\end{equation*}%
We have:%
\begin{equation*}
\mu \left( x^{\star }\right) =\mu ^{\top }x^{\star }=\frac{\mu ^{\top
}\Sigma ^{-1}\left( \mu -r\mathbf{1}\right) }{\mathbf{1}^{\top }\Sigma
^{-1}\left( \mu -r\mathbf{1}\right) }
\end{equation*}%
and:%
\begin{equation*}
\mu \left( x^{\star }\right) -r=\frac{\left( \mu -r\mathbf{1}\right) ^{\top
}\Sigma ^{-1}\left( \mu -r\mathbf{1}\right) }{\mathbf{1}^{\top }\Sigma
^{-1}\left( \mu -r\mathbf{1}\right) }
\end{equation*}%
The return variance of $x^{\star }$ is also:%
\begin{eqnarray*}
\sigma ^{2}\left( x^{\star }\right)  &=&\frac{1}{\mathbf{1}^{\top }\Sigma
^{-1}\left( \mu -r\mathbf{1}\right) }\cdot \frac{\left( \mu -r\mathbf{1}%
\right) ^{\top }\Sigma ^{-1}\left( \mu -r\mathbf{1}\right) }{\mathbf{1}%
^{\top }\Sigma ^{-1}\left( \mu -r\mathbf{1}\right) } \\
&=&\frac{1}{\mathbf{1}^{\top }\Sigma ^{-1}\left( \mu -r\mathbf{1}\right) }%
\left( \mu \left( x^{\star }\right) -r\right)
\end{eqnarray*}%
It follows that:%
\begin{eqnarray*}
\frac{\partial _{x}\,\sigma \left( x^{\star }\right) }{\sigma \left(
x^{\star }\right) } &=&\frac{\Sigma x^{\star }}{\sigma ^{2}\left( x^{\star
}\right) } \\
&=&\frac{\mathbf{1}^{\top }\Sigma ^{-1}\left( \mu -r\mathbf{1}\right) \Sigma
x^{\star }}{\left( \mu \left( x^{\star }\right) -r\right) } \\
&=&\frac{\mathbf{1}^{\top }\Sigma ^{-1}\Sigma x^{\star }}{\left( \mu \left(
x^{\star }\right) -r\right) }\left( \mu -r\mathbf{1}\right)  \\
&=&\frac{\mu -r\mathbf{1}}{\mu \left( x^{\star }\right) -r} \\
&=&\frac{\partial _{x}\,\mu \left( x^{\star }\right) -r\mathbf{1}}{\mu
\left( x^{\star }\right) -r}
\end{eqnarray*}%
$x^{\star }$ satisfies then the first-order condition.

\item The first-order condition is $\left( \mu -r\mathbf{1}\right) -\phi
\Sigma x=\mathbf{0}$. The solution is then:%
\begin{equation*}
x^{\star }=\frac{1}{\phi }\Sigma ^{-1}\left( \mu -r\mathbf{1}\right)
\end{equation*}%
The value of the utility function at the optimum is:%
\begin{eqnarray*}
\mathcal{U}\left( x^{\star }\right)  &=&\frac{1}{\phi }\left( \mu -r\mathbf{1%
}\right) ^{\top }\Sigma ^{-1}\left( \mu -r\mathbf{1}\right) - \\
&&\frac{\phi }{2}\frac{\left( \mu -r\mathbf{1}\right) ^{\top }\Sigma
^{-1}\Sigma \Sigma ^{-1}\left( \mu -r\mathbf{1}\right) }{\phi ^{2}} \\
&=&\frac{1}{2\phi }\left( \mu -r\mathbf{1}\right) ^{\top }\Sigma ^{-1}\left(
\mu -r\mathbf{1}\right)
\end{eqnarray*}
We also have:%
\begin{eqnarray*}
\limfunc{SR}\left( x^{\star }\mid r\right)  &=&\frac{\left( \mu -r\mathbf{1}%
\right) ^{\top }x^{\star }}{\sqrt{x^{\star ^{\top }}\Sigma x^{\star }}} \\
&=&\sqrt{\left( \mu -r\mathbf{1}\right) ^{\top }\Sigma ^{-1}\left( \mu -r%
\mathbf{1}\right) }
\end{eqnarray*}%
We obtain:%
\begin{equation*}
\mathcal{U}\left( x^{\star }\right) =\frac{1}{2\phi }\limfunc{SR}%
\nolimits^{2}\left( x^{\star }\mid r\right)
\end{equation*}%
Maximizing the utility function is then equivalent to maximizing the Sharpe
ratio. In fact, the tangency portfolio corresponds to the value of $\phi $
such that $\mathbf{1}^{\top }x^{\star }=1$ (no cash in the portfolio). We
have:%
\begin{equation*}
\mathbf{1}^{\top }x^{\star }=\frac{1}{\phi }\mathbf{1}^{\top }\Sigma
^{-1}\left( \mu -r\mathbf{1}\right)
\end{equation*}%
It follows that:%
\begin{equation*}
\phi =\frac{1}{\mathbf{1}^{\top }\Sigma ^{-1}\left( \mu -r\mathbf{1}\right) }
\end{equation*}%
We deduce that the tangency portfolio is equal to:%
\begin{equation*}
x^{\star }=\frac{\Sigma ^{-1}\left( \mu -r\mathbf{1}\right) }{\mathbf{1}%
^{\top }\Sigma ^{-1}\left( \mu -r\mathbf{1}\right) }
\end{equation*}
\end{enumerate}

\item

\begin{enumerate}
\item We obtain the following results:
\begin{equation*}
\begin{tabular}{|c|cccc|}
\hline
$i$ & $x^{\star}_{i}$ & $\beta \left( \mathbf{e}_{i}\mid x^{\star }\right) $ & $\mathcal{MR}_{i}\left( x^{\star}\right) $ &
$\pi \left( \mathbf{e}_{i}\mid x^{\star}\right) $  \\ \hline
1 &     $-13.27\%$ & $1.77\%$ & $6.46\%$ & $5.00\%$ \\
2 & ${\bN}21.27\%$ & $1.77\%$ & $6.46\%$ & $5.00\%$ \\
3 & ${\bN}62.84\%$ & $0.71\%$ & $2.58\%$ & $2.00\%$ \\
4 & ${\bN}29.16\%$ & $1.42\%$ & $5.17\%$ & $4.00\%$ \\ \hline
\end{tabular}%
\end{equation*}
We verify that $\pi \left( \mathbf{e}_{i}\mid x\right) =\mu _{i}-r$.

\item We obtain the following results:
\begin{equation*}
\begin{tabular}{|c|ccccc|}
\hline
$i$ & $x_{i}$ & $\beta \left( \mathbf{e}_{i}\mid x\right) $ & $\mathcal{MR}_{i}\left( x\right) $ &
$\pi \left( \mathbf{e}_{i}\mid x\right) $ & $\delta_{i}\left( x^{\star },x\right) $ \\ \hline
1 & ${\bP}0.00\%$ & $2.96\%$ &     $11.79\%$ & $8.38\%$ &     $-3.38\%$ \\
2 & ${\bP}9.08\%$ & $1.77\%$ & ${\bP}7.04\%$ & $5.00\%$ & ${\bN}0.00\%$ \\
3 &     $63.24\%$ & $0.71\%$ & ${\bP}2.82\%$ & $2.00\%$ & ${\bN}0.00\%$ \\
4 &     $27.68\%$ & $1.42\%$ & ${\bP}5.63\%$ & $4.00\%$ & ${\bN}0.00\%$ \\ \hline
\end{tabular}%
\end{equation*}
Even if $x$ is a tangency portfolio, the beta return differs from the risk
premium because of the constraints. By imposing that $x_{i}\geq 0$, we
overestimate the beta of the first asset and its beta return. This explains
that $\delta _{1}\left( x^{\star },x\right) <0$.

\item We obtain the following results:
\begin{equation*}
\begin{tabular}{|c|ccccc|}
\hline
$i$ & $x_{i}$ & $\beta \left( \mathbf{e}_{i}\mid x\right) $ & $\mathcal{MR}_{i}\left( x\right) $ &
$\pi \left( \mathbf{e}_{i}\mid x\right) $ & $\delta_{i}\left( x^{\star },x\right) $ \\ \hline
1 & $10.00\%$ & $3.04$ &     $16.26\%$ & $9.82\%$ &     $-4.82\%$ \\
2 & $10.00\%$ & $1.83$ & ${\bP}9.79\%$ & $5.91\%$ &     $-0.91\%$ \\
3 & $48.55\%$ & $0.40$ & ${\bP}2.13\%$ & $1.28\%$ & ${\bN}0.72\%$ \\
4 & $31.45\%$ & $1.02$ & ${\bP}5.44\%$ & $3.28\%$ & ${\bN}0.72\%$ \\ \hline
\end{tabular}%
\end{equation*}

\item We consider the portfolio $x=\left( 0\%,0\%,50\%,50\%\right) $. We
obtain the following results:%
\begin{equation*}
\begin{tabular}{|c|ccccc|}
\hline
$i$ & $x_{i}$ & $\beta \left( \mathbf{e}_{i}\mid x\right) $ & $\mathcal{MR}_{i}\left( x\right) $ &
$\pi \left( \mathbf{e}_{i}\mid x\right) $ & $\delta_{i}\left( x^{\star },x\right) $ \\ \hline
1 & ${\bP}0.00\%$ & $1.24$ & $6.09\%$ & $3.71\%$ & ${\bN}1.29\%$ \\
2 & ${\bP}0.00\%$ & $0.25$ & $1.22\%$ & $0.74\%$ & ${\bN}4.26\%$ \\
3 &     $50.00\%$ & $0.33$ & $1.62\%$ & $0.99\%$ & ${\bN}1.01\%$ \\
4 &     $50.00\%$ & $1.67$ & $8.22\%$ & $5.01\%$ &     $-1.01\%$ \\ \hline
\end{tabular}%
\end{equation*}
\end{enumerate}
\end{enumerate}

\section{Generalization of the Jagannathan-Ma shrinkage approach}

\begin{enumerate}
\item
\begin{enumerate}
\item Jagannathan and Ma (2003) show that the constrained portfolio is
the solution of the unconstrained problem (TR-RPB, page 66):%
\begin{equation*}
\tilde{x}=x^{\star }\left( \tilde{\mu},\tilde{\Sigma}\right)
\end{equation*}%
with:%
\begin{equation*}
\left\{
\begin{array}{l}
\tilde{\mu}=\mu  \\
\tilde{\Sigma}=\Sigma +\left( \lambda ^{+}-\lambda ^{-}\right) \mathbf{1}%
^{\top }+\mathbf{1}\left( \lambda ^{+}-\lambda ^{-}\right) ^{\top }%
\end{array}%
\right.
\end{equation*}%
where $\lambda ^{-}$ and $\lambda ^{+}$ are the vectors of Lagrange coefficients
associated to the lower and upper bounds.

\item The unconstrained MV portfolio is:%
\begin{equation*}
x^{\star }=\left(
\begin{array}{r}
50.581\% \\
 1.193\% \\
-6.299\% \\
-3.054\% \\
57.579\%
\end{array}%
\right)
\end{equation*}

\item The constrained MV portfolio is:%
\begin{equation*}
\tilde{x}=\left(
\begin{array}{r}
40.000\% \\
16.364\% \\
 3.636\% \\
 0.000\% \\
40.000\%
\end{array}%
\right)
\end{equation*}%
The Lagrange coefficients are:
\begin{equation*}
\lambda ^{-}=\left(
\begin{array}{r}
0.000\% \\
0.000\% \\
0.000\% \\
0.118\% \\
0.000\%
\end{array}%
\right) \text{\quad and\quad }\lambda ^{+}=\left(
\begin{array}{r}
0.345\% \\
0.000\% \\
0.000\% \\
0.000\% \\
0.290\%
\end{array}%
\right)
\end{equation*}
The implied volatilities are $17.14\%$, $20.00\%$, $25.00\%$, $24.52\%$ and $16.82\%$.
For the implied shrinkage correlation matrix, we obtain:
\begin{equation*}
\tilde{\rho} =\left(
\begin{array}{rrrrr}
100.00\% &          &          &          &           \\
 53.80\% & 100.00\% &          &          &           \\
 34.29\% &  20.00\% & 100.00\% &          &           \\
 49.98\% &  38.37\% &  79.63\% & 100.00\% &           \\
 53.21\% &  53.20\% &  69.31\% &  49.62\% & 100.00\%
\end{array}%
\right)
\end{equation*}

\item If we impose that $3\%\leq x_{i}\leq 40\%$, the optimal solution
becomes:%
\begin{equation*}
\tilde{x}=\left(
\begin{array}{r}
40.000\% \\
14.000\% \\
 3.000\% \\
 3.000\% \\
40.000\%
\end{array}%
\right)
\end{equation*}%
The Lagrange function of the optimization problem is (TR-RPB, page 66):%
\begin{eqnarray*}
\mathcal{L}\left( x;\lambda _{0},\lambda _{1},\lambda ^{-},\lambda
^{+}\right)  &=&\frac{1}{2}x^{\top }\Sigma x- \\
& & \lambda _{0}\left( \mathbf{1}%
^{\top }x-1\right) -\lambda _{1}\left( \mu ^{\top }x-\mu ^{\star }\right) -
\\
&&\lambda ^{-^{\top }}\left( x-x^{-}\right) -\lambda ^{+^{\top }}\left(
x^{+}-x\right)
\end{eqnarray*}%
The first-order condition is:%
\begin{equation*}
\sigma \left( x\right) \frac{\partial \,\sigma \left( x\right) }{\partial \,x%
}-\lambda _{0}\mathbf{1}-\lambda _{1}\mu -\lambda ^{-}+\lambda ^{+}=\mathbf{0%
}
\end{equation*}%
It follows that:%
\begin{equation*}
\frac{\partial \,\sigma \left( x\right) }{\partial \,x}=\frac{\lambda _{0}%
\mathbf{1}+\lambda _{1}\mu +\lambda ^{-}-\lambda ^{+}}{\sigma \left(
x\right) }
\end{equation*}
We deduce that:%
\begin{equation*}
\sigma \left( x+\Delta x\right) \simeq \sigma \left( x\right) +\Delta
x^{\top }\frac{\partial \,\sigma \left( x\right) }{\partial \,x}
\end{equation*}%
Using the portfolio obtained in Question 1(c), we have $\sigma \left(
x\right) =12.79\%$ whereas the marginal volatilities
$\partial_x \,\sigma \left( x\right)$ are equal to:%
\begin{equation*}
\frac{1}{12.79}%
\left( 1.891 + \left(
\begin{array}{c}
0.000 \\
0.000 \\
0.000 \\
0.118 \\
0.000
\end{array}%
\right)
- \left(
\begin{array}{c}
0.345 \\
0.000 \\
0.000 \\
0.000 \\
0.290
\end{array}%
\right) \right) =\left(
\begin{array}{r}
12.09\% \\
14.78\% \\
14.78\% \\
15.70\% \\
12.51\%
\end{array}%
\right)
\end{equation*}%
It follows that the approximated value of the portfolio volatility is $12.82\%$
whereas the exact value is $12.84\%$.
\end{enumerate}
\item

\begin{enumerate}
\item The Lagrange function of the unconstrained problem is:%
\begin{equation*}
\mathcal{L}\left( x;\lambda _{0},\lambda _{1}\right) =\frac{1}{2}x^{\top
}\Sigma x-\lambda _{0}\left( \mathbf{1}^{\top }x-1\right) -\lambda
_{1}\left( \mu ^{\top }x-\mu ^{\star }\right)
\end{equation*}%
with $\lambda _{0}\geq 0$ and $\lambda _{1}\geq 0$. The unconstrained
solution $x^{\star }$ satisfies the following first-order conditions:%
\begin{equation*}
\left\{
\begin{array}{l}
\Sigma x^{\star }-\lambda _{0}\mathbf{1}-\lambda _{1}\mu =\mathbf{0} \\
\mathbf{1}^{\top }x^{\star }-1=0 \\
\mu ^{\top }x^{\star }-\mu ^{\star }=0%
\end{array}%
\right.
\end{equation*}%
If we now consider the constraints $Cx\geq D$, we have:%
\begin{eqnarray*}
\mathcal{L}\left( x;\lambda _{0},\lambda _{1},\lambda ^{-},\lambda
^{+}\right) & = & \frac{1}{2}x^{\top }\Sigma x-\lambda _{0}\left( \mathbf{1}%
^{\top }x-1\right) - \\
& & \lambda _{1}\left( \mu ^{\top }x-\mu ^{\star }\right) - \lambda ^{\top }\left( Cx-D\right)
\end{eqnarray*}%
with $\lambda _{0}\geq 0$, $\lambda _{1}\geq 0$ and $\lambda \geq \mathbf{0}$.
In this case, the constrained solution $\tilde{x}$ satisfies the following
Kuhn-Tucker conditions:%
\begin{equation*}
\left\{
\begin{array}{l}
\Sigma \tilde{x}-\lambda _{0}\mathbf{1}-\lambda _{1}\mu -C^{\top }\lambda =%
\mathbf{0} \\
\mathbf{1}^{\top }\tilde{x}-1=0 \\
\mu ^{\top }\tilde{x}-\mu ^{\star }=0 \\
\min \left( \lambda ,C\tilde{x}-D\right) =\mathbf{0}%
\end{array}%
\right.
\end{equation*}%
To show that $x^{\star }\left( \mu ,\tilde{\Sigma}\right) $ is the solution
of the constrained problem, we follow the same approach used in the case of
lower and upper bounds (TR-RPB, page 67). We have:%
\begin{eqnarray*}
\tilde{\Sigma}\tilde{x} &=&\Sigma \tilde{x}-\left( C^{\top }\lambda \mathbf{1%
}^{\top }+\mathbf{1}\lambda ^{\top }C\right) \tilde{x} \\
&=&\lambda _{0}\mathbf{1}+\lambda _{1}\mu +C^{\top }\lambda -C^{\top
}\lambda \mathbf{1}^{\top }\tilde{x}-\mathbf{1}\lambda ^{\top }C\tilde{x}
\end{eqnarray*}%
The Kuhn-Tucker condition $\min \left( \lambda ,C\tilde{x}-D\right) =\mathbf{%
0}$ implies that $\lambda ^{\top }\left( C\tilde{x}-D\right) =0$. We deduce
that:
\begin{eqnarray*}
\tilde{\Sigma}\tilde{x} &=&\lambda _{0}\mathbf{1}+\lambda _{1}\mu +C^{\top
}\lambda -C^{\top }\lambda -\mathbf{1}\lambda ^{\top }D \\
&=&\left( \lambda _{0}-\lambda ^{\top }D\right) \mathbf{1}+\lambda _{1}\mu
\end{eqnarray*}%
It proves that $\tilde{x}$ is the solution of the unconstrained optimization
problem with the following unconstrained Lagrange coefficients $\lambda
_{0}^{\star }=\lambda _{0}-\lambda ^{\top }D$ and $\lambda _{0}^{\star
}=\lambda _{1}$.

\item Because $\left( AB\right) ^{\top }=B^{\top }A^{\top }$, we have:%
\begin{eqnarray*}
\tilde{\Sigma}^{\top } &=&\Sigma ^{\top }-\left( C^{\top }\lambda \mathbf{1}%
^{\top }+\mathbf{1}\lambda ^{\top }C\right) ^{\top } \\
&=&\Sigma -\left( \mathbf{1}\lambda ^{\top }C+C^{\top }\lambda \mathbf{1}%
^{\top }\right)  \\
&=&\tilde{\Sigma}
\end{eqnarray*}%
This proves that $\tilde{\Sigma}$ is a symmetric matrix. For any vector $x$,
we have:%
\begin{eqnarray*}
x^{\top }\tilde{\Sigma}x &=&x^{\top }\left( \Sigma -\left( C^{\top }\lambda
\mathbf{1}^{\top }+\mathbf{1}\lambda ^{\top }C\right) \right) x \\
&=&x^{\top }\Sigma x-x^{\top }C^{\top }\lambda \mathbf{1}^{\top }x-x^{\top }%
\mathbf{1}\lambda ^{\top }Cx \\
&=&x^{\top }\Sigma x-2x^{\top }C^{\top }\lambda \mathbf{1}^{\top }x
\end{eqnarray*}%
We first consider the case where the constraint $\mu ^{\top }x\geq \mu
^{\star }$ vanishes and the optimization program corresponds to the minimum
variance problem. The first-order condition is:%
\begin{equation*}
C^{\top }\lambda =\Sigma \tilde{x}-\lambda _{0}\mathbf{1}
\end{equation*}%
It follows that:%
\begin{equation*}
\left( x^{\top }\mathbf{1}\right) \cdot \left( x^{\top }C^{\top }\lambda
\right) =\left( x^{\top }\mathbf{1}\right) \cdot \left( x^{\top }\Sigma
\tilde{x}\right) -\lambda _{0}\left( x^{\top }\mathbf{1}\right) ^{2}
\end{equation*}%
Due to Cauchy-Schwarz inequality, we also have:%
\begin{eqnarray*}
\left\vert \left( x^{\top }\mathbf{1}\right) \cdot \left( x^{\top }\Sigma
\tilde{x}\right) \right\vert  &=&\left\vert \left( x^{\top }\mathbf{1}%
\right) \cdot \left( x^{\top }\Sigma ^{1/2}\Sigma ^{1/2}\tilde{x}\right)
\right\vert  \\
&\leq &\left\vert \left( x^{\top }\mathbf{1}\right) \right\vert \cdot \left(
x^{\top }\Sigma x\right) ^{1/2}\cdot \left( \tilde{x}^{\top }\Sigma \tilde{x}%
\right) ^{1/2}
\end{eqnarray*}%
Using the Kuhn-Tucker condition\footnote{%
We have:%
\begin{eqnarray*}
\tilde{x}^{\top }\Sigma \tilde{x} &=&\tilde{x}^{\top }\left( C^{\top
}\lambda +\lambda _{0}\mathbf{1}\right)  \\
&=&\tilde{x}^{\top }C^{\top }\lambda +\lambda _{0}\tilde{x}^{\top }\mathbf{1}
\\
&=&\lambda ^{\top }\left( C\tilde{x}\right) +\lambda _{0} \\
&=&\lambda ^{\top }D+\lambda _{0}
\end{eqnarray*}%
}, we obtain:%
\begin{eqnarray*}
x^{\top }\tilde{\Sigma}x &=&x^{\top }\Sigma x-2\left( x^{\top }\mathbf{1}%
\right) \cdot \left( x^{\top }\Sigma \tilde{x}\right) +2\lambda _{0}\left(
x^{\top }\mathbf{1}\right) ^{2} \\
&\geq &x^{\top }\Sigma x-2\left\vert \left( x^{\top }\mathbf{1}\right) \cdot
\left( x^{\top }\Sigma \tilde{x}\right) \right\vert +2\lambda _{0}\left(
x^{\top }\mathbf{1}\right) ^{2} \\
&\geq &x^{\top }\Sigma x-2\left\vert \left( x^{\top }\mathbf{1}\right)
\right\vert \cdot \left( x^{\top }\Sigma x\right) ^{1/2}\cdot \left( \tilde{x%
}^{\top }\Sigma \tilde{x}\right) ^{1/2}+ \\
& & 2\lambda _{0}\left( x^{\top }\mathbf{1}\right) ^{2} \\
& = & x^{\top }\Sigma x-2\left\vert \left( x^{\top }\mathbf{1}\right)
\right\vert \cdot \left( x^{\top }\Sigma x\right) ^{1/2}\cdot \left( \lambda
^{\top }D+\lambda _{0}\right) ^{1/2}+\\
& & 2\lambda _{0}\left( x^{\top }\mathbf{1}\right) ^{2}
\end{eqnarray*}%
We deduce that:%
\begin{eqnarray*}
x^{\top }\tilde{\Sigma}x &\geq &x^{\top }\Sigma x-2\left\vert \left( x^{\top
}\mathbf{1}\right) \right\vert \cdot \left( x^{\top }\Sigma x\right)
^{1/2}\cdot \left( \lambda ^{\top }D+\lambda _{0}\right) ^{1/2}+ \\
& & \lambda_{0}\left( x^{\top }\mathbf{1}\right) ^{2} +
\left( \lambda ^{\top }D+\lambda _{0}\right) \left( x^{\top }\mathbf{1}%
\right) ^{2}-\left( \lambda ^{\top }D\right) \left( x^{\top }\mathbf{1}%
\right) ^{2} \\
& = &\left( a-b\right) ^{2}+\left( \lambda _{0}-\lambda ^{\top }D\right)
\left( x^{\top }\mathbf{1}\right) ^{2}
\end{eqnarray*}%
where $a=\sqrt{x^{\top }\Sigma x}$ and $b=\left\vert \left( x^{\top }\mathbf{%
1}\right) \right\vert \cdot \left( \lambda ^{\top }D+\lambda _{0}\right)
^{1/2}$. If $\lambda _{0}\geq \lambda ^{\top }D$, then $x^{\top }\tilde{%
\Sigma}x\geq 0$ and $\tilde{\Sigma}$ is a positive semi-definite matrix. If $%
\lambda _{0}<\lambda ^{\top }D$, the matrix $\tilde{\Sigma}$ may be
indefinite. Let us consider a universe of three assets. Their volatilities
are equal to $15\%$, $15\%$ and $5\%$ whereas the correlation matrix of
asset returns is:%
\begin{equation*}
\rho =\left(
\begin{array}{rrr}
100\% &  &  \\
50\% & 100\% &  \\
20\% & 20\% & 100\%%
\end{array}%
\right)
\end{equation*}%
If $\mathcal{C}=\left\{ 20\%\leq x_{i}\leq 80\%\right\} $, the minimum
variance portfolio is $\left( 20\%,20\%,60\%\right) $ and the implied
covariance matrix $\tilde{\Sigma}$ is not positive semi-definite. If $%
\mathcal{C}=\left\{ 20\%\leq x_{i}\leq 50\%\right\} $, the minimum variance
portfolio is $\left( 25\%,25\%,50\%\right) $ and the implied covariance
matrix $\tilde{\Sigma}$ is positive semi-definite. The extension to the case
$\mu ^{\top }x\geq \mu ^{\star }$ is straightforward because this constraint
may be encompassed in the restriction set
$\mathcal{C}=\left\{ x\in \mathbb{R}^{n}:Cx\geq D\right\} $.

\item We have $x\geq x^{-}$ and $x\leq x^{+}$. Imposing lower and upper
bounds is then equivalent to:%
\begin{equation*}
\left(
\begin{array}{r}
I_{n} \\
-I_{n}%
\end{array}%
\right) x\geq \left(
\begin{array}{r}
x^{-} \\
-x^{+}%
\end{array}%
\right)
\end{equation*}%
Let $\lambda =\left( \lambda ^{-},\lambda ^{+}\right) $ be the lagrange
coefficients associated with the constraint $Cx\geq D$. We have:%
\begin{eqnarray*}
C^{\top }\lambda  &=&\left(
\begin{array}{cc}
I_{n} & -I_{n}%
\end{array}%
\right) \left(
\begin{array}{r}
\lambda ^{-} \\
\lambda ^{+}%
\end{array}%
\right)  \\
&=&\lambda ^{-}-\lambda ^{+}
\end{eqnarray*}%
We deduce that the implied shrinkage covariance matrix is:%
\begin{eqnarray*}
\tilde{\Sigma} &=&\Sigma -\left( C^{\top }\lambda \mathbf{1}^{\top }+\mathbf{%
1}\lambda ^{\top }C\right)  \\
&=&\Sigma -\left( \lambda ^{-}-\lambda ^{+}\right) \mathbf{1}^{\top }-%
\mathbf{1}\left( \lambda ^{-}-\lambda ^{+}\right) ^{\top } \\
&=&\Sigma +\left( \lambda ^{+}-\lambda ^{-}\right) \mathbf{1}^{\top }+%
\mathbf{1}\left( \lambda ^{+}-\lambda ^{-}\right) ^{\top }
\end{eqnarray*}%
We retrieve the results of Jagannathan and Ma (2003).

\item We write the constraints as follows:%
\begin{equation*}
\left(
\begin{array}{rrrrr}
-1 & -1 & 0 & 0 & 0 \\
0 & 0 & 0 & 1 & 0%
\end{array}%
\right) x\geq \left(
\begin{array}{r}
-0.40 \\
0.10%
\end{array}%
\right)
\end{equation*}%
We obtain the following composition for the minimum variance portfolio:%
\begin{equation*}
\tilde{x}=\left(
\begin{array}{r}
 44.667\% \\
 -4.667\% \\
-19.195\% \\
 10.000\% \\
 69.195\%
\end{array}%
\right)
\end{equation*}%
The lagrange coefficients are $0.043\%$ and $0.134\%$. The implied
volatilities are $15.29\%$, $20.21\%$, $25.00\%$, $24.46\%$ and $15.00\%$.
For the implied shrinkage correlation matrix, we obtain:%
\begin{equation*}
\tilde{\rho} =\left(
\begin{array}{rrrrr}
100.00\% &          &          &          &           \\
 51.34\% & 100.00\% &          &          &           \\
 30.57\% &  20.64\% & 100.00\% &          &           \\
 47.72\% &  38.61\% &  79.58\% & 100.00\% &           \\
 41.14\% &  50.89\% &  70.00\% &  47.45\% & 100.00\%
\end{array}%
\right)
\end{equation*}
\end{enumerate}

\item

\begin{enumerate}
\item We consider the same technique used in QP problems (TR-RPB, page 302):%
\begin{eqnarray*}
Ax=B &\Leftrightarrow &\left\{
\begin{array}{c}
Ax\geq B \\
Ax\leq B%
\end{array}%
\right.  \\
&\Leftrightarrow &\left(
\begin{array}{r}
A \\
-A%
\end{array}%
\right) x\geq \left(
\begin{array}{r}
B \\
-B%
\end{array}%
\right)
\end{eqnarray*}%
We can then use the previous framework with:%
\begin{equation*}
C=\left(
\begin{array}{r}
A \\
-A%
\end{array}%
\right) \text{\quad and\quad }D=\left(
\begin{array}{r}
B \\
-B%
\end{array}%
\right)
\end{equation*}

\item We write the constraints as follows:%
\begin{equation*}
\left(
\begin{array}{rrrrr}
-1 & -1 & 0 & 0 & 0 \\
0 & 0 & 0 & 1 & -1 \\
0 & 0 & 0 & -1 & 1%
\end{array}%
\right) x\geq \left(
\begin{array}{r}
-0.50 \\
0.00 \\
0.00%
\end{array}%
\right)
\end{equation*}%
We obtain the following composition for the minimum variance portfolio:%
\begin{equation*}
\tilde{x}=\left(
\begin{array}{r}
 46.033\% \\
  3.967\% \\
-13.298\% \\
 31.649\% \\
 31.649\%
\end{array}%
\right)
\end{equation*}%
The lagrange coefficients are $0.316\%$, $0.709\%$ and $0.$ The implied
volatilities are $16.97\%$, $21.52\%$, $25.00\%$, $21.98\%$ and $19.15\%$.
For the implied shrinkage correlation matrix, we obtain:%
\begin{equation*}
\tilde{\rho} =\left(
\begin{array}{rrrrr}
100.00\% &          &          &          &           \\
 58.35\% & 100.00\% &          &          &           \\
 33.95\% &  24.46\% & 100.00\% &          &           \\
 39.70\% &  33.96\% &  78.08\% & 100.00\% &           \\
 59.21\% &  61.26\% &  69.63\% &  44.54\% & 100.00\%
\end{array}%
\right)
\end{equation*}
\end{enumerate}
\end{enumerate}

\begin{remark}
The original model of Jagannathan and Ma (2003) concerns the minimum variance portfolio.
Extension to mean-variance portfolios is straightforward if we consider
the Markowitz constraint $\mu\left(x\right) \geq \mu ^{\star }$ as a special case of
the general constraint $C x \geq \ D$ treated in this exercise.
\end{remark}

\chapter{Exercises related to the risk budgeting approach}

\section{Risk measures}

\begin{enumerate}
\item
\begin{enumerate}
\item We have (TR-RPB, page 74):%
\begin{equation*}
\limfunc{VaR}\left( \alpha \right) =\inf \left\{ \ell :\Pr \left\{ L\geq
\ell\right\} \geq \alpha \right\}
\end{equation*}%
and:%
\begin{equation*}
\func{ES}\left( \alpha \right) =\mathbb{E}\left[ \left. L\right\vert L\geq
\limfunc{VaR}\left( \alpha \right) \right]
\end{equation*}

\item We assume that $\mathbf{F}$ is continuous. It follows that $\limfunc{VaR}%
\left( \alpha \right) =\mathbf{F}^{-1}\left( \alpha \right) $. We deduce
that:%
\begin{eqnarray*}
\func{ES}\left( \alpha \right)  &=&\mathbb{E}\left[ \left. L\right\vert
L\geq \mathbf{F}^{-1}\left( \alpha \right) \right]  \\
&=&\int_{\mathbf{F}^{-1}\left( \alpha \right) }^{\infty }x\frac{f\left(
x\right) }{1-\mathbf{F}\left( \mathbf{F}^{-1}\left( \alpha \right) \right) }%
\,\mathrm{d}x \\
&=&\frac{1}{1-\alpha }\int_{\mathbf{F}^{-1}\left( \alpha \right) }^{\infty
}xf\left( x\right) \,\mathrm{d}x
\end{eqnarray*}%
We consider the change of variable $t=\mathbf{F}\left( x\right) $. Because $%
\mathrm{d}t=f\left( x\right) \,\mathrm{d}x$ and $\mathbf{F}\left( \infty
\right) =1$, we obtain:%
\begin{equation*}
\func{ES}\left( \alpha \right) =\frac{1}{1-\alpha }\int_{\alpha }^{1}\mathbf{%
F}^{-1}\left( t\right) \,\mathrm{d}t
\end{equation*}

\item We have:%
\begin{equation*}
f\left( x\right) =\theta \frac{x^{-\left( \theta +1\right) }}{x_{-}^{-\theta
}}
\end{equation*}%
The non-centered moment of order $n$ is%
\footnote{The moment exists if $n \neq \theta$.}:%
\begin{eqnarray*}
\mathbb{E}\left[ L^{n}\right]  &=&\int_{x_{-}}^{\infty }x^{n}\theta \frac{%
x^{-\left( \theta +1\right) }}{x_{-}^{-\theta }}\,\mathrm{d}x \\
&=&\frac{\theta }{x_{-}^{-\theta }}\int_{x_{-}}^{\infty }x^{n-\theta -1}\,%
\mathrm{d}x \\
&=&\frac{\theta }{x_{-}^{-\theta }}\left[ \frac{x^{n-\theta }}{n-\theta }%
\right] _{x_{-}}^{\infty } \\
&=&\frac{\theta }{\theta -n}x_{-}^{n}
\end{eqnarray*}%
We deduce that:%
\begin{equation*}
\mathbb{E}\left[ L\right] =\frac{\theta }{\theta -1}x_{-}
\end{equation*}%
and:%
\begin{equation*}
\mathbb{E}\left[ L^{2}\right] =\frac{\theta }{\theta -2}x_{-}^{2}
\end{equation*}%
The variance of the loss is then:%
\begin{equation*}
\limfunc{var}\left( L\right) =\mathbb{E}\left[ L^{2}\right] -\mathbb{E}^{2}%
\left[ L\right] =\frac{\theta }{\left( \theta -1\right) ^{2}\left( \theta
-2\right) }x_{-}^{2}
\end{equation*}%
$x_{-}$ is a scale parameter whereas $\theta $ is a parameter to control the
distribution tail. We have:%
\begin{equation*}
1-\left( \frac{\mathbf{F}^{-1}\left( \alpha \right) }{x_{-}}\right)
^{-\theta }=\alpha
\end{equation*}%
We deduce that:%
\begin{equation*}
\func{VaR}\left( \alpha \right) =\mathbf{F}^{-1}\left( \alpha \right)
=x_{-}\left( 1-\alpha \right) ^{-\theta ^{-1}}
\end{equation*}%
We also obtain:%
\begin{eqnarray*}
\func{ES}\left( \alpha \right)  &=&\frac{1}{1-\alpha }\int_{\alpha
}^{1}x_{-}\left( 1-t\right) ^{-\theta ^{-1}}\,\mathrm{d}t \\
&=&\frac{x_{-}}{1-\alpha }\left[ -\frac{1}{1-\theta ^{-1}}\left( 1-t\right)
^{1-\theta ^{-1}}\right] _{\alpha }^{1} \\
&=&\frac{\theta }{\theta -1}x_{-}\left( 1-\alpha \right) ^{-\theta ^{-1}} \\
&=&\frac{\theta }{\theta -1}\func{VaR}\left( \alpha \right)
\end{eqnarray*}%
Because $\theta >1$, we have $\frac{\theta }{\theta -1}>1$ and:%
\begin{equation*}
\func{ES}\left( \alpha \right) >\func{VaR}\left( \alpha \right)
\end{equation*}

\item We have:%
\begin{equation*}
\func{ES}\left( \alpha \right) =\frac{1}{1-\alpha }\int_{\mu +\sigma \Phi
^{-1}\left( \alpha \right) }^{\infty }x\frac{1}{\sigma \sqrt{2\pi }}\exp
\left( -\frac{1}{2}\left( \frac{x-\mu }{\sigma }\right) ^{2}\right) \,%
\mathrm{d}x
\end{equation*}%
By considering the change of variable $t=\sigma ^{-1}\left( x-\mu \right) $,
we obtain (TR-RPB, page 75):%
\begin{eqnarray*}
\func{ES}\left( \alpha \right)  &=&\frac{1}{1-\alpha }\int_{\Phi ^{-1}\left(
\alpha \right) }^{\infty }\left( \mu +\sigma t\right) \frac{1}{\sqrt{2\pi }}%
\exp \left( -\frac{1}{2}t^{2}\right) \,\mathrm{d}t \\
&=&\frac{\mu }{1-\alpha }\left[ \Phi \left( t\right) \right] _{\Phi
^{-1}\left( \alpha \right) }^{\infty }+ \\
& & \frac{\sigma }{\left( 1-\alpha
\right) \sqrt{2\pi }}\int_{\Phi ^{-1}\left( \alpha \right) }^{\infty }t\exp
\left( -\frac{1}{2}t^{2}\right) \,\mathrm{d}t \\
&=&\mu +\frac{\sigma }{\left( 1-\alpha \right) \sqrt{2\pi }}\left[ -\exp
\left( -\frac{1}{2}t^{2}\right) \right] _{\Phi ^{-1}\left( \alpha \right)
}^{\infty } \\
&=&\mu +\frac{\sigma }{\left( 1-\alpha \right) \sqrt{2\pi }}\exp \left( -%
\frac{1}{2}\left[ \Phi ^{-1}\left( \alpha \right) \right] ^{2}\right)  \\
&=&\mu +\frac{\sigma }{\left( 1-\alpha \right) }\phi \left( \Phi ^{-1}\left(
\alpha \right) \right)
\end{eqnarray*}%
Because $\phi ^{\prime }\left( x\right) =-x\phi \left( x\right) $, we have:%
\begin{eqnarray*}
1-\Phi \left( x\right)  &=&\int_{x}^{\infty }\phi \left( t\right) \,\mathrm{d%
}t \\
&=&\int_{x}^{\infty }\left( -\frac{1}{t}\right) \left( -t\phi \left(
t\right) \right) \,\mathrm{d}t \\
&=&\int_{x}^{\infty }\left( -\frac{1}{t}\right) \phi ^{\prime }\left(
t\right) \,\mathrm{d}t
\end{eqnarray*}%
We consider the integration by parts with $u\left( t\right) =-t^{-1}$ and $%
v^{\prime }\left( t\right) =\phi \left( t\right) $:%
\begin{eqnarray*}
1-\Phi \left( x\right)  &=&\left[ -\frac{\phi \left( t\right) }{t}\right]
_{x}^{\infty }-\int_{x}^{\infty }\frac{1}{t^{2}}\phi \left( t\right) \,%
\mathrm{d}t \\
&=&\frac{\phi \left( x\right) }{x}+\int_{x}^{\infty }\frac{1}{t^{3}}\left(
-t\phi \left( t\right) \right) \,\mathrm{d}t \\
&=&\frac{\phi \left( x\right) }{x}+\int_{x}^{\infty }\frac{1}{t^{3}}\phi
^{\prime }\left( t\right) \,\mathrm{d}t
\end{eqnarray*}%
We consider another integration by parts with $u\left( t\right) =t^{-3}$ and
$v^{\prime }\left( t\right) =\phi \left( t\right) $:%
\begin{eqnarray*}
1-\Phi \left( x\right)  &=&\frac{\phi \left( x\right) }{x}+\left[ \frac{\phi
\left( t\right) }{t^{3}}\right] _{x}^{\infty }-\int_{x}^{\infty }-\frac{3}{%
t^{4}}\phi \left( t\right) \,\mathrm{d}t \\
&=&\frac{\phi \left( x\right) }{x}-\frac{\phi \left( x\right) }{x^{3}}%
-\int_{x}^{\infty }\frac{3}{t^{5}}\phi ^{\prime }\left( t\right) \,\mathrm{d}%
t
\end{eqnarray*}%
We continue to use the integration by parts with $v^{\prime }\left( t\right)
=\phi \left( t\right) $. At the end, we obtain:%
\begin{eqnarray*}
1-\Phi \left( x\right)  &=&\frac{\phi \left( x\right) }{x}-\frac{\phi \left(
x\right) }{x^{3}}+3\frac{\phi \left( x\right) }{x^{5}}-3\cdot 5\frac{\phi
\left( x\right) }{x^{7}}+ \\
&& 3\cdot 5\cdot 7\frac{\phi \left( x\right) }{x^{9}}-\ldots \\
&=&\frac{\phi \left( x\right) }{x}+\frac{1}{x^{2}}\sum_{n=1}^{\infty }\left(
-1\right) ^{n}\left( \prod_{i=1}^{n}\left( 2i-1\right) \right) \frac{\phi
\left( x\right) }{x^{2n-1}} \\
&=&\frac{\phi \left( x\right) }{x}+\frac{\Psi \left( x\right) }{x^{2}}
\end{eqnarray*}%
We have represented the approximation in Figure \ref{fig:app2-2-1-1}.
\begin{figure}[tb]
\centering
\includegraphics[width = \figurewidth, height = \figureheight]{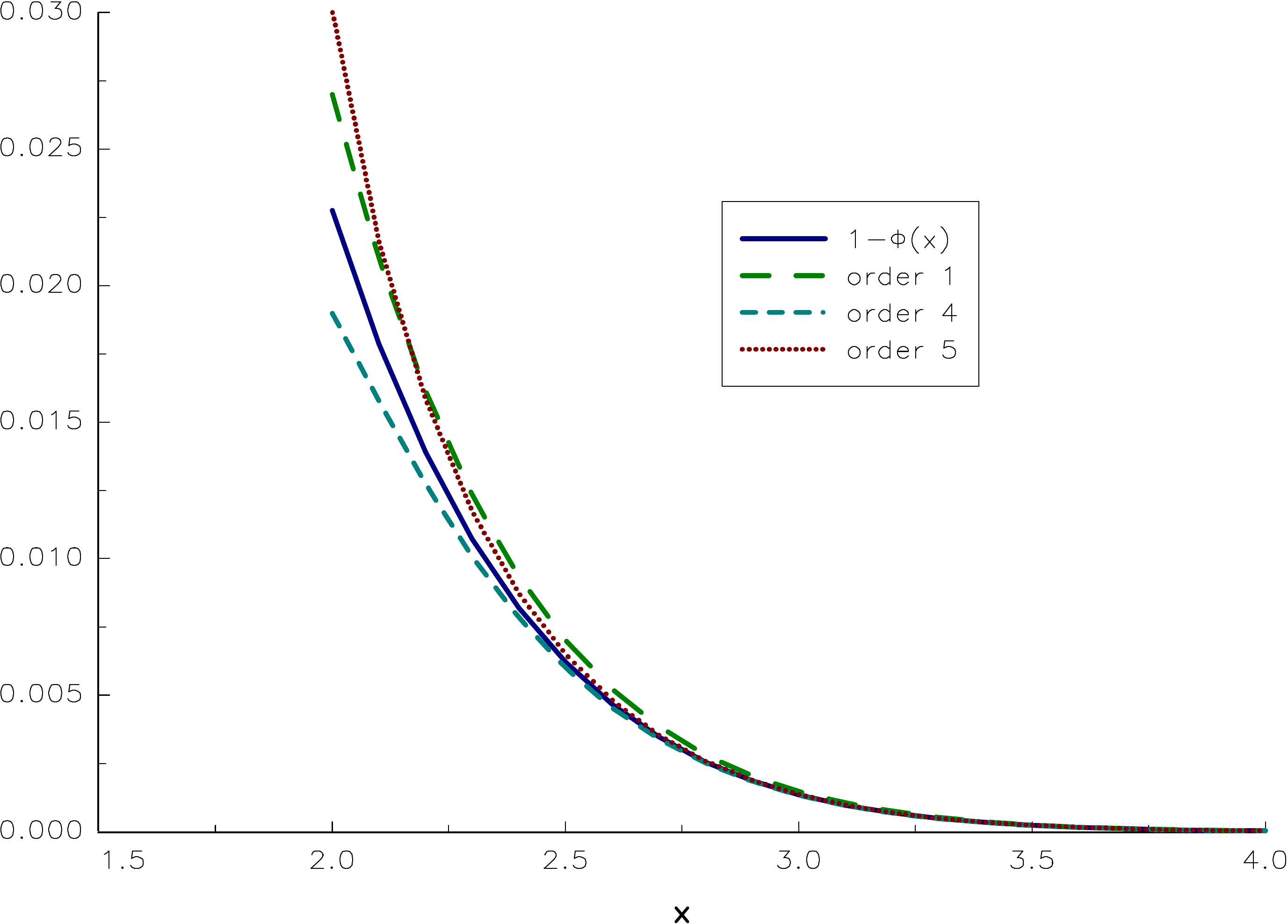}
\caption{Approximation of $1-\Phi \left( x\right) $}
\label{fig:app2-2-1-1}
\end{figure}
We finally deduce that:%
\begin{equation*}
\phi \left( x\right) =x\left( 1-\Phi \left( x\right) \right) -\frac{\Psi
\left( x\right) }{x}
\end{equation*}%
By using the previous expression of $\func{ES}\left( \alpha \right) $, we
obtain with $x=\Phi ^{-1}\left( \alpha \right) $:%
\begin{eqnarray*}
\func{ES}\left( \alpha \right)  &=&\mu +\frac{\sigma }{\left( 1-\alpha
\right) }\phi \left( \Phi ^{-1}\left( \alpha \right) \right)  \\
&=&\mu +\frac{\sigma }{\left( 1-\alpha \right) }\phi \left( x\right)  \\
&=&\mu +\frac{\sigma }{\left( 1-\alpha \right) }\left( \Phi ^{-1}\left(
\alpha \right) \left( 1-\alpha \right) -\frac{\Psi \left( \Phi ^{-1}\left(
\alpha \right) \right) }{\Phi ^{-1}\left( \alpha \right) }\right)  \\
&=&\mu +\sigma \Phi ^{-1}\left( \alpha \right) -\sigma \frac{\Psi \left(
\Phi ^{-1}\left( \alpha \right) \right) }{\left( 1-\alpha \right) \Phi
^{-1}\left( \alpha \right) } \\
&=&\func{VaR}\left( \alpha \right) -\sigma \frac{\Psi \left( \Phi
^{-1}\left( \alpha \right) \right) }{\left( 1-\alpha \right) \Phi
^{-1}\left( \alpha \right) }
\end{eqnarray*}%
We deduce that $\func{ES}\left( \alpha \right) \rightarrow \func{VaR}\left(
\alpha \right) $ because:%
\begin{equation*}
\lim_{\alpha \rightarrow 1}\frac{\Psi \left( \Phi ^{-1}\left( \alpha \right)
\right) }{\left( 1-\alpha \right) \Phi ^{-1}\left( \alpha \right) }=0
\end{equation*}

\item For the Gaussian distribution, the expected shortfall and the
value-at-risk coincide for high confidence level $\alpha $. It is not the
case with the Pareto distribution, which has a fat tail. The use of the
Pareto distribution can then produce risk measures which may be much higher
than those based on the Gaussian distribution.
\end{enumerate}

\item
\begin{enumerate}
\item We have (TR-RPB, page 73):%
\begin{eqnarray*}
\mathcal{R}\left( L_{1}+L_{2}\right) &=&\mathbb{E}\left[ L_{1}+L_{2}\right] =%
\mathbb{E}\left[ L_{1}\right] +\mathbb{E}\left[ L_{2}\right] =\mathcal{R}%
\left( L_{1}\right) +\mathcal{R}\left( L_{2}\right)  \\
\mathcal{R}\left( \lambda L\right) &=&\mathbb{E}\left[ \lambda L\right]
=\lambda \mathbb{E}\left[ L\right] =\lambda \mathcal{R}\left( L\right)  \\
\mathcal{R}\left( L+m\right) &=&\mathbb{E}\left[ L-m\right] =\mathbb{E}\left[
L\right] -m=\mathcal{R}\left( L\right) -m
\end{eqnarray*}%
We notice that:%
\begin{equation*}
\mathbb{E}\left[ L\right] =\int_{-\infty }^{\infty }x\,\mathrm{d}\mathbf{F}%
\left( x\right) =\int_{0}^{1}\mathbf{F}^{-1}\left( t\right) \,\mathrm{d}t
\end{equation*}%
We deduce that if $\mathbf{F}_{1}\left( x\right) \geq \mathbf{F}_{2}\left(
x\right) $, then $\mathbf{F}_{1}^{-1}\left( t\right) \leq \mathbf{F}%
_{2}^{-1}\left( t\right) $ and $\mathbb{E}\left[ L_{1}\right] \leq \mathbb{E}%
\left[ L_{2}\right] $. We conclude that $\mathcal{R}$ is a coherent risk
measure.

\item We have:%
\begin{eqnarray*}
\mathcal{R}\left( L_{1}+L_{2}\right)  &=&\mathbb{E}\left[ L_{1}+L_{2}\right]
+\mathbb{\sigma }\left( L_{1}+L_{2}\right)  \\
&=&\mathbb{E}\left[ L_{1}\right] +\mathbb{E}\left[ L_{2}\right] + \\
&&\sqrt{\mathbb{\sigma }^{2}\left( L_{1}\right) +\mathbb{\sigma }^{2}\left(
L_{2}\right) +2\rho \left( L_{1},L_{2}\right) \mathbb{\sigma }\left(
L_{1}\right) \mathbb{\sigma }\left( L_{2}\right) }
\end{eqnarray*}%
Because $\rho \left( L_{1},L_{2}\right) \leq 1$, we deduce that:%
\begin{eqnarray*}
\mathcal{R}\left( L_{1}+L_{2}\right)  &\leq &\mathbb{E}\left[ L_{1}\right] +%
\mathbb{E}\left[ L_{2}\right] + \\
&&\sqrt{\mathbb{\sigma }^{2}\left( L_{1}\right) +\mathbb{\sigma }^{2}\left(
L_{2}\right) +2\mathbb{\sigma }\left( L_{1}\right) \mathbb{\sigma }\left(
L_{2}\right) } \\
&\leq &\mathbb{E}\left[ L_{1}\right] +\mathbb{E}\left[ L_{2}\right] +\mathbb{%
\sigma }\left( L_{1}\right) +\mathbb{\sigma }\left( L_{2}\right)  \\
&\leq &\mathcal{R}\left( L_{1}\right) +\mathcal{R}\left( L_{2}\right)
\end{eqnarray*}%
We have:%
\begin{eqnarray*}
\mathcal{R}\left( \lambda L\right)  &=&\mathbb{E}\left[ \lambda L\right] +%
\mathbb{\sigma }\left( \lambda L\right)  \\
&=&\lambda \mathbb{E}\left[ L\right] +\lambda \mathbb{\sigma }\left(
L\right)  \\
&=&\lambda \mathcal{R}\left( L\right)
\end{eqnarray*}%
and:%
\begin{eqnarray*}
\mathcal{R}\left( L+m\right)  &=&\mathbb{E}\left[ L-m\right] +\mathbb{\sigma
}\left( L-m\right)  \\
&=&\mathbb{E}\left[ L\right] -m+\mathbb{\sigma }\left( L\right)  \\
&=&\mathcal{R}\left( L\right) -m
\end{eqnarray*}%
If we consider the convexity property, we notice that (TR-RPB, page 73):%
\begin{eqnarray*}
\mathcal{R}\left( \lambda L_{1}+\left( 1-\lambda \right) L_{2}\right)  &\leq
&\mathcal{R}\left( \lambda L_{1}\right) +\mathcal{R}\left( \left( 1-\lambda
\right) L_{2}\right)  \\
&\leq &\lambda \mathcal{R}\left( L_{1}\right) +\left( 1-\lambda \right)
\mathcal{R}\left( L_{2}\right)
\end{eqnarray*}%
We conclude that $\mathcal{R}$ is a convex risk measure.
\end{enumerate}

\item We have:%
\begin{equation*}
\begin{tabular}{|c|ccccccccc|}
\hline
$\ell _{i}$ & $0$ & $1$ & $2$ & $3$ & $4$ & $5$ & $6$ & $7$ & $8$ \\
\hline
$\Pr \left\{ L=\ell _{i}\right\} $ & $0.2$ & $0.1$ & $0.1$ & $0.1$ & $0.1$ & $0.1$ & $0.1$ & $0.1$ & $0.1$ \\
$\Pr \left\{ L\leq \ell _{i}\right\} $ & $0.2$ & $0.3$ & $0.4$ & $0.5$ & $0.6$ & $0.7$ & $0.8$ & $0.9$ & $1.0$ \\
\hline
\end{tabular}%
\end{equation*}

\begin{enumerate}
\item We have $\limfunc{VaR}\left( 50\%\right) =3$, $\limfunc{VaR}\left(
75\%\right) =6$, $\limfunc{VaR}\left( 90\%\right) =7$ and:%
\begin{eqnarray*}
\func{ES}\left( 50\%\right)  &=&\frac{3\times 10\%+\ldots +8\times 10\%}{60\%%
}=5.5 \\
\func{ES}\left( 75\%\right)  &=&\frac{6\times 10\%+\ldots +8\times 10\%}{30\%%
}=7.0 \\
\func{ES}\left( 90\%\right)  &=&\frac{7\times 10\%+8\times 10\%}{20\%}=7.5
\end{eqnarray*}

\item We have to build a bivariate distribution such that (TR-RPB, page 73):%
\begin{equation*}
\mathbf{F}_{1}^{-1}\left( \alpha \right) +\mathbf{F}_{2}^{-1}\left( \alpha
\right) <\mathbf{F}_{1+2}^{-1}\left( \alpha \right)
\end{equation*}%
To this end, we may use the Makarov inequalities. For instance, we may
consider an ordinal sum of the copula $\mathbf{C}^{+}$ for $\left(
u_{1},u_{2}\right) \leq \left( \alpha ,\alpha \right) $ and another copula $%
\mathbf{C}_{\alpha }$ for $\left( u_{1},u_{2}\right) >\left( \alpha ,\alpha
\right) $ to produce a bivariate distribution which does not satisfy the
subadditivity property. By taking for example $\alpha =70\%$ and $\mathbf{C}%
_{\alpha }=\mathbf{C}^{-}$, we obtain the following bivariate distribution%
\footnote{We have $p_{1,i} = \Pr \left\{ L_{1}=\ell _{i}\right\} $
and $p_{2,i} = \Pr \left\{ L_{2}=\ell _{i}\right\} $.}:%
\begin{equation*}
\begin{tabular}{|c|ccccccccc|c|}
\hline
$\ell _{i}$ & $0$ & $1$ & $2$ & $3$ & $4$ & $5$ & $6$ & $7$ & $8$ &
$p_{2,i} $ \\ \hline
$0$ & $0.2$ &  &  &  &  &  &  &  &  & $0.2$ \\
$1$ &  & $0.1$ &  &  &  &  &  &  &  & $0.1$ \\
$2$ &  &  & $0.1$ &  &  &  &  &  &  & $0.1$ \\
$3$ &  &  &  & $0.1$ &  &  &  &  &  & $0.1$ \\
$4$ &  &  &  &  & $0.1$ &  &  &  &  & $0.1$ \\
$5$ &  &  &  &  &  & $0.1$ &  &  &  & $0.1$ \\ \cline{8-10}
$6$ &  &  &  &  &  &  & \multicolumn{1}{|c}{} &  & $\mathbf{0,1}$ & $0.1$ \\
$7$ &  &  &  &  &  &  & \multicolumn{1}{|c}{} & $\mathbf{0,1}$ &  & $0.1$ \\
$8$ &  &  &  &  &  &  & \multicolumn{1}{|c}{$\mathbf{0,1}$} &  &  & $0.1$ \\ \hline
$p_{1,i}$    & $0.2$ & $0.1$ & $0.1$ & $0.1 $ & $0.1$ & $0.1$ & $0.1$ & $0.1$ & $0.1$ & \\
\hline
\end{tabular}%
\end{equation*}%
We then have:%
\begin{equation*}
\begin{tabular}{c|ccccccc}
$\ell _{i}$ & $0$ & $2$ & $4$ & $6$ & $8$ & $10$ & $14$ \\ \hline
$\Pr \left\{ L_{1}+L_{2}=\ell _{i}\right\} $ & $0.2$ & $0.1$ & $0.1$ &
$0.1$ & $0.1$ & $0.1$ & $0.3$ \\
$\Pr \left\{ L_{1}+L_{2}\leq \ell _{i}\right\} $ & $0.2$ & $0.3$ & $0.4
$ & $0.5$ & $0.6$ & $0.7$ & $1.0$%
\end{tabular}%
\end{equation*}%
Because $\mathbf{F}_{1}^{-1}\left( 80\%\right) =\mathbf{F}_{2}^{-1}\left(
80\%\right) =6$ and $\mathbf{F}_{1+2}^{-1}\left( 80\%\right) =14$, we obtain:%
\begin{equation*}
\mathbf{F}_{1}^{-1}\left( 80\%\right) +\mathbf{F}_{2}^{-1}\left( 80\%\right)
<\mathbf{F}_{1+2}^{-1}\left( 80\%\right)
\end{equation*}
\end{enumerate}
\end{enumerate}

\section{Weight concentration of a portfolio}

\begin{enumerate}
\item
\begin{enumerate}
\item We have represented the function $y=\mathcal{L}\left( x\right) $ in
Figure \ref{fig:app2-2-2-1}. It verifies $\mathcal{L}\left( x\right) \geq x$
and $\mathcal{L}\left( x\right) \leq 1$. The Gini coefficient is defined as
follows (TR-RPB, page 127):%
\begin{eqnarray*}
G &=&\frac{A}{A+B} \\
&=&\left. \left( \int_{0}^{1}\mathcal{L}\left( x\right) \,\mathrm{d}x-\frac{1%
}{2}\right) \right/ \frac{1}{2} \\
&=&2\int_{0}^{1}\mathcal{L}\left( x\right) \,\mathrm{d}x-1
\end{eqnarray*}

\item If $\alpha \geq 0$, the function $\mathcal{L}_{\alpha }\left(
x\right) =x^{\alpha }$ is increasing. We have $\mathcal{L}_{\alpha }\left(
1\right) =1$, $\mathcal{L}_{\alpha }\left( x\right) \leq 1$ and $\mathcal{L}%
_{\alpha }\left( x\right) \geq x$. We deduce that $\mathcal{L}_{\alpha }$ is
a Lorenz curve. For the Gini index, we have:%
\begin{eqnarray*}
\mathcal{G}\left( \alpha \right)  &=&2\int_{0}^{1}x^{\alpha }\,\mathrm{d}x-1
\\
&=&2\left[ \frac{x^{\alpha +1}}{\alpha +1}\right] _{0}^{1}-1 \\
&=&\frac{1-\alpha }{1+\alpha }
\end{eqnarray*}%
We deduce that $\mathcal{G}\left( 0\right) =1$, $\mathcal{G}\left( \frac{1}{2%
}\right) =\nicefrac{1}{3}$ et $\mathcal{G}\left( 1\right) =0$. $\alpha =0$
corresponds to the perfect concentration whereas $\alpha =1$ corresponds to
the perfect equality.
\end{enumerate}

\begin{figure}[tbph]
\centering
\includegraphics[width = \figurewidth, height = \figureheight]{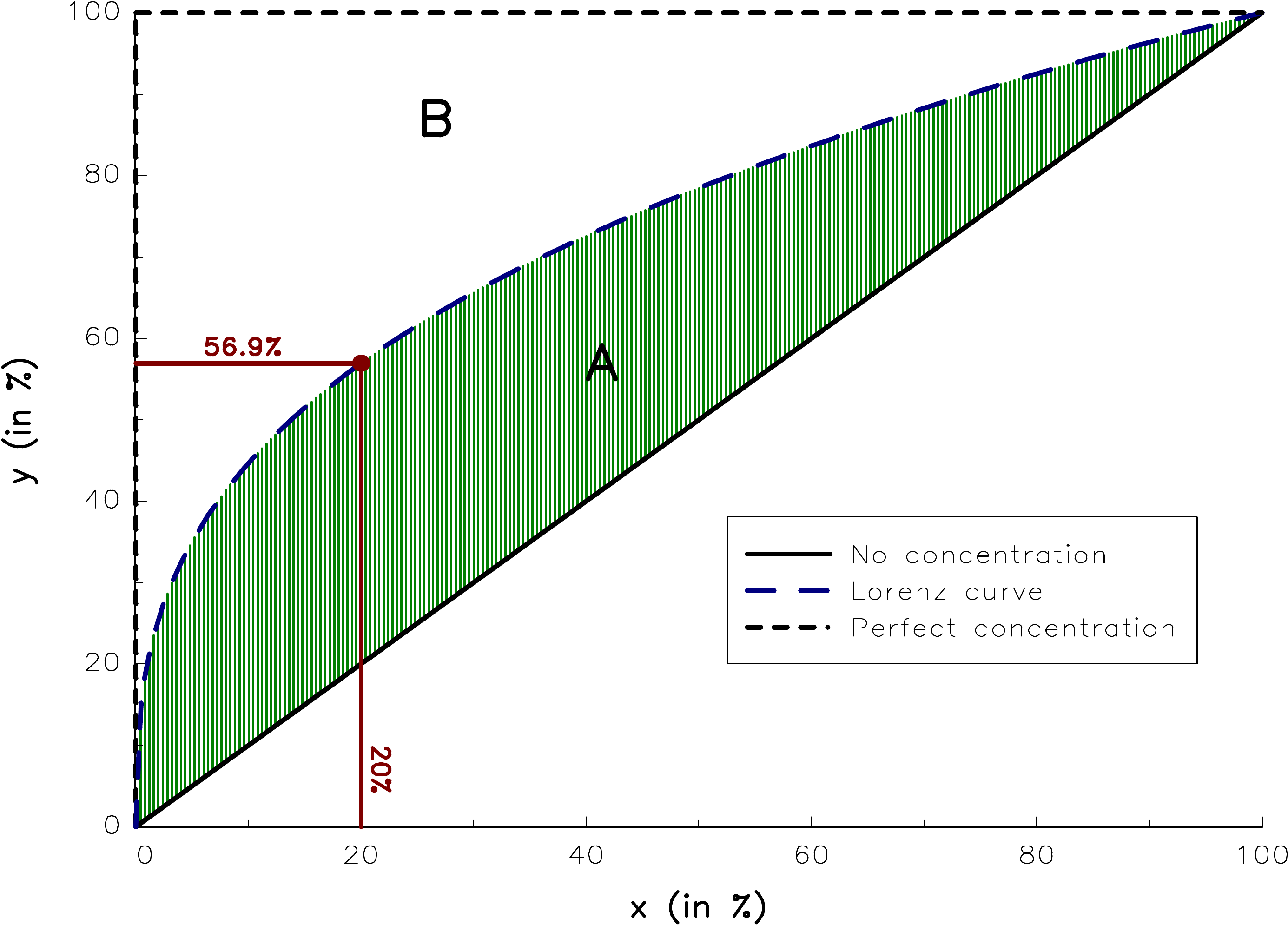}
\caption{Lorenz curve}
\label{fig:app2-2-2-1}
\end{figure}

\begin{figure}[tbph]
\centering
\includegraphics[width = \figurewidth, height = \figureheight]{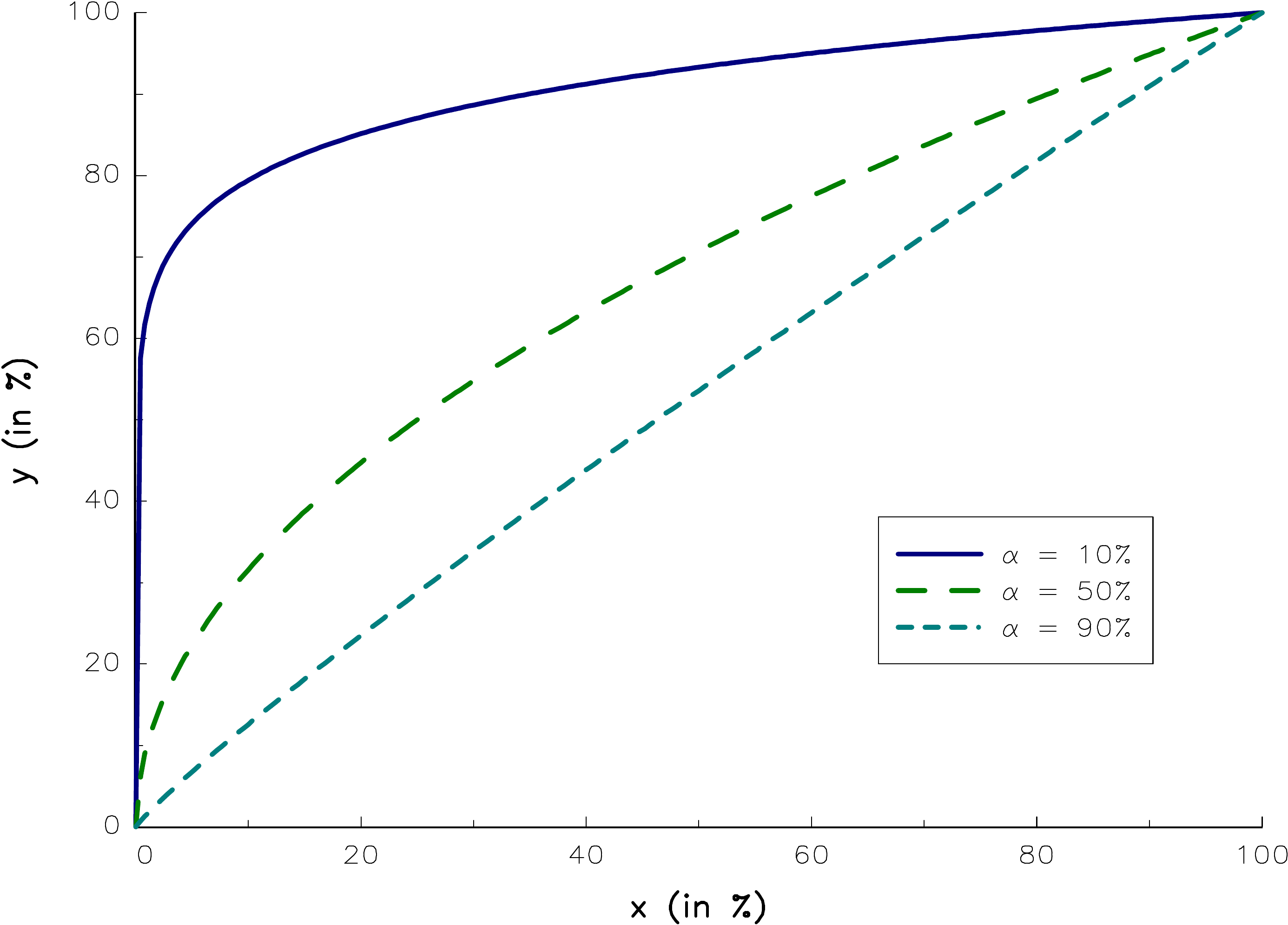}
\caption{Function $y=x^{\alpha }$}
\label{fig:app2-2-2-2}
\end{figure}

\item

\begin{enumerate}
\item We have $\mathcal{L}_{w}\left( 0\right) =0$ and $\mathcal{L}_{w}\left(
1\right) =\sum_{j=1}^{n}w_{j}=1$. If $x_{2}\geq x_{1}$, we have $\mathcal{L}%
_{w}\left( x_{2}\right) \geq \mathcal{L}_{w}\left( x_{2}\right) $. $\mathcal{%
L}_{w}$ is then a Lorenz curve. The Gini coefficient is equal to:%
\begin{eqnarray*}
\mathcal{G} &=&2\int_{0}^{1}\mathcal{L}\left( x\right) \,\mathrm{d}x-1 \\
&=&\frac{2}{n}\sum_{i=1}^{n}\sum_{j=1}^{i}w_{j}-1
\end{eqnarray*}%
If $w_{j}=n^{-1}$, we have:%
\begin{eqnarray*}
\underset{n\rightarrow \infty }{\lim }\mathcal{G} &=&\underset{n\rightarrow
\infty }{\lim }\frac{2}{n}\sum_{i=1}^{n}\frac{i}{n}-1 \\
&=&\underset{n\rightarrow \infty }{\lim }\frac{2}{n}\cdot \frac{n\left(
n+1\right) }{2n}-1 \\
&=&\underset{n\rightarrow \infty }{\lim }\frac{1}{n} = 0 %\\
%&=&0
\end{eqnarray*}%
If $w_{1}=1$, we have:%
\begin{eqnarray*}
\underset{n\rightarrow \infty }{\lim }\mathcal{G} &=&\underset{n\rightarrow
\infty }{\lim }1-\frac{1}{n} \\
&=&1
\end{eqnarray*}%
We note that that the perfect equality does not correspond to the case $%
\mathcal{G}=0$ except in the asymptotic case. This is why we may
slightly modify the definition of $\mathcal{L}_{w}\left( x\right) $:%
\begin{equation*}
\mathcal{L}_{w}\left( x\right) =\left\{
\begin{array}{lcl}
\sum_{j=1}^{i}w_{j} & \text{if} & x=n^{-1}i \\
\sum_{j=1}^{i}w_{j}+w_{i+1}\left( nx-i\right)  & \text{if} &
n^{-1}i<x<n^{-1}\left( i+1\right)
\end{array}%
\right.
\end{equation*}%
While the previous definition corresponds to a constant piecewise
function, this one defines an affine piecewise function. In this case, the
computation of the Gini index is done using a trapezoidal integration:
\begin{equation*}
\mathcal{G}=\frac{2}{n}\left( \sum_{i=1}^{n-1}\sum_{j=1}^{i}w_{j}+\frac{1}{2}%
\right) -1
\end{equation*}

\item The Herfindahl index is equal to $1$ if the portfolio is concentrated
in only one asset. We seek to minimize $\mathcal{H}=\sum_{i=1}^{n}w_{i}^{2}$
under the constraint $\sum_{i=1}^{n}w_{i}=1$. The Lagrange function is then:%
\begin{equation*}
f\left( w_{1},\ldots ,w_{n};\lambda \right) =\sum_{i=1}^{n}w_{i}^{2}-\lambda
\left( \sum_{i=1}^{n}w_{i}-1\right)
\end{equation*}%
The first-order conditions are $2w_{i}-\lambda =0$. We deduce that $%
w_{i}=w_{j}$. $\mathcal{H}$ reaches its minimum when $w_{i}=n^{-1}$. It
corresponds to the equally weighted portfolio. In this case, we have:%
\begin{equation*}
\mathcal{H}=\frac{1}{n}
\end{equation*}

\item The statistic $\mathcal{N}$ is the degree of freedom or the equivalent
number of equally weighted assets. For instance, if $\mathcal{H}=0.5$,
then $\mathcal{N}=2$. It is a portfolio equivalent to two equally weighted
assets.
\end{enumerate}

\item
\begin{enumerate}
\item The minimum variance portfolio is $w_{1}^{\left( 4\right) }=82.342\%$%
, $w_{2}^{\left( 4\right) }=13.175\%$, $w_{3}^{\left( 4\right) }=3.294\%$%
, $w_{4}^{\left( 4\right) }=0.823\%$ and $w_{5}^{\left( 4\right) }=0.%
366\%$.

\item For each portfolio, we sort the weights in descending order.
For the portfolio $w^{\left( 1\right) }$, we have $w_{1}^{\left( 1\right)
}=40\%$, $w_{2}^{\left( 1\right) }=30\%$, $w_{3}^{\left( 1\right) }=20\%$, $%
w_{4}^{\left( 1\right) }=10\%$ and $w_{5}^{\left( 1\right) }=0\%$. It follows
that:%
\begin{eqnarray*}
\mathcal{H}\left( w^{\left( 1\right) }\right)  &=&\sum_{i=1}^{5}\left(
w_{i}^{\left( 1\right) }\right) ^{2} \\
&=&0.10^{2}+0.20^{2}+0.30^{2}+0.40^{2} \\
&=&0.30
\end{eqnarray*}%
We also have:%
\begin{eqnarray*}
\mathcal{G}\left( w^{\left( 1\right) }\right)  &=&\frac{2}{5}\left(
\sum_{i=1}^{4}\sum_{j=1}^{i}\tilde{w}_{j}^{\left( 1\right) }+\frac{1}{2}%
\right) -1 \\
&=&\frac{2}{5}\left( 0.40+0.70+0.90+1.00+\frac{1}{2}\right) -1 \\
&=&0.40
\end{eqnarray*}%
For the portfolios $w^{\left( 2\right) }$, $w^{\left( 3\right) }$ and $%
w^{\left( 4\right) }$, we obtain $\mathcal{H}\left( w^{\left( 2\right)
}\right) =0.30$, $\mathcal{H}\left( w^{\left( 3\right) }\right) =0.25$, $%
\mathcal{H}\left( w^{\left( 4\right) }\right) =0.70$, $\mathcal{G}\left(
w^{\left( 2\right) }\right) =0.40$, $\mathcal{G}\left( w^{\left( 3\right)
}\right) =0.28$ and  $\mathcal{G}\left( w^{\left( 4\right) }\right) =0.71
$. We have $\mathcal{N}\left( w^{\left( 2\right) }\right) =\mathcal{N}\left(
w^{\left( 1\right) }\right) =3.33$, $\mathcal{N}\left( w^{\left( 3\right)
}\right) =4.00$ and $\mathcal{N}\left( w^{\left( 4\right) }\right) =1.44$%
.

\item All the statistics show that the least concentrated portfolio is $%
w^{\left( 3\right) }$. The most concentrated portfolio is
paradoxically the minimum variance portfolio $w^{\left( 4\right) }$.
We generally assimilate variance optimization to diversification
optimization. We show in this example that diversifying in the
Markowitz sense does not permit to minimize the concentration.
\end{enumerate}
\end{enumerate}

\section{ERC portfolio}

\begin{enumerate}
\item We note $\Sigma $ the covariance matrix of asset returns.

\begin{enumerate}
\item Let $\mathcal{R}\left( x\right) $ be a risk measure of the portfolio $x
$. If this risk measure satisfies the Euler principle, we have (TR-RPB, page
78):%
\begin{equation*}
\mathcal{R}\left( x\right) =\sum_{i=1}^{n}x_{i}\frac{\partial \,\mathcal{R}%
\left( x\right) }{\partial \,x_{i}}
\end{equation*}%
We can then decompose the risk measure as a sum of asset contributions.
This is why we define the risk contribution $\mathcal{RC}_{i}$ of asset $i$
as the product of the weight by the marginal risk:%
\begin{equation*}
\mathcal{RC}_{i}=x_{i}\frac{\partial \,\mathcal{R}\left( x\right) }{\partial
\,x_{i}}
\end{equation*}%
When the risk measure is the volatility $\sigma \left( x\right) $, it follows
that:%
\begin{eqnarray*}
\mathcal{RC}_{i} &=&x_{i}\frac{\left( \Sigma x\right) _{i}}{\sqrt{x^{\top
}\Sigma x}} \\
&=&\frac{x_{i}\left( \sum_{k=1}^{n}\rho _{i,k}\sigma _{i}\sigma
_{k}x_{k}\right) }{\sigma \left( x\right) }
\end{eqnarray*}

\item The ERC portfolio corresponds to the risk budgeting portfolio when the
risk measure is the return volatility $\sigma \left( x\right) $ and when the
risk budgets are the same for all the assets (TR-RPB, page 119). It means
that $\mathcal{RC}_{i}=\mathcal{RC}_{j}$, that is:%
\begin{equation*}
x_{i}\frac{\partial \,\sigma \left( x\right) }{\partial \,x_{i}}=x_{j}\frac{%
\partial \,\sigma \left( x\right) }{\partial \,x_{j}}
\end{equation*}

\item We have:%
\begin{eqnarray*}
\overline{\mathcal{RC}} &=&\frac{1}{n}\sum_{i=1}^{n}\mathcal{RC}_{i} \\
&=&\frac{1}{n}\sigma \left( x\right)
\end{eqnarray*}%
It follows that:%
\begin{eqnarray*}
\func{var}\left( \mathcal{RC}\right)  &=&\frac{1}{n}\sum_{i=1}^{n}\left(
\mathcal{RC}_{i}-\overline{\mathcal{RC}}\right) ^{2} \\
&=&\frac{1}{n}\sum_{i=1}^{n}\left( \mathcal{RC}_{i}-\frac{1}{n}\sigma \left(
x\right) \right) ^{2} \\
&=&\frac{1}{n^{2}\sigma \left( x\right) }\sum_{i=1}^{n}\left( nx_{i}\left(
\Sigma x\right) _{i}-\sigma ^{2}\left( x\right) \right) ^{2}
\end{eqnarray*}%
To compute the ERC portfolio, we may consider the following optimization
program:%
\begin{equation*}
x^{\star }=\arg \min \sum_{i=1}^{n}\left( nx_{i}\left( \Sigma x\right)
_{i}-\sigma ^{2}\left( x\right) \right) ^{2}
\end{equation*}%
Because we know that the ERC portfolio always exists (TR-RPB, page 108), the objective
function at the optimum $x^{\star }$ is necessarily equal to $0$. Another
equivalent optimization program is to consider the $L^{2}$ norm. In this
case, we have (TR-RPB, page 102):%
\begin{equation*}
x^{\star }=\arg \min \sum_{i=1}^{n}\sum_{j=1}^{n}\left( x_{i}\cdot \left(
\Sigma x\right) _{i}-x_{j}\cdot \left( \Sigma x\right) _{j}\right) ^{2}
\end{equation*}

\item We have:%
\begin{eqnarray*}
\beta _{i}\left( x\right)  &=&\frac{\left( \Sigma x\right) _{i}}{x^{\top
}\Sigma x} \\
&=&\frac{\mathcal{MR}_{i}}{\sigma \left( x\right) }
\end{eqnarray*}%
We deduce that:%
\begin{eqnarray*}
\mathcal{RC}_{i} &=&x_{i}\cdot \mathcal{MR}_{i} \\
&=&x_{i}\beta _{i}\left( x\right) \sigma \left( x\right)
\end{eqnarray*}%
The relationship $\mathcal{RC}_{i}=\mathcal{RC}_{j}$ becomes:%
\begin{equation*}
x_{i}\beta _{i}\left( x\right) =x_{j}\beta _{j}\left( x\right)
\end{equation*}%
It means that the weight is inversely proportional to the beta:%
\begin{equation*}
x_{i}\propto \frac{1}{\beta _{i}\left( x\right) }
\end{equation*}%
We can use the Jacobi power algorithm (TR-RPB, page 308). Let $x^{\left(
k\right) }$ be the portfolio at iteration $k$. We define the portfolio $%
x^{\left( k+1\right) }$ as follows:%
\begin{equation*}
x^{\left( k+1\right) }=\frac{\beta _{i}^{-1}\left( x^{\left( k\right)
}\right) }{\sum_{j=1}^{n}\beta _{j}^{-1}\left( x^{\left( k\right) }\right) }
\end{equation*}%
Starting from an initial portfolio $x^{\left( 0\right) }$, the limit
portfolio is the ERC portfolio if the algorithm converges:%
\begin{equation*}
\lim_{k\rightarrow \infty }x^{\left( k\right) }=x_{\mathrm{erc}}
\end{equation*}

\item Starting from the EW portfolio, we obtain for the first five iterations:%
\begin{equation*}
\begin{footnotesize}
\begin{tabular}{|c|cccccc|}
\hline
$k$                                             &         $0$ &         $1$ &         $2$ &         $3$ &         $4$  &         $5$ \\ \hline
$x_{1}^{\left( k\right) }$ (in \%)              &     $33.3333$ &     $43.1487$ &     $40.4122$ &     $41.2314$ &     $40.9771$  &     $41.0617$ \\
$x_{2}^{\left( k\right) }$ (in \%)              &     $33.3333$ &     $32.3615$ &     $31.9164$ &     $32.3529$ &     $32.1104$  &     $32.2274$ \\
$x_{3}^{\left( k\right) }$ (in \%)              &     $33.3333$ &     $24.4898$ &     $27.6714$ &     $26.4157$ &     $26.9125$  &     $26.7109$ \\ \hdashline
$\beta _{1}\left( x^{\left( k\right) }\right) $ & ${\bP}0.7326$ & ${\bP}0.8341$ & ${\bP}0.8046$ & ${\bP}0.8147$ & ${\bP}0.8113$  & ${\bP}0.8126$ \\
$\beta _{2}\left( x^{\left( k\right) }\right) $ & ${\bP}0.9767$ & ${\bP}1.0561$ & ${\bP}1.0255$ & ${\bP}1.0397$ & ${\bP}1.0337$  & ${\bP}1.0363$ \\
$\beta _{3}\left( x^{\left( k\right) }\right) $ & ${\bP}1.2907$ & ${\bP}1.2181$ & ${\bP}1.2559$ & ${\bP}1.2405$ & ${\bP}1.2472$  & ${\bP}1.2444$ \\
\hline
\end{tabular}%
\end{footnotesize}
\end{equation*}
The next iterations give the following results:
\begin{equation*}
\begin{footnotesize}
\begin{tabular}{|c|cccccc|}
\hline
$k$                                             &         $6$ &         $7$ &         $8$ &         $9$ &         $10$  &        $11$ \\ \hline
$x_{1}^{\left( k\right) }$ (in \%)              &     $41.0321$ &     $41.0430$ &     $41.0388$ &     $41.0405$ &     $41.0398$  &     $41.0401$ \\
$x_{2}^{\left( k\right) }$ (in \%)              &     $32.1746$ &     $32.1977$ &     $32.1878$ &     $32.1920$ &     $32.1902$  &     $32.1909$ \\
$x_{3}^{\left( k\right) }$ (in \%)              &     $26.7933$ &     $26.7593$ &     $26.7734$ &     $26.7676$ &     $26.7700$  &     $26.7690$ \\ \hdashline
$\beta _{1}\left( x^{\left( k\right) }\right) $ & ${\bP}0.8121$ & ${\bP}0.8123$ & ${\bP}0.8122$ & ${\bP}0.8122$ & ${\bP}0.8122$  & ${\bP}0.8122$ \\
$\beta _{2}\left( x^{\left( k\right) }\right) $ & ${\bP}1.0352$ & ${\bP}1.0356$ & ${\bP}1.0354$ & ${\bP}1.0355$ & ${\bP}1.0355$  & ${\bP}1.0355$ \\
$\beta _{3}\left( x^{\left( k\right) }\right) $ & ${\bP}1.2456$ & ${\bP}1.2451$ & ${\bP}1.2453$ & ${\bP}1.2452$ & ${\bP}1.2452$  & ${\bP}1.2452$ \\
\hline
\end{tabular}%
\end{footnotesize}
\end{equation*}
Finally, the algorithm converges after $14$ iterations with the following stopping criteria:
\begin{equation*}
\sup_{i}\left\vert x_{i}^{\left( k+1\right) }-x_{i}^{\left( k\right)
}\right\vert \leq 10^{-6}
\end{equation*}
and we obtain the following results:
\begin{equation*}
\begin{tabular}{|c|cccc|}
\hline
Asset & $x_{i}$ & $\mathcal{MR}_{i}$ & $\mathcal{RC}_{i}$ & $\mathcal{RC}_{i}^{\star }$ \\ \hline
1 & $41.04\%$ & $12.12\%$ & $4.97\%$ & $33.33\%$ \\
2 & $32.19\%$ & $15.45\%$ & $4.97\%$ & $33.33\%$ \\
3 & $26.77\%$ & $18.58\%$ & $4.97\%$ & $33.33\%$ \\
\hline
\end{tabular}%
\end{equation*}
\end{enumerate}

\item

\begin{enumerate}
\item We have:%
\begin{equation*}
\Sigma =\beta \beta ^{\top }\sigma_{m}^{2}+\limfunc{diag}\left( \tilde{\sigma}
_{1}^{2},\ldots ,\tilde{\sigma} _{n}^{2}\right)
\end{equation*}%
We deduce that:%
\begin{eqnarray*}
\mathcal{RC}_{i} &=&\frac{x_{i}\left( \sum_{k=1}^{n}\beta _{i}\beta
_{k}\sigma_{m}^{2}x_{k}+\tilde{\sigma} _{i}^{2}x_{i}\right) }{\tilde{\sigma} \left( x\right) }
\\
&=&\frac{x_{i}\beta _{i}B+x_{i}^{2}\tilde{\sigma} _{i}^{2}}{\sigma \left( x\right) }
\end{eqnarray*}%
with:%
\begin{equation*}
B=\sum_{k=1}^{n}x_{k}\beta _{k}\sigma_{m}^{2}
\end{equation*}%
The ERC portfolio satisfies then:%
\begin{equation*}
x_{i}\beta _{i}B+x_{i}^{2}\tilde{\sigma} _{i}^{2}=x_{j}\beta _{j}B+x_{j}^{2}\tilde{\sigma}
_{j}^{2}
\end{equation*}%
or:%
\begin{equation*}
\left( x_{i}\beta _{i}-x_{j}\beta _{j}\right) B=\left( x_{j}^{2}\tilde{\sigma}
_{j}^{2}-x_{i}^{2}\tilde{\sigma} _{i}^{2}\right)
\end{equation*}

\item If $\beta _{i}=\beta _{j}=\beta $, we have:%
\begin{equation*}
\left( x_{i}-x_{j}\right) \beta B=\left( x_{j}^{2}\tilde{\sigma}
_{j}^{2}-x_{i}^{2}\tilde{\sigma} _{i}^{2}\right)
\end{equation*}%
Because $\beta >0$, we deduce that:%
\begin{eqnarray*}
x_{i}>x_{j} &\Leftrightarrow &x_{j}^{2}\tilde{\sigma} _{j}^{2}-x_{i}^{2}\tilde{\sigma}
_{i}^{2}>0 \\
&\Leftrightarrow &x_{j}\tilde{\sigma} _{j}>x_{i}\tilde{\sigma} _{i} \\
&\Leftrightarrow &\tilde{\sigma} _{i} < \tilde{\sigma} _{j}
\end{eqnarray*}%
We conclude that the weight $x_{i}$ is a decreasing function of the specific
volatility $\tilde{\sigma} _{i}$.

\item If $\tilde{\sigma} _{i}=\tilde{\sigma} _{j}=\tilde{\sigma} $, we have:%
\begin{equation*}
\left( x_{i}\beta _{i}-x_{j}\beta _{j}\right) B=\left(
x_{j}^{2}-x_{i}^{2}\right) \tilde{\sigma} ^{2}
\end{equation*}%
We deduce that:%
\begin{eqnarray*}
x_{i}>x_{j} &\Leftrightarrow &\left( x_{i}\beta _{i}-x_{j}\beta _{j}\right)
B<0 \\
&\Leftrightarrow &x_{i}\beta _{i}<x_{j}\beta _{j} \\
&\Leftrightarrow &\beta _{i}<\beta _{j}
\end{eqnarray*}%
We conclude that the weight $x_{i}$ is a decreasing function of the
sensitivity $\beta _{i}$.

\item We obtain the following results:
\begin{equation*}
\begin{tabular}{|c|cccc|}
\hline
Asset & $x_{i}$ & $\mathcal{MR}_{i}$ & $\mathcal{RC}_{i}$ & $\mathcal{RC}_{i}^{\star }$ \\ \hline
1 & $21.92\%$ & $19.73\%$ & $4.32\%$ & $25.00\%$ \\
2 & $24.26\%$ & $17.83\%$ & $4.32\%$ & $25.00\%$ \\
3 & $25.43\%$ & $17.00\%$ & $4.32\%$ & $25.00\%$ \\
4 & $28.39\%$ & $15.23\%$ & $4.32\%$ & $25.00\%$ \\
\hline
\end{tabular}%
\end{equation*}

\end{enumerate}
\end{enumerate}

\section{Computing the Cornish-Fisher value-at-risk}

\begin{enumerate}
\item We have:%
\begin{eqnarray*}
\mathbb{E}\left[ X^{2n}\right] &=&\int_{-\infty }^{+\infty }x^{2n}\phi
\left( x\right) \,\mathrm{d}x \\
&=&\int_{-\infty }^{+\infty }x^{2n-1}x\phi \left( x\right) \,\mathrm{d}x
\end{eqnarray*}%
Using the integration by parts formula, we obtain\footnote{%
because $\phi ^{\prime }\left( x\right) =-x\phi \left( x\right) $.}:%
\begin{eqnarray*}
\mathbb{E}\left[ X^{2n}\right] &=&\left[ -x^{2n-1}\phi \left( x\right) %
\right] _{-\infty }^{+\infty }+\left( 2n-1\right) \int_{-\infty }^{+\infty
}x^{2n-2}\phi \left( x\right) \,\mathrm{d}x \\
&=&\left( 2n-1\right) \int_{-\infty }^{+\infty }x^{2n-2}\phi \left( x\right)
\,\mathrm{d}x \\
&=&\left( 2n-1\right) \mathbb{E}\left[ X^{2n-2}\right]
\end{eqnarray*}%
We deduce that $\mathbb{E}\left[ X^{2}\right] =1$, $\mathbb{E}\left[ X^{4}%
\right] =\left( 2\times 2-1\right) \mathbb{E}\left[ X^{2}\right] =3$, $%
\mathbb{E}\left[ X^{6}\right] =\left( 2\times 3-1\right) \mathbb{E}\left[
X^{4}\right] =15$\ and $\mathbb{E}\left[ X^{8}\right] =\left( 2\times
4-1\right) \mathbb{E}\left[ X^{4}\right] =105$. For the odd moments, we
obtain:%
\begin{eqnarray*}
\mathbb{E}\left[ X^{2n+1}\right] &=&\int_{-\infty }^{+\infty }x^{2n+1}\phi
\left( x\right) \,\mathrm{d}x \\
&=&0
\end{eqnarray*}%
because $x^{2n+1}\phi \left( x\right) $ is an odd function.

\item Let $C_{t}$ be the value of the call option at time $t$. The PnL is
equal to:
\begin{equation*}
\Pi =C_{t+1}-C_{t}
\end{equation*}%
We also have $S_{t+1}=\left( 1+R_{t+1}\right) S_{t}$ with $R_{t+1}$ the
daily asset return. We notice that the daily volatility is equal to:%
\begin{equation*}
\sigma =\frac{32.25\%}{\sqrt{260}}=2\%
\end{equation*}%
We deduce that $R_{t+1}\sim \mathcal{N}\left( 0,2\%\right) $.

\begin{enumerate}
\item We have:%
\begin{eqnarray*}
\Pi &\simeq &\Delta \left( S_{t+1}-S_{t}\right) \\
&=&\Delta R_{t+1}S_{t}
\end{eqnarray*}%
It follows that $\Pi \sim \mathcal{N}\left( 0,\Delta \sigma S_{t}\right) $ and:%
\begin{equation*}
\limfunc{VaR}\nolimits_{\alpha }=\Phi ^{-1}\left( \alpha \right) \Delta
\sigma S_{t}
\end{equation*}%
The numerical application gives $\limfunc{VaR}\nolimits_{\alpha }=2.33$
dollars.

\item In the case of the delta-gamma approximation, we obtain:%
\begin{eqnarray*}
\Pi &\simeq &\Delta \left( S_{t+1}-S_{t}\right) +\frac{1}{2}\Gamma \left(
S_{t+1}-S_{t}\right) ^{2} \\
&=&\Delta R_{t+1}S_{t}+\frac{1}{2}\Gamma R_{t+1}^{2}S_{t}^{2}
\end{eqnarray*}%
We deduce that:%
\begin{eqnarray*}
\mathbb{E}\left[ \Pi \right] &=&\mathbb{E}\left[ \Delta R_{t+1}S_{t}+\frac{1%
}{2}\Gamma R_{t+1}^{2}S_{t}^{2}\right] \\
&=&\frac{1}{2}\Gamma S_{t}^{2}\mathbb{E}\left[ R_{t+1}^{2}\right] \\
&=&\frac{1}{2}\Gamma \mathbb{\sigma }^{2}S_{t}^{2}
\end{eqnarray*}%
and:%
\begin{eqnarray*}
\mathbb{E}\left[ \Pi ^{2}\right] &=&\mathbb{E}\left[ \left( \Delta
R_{t+1}S_{t}+\frac{1}{2}\Gamma R_{t+1}^{2}S_{t}^{2}\right) ^{2}\right] \\
&=&\mathbb{E}\left[ \Delta ^{2}R_{t+1}^{2}S_{t}^{2}+\Delta \Gamma
R_{t+1}^{3}S_{t}^{3}+\frac{1}{4}\Gamma ^{2}R_{t+1}^{4}S_{t}^{4}\right]
\end{eqnarray*}%
We have $R_{t+1}=\sigma X$ with $X\sim \mathcal{N}\left( 0,1\right) $. It
follows that:%
\begin{equation*}
\mathbb{E}\left[ \Pi ^{2}\right] =\Delta ^{2}\sigma ^{2}S_{t}^{2}+\frac{3}{4}%
\Gamma ^{2}\sigma ^{4}S_{t}^{4}
\end{equation*}%
because $\mathbb{E}\left[ X\right] =0$, $\mathbb{E}\left[ X^{2}\right] =1$, $%
\mathbb{E}\left[ X^{3}\right] =0$ and $\mathbb{E}\left[ X^{4}\right] =3$.
The standard deviation of the PnL is then:%
\begin{eqnarray*}
\sigma \left( \Pi \right) &=&\sqrt{\Delta ^{2}\sigma ^{2}S_{t}^{2}+\frac{3}{4%
}\Gamma ^{2}\sigma ^{4}S_{t}^{4}-\left( \frac{1}{2}\Gamma \mathbb{\sigma }%
^{2}S_{t}^{2}\right) ^{2}} \\
&=&\sqrt{\Delta ^{2}\sigma ^{2}S_{t}^{2}+\frac{1}{2}\Gamma ^{2}\sigma
^{4}S_{t}^{4}}
\end{eqnarray*}%
Therefore, the Gaussian approximation of the PnL is:%
\begin{equation*}
\Pi \sim \mathcal{N}\left( \frac{1}{2}\Gamma \mathbb{\sigma }^{2}S_{t}^{2},%
\sqrt{\Delta ^{2}\sigma ^{2}S_{t}^{2}+\frac{1}{2}\Gamma ^{2}\sigma
^{4}S_{t}^{4}}\right)
\end{equation*}%
We deduce that the Gaussian value-at-risk is:%
\begin{equation*}
\limfunc{VaR}\nolimits_{\alpha }=-\frac{1}{2}\Gamma \mathbb{\sigma }%
^{2}S_{t}^{2}+\Phi ^{-1}\left( \alpha \right) \sqrt{\Delta ^{2}\sigma
^{2}S_{t}^{2}+\frac{1}{2}\Gamma ^{2}\sigma ^{4}S_{t}^{4}}
\end{equation*}%
The numerical application gives $\limfunc{VaR}\nolimits_{\alpha }=2.29$
dollars.

\item Let $L=-\Pi $ be the loss. We recall that the Cornish-Fisher
value-at-risk is equal to (TR-RPB, page 94):%
\begin{equation*}
\limfunc{VaR}\left( \alpha \right) =\mu \left( L\right) +z_{\alpha }\left(
\gamma _{1},\gamma _{2}\right) \cdot \sigma \left( L\right)
\end{equation*}%
with:
\begin{eqnarray*}
z_{\alpha }\left( \gamma _{1},\gamma _{2}\right) & = & z_{\alpha }+\frac{1}{6}%
\left( z_{\alpha }^{2}-1\right) \gamma _{1}+\frac{1}{24}\left( z_{\alpha
}^{3}-3z_{\alpha }\right) \gamma _{2}- \\
& & \frac{1}{36}\left( 2z_{\alpha
}^{3}-5z_{\alpha }\right) \gamma _{1}^{2}+\cdots
\end{eqnarray*}
and $z_{\alpha }=\Phi ^{-1}\left( \alpha \right) $. $\gamma _{1}$ et $\gamma
_{2}$ are the skewness and excess kurtosis of the loss $L$. We have seen
that:%
\begin{equation*}
\Pi =\Delta \sigma S_{t}X+\frac{1}{2}\Gamma \sigma ^{2}S_{t}^{2}X^{2}
\end{equation*}%
with $X\sim \mathcal{N}\left( 0,1\right) $. Using the results in Question 1,
we have $\mathbb{E}\left[ X\right] =\mathbb{E}\left[ X^{3}\right] =\mathbb{E}%
\left[ X^{5}\right] =\mathbb{E}\left[ X^{7}\right] =0$, $\mathbb{E}\left[
X^{2}\right] =1$, $\mathbb{E}\left[ X^{4}\right] =3$, $\mathbb{E}\left[ X^{6}%
\right] =15$\ and $\mathbb{E}\left[ X^{8}\right] =105$. We deduce that:%
\begin{eqnarray*}
\mathbb{E}\left[ \Pi ^{3}\right]
% &=&\mathbb{E}\left[ \left( \Delta \sigma
% S_{t}X+\frac{1}{2}\Gamma \sigma ^{2}S_{t}^{2}X^{2}\right) ^{3}\right]  \\
&=&\mathbb{E}\left[ \Delta ^{3}\sigma ^{3}S_{t}^{3}X^{3}+\frac{3}{2}\Delta
^{2}\Gamma \sigma ^{4}S_{t}^{4}X^{4}\right] + \\
& &\mathbb{E}\left[\frac{3}{4}\Delta \Gamma ^{2}\sigma
^{5}S_{t}^{5}X^{5}+\frac{1}{8}\Gamma ^{3}\sigma ^{6}S_{t}^{6}X^{6}\right]  \\
&=&\frac{9}{2}\Delta ^{2}\Gamma \sigma ^{4}S_{t}^{4}+\frac{15}{8}\Gamma
^{3}\sigma ^{6}S_{t}^{6}
\end{eqnarray*}%
and:%
\begin{eqnarray*}
\mathbb{E}\left[ \Pi ^{4}\right]  &=&\mathbb{E}\left[ \left( \Delta \sigma
S_{t}X+\frac{1}{2}\Gamma \sigma ^{2}S_{t}^{2}X^{2}\right) ^{4}\right]  \\
&=&3\Delta ^{4}\sigma ^{4}S_{t}^{4}+\frac{45}{2}\Delta ^{2}\Gamma ^{2}\sigma
^{6}S_{t}^{6}+\frac{105}{16}\Gamma ^{4}\sigma ^{8}S_{t}^{8}
\end{eqnarray*}%
The centered moments are then:%
\begin{eqnarray*}
\mathbb{E}\left[ \left( \Pi -\mathbb{E}\left[ \Pi \right] \right) ^{3}\right]
&=&\mathbb{E}\left[ \Pi ^{3}\right] -3\mathbb{E}\left[ \Pi \right] \mathbb{E}%
\left[ \Pi ^{2}\right] +2\mathbb{E}^{3}\left[ \Pi \right]  \\
&=&\frac{9}{2}\Delta ^{2}\Gamma \sigma ^{4}S_{t}^{4}+\frac{15}{8}\Gamma
^{3}\sigma ^{6}S_{t}^{6}-\frac{3}{2}\Delta ^{2}\Gamma \mathbb{\sigma }%
^{4}S_{t}^{4}- \\
& & \frac{9}{8}\Gamma ^{3}\sigma ^{6}S_{t}^{6}+\frac{2}{8}\Gamma
^{3}\mathbb{\sigma }^{6}S_{t}^{6} \\
&=&3\Delta ^{2}\Gamma \sigma ^{4}S_{t}^{4}+\Gamma ^{3}\sigma ^{6}S_{t}^{6}
\end{eqnarray*}%
and:%
\begin{eqnarray*}
\mathbb{E}\left[ \left( \Pi -\mathbb{E}\left[ \Pi \right] \right) ^{4}\right]
&=&\mathbb{E}\left[ \Pi ^{4}\right] -4\mathbb{E}\left[ \Pi \right] \mathbb{E}%
\left[ \Pi ^{3}\right] +6\mathbb{E}^{2}\left[ \Pi \right] \mathbb{E}\left[
\Pi ^{2}\right] - \\
& & 3\mathbb{E}^{4}\left[ \Pi \right]  \\
&=&3\Delta ^{4}\sigma ^{4}S_{t}^{4}+\frac{45}{2}\Delta ^{2}\Gamma ^{2}\sigma
^{6}S_{t}^{6}+\frac{105}{16}\Gamma ^{4}\sigma ^{8}S_{t}^{8}- \\
& & 9\Delta
^{2}\Gamma ^{2}\mathbb{\sigma }^{6}S_{t}^{6}-\frac{15}{4}\Gamma ^{4}\mathbb{%
\sigma }^{8}S_{t}^{8}+ \\
&&\frac{3}{2}\Delta ^{2}\Gamma ^{2}\mathbb{\sigma }^{6}S_{t}^{6}+\frac{9}{8}%
\Gamma ^{4}\mathbb{\sigma }^{8}S_{t}^{8}-\frac{3}{16}\Gamma ^{4}\mathbb{%
\sigma }^{8}S_{t}^{8} \\
&=&3\Delta ^{4}\sigma ^{4}S_{t}^{4}+15\Delta ^{2}\Gamma ^{2}\sigma
^{6}S_{t}^{6}+\frac{15}{4}\Gamma ^{4}\sigma ^{8}S_{t}^{8}
\end{eqnarray*}%
It follows that the skewness is:%
\begin{eqnarray*}
\gamma _{1}\left( L\right)  &=&-\gamma _{1}\left( \Pi \right)  \\
&=&-\frac{\mathbb{E}\left[ \left( \Pi -\mathbb{E}\left[ \Pi \right] \right)
^{3}\right] }{\sigma ^{3}\left( \Pi \right) } \\
&=&-\frac{3\Delta ^{2}\Gamma \sigma ^{4}S_{t}^{4}+\Gamma ^{3}\sigma
^{6}S_{t}^{6}}{\left( \Delta ^{2}\sigma ^{2}S_{t}^{2}+\frac{1}{2}\Gamma
^{2}\sigma ^{4}S_{t}^{4}\right) ^{3/2}} \\
&=&-\frac{6\sqrt{2}\Delta ^{2}\Gamma \sigma ^{4}S_{t}^{4}+2\sqrt{2}\Gamma
^{3}\sigma ^{6}S_{t}^{6}}{\left( 2\Delta ^{2}\sigma ^{2}S_{t}^{2}+\Gamma
^{2}\sigma ^{4}S_{t}^{4}\right) ^{3/2}}
\end{eqnarray*}%
whereas the excess kurtosis is:%
\begin{eqnarray*}
\gamma _{2}\left( L\right)  &=&\gamma _{2}\left( \Pi \right)  \\
&=&\frac{\mathbb{E}\left[ \left( \Pi -\mathbb{E}\left[ \Pi \right] \right)
^{4}\right] }{\sigma ^{4}\left( \Pi \right) }-3 \\
&=&\frac{3\Delta ^{4}\sigma ^{4}S_{t}^{4}+15\Delta ^{2}\Gamma ^{2}\sigma
^{6}S_{t}^{6}+\frac{15}{4}\Gamma ^{4}\sigma ^{8}S_{t}^{8}}{\left( \Delta
^{2}\sigma ^{2}S_{t}^{2}+\frac{1}{2}\Gamma ^{2}\sigma ^{4}S_{t}^{4}\right)
^{2}}-3 \\
&=&\frac{12\Delta ^{2}\Gamma ^{2}\sigma ^{6}S_{t}^{6}+3\Gamma ^{4}\sigma
^{8}S_{t}^{8}}{\left( \Delta ^{2}\sigma ^{2}S_{t}^{2}+\frac{1}{2}\Gamma
^{2}\sigma ^{4}S_{t}^{4}\right) ^{2}}
\end{eqnarray*}%
Using the numerical values, we obtain $\mu \left( L\right) =-0.0400$, $%
\sigma \left( L\right) =1.0016$, $\gamma _{1}\left( L\right) =-0.2394$, $%
\gamma _{2}\left( L\right) =0.0764$, $z_{\alpha }\left( \gamma _{1},\gamma
_{2}\right) =2.1466$ and $\limfunc{VaR}\nolimits_{\alpha }=2.11$ dollars.
The value-at-risk is reduced with the Cornish-Fisher approximation because
the skewness is negative whereas the excess kurtosis is very small.
\end{enumerate}

\item

\begin{enumerate}
\item We have:%
\begin{eqnarray*}
Y &=&X^{\top }AX \\
&=&\left( \Sigma ^{-1/2}X\right) ^{\top }\Sigma ^{1/2}A\Sigma ^{1/2}\left(
\Sigma ^{-1/2}X\right)  \\
&=&\tilde{X}^{\top }\tilde{A}\tilde{X}
\end{eqnarray*}%
with $\tilde{A}=\Sigma ^{1/2}A\Sigma ^{1/2}$, $\tilde{X}\sim \mathcal{N}%
\left( \tilde{\mu},\tilde{\Sigma}\right) $, $\tilde{\mu}=\Sigma ^{-1/2}\mu $
and $\tilde{\Sigma}=I$. We deduce that:%
\begin{eqnarray*}
\mathbb{E}\left[ Y\right]  &=&\tilde{\mu}^{\top }\tilde{A}\tilde{\mu}+\func{%
tr}\left( \tilde{A}\right)  \\
&=&\mu ^{\top }A\mu +\func{tr}\left( \Sigma ^{1/2}A\Sigma ^{1/2}\right)  \\
&=&\mu ^{\top }A\mu +\func{tr}\left( A\Sigma \right)
\end{eqnarray*}%
and:%
\begin{eqnarray*}
\func{var}\left( Y\right)  &=&\mathbb{E}\left[ Y^{2}\right] -\mathbb{E}^{2}%
\left[ Y\right]  \\
&=&4\tilde{\mu}^{\top }\tilde{A}^{2}\tilde{\mu}+2\func{tr}\left( \tilde{A}%
^{2}\right)  \\
&=&4\mu ^{\top }A\Sigma A\mu +2\func{tr}\left( \Sigma ^{1/2}A\Sigma A\Sigma
^{1/2}\right)  \\
&=&4\mu ^{\top }A\Sigma A\mu +2\func{tr}\left( \left( A\Sigma \right)
^{2}\right)
\end{eqnarray*}

\item For the moments, we obtain:%
\begin{eqnarray*}
\mathbb{E}\left[ Y\right]  &=&\func{tr}\left( A\Sigma \right)  \\
\mathbb{E}\left[ Y^{2}\right]  &=&\left( \func{tr}\left( A\Sigma \right)
\right) ^{2}+2\func{tr}\left( \left( A\Sigma \right) ^{2}\right)  \\
\mathbb{E}\left[ Y^{3}\right]  &=&\left( \func{tr}\left( A\Sigma \right)
\right) ^{3}+6\func{tr}\left( A\Sigma \right) \func{tr}\left( \left( A\Sigma
\right) ^{2}\right) +8\func{tr}\left( \left( A\Sigma \right) ^{3}\right)  \\
\mathbb{E}\left[ Y^{4}\right]  &=&\left( \func{tr}\left( A\Sigma \right)
\right) ^{4}+32\func{tr}\left( A\Sigma \right) \func{tr}\left( \left(
A\Sigma \right) ^{3}\right) + \\
& & 12\left( \func{tr}\left( \left( A\Sigma \right)
^{2}\right) \right) ^{2} + 12\left( \func{tr}\left( A\Sigma \right) \right) ^{2}\func{tr}\left(
\left( A\Sigma \right) ^{2}\right) + \\
& & 48\func{tr}\left( \left( A\Sigma \right)^{4}\right)
\end{eqnarray*}%
It follows that the first and second centered moments are $\mu \left( Y\right)
=\func{tr}\left( A\Sigma \right) $ and $\func{var}\left( Y\right) =2\func{tr}%
\left( \left( A\Sigma \right) ^{2}\right) $. For the third centered moment,
we have:%
\begin{eqnarray*}
\mathbb{E}\left[ \left( Y-\mathbb{E}\left[ Y\right] \right) ^{3}\right]  &=&%
\mathbb{E}\left[ Y^{3}\right] -3\mathbb{E}\left[ Y^{2}\right] \mathbb{E}%
\left[ Y\right] +2\mathbb{E}^{3}\left[ Y\right]  \\
&=&\left( \func{tr}\left( A\Sigma \right) \right) ^{3}+6\func{tr}\left(
A\Sigma \right) \func{tr}\left( \left( A\Sigma \right) ^{2}\right) + \\
& & 8\func{tr}\left( \left( A\Sigma \right) ^{3}\right) -
3\left( \func{tr}\left( A\Sigma \right) \right) ^{3}-\\
& & 6\func{tr}\left(
\left( A\Sigma \right) ^{2}\right) \func{tr}\left( A\Sigma \right) +2\left(
\func{tr}\left( A\Sigma \right) \right) ^{3} \\
&=&8\func{tr}\left( \left( A\Sigma \right) ^{3}\right)
\end{eqnarray*}%
whereas we obtain for the fourth centered moment:%
\begin{eqnarray*}
\mathbb{E}\left[ \left( Y-\mathbb{E}\left[ Y\right] \right) ^{4}\right]  &=&%
\mathbb{E}\left[ Y^{4}\right] -4\mathbb{E}\left[ Y^{3}\right] \mathbb{E}%
\left[ Y\right] +6\mathbb{E}\left[ Y^{2}\right] \mathbb{E}^{2}\left[ Y\right]
- \\
&&3\mathbb{E}^{4}\left[ Y\right]  \\
&=&\left( \func{tr}\left( A\Sigma \right) \right) ^{4}+32\func{tr}\left(
A\Sigma \right) \func{tr}\left( \left( A\Sigma \right) ^{3}\right) + \\
&&12\left( \func{tr}\left( \left( A\Sigma \right) ^{2}\right) \right) ^{2}+48%
\func{tr}\left( \left( A\Sigma \right) ^{4}\right)  \\
&&12\left( \func{tr}\left( A\Sigma \right) \right) ^{2}\func{tr}\left(
\left( A\Sigma \right) ^{2}\right) -4\left( \func{tr}\left( A\Sigma \right)
\right) ^{4}- \\
&&24\left( \func{tr}\left( A\Sigma \right) \right) ^{2}\func{tr}\left(
\left( A\Sigma \right) ^{2}\right) - \\
&&32\func{tr}\left( \left( A\Sigma \right) ^{3}\right) \func{tr}\left(
A\Sigma \right) +6\left( \func{tr}\left( A\Sigma \right) \right) ^{4}+ \\
&&12\func{tr}\left( \left( A\Sigma \right) ^{2}\right) \left( \func{tr}%
\left( A\Sigma \right) \right) ^{2}-3\left( \func{tr}\left( A\Sigma \right)
\right) ^{4} \\
&=&12\left( \func{tr}\left( \left( A\Sigma \right) ^{2}\right) \right)
^{2}+48\func{tr}\left( \left( A\Sigma \right) ^{4}\right)
\end{eqnarray*}
The skewness is then equal to:%
\begin{eqnarray*}
\gamma _{1}\left( Y\right)  &=&\frac{8\func{tr}\left( \left( A\Sigma \right)
^{3}\right) }{\left( 2\func{tr}\left( \left( A\Sigma \right) ^{2}\right)
\right) ^{3/2}} \\
&=&\frac{2\sqrt{2}\func{tr}\left( \left( A\Sigma \right) ^{3}\right) }{%
\left( \func{tr}\left( \left( A\Sigma \right) ^{2}\right) \right) ^{3/2}}
\end{eqnarray*}%
For the excess kurtosis, we obtain:%
\begin{eqnarray*}
\gamma _{2}\left( Y\right)  &=&\frac{12\left( \func{tr}\left( \left( A\Sigma
\right) ^{2}\right) \right) ^{2}+48\func{tr}\left( \left( A\Sigma \right)
^{4}\right) }{\left( 2\func{tr}\left( \left( A\Sigma \right) ^{2}\right)
\right) ^{2}}-3 \\
&=&\frac{12\func{tr}\left( \left( A\Sigma \right) ^{4}\right) }{\left( \func{%
tr}\left( \left( A\Sigma \right) ^{2}\right) \right) ^{2}}
\end{eqnarray*}
\end{enumerate}

\item We have:%
\begin{equation*}
\Pi =x^{\top }\left( C_{t+1}-C_{t}\right)
\end{equation*}%
where $C_{t}$ is the vector of option prices.

\begin{enumerate}
\item The expression of the PnL is:%
\begin{eqnarray*}
\Pi  &\simeq &x^{\top }\left( \Delta \circ \left( S_{t+1}-S_{t}\right)
\right)  \\
&=&x^{\top }\left( \left( \Delta \circ S_{t}\right) \circ R_{t+1}\right)  \\
&=&\tilde{\Delta}^{\top }R_{t+1}
\end{eqnarray*}%
with $\tilde{\Delta}$ the vector of delta exposures in dollars:%
\begin{equation*}
\tilde{\Delta}_{i}=x_{i}\Delta _{i}S_{i,t}
\end{equation*}%
Because $R_{t+1}\sim \mathcal{N}\left( \mathbf{0},\Sigma \right) $, it follows
that $\Pi \sim \mathcal{N}\left( 0,\sqrt{\tilde{\Delta}^{\top }\Sigma \tilde{%
\Delta}}\right) $. We deduce that the Gaussian value-at-risk is:%
\begin{equation*}
\limfunc{VaR}\nolimits_{\alpha }=\Phi ^{-1}\left( \alpha \right) \sqrt{%
\tilde{\Delta}^{\top }\Sigma \tilde{\Delta}}
\end{equation*}%
The risk contribution of option $i$ is then equal to:%
\begin{eqnarray*}
\mathcal{RC}_{i} &=&x_{i}\frac{\Phi ^{-1}\left( \alpha \right) \left( \Sigma
\tilde{\Delta}\right) _{i}\Delta _{i}S_{i,t}}{\sqrt{\tilde{\Delta}^{\top
}\Sigma \tilde{\Delta}}} \\
&=&\Phi ^{-1}\left( \alpha \right) \frac{\tilde{\Delta}_{i}\cdot \left(
\Sigma \tilde{\Delta}\right) _{i}}{\sqrt{\tilde{\Delta}^{\top }\Sigma \tilde{%
\Delta}}}
\end{eqnarray*}

\item In the case of the delta-gamma approximation, we obtain:%
\begin{eqnarray*}
\Pi  &\simeq &x^{\top }\left( \Delta \circ \left( S_{t+1}-S_{t}\right)
\right) + \\
& & \frac{1}{2}x^{\top }\left( \Gamma \circ \left( S_{t+1}-S_{t}\right)
\circ \left( S_{t+1}-S_{t}\right) ^{\top }\right) x \\
&=&\tilde{\Delta}^{\top }R_{t+1}+\frac{1}{2}R_{t+1}^{\top }\tilde{\Gamma}%
R_{t+1}
\end{eqnarray*}%
with $\tilde{\Gamma}$ the matrix of gamma exposures in dollars:%
\begin{equation*}
\tilde{\Gamma}_{i,j}=x_{i}x_{j}\Gamma _{i,j}S_{i,t}S_{j,t}
\end{equation*}%
We deduce that:%
\begin{eqnarray*}
\mathbb{E}\left[ \Pi \right]  &=&\mathbb{E}\left[ \tilde{\Delta}^{\top
}R_{t+1}+\frac{1}{2}R_{t+1}^{\top }\tilde{\Gamma}R_{t+1}\right]  \\
&=&\frac{1}{2}\mathbb{E}\left[ R_{t+1}^{\top }\tilde{\Gamma}R_{t+1}\right]
\\
&=&\frac{1}{2}\func{tr}\left( \tilde{\Gamma}\Sigma \right)
\end{eqnarray*}%
and:%
\begin{eqnarray*}
\func{var}\left( \Pi \right)  &=&\mathbb{E}\left[ \left( \Pi -\mathbb{E}%
\left[ \Pi \right] \right) ^{2}\right]  \\
&=&\mathbb{E}\left[ \left( \tilde{\Delta}^{\top }R_{t+1}+\frac{1}{2}%
R_{t+1}^{\top }\tilde{\Gamma}R_{t+1}-\frac{1}{2}\func{tr}\left( \tilde{\Gamma%
}\Sigma \right) \right) ^{2}\right]  \\
&=&\mathbb{E}\left[ \left( \tilde{\Delta}^{\top }R_{t+1}\right) ^{2}\right] +%
\frac{1}{4}\mathbb{E}\left[ \left( R_{t+1}^{\top }\tilde{\Gamma}R_{t+1}-%
\func{tr}\left( \tilde{\Gamma}\Sigma \right) \right) ^{2}\right] + \\
&&\mathbb{E}\left[ \left( \tilde{\Delta}^{\top }R_{t+1}\right) \left(
R_{t+1}^{\top }\tilde{\Gamma}R_{t+1}-\func{tr}\left( \tilde{\Gamma}\Sigma
\right) \right) \right]  \\
&=&\mathbb{E}\left[ \left( \tilde{\Delta}^{\top }R_{t+1}\right) ^{2}\right] +%
\frac{1}{4}\func{var}\left( R_{t+1}^{\top }\tilde{\Gamma}R_{t+1}\right)  \\
&=&\tilde{\Delta}^{\top }\Sigma \tilde{\Delta}+\frac{1}{2}\func{tr}\left(
\left( \tilde{\Gamma}\Sigma \right) ^{2}\right)
\end{eqnarray*}
Therefore, the Gaussian approximation of the PnL is:%
\begin{equation*}
\Pi \sim \mathcal{N}\left( \frac{1}{2}\func{tr}\left( \tilde{\Gamma}\Sigma
\right) ,\sqrt{\tilde{\Delta}^{\top }\Sigma \tilde{\Delta}+\frac{1}{2}\func{%
tr}\left( \left( \tilde{\Gamma}\Sigma \right) ^{2}\right) }\right)
\end{equation*}%
We deduce that the Gaussian value-at-risk is:%
\begin{equation*}
\limfunc{VaR}\nolimits_{\alpha }=-\frac{1}{2}\func{tr}\left( \tilde{\Gamma}%
\Sigma \right) +\Phi ^{-1}\left( \alpha \right) \sqrt{\tilde{\Delta}^{\top
}\Sigma \tilde{\Delta}+\frac{1}{2}\func{tr}\left( \left( \tilde{\Gamma}%
\Sigma \right) ^{2}\right) }
\end{equation*}

\item If the portfolio is delta neutral, $\Delta $ is equal to zero and we
have:%
\begin{equation*}
\Pi \simeq \frac{1}{2}R_{t+1}^{\top }\tilde{\Gamma}R_{t+1}
\end{equation*}%
Let $L=-\Pi $ be the loss. Using the formulas of Question 3(b), we
obtain:%
\begin{equation*}
\mu \left( L\right) =-\frac{1}{2}\func{tr}\left( \tilde{\Gamma}\Sigma
\right)
\end{equation*}%
\begin{equation*}
\sigma \left( L\right) =\sqrt{\frac{1}{2}\func{tr}\left( \left( \tilde{\Gamma%
}\Sigma \right) ^{2}\right) }
\end{equation*}%
\begin{equation*}
\gamma _{1}\left( L\right) =-\frac{2\sqrt{2}\func{tr}\left( \left( \tilde{%
\Gamma}\Sigma \right) ^{3}\right) }{\left( \func{tr}\left( \left( \tilde{%
\Gamma}\Sigma \right) ^{2}\right) \right) ^{3/2}}
\end{equation*}%
\begin{equation*}
\gamma _{2}\left( L\right) =\frac{12\func{tr}\left( \left( \tilde{\Gamma}%
\Sigma \right) ^{4}\right) }{\left( \func{tr}\left( \left( \tilde{\Gamma}%
\Sigma \right) ^{2}\right) \right) ^{2}}
\end{equation*}%
We have all the statistics to compute the Cornish-Fisher value-at-risk.

\item We notice that the previous formulas obtained in the multivariate case
are perfectly coherent with those obtained in the univariate case. When the
portfolio is not delta neutral, we could then postulate that the skewness is%
\footnote{%
You may easily verify that we obtained this formula in the case $n=2$ by
developing the different polynomials.}:%
\begin{equation*}
\gamma _{1}\left( L\right) =-\frac{6\sqrt{2}\tilde{\Delta}^{\top }\Sigma
\Gamma \Sigma \tilde{\Delta}+2\sqrt{2}\func{tr}\left( \left( \tilde{\Gamma}%
\Sigma \right) ^{3}\right) }{\left( 2\tilde{\Delta}^{\top }\Sigma \tilde{%
\Delta}+\func{tr}\left( \left( \tilde{\Gamma}\Sigma \right) ^{2}\right)
\right) ^{3/2}}
\end{equation*}%
In fact, it is the formula obtained by Britten-Jones and Schaeffer (1999)%
\footnote{\textsc{Britten-Jones} M. and \textsc{Schaeffer} S.M. (1999),
Non-Linear Value-at-Risk, \textit{European Finance Review}, 2(2), pp.
167-187.}.
\end{enumerate}

\item

\begin{enumerate}
\item Using the numerical values, we obtain $\mu \left( L\right) =-78.65$, $%
\sigma \left( L\right) =88.04$, $\gamma _{1}\left( L\right) =-2.5583$ and $%
\gamma _{2}\left( L\right) =10.2255$. The value-at-risk is then equal to $0$
for the delta approximation, $126.16$ for the delta-gamma approximation and $%
-45.85$ for the Cornish-Fisher approximation. We notice that we obtain an
absurd result in the last case, because the distribution is far from the
Gaussian distribution (high skewness and kurtosis). If we consider a smaller
order expansion:%
\begin{equation*}
_{\alpha }\left( \gamma _{1},\gamma _{2}\right) =z_{\alpha }+\frac{1}{6}%
\left( z_{\alpha }^{2}-1\right) \gamma _{1}+\frac{1}{24}\left( z_{\alpha
}^{3}-3z_{\alpha }\right) \gamma _{2}
\end{equation*}%
the value-at-risk is equal to $171.01$.

\item In this case, we obtain $126.24$ for the delta approximation, $161.94$
for the delta-gamma approximation and $-207.84$ for the Cornish-Fisher
approximation. For the delta approximation, the risk decomposition is:%
\begin{equation*}
\begin{tabular}{|c|cccc|}
\hline Option & $x_{i}$  & $\mathcal{MR}_{i}$ & $\mathcal{RC}_{i}$ & $\mathcal{RC}^{\star}_{i}$ \\
\hline
$1$  & $50.00$ & $0.86$ & ${\bP}42.87$ & ${\bP}33.96\%$ \\
$2$  & $20.00$ & $0.77$ & ${\bP}15.38$ & ${\bP}12.19\%$ \\
$3$  & $30.00$ & $2.27$ & ${\bP}67.98$ & ${\bP}53.85\%$ \\ \hdashline
$\limfunc{VaR}\nolimits_{\alpha }$ &  &  & $126.24$ & $100.00\%$ \\
\hline
\end{tabular}
\end{equation*}
For the delta-gamma approximation, we have:%
\begin{equation*}
\begin{tabular}{|c|cccc|}
\hline Option & $x_{i}$  & $\mathcal{MR}_{i}$ & $\mathcal{RC}_{i}$ & $\mathcal{RC}^{\star}_{i}$ \\
\hline
$1$  & $50.00$ & $4.06$ &     $202.92$ &     $125.31\%$ \\
$2$  & $20.00$ & $1.18$ & ${\bP}23.62$ & ${\bP}14.59\%$ \\
$3$  & $30.00$ & $1.04$ & ${\bP}31.10$ & ${\bP}19.21\%$ \\ \hdashline
$\limfunc{VaR}\nolimits_{\alpha }$ &  &  & $161.94$ & $159.10\%$ \\
\hline
\end{tabular}
\end{equation*}
We notice that the delta-gamma approximation does not satisfy the Euler
decomposition. Finally, we obtain the following ERC portfolio:%
\begin{equation*}
\begin{tabular}{|c|cccc|}
\hline Option & $x_{i}$  & $\mathcal{MR}_{i}$ & $\mathcal{RC}_{i}$ & $\mathcal{RC}^{\star}_{i}$ \\
\hline
$1$  & $42.38$ & $0.90$ & ${\bP}38.35$ & ${\bP}33.33\%$ \\
$2$  & $37.16$ & $1.03$ & ${\bP}38.35$ & ${\bP}33.33\%$ \\
$3$  & $20.46$ & $1.87$ & ${\bP}38.35$ & ${\bP}33.33\%$ \\ \hdashline
$\limfunc{VaR}\nolimits_{\alpha }$ &  &  & $115.05$ &   \\
\hline
\end{tabular}
\end{equation*}
\end{enumerate}
\end{enumerate}

\section{Risk budgeting when risk budgets are not strictly positive}

\begin{enumerate}
\item
\begin{enumerate}
\item We obtain the following portfolio:%
\begin{equation*}
\hspace*{-1cm}
\begin{small}
\begin{tabular}{|cc|rrrrrr:c|}
\hline
\multicolumn{2}{|c|}{Solution} & \multicolumn{1}{c}{1} & \multicolumn{1}{c}{2} & \multicolumn{1}{c}{3} &
                                 \multicolumn{1}{c}{4} & \multicolumn{1}{c}{5} & \multicolumn{1}{c:}{6} & $\sigma \left( x\right) $ \\
\hline
\multirow{4}{*}{$\mathcal{S}_{1}$}
                  & $x_{i}$                                      & $20.29\%$ & $15.95\%$ & $20.82\%$ & $14.88\%$ & $9.97\%$ & $18.08\%$ & \multirow{4}{*}{$8.55\%$} \\
                  & $\mathcal{MR}_{i}$                           & $16.85\%$ & $16.07\%$ & $12.31\%$ & $ 0.00\%$ & $0.00\%$ & $ 0.00\%$ &          \\
                  & $\mathcal{RC}_{i}$                           & $ 3.42\%$ & $ 2.56\%$ & $ 2.56\%$ & $ 0.00\%$ & $0.00\%$ & $ 0.00\%$ &          \\
                  & $\mathcal{RC}_{i}^{\star}$                   & $40.00\%$ & $30.00\%$ & $30.00\%$ & $ 0.00\%$ & $0.00\%$ & $ 0.00\%$ &          \\
\hline
\end{tabular}
\end{small}
\end{equation*}%
We notice that the last three assets have a
positive weight ($x_{i}>0$) and a marginal risk equal to zero ($%
\mathcal{MR}_{i}=0$). We deduce that the number of solutions is $2^{3}=8$
(TR-RPB, page 110).

\item We obtain the following portfolio:%
\begin{center}
\begin{equation*}
\hspace*{-1.75cm}
\begin{small}
\begin{tabular}{|cc|rrrrrr:c|}
\hline
\multicolumn{2}{|c|}{Solution} & \multicolumn{1}{c}{1} & \multicolumn{1}{c}{2} & \multicolumn{1}{c}{3} &
                                 \multicolumn{1}{c}{4} & \multicolumn{1}{c}{5} & \multicolumn{1}{c:}{6} & $\sigma \left( x\right) $ \\
\hline
\multirow{4}{*}{$\mathcal{S}_{2}$}
                  & $x_{i}$                                      & $33.77\%$ & $29.05\%$ & $37.18\%$ & $ 0.00\%$ & $ 0.00\%$ & $ 0.00\%$ & \multirow{4}{*}{$16.07\%$} \\
                  & $\mathcal{MR}_{i}$                           & $19.03\%$ & $16.59\%$ & $12.96\%$ & $-4.54\%$ & $-1.55\%$ & $-2.39\%$ &          \\
                  & $\mathcal{RC}_{i}$                           & $ 6.43\%$ & $ 4.82\%$ & $ 4.82\%$ & $ 0.00\%$ & $ 0.00\%$ & $ 0.00\%$ &          \\
                  & $\mathcal{RC}_{i}^{\star}$                   & $40.00\%$ & $30.00\%$ & $30.00\%$ & $ 0.00\%$ & $ 0.00\%$ & $ 0.00\%$ &          \\
\hline
\end{tabular}
\end{small}
\end{equation*}%
\end{center}
We notice that the marginal risk of the assets that have a nil weight is
negative. We confirm that that the number of solutions is $2^{3}=8$ (TR-RPB,
page 110).

\item We obtain the six other solutions reported in Table \ref{tab:app2-2-5-1}.
\begin{table}[tbh]
\centering
\caption{The six other RB portfolios}
\label{tab:app2-2-5-1}
\tableskip
\begin{small}
\begin{tabular}{|cc|rrrrrr:c|}
\hline
\multicolumn{2}{|c|}{Solution} & \multicolumn{1}{c}{1} & \multicolumn{1}{c}{2} & \multicolumn{1}{c}{3} &
                                 \multicolumn{1}{c}{4} & \multicolumn{1}{c}{5} & \multicolumn{1}{c:}{6} & $\sigma \left( x\right) $ \\
\hline
\multirow{4}{*}{$\mathcal{S}_{3}$}
& $x_{i}$                     & $25.57\%$ & $21.56\%$ & $28.38\%$ & $24.49\%$ & $ 0.00\%$ & $ 0.00\%$ & \multirow{4}{*}{$11.55\%$}  \\
& $\mathcal{MR}_{i}$          & $18.07\%$ & $16.08\%$ & $12.21\%$ & $ 0.00\%$ & $-0.99\%$ & $-1.88\%$ &                             \\
& $\mathcal{RC}_{i}$          & $ 4.62\%$ & $ 3.47\%$ & $ 3.47\%$ & $ 0.00\%$ & $ 0.00\%$ & $ 0.00\%$ &                             \\
& $\mathcal{RC}_{i}^{\star}$  & $40.00\%$ & $30.00\%$ & $30.00\%$ & $ 0.00\%$ & $ 0.00\%$ & $ 0.00\%$ &                             \\
\hline
\multirow{4}{*}{$\mathcal{S}_{4}$}
& $x_{i}$                     & $27.27\%$ & $23.15\%$ & $29.60\%$ & $ 0.00\%$ & $19.98\%$ & $ 0.00\%$ & \multirow{4}{*}{$12.71\%$}  \\
& $\mathcal{MR}_{i}$          & $18.64\%$ & $16.47\%$ & $12.88\%$ & $-4.12\%$ & $ 0.00\%$ & $-2.43\%$ &                             \\
& $\mathcal{RC}_{i}$          & $ 5.08\%$ & $ 3.81\%$ & $ 3.81\%$ & $ 0.00\%$ & $ 0.00\%$ & $ 0.00\%$ &                             \\
& $\mathcal{RC}_{i}^{\star}$  & $40.00\%$ & $30.00\%$ & $30.00\%$ & $ 0.00\%$ & $ 0.00\%$ & $ 0.00\%$ &                             \\
\hline
\multirow{4}{*}{$\mathcal{S}_{5}$}
& $x_{i}$                     & $25.42\%$ & $20.34\%$ & $25.95\%$ & $ 0.00\%$ & $ 0.00\%$ & $28.29\%$ & \multirow{4}{*}{$11.22\%$}  \\
& $\mathcal{MR}_{i}$          & $17.66\%$ & $16.55\%$ & $12.98\%$ & $-3.93\%$ & $-1.62\%$ & $ 0.00\%$ &                             \\
& $\mathcal{RC}_{i}$          & $ 4.49\%$ & $ 3.37\%$ & $ 3.37\%$ & $ 0.00\%$ & $ 0.00\%$ & $ 0.00\%$ &                             \\
& $\mathcal{RC}_{i}^{\star}$  & $40.00\%$ & $30.00\%$ & $30.00\%$ & $ 0.00\%$ & $ 0.00\%$ & $ 0.00\%$ &                             \\
\hline
\multirow{4}{*}{$\mathcal{S}_{6}$}
& $x_{i}$                     & $23.31\%$ & $19.49\%$ & $25.61\%$ & $20.76\%$ & $10.84\%$ & $ 0.00\%$ & \multirow{4}{*}{$10.42\%$}  \\
& $\mathcal{MR}_{i}$          & $17.88\%$ & $16.03\%$ & $12.21\%$ & $ 0.00\%$ & $ 0.00\%$ & $-1.94\%$ &                             \\
& $\mathcal{RC}_{i}$          & $ 4.17\%$ & $ 3.13\%$ & $ 3.13\%$ & $ 0.00\%$ & $ 0.00\%$ & $ 0.00\%$ &                             \\
& $\mathcal{RC}_{i}^{\star}$  & $40.00\%$ & $30.00\%$ & $30.00\%$ & $ 0.00\%$ & $ 0.00\%$ & $ 0.00\%$ &                             \\
\hline
\multirow{4}{*}{$\mathcal{S}_{7}$}
& $x_{i}$                     & $22.14\%$ & $17.61\%$ & $23.05\%$ & $17.89\%$ & $ 0.00\%$ & $19.32\%$ & \multirow{4}{*}{${\bP}9.46\%$} \\
& $\mathcal{MR}_{i}$          & $17.09\%$ & $16.11\%$ & $12.31\%$ & $ 0.00\%$ & $-1.11\%$ & $ 0.00\%$ &                                \\
& $\mathcal{RC}_{i}$          & $ 3.78\%$ & $ 2.84\%$ & $ 2.84\%$ & $ 0.00\%$ & $ 0.00\%$ & $ 0.00\%$ &                                \\
& $\mathcal{RC}_{i}^{\star}$  & $40.00\%$ & $30.00\%$ & $30.00\%$ & $ 0.00\%$ & $ 0.00\%$ & $ 0.00\%$ &                                \\
\hline
\multirow{4}{*}{$\mathcal{S}_{8}$}
& $x_{i}$                     & $21.71\%$ & $17.09\%$ & $21.77\%$ & $ 0.00\%$ & $15.37\%$ & $24.05\%$ & \multirow{4}{*}{${\bP}9.36\%$} \\
& $\mathcal{MR}_{i}$          & $17.25\%$ & $16.44\%$ & $12.90\%$ & $-3.49\%$ & $ 0.00\%$ & $ 0.00\%$ &                                \\
& $\mathcal{RC}_{i}$          & $ 3.75\%$ & $ 2.81\%$ & $ 2.81\%$ & $ 0.00\%$ & $ 0.00\%$ & $ 0.00\%$ &                                \\
& $\mathcal{RC}_{i}^{\star}$  & $40.00\%$ & $30.00\%$ & $30.00\%$ & $ 0.00\%$ & $ 0.00\%$ & $ 0.00\%$ &                                \\
\hline
\end{tabular}
\end{small}
\end{table}
\end{enumerate}

\item

\begin{enumerate}
\item We obtain the following portfolio:%
\begin{equation*}
\hspace*{-1.25cm}
\begin{small}
\begin{tabular}{|cc|rrrrrr:c|}
\hline
\multicolumn{2}{|c|}{Solution} & \multicolumn{1}{c}{1} & \multicolumn{1}{c}{2} & \multicolumn{1}{c}{3} &
                                 \multicolumn{1}{c}{4} & \multicolumn{1}{c}{5} & \multicolumn{1}{c:}{6} & $\sigma \left( x\right) $ \\
\hline
\multirow{4}{*}{$\mathcal{S}_{1}$}
                  & $x_{i}$                                      & $33.77\%$ & $29.05\%$ & $37.18\%$ & $0.00\%$ & $ 0.00\%$ & $0.00\%$ & \multirow{4}{*}{$16.07\%$} \\
                  & $\mathcal{MR}_{i}$                           & $19.03\%$ & $16.59\%$ & $12.96\%$ & $4.54\%$ & $-1.55\%$ & $2.39\%$ &          \\
                  & $\mathcal{RC}_{i}$                           & $ 6.43\%$ & $ 4.82\%$ & $ 4.82\%$ & $0.00\%$ & $ 0.00\%$ & $0.00\%$ &          \\
                  & $\mathcal{RC}_{i}^{\star}$                   & $40.00\%$ & $30.00\%$ & $30.00\%$ & $0.00\%$ & $ 0.00\%$ & $0.00\%$ &          \\
\hline
\end{tabular}
\end{small}
\end{equation*}%
There is only one asset such that that the weight is zero ($x_{i}=0$) and
the marginal risk is negative ($\mathcal{MR}_{i}<0$). We deduce that the
number of solutions is $2^{1}=2$ (TR-RPB, page 110).

\item The second solution is:%
\begin{equation*}
\hspace*{-1.05cm}
\begin{small}
\begin{tabular}{|cc|rrrrrr:c|}
\hline
\multicolumn{2}{|c|}{Solution} & \multicolumn{1}{c}{1} & \multicolumn{1}{c}{2} & \multicolumn{1}{c}{3} &
                                 \multicolumn{1}{c}{4} & \multicolumn{1}{c}{5} & \multicolumn{1}{c:}{6} & $\sigma \left( x\right) $ \\
\hline
\multirow{4}{*}{$\mathcal{S}_{2}$}
                  & $x_{i}$                                      & $27.27\%$ & $23.15\%$ & $29.60\%$ & $0.00\%$ & $19.98\%$ & $0.00\%$ & \multirow{4}{*}{$12.71\%$} \\
                  & $\mathcal{MR}_{i}$                           & $18.64\%$ & $16.47\%$ & $12.88\%$ & $4.12\%$ & $ 0.00\%$ & $2.43\%$ &          \\
                  & $\mathcal{RC}_{i}$                           & $ 5.08\%$ & $ 3.81\%$ & $ 3.81\%$ & $0.00\%$ & $ 0.00\%$ & $0.00\%$ &          \\
                  & $\mathcal{RC}_{i}^{\star}$                   & $40.00\%$ & $30.00\%$ & $30.00\%$ & $0.00\%$ & $ 0.00\%$ & $0.00\%$ &          \\
\hline
\end{tabular}
\end{small}
\end{equation*}
\end{enumerate}

\item

\begin{enumerate}
\item All the correlations are positive. We deduce that there is only one
solution (TR-RPB, page 101).

\item We obtain the following portfolio:%
\begin{equation*}
\hspace*{-0.90cm}
\begin{small}
\begin{tabular}{|cc|rrrrrr:c|}
\hline
\multicolumn{2}{|c|}{Solution} & \multicolumn{1}{c}{1} & \multicolumn{1}{c}{2} & \multicolumn{1}{c}{3} &
                                 \multicolumn{1}{c}{4} & \multicolumn{1}{c}{5} & \multicolumn{1}{c:}{6} & $\sigma \left( x\right) $ \\
\hline
\multirow{4}{*}{$\mathcal{S}_{1}$}
                  & $x_{i}$                                      & $33.78\%$ & $29.05\%$ & $37.17\%$ & $0.00\%$ & $0.00\%$ & $0.00\%$ & \multirow{4}{*}{$16.07\%$} \\
                  & $\mathcal{MR}_{i}$                           & $19.03\%$ & $16.59\%$ & $12.96\%$ & $4.54\%$ & $1.55\%$ & $2.39\%$ &          \\
                  & $\mathcal{RC}_{i}$                           & $ 6.43\%$ & $ 4.82\%$ & $ 4.82\%$ & $0.00\%$ & $0.00\%$ & $0.00\%$ &          \\
                  & $\mathcal{RC}_{i}^{\star}$                   & $40.00\%$ & $30.00\%$ & $30.00\%$ & $0.00\%$ & $0.00\%$ & $0.00\%$ &          \\
\hline
\end{tabular}
\end{small}
\end{equation*}%
\end{enumerate}

\item We obtain now:%
\begin{equation*}
\hspace*{-0.80cm}
\begin{small}
\begin{tabular}{|cc|rrrrrr:c|}
\hline
\multicolumn{2}{|c|}{Solution} & \multicolumn{1}{c}{1} & \multicolumn{1}{c}{2} & \multicolumn{1}{c}{3} &
                                 \multicolumn{1}{c}{4} & \multicolumn{1}{c}{5} & \multicolumn{1}{c:}{6} & $\sigma \left( x\right) $ \\
\hline
\multirow{4}{*}{$\mathcal{S}_{1}$}
                  & $x_{i}$                                      & $33.77\%$ & $29.05\%$ & $37.17\%$ & $ 0.00\%$ & $ 0.00\%$ & $ 0.00\%$ & \multirow{4}{*}{$16.07\%$} \\
                  & $\mathcal{MR}_{i}$                           & $19.03\%$ & $16.59\%$ & $12.96\%$ & $-0.61\%$ & $-0.31\%$ & $-0.48\%$ &          \\
                  & $\mathcal{RC}_{i}$                           & $ 6.43\%$ & $ 4.82\%$ & $ 4.82\%$ & $ 0.00\%$ & $ 0.00\%$ & $ 0.00\%$ &          \\
                  & $\mathcal{RC}_{i}^{\star}$                   & $40.00\%$ & $30.00\%$ & $30.00\%$ & $ 0.00\%$ & $ 0.00\%$ & $ 0.00\%$ &          \\
\hline
\end{tabular}
\end{small}
\end{equation*}%
We deduce that there are many solutions.
\end{enumerate}

\begin{remark}
This last question has been put in the wrong way. In fact, we wanted to show that
the number of solutions depends on the correlation coefficients, but also on the values taken by the
volatilities. If one or more assets which are not risk budgeted ($b_{i}=0$)
present a negative correlation with the assets which are risk budgeted ($%
b_{i}>0$), the solution may not be unique. The number of solutions will
depend on the anti-correlation strength and on the volatility level. If the
volatilities of the assets which are not risk budgeted are very different
from the other volatilities, the
solution may be unique, because the diversification effect is small.
\end{remark}

\section{Risk parity and factor models}

\begin{enumerate}
\item

\begin{enumerate}
\item We have:%
\begin{eqnarray*}
\sigma _{i}^{2} &=&\left( A\Omega A^{\top }+D\right) _{i,i} \\
&=&\sum_{j=1}^{3}A_{i,j}^{2}\omega _{j}^{2}+\tilde{\sigma}_{i}^{2}
\end{eqnarray*}%
The normalized risk decomposition with common and specific factors is then:%
\begin{equation*}
\sum_{j=1}^{3}c_{i,j}+\tilde{c}_{i}=1
\end{equation*}%
with $c_{i,j}=A_{i,j}^{2}\omega _{j}^{2}/\sigma _{i}^{2}$ and $\tilde{c}_{i}=%
\tilde{\sigma}_{i}^{2}/\sigma _{i}^{2}$. We obtain the following results:%
\begin{equation*}
\begin{tabular}{|c|cccc:cc|}
\hline
$i$ & $\sigma _{i}$ & $c_{i,1}$ & $c_{i,2}$ & $c_{i,3}$ & $%
\sum_{j=1}^{3}c_{i,j}$ & $\tilde{c}_{i}$ \\
\hline
$1$ & $18.89\%$ & $90.76\%$ & $0.28\%$ & $ 4.48\%$ & $95.52\%$ & $ 4.48\%$ \\
$2$ & $22.83\%$ & $92.90\%$ & $1.20\%$ & $ 1.11\%$ & $95.20\%$ & $ 4.80\%$ \\
$3$ & $25.39\%$ & $89.33\%$ & $0.35\%$ & $ 0.40\%$ & $90.07\%$ & $ 9.93\%$ \\
$4$ & $17.68\%$ & $81.89\%$ & $0.08\%$ & $10.03\%$ & $92.00\%$ & $ 8.00\%$ \\
$5$ & $14.91\%$ & $44.99\%$ & $2.81\%$ & $ 7.20\%$ & $55.01\%$ & $44.99\%$ \\
$6$ & $28.98\%$ & $93.38\%$ & $0.48\%$ & $ 0.30\%$ & $94.16\%$ & $ 5.84\%$ \\ \hline
\end{tabular}%
\end{equation*}%
We notice that individual risks are mainly concentrated in the first factor. We may then
assimilate this factor as a market risk factor.
The expression of the correlation is:%
\begin{eqnarray*}
\rho _{i,j} &=&\frac{\left( A\Omega A^{\top }+D\right) _{i,j}}{\sigma
_{i}\sigma _{j}} \\
&=&\frac{\sum_{k=1}^{3}A_{i,k}A_{j,k}\omega _{k}^{2}}{\sqrt{\left(
\sum_{k=1}^{3}A_{i,k}^{2}\omega _{k}^{2}+\tilde{\sigma}_{i}^{2}\right)
\left( \sum_{k=1}^{3}A_{j,k}^{2}\omega _{k}^{2}+\tilde{\sigma}%
_{j}^{2}\right) }}
\end{eqnarray*}%
We obtain:%
\begin{equation*}
\rho =\left(
\begin{array}{rrrrrr}
100.0\% &         &         &         &         &         \\
 90.2\% & 100.0\% &         &         &         &         \\
 91.7\% &  91.1\% & 100.0\% &         &         &         \\
 79.4\% &  90.2\% &  83.4\% & 100.0\% &         &         \\
 68.7\% &  60.0\% &  64.1\% &  52.7\% & 100.0\% &         \\
 90.5\% &  93.0\% &  90.6\% &  89.4\% &  64.5\% & 100.0\%
\end{array}%
\right)
\end{equation*}

\item We obtain:%
\begin{equation*}
A^{+}=\left(
\begin{array}{rrrrrr}
 0.152 &  0.150 &  0.188 &  0.112 & 0.113 &  0.234 \\
-0.266 & -0.567 & -0.355 &  0.220 & 0.608 &  0.578 \\
 0.482 & -0.185 &  0.237 & -0.598 & 0.406 & -0.171
\end{array}%
\right)
\end{equation*}%
The number of assets $n$ is larger than the number of risk factors $m$. We
deduce that the Moore-Penrose inverse $A^{+}$ can be written as the OLS
projector:%
\begin{equation*}
A^{+}=\left( A^{\top }A\right) ^{-1}A^{\top }
\end{equation*}%
We obtain:%
\begin{equation*}
B^{+}=\left(
\begin{array}{rrr}
0.152 & -0.266 &  0.482 \\
0.150 & -0.567 & -0.185 \\
0.188 & -0.355 &  0.237 \\
0.112 &  0.220 & -0.598 \\
0.113 &  0.608 &  0.406 \\
0.234 &  0.578 & -0.171
\end{array}%
\right)
\end{equation*}%
\begin{equation*}
\tilde{B}=\left(
\begin{array}{rrrrrr}
0.579 & -0.385 & -0.074 &  0.624 & -0.045 & -0.347 \\
0.064 &  0.587 & -0.519 &  0.109 &  0.525 & -0.307 \\
0.480 &  0.061 & -0.588 & -0.262 & -0.399 &  0.439
\end{array}%
\right)
\end{equation*}%
\begin{equation*}
\tilde{B}^{+}=\left(
\begin{array}{rrr}
 0.579 &  0.064 &  0.480 \\
-0.385 &  0.587 &  0.061 \\
-0.074 & -0.519 & -0.588 \\
 0.624 &  0.109 & -0.262 \\
-0.045 &  0.525 & -0.399 \\
-0.347 & -0.307 &  0.439
\end{array}%
\right)
\end{equation*}%
By construction, we have:%
\begin{equation*}
B^{+}=B^{\top }\left( BB^{\top }\right) ^{-1}
\end{equation*}

\item The weights $y$ and $\tilde{y}$ are equal to:%
\begin{eqnarray*}
y &=&A^{\top }x \\
&=&\left(
\begin{array}{r}
94.000\% \\
10.000\% \\
17.500\%
\end{array}%
\right)
\end{eqnarray*}%
and:%
\begin{eqnarray*}
\tilde{y} &=&\tilde{B}x \\
&=&\left(
\begin{array}{r}
 1.445\% \\
 9.518\% \\
-3.091\%
\end{array}%
\right)
\end{eqnarray*}%
We have (TR-RPB, page 141):%
\begin{equation*}
x=x_{c}+x_{s}
\end{equation*}%
where $x_{c}=B^{+}y$ is the exposure to common factors and $x_{s}=\tilde{B}%
^{+}\tilde{y}$ is the exposure to specific factors. We then obtain:%
\begin{equation*}
\begin{tabular}{|c|rrr|}
\hline
$i$ & \multicolumn{1}{c}{$x_{i}$} & \multicolumn{1}{c}{$x_{i,c}$} & \multicolumn{1}{c|}{$x_{i,s}$} \\
\hline
$1$ & $20.000\%$ & $20.033\%$ & $-0.033\%$ \\
$2$ & $10.000\%$ & $ 5.159\%$ & $ 4.841\%$ \\
$3$ & $15.000\%$ & $18.234\%$ & $-3.234\%$ \\
$4$ & $ 5.000\%$ & $ 2.256\%$ & $ 2.744\%$ \\
$5$ & $30.000\%$ & $23.839\%$ & $ 6.161\%$ \\
$6$ & $20.000\%$ & $24.779\%$ & $-4.779\%$ \\
\hline
\end{tabular}%
\end{equation*}
\item We recall that (TR-RPB, page 142):%
\begin{eqnarray*}
\mathcal{MR}\left( \mathcal{F}_{j}\right)  &=&\left( A^{+}\frac{\partial
\,\sigma \left( x\right) }{\partial \,x}\right) _{j} \\
\mathcal{MR}\left( \mathcal{\tilde{F}}_{j}\right)  &=&\left( \tilde{B}\frac{%
\partial \,\sigma \left( x\right) }{\partial \,x}\right) _{j}
\end{eqnarray*}%
We deduce that:%
\begin{equation*}
\begin{tabular}{|c|rrrr|}
\hline
Factor & \multicolumn{1}{c}{$y_{j}$} & \multicolumn{1}{c}{$\mathcal{MR}_{j}$} &
\multicolumn{1}{c}{$\mathcal{RC}_{j}$} & \multicolumn{1}{c|}{$\mathcal{RC}_{j}^{\star }$} \\
\hline
$\mathcal{F}_{1}$         & $94.00\%$ & $20.02\%$ & $18.81\%$ & $97.95\%$  \\
$\mathcal{F}_{2}$         & $10.00\%$ & $ 1.09\%$ & $ 0.11\%$ & $ 0.57\%$  \\
$\mathcal{F}_{3}$         & $17.50\%$ & $ 1.27\%$ & $ 0.22\%$ & $ 1.15\%$  \\ \hdashline
$\mathcal{\tilde{F}}_{1}$ & $ 1.44\%$ & $-0.20\%$ & $ 0.00\%$ & $-0.01\%$  \\
$\mathcal{\tilde{F}}_{2}$ & $ 9.52\%$ & $ 0.50\%$ & $ 0.05\%$ & $ 0.25\%$  \\
$\mathcal{\tilde{F}}_{3}$ & $-3.09\%$ & $-0.62\%$ & $ 0.02\%$ & $ 0.10\%$  \\ \hdashline
$\sigma\left(x\right)$    &           &             & $19.21\%$ &          \\
\hline
\end{tabular}%
\end{equation*}

\item We have:%
\begin{eqnarray*}
R_{t} &=&A\mathcal{F}_{t}+\varepsilon _{t} \\
&=&\left(
\begin{array}{cc}
A & I_{n}%
\end{array}%
\right) \left(
\begin{array}{c}
\mathcal{F}_{t} \\
\varepsilon _{t}%
\end{array}%
\right)  \\
&=&A^{\prime }\mathcal{F}_{t}^{\prime }+\varepsilon _{t}^{\prime }
\end{eqnarray*}%
with $D^{\prime }=\mathbf{0}$ and:%
\begin{equation*}
\Omega ^{\prime }=\left(
\begin{array}{cc}
\Omega  & \mathbf{0} \\
\mathbf{0} & D%
\end{array}%
\right)
\end{equation*}%
We obtain:%
\begin{equation*}
\begin{tabular}{|c|rrrr|}
\hline
Factor & \multicolumn{1}{c}{$y_{j}$} & \multicolumn{1}{c}{$\mathcal{MR}_{j}$} &
\multicolumn{1}{c}{$\mathcal{RC}_{j}$} & \multicolumn{1}{c|}{$\mathcal{RC}_{j}^{\star }$} \\
\hline
$\mathcal{F}_{1}$         & $94.00\%$ & $17.23\%$ & $16.19\%$ & $84.30\%$  \\
$\mathcal{F}_{2}$         & $10.00\%$ & $ 0.22\%$ & $ 0.02\%$ & $ 0.11\%$  \\
$\mathcal{F}_{3}$         & $17.50\%$ & $ 0.42\%$ & $ 0.07\%$ & $ 0.38\%$  \\ \hdashline
$\mathcal{F}_{4}$         & $20.00\%$ & $ 2.38\%$ & $ 0.48\%$ & $ 2.48\%$  \\
$\mathcal{F}_{5}$         & $10.00\%$ & $ 2.71\%$ & $ 0.27\%$ & $ 1.41\%$  \\
$\mathcal{F}_{6}$         & $15.00\%$ & $ 3.37\%$ & $ 0.51\%$ & $ 2.64\%$  \\
$\mathcal{F}_{7}$         & $ 5.00\%$ & $ 1.82\%$ & $ 0.09\%$ & $ 0.47\%$  \\
$\mathcal{F}_{8}$         & $30.00\%$ & $ 2.77\%$ & $ 0.83\%$ & $ 4.33\%$  \\
$\mathcal{F}_{9}$         & $20.00\%$ & $ 3.73\%$ & $ 0.75\%$ & $ 3.88\%$  \\ \hdashline
$\sigma\left(x\right)$    &           &             & $19.21\%$ &          \\
\hline
\end{tabular}%
\end{equation*}
\end{enumerate}
We don't find the same results for the risk decomposition with respect
to the common factors. This is normal because we face an identification problem.
Other parameterizations may induce other results. By considering the specific factors
as common factors, we reduce the part explained by the common factors. Indeed, the
identification problem becomes less and less important when $n/m$ tends to $\infty$.

\item

\begin{enumerate}
\item If we consider the optimization problem defined in TR-RPB on page
144, we obtain:%
\begin{equation*}
\begin{tabular}{|c|rrrr|}
\hline
Factor & \multicolumn{1}{c}{$y_{j}$} & \multicolumn{1}{c}{$\mathcal{MR}_{j}$} &
\multicolumn{1}{c}{$\mathcal{RC}_{j}$} & \multicolumn{1}{c|}{$\mathcal{RC}_{j}^{\star }$} \\
\hline
$\mathcal{F}_{1}$         & $ 70.11\%$ & $18.05\%$ & $12.66\%$ & $78.72\%$  \\
$\mathcal{F}_{2}$         & $ 37.06\%$ & $ 3.39\%$ & $ 1.26\%$ & $ 7.81\%$  \\
$\mathcal{F}_{3}$         & $ 39.44\%$ & $ 3.30\%$ & $ 1.30\%$ & $ 8.10\%$  \\ \hdashline
$\mathcal{\tilde{F}}_{1}$ & $  1.11\%$ & $-0.26\%$ & $ 0.00\%$ & $-0.02\%$  \\
$\mathcal{\tilde{F}}_{2}$ & $ 32.42\%$ & $ 2.11\%$ & $ 0.68\%$ & $ 4.25\%$  \\
$\mathcal{\tilde{F}}_{3}$ & $-12.88\%$ & $-1.42\%$ & $ 0.18\%$ & $ 1.14\%$  \\ \hdashline
$\sigma\left(x\right)$    &           &            & $16.08\%$ &            \\
\hline
\end{tabular}%
\end{equation*}%
We see that it is not possible to target a risk contribution of $10\%$ for
the second and third risk factors, because the first factor
explains most of the risk of long-only portfolios. To have a
smaller sensibility to the first risk factor, we need to consider a long-short
portfolio.

\item In terms of risk factors, we obtain:%
\begin{equation*}
\begin{tabular}{|c|rrrr|}
\hline
Factor & \multicolumn{1}{c}{$y_{j}$} & \multicolumn{1}{c}{$\mathcal{MR}_{j}$} &
\multicolumn{1}{c}{$\mathcal{RC}_{j}$} & \multicolumn{1}{c|}{$\mathcal{RC}_{j}^{\star }$} \\
\hline
$\mathcal{F}_{1}$         & $ 27.63\%$ & $ 6.72\%$ & $ 1.86\%$ & $10.00\%$  \\
$\mathcal{F}_{2}$         & $107.39\%$ & $ 6.92\%$ & $ 7.43\%$ & $40.00\%$  \\
$\mathcal{F}_{3}$         & $107.23\%$ & $ 6.93\%$ & $ 7.43\%$ & $40.00\%$  \\ \hdashline
$\mathcal{\tilde{F}}_{1}$ & $ 24.27\%$ & $-0.09\%$ & $-0.02\%$ & $-0.12\%$  \\
$\mathcal{\tilde{F}}_{2}$ & $ 38.77\%$ & $ 3.58\%$ & $ 1.39\%$ & $ 7.47\%$  \\
$\mathcal{\tilde{F}}_{3}$ & $-21.45\%$ & $-2.29\%$ & $ 0.49\%$ & $ 2.65\%$  \\ \hdashline
$\sigma\left(x\right)$    &           &            & $18.57\%$ &            \\
\hline
\end{tabular}%
\end{equation*}%
We deduce the following RB portfolio:%
\begin{equation*}
\begin{tabular}{|c|rrrr|}
\hline
Asset & \multicolumn{1}{c}{$x_{i}$} & \multicolumn{1}{c}{$\mathcal{MR}_{i}$} &
\multicolumn{1}{c}{$\mathcal{RC}_{i}$} & \multicolumn{1}{c|}{$\mathcal{RC}_{i}^{\star }$} \\
\hline
$1$                       & $  33.52\%$ & $ 7.21\%$ & $ 2.42\%$ & $ 13.01\%$  \\
$2$                       & $ -64.47\%$ & $ 3.85\%$ & $-2.48\%$ & $-13.36\%$  \\
$3$                       & $ -16.81\%$ & $ 6.87\%$ & $-1.16\%$ & $ -6.22\%$  \\
$4$                       & $ -12.44\%$ & $ 2.15\%$ & $-0.27\%$ & $ -1.44\%$  \\
$5$                       & $ 139.75\%$ & $13.08\%$ & $18.27\%$ & $ 98.42\%$  \\
$6$                       & $  20.45\%$ & $ 8.71\%$ & $ 1.78\%$ & $  9.60\%$  \\ \hdashline
$\sigma\left(x\right)$    &             &           & $18.57\%$ &             \\
\hline
\end{tabular}%
\end{equation*}%

\item In terms of risk factors, we obtain:%
\begin{equation*}
\begin{tabular}{|c|rrrr|}
\hline
Factor & \multicolumn{1}{c}{$y_{j}$} & \multicolumn{1}{c}{$\mathcal{MR}_{j}$} &
\multicolumn{1}{c}{$\mathcal{RC}_{j}$} & \multicolumn{1}{c|}{$\mathcal{RC}_{j}^{\star }$} \\
\hline
$\mathcal{F}_{1}$         & $ 23.53\%$ & $ 4.87\%$ & $ 1.15\%$ & $ 5.00\%$  \\
$\mathcal{F}_{2}$         & $220.92\%$ & $ 9.33\%$ & $20.62\%$ & $90.00\%$  \\
$\mathcal{F}_{3}$         & $ 40.05\%$ & $ 2.86\%$ & $ 1.15\%$ & $ 5.00\%$  \\ \hdashline
$\mathcal{\tilde{F}}_{1}$ & $ 71.12\%$ & $ 0.27\%$ & $ 0.19\%$ & $ 0.83\%$  \\
$\mathcal{\tilde{F}}_{2}$ & $-12.15\%$ & $ 2.59\%$ & $-0.32\%$ & $-1.38\%$  \\
$\mathcal{\tilde{F}}_{3}$ & $-12.20\%$ & $-1.04\%$ & $ 0.13\%$ & $ 0.56\%$  \\ \hdashline
$\sigma\left(x\right)$    &           &            & $22.91\%$ &            \\
\hline
\end{tabular}%
\end{equation*}%
We deduce the following RB portfolio:%
\begin{equation*}
\begin{tabular}{|c|rrrr|}
\hline
Asset & \multicolumn{1}{c}{$x_{i}$} & \multicolumn{1}{c}{$\mathcal{MR}_{i}$} &
\multicolumn{1}{c}{$\mathcal{RC}_{i}$} & \multicolumn{1}{c|}{$\mathcal{RC}_{i}^{\star }$} \\
\hline
$1$                       & $  -1.43\%$ & $ 3.76\%$ & $-0.05\%$ & $-0.24\%$  \\
$2$                       & $-164.36\%$ & $ 1.18\%$ & $-1.95\%$ & $-8.50\%$  \\
$3$                       & $ -56.24\%$ & $ 2.86\%$ & $-1.61\%$ & $-7.02\%$  \\
$4$                       & $  73.59\%$ & $ 3.55\%$ & $ 2.61\%$ & $11.39\%$  \\
$5$                       & $ 148.43\%$ & $10.30\%$ & $15.28\%$ & $66.69\%$  \\
$6$                       & $ 100.02\%$ & $ 8.63\%$ & $ 8.63\%$ & $37.67\%$  \\ \hdashline
$\sigma\left(x\right)$    &             &           & $22.91\%$ &             \\
\hline
\end{tabular}%
\end{equation*}%

\item In terms of risk factors, we obtain:%
\begin{equation*}
\begin{tabular}{|c|rrrr|}
\hline
Factor & \multicolumn{1}{c}{$y_{j}$} & \multicolumn{1}{c}{$\mathcal{MR}_{j}$} &
\multicolumn{1}{c}{$\mathcal{RC}_{j}$} & \multicolumn{1}{c|}{$\mathcal{RC}_{j}^{\star }$} \\
\hline
$\mathcal{F}_{1}$         & $ 26.73\%$ & $ 4.70\%$ & $ 1.26\%$ & $ 5.00\%$  \\
$\mathcal{F}_{2}$         & $ 43.70\%$ & $ 2.87\%$ & $ 1.26\%$ & $ 5.00\%$  \\
$\mathcal{F}_{3}$         & $239.40\%$ & $ 9.44\%$ & $22.60\%$ & $90.00\%$  \\ \hdashline
$\mathcal{\tilde{F}}_{1}$ & $ 70.84\%$ & $ 0.27\%$ & $ 0.19\%$ & $ 0.77\%$  \\
$\mathcal{\tilde{F}}_{2}$ & $ 14.03\%$ & $ 1.92\%$ & $ 0.27\%$ & $ 1.07\%$  \\
$\mathcal{\tilde{F}}_{3}$ & $ 26.81\%$ & $-1.72\%$ & $-0.46\%$ & $-1.84\%$  \\ \hdashline
$\sigma\left(x\right)$    &           &            & $25.12\%$ &            \\
\hline
\end{tabular}%
\end{equation*}%
We deduce the following RB portfolio:%
\begin{equation*}
\begin{tabular}{|c|rrrr|}
\hline
Asset & \multicolumn{1}{c}{$x_{i}$} & \multicolumn{1}{c}{$\mathcal{MR}_{i}$} &
\multicolumn{1}{c}{$\mathcal{RC}_{i}$} & \multicolumn{1}{c|}{$\mathcal{RC}_{i}^{\star }$} \\
\hline
$1$                       & $162.53\%$ & $ 7.83\%$ & $12.73\%$ & $50.67\%$  \\
$2$                       & $-82.45\%$ & $ 1.81\%$ & $-1.50\%$ & $-5.96\%$  \\
$3$                       & $ 17.99\%$ & $ 6.66\%$ & $ 1.20\%$ & $ 4.77\%$  \\
$4$                       & $-91.93\%$ & $-1.74\%$ & $ 1.60\%$ & $ 6.35\%$  \\
$5$                       & $120.33\%$ & $10.19\%$ & $12.26\%$ & $48.80\%$  \\
$6$                       & $-26.47\%$ & $ 4.40\%$ & $-1.16\%$ & $-4.64\%$  \\ \hdashline
$\sigma\left(x\right)$    &             &          & $25.12\%$ &             \\
\hline
\end{tabular}%
\end{equation*}%

\item To be exposed to the second factor, we use a portfolio which is long
on the first three assets and short on the last three assets. This result is
coherent with the matrix $A$. We obtain a similar result if we want to be
exposed to the third asset. Nevertheless, the figures of risk contributions
may be confusing. We might have thought that the risk was split between
the long leg and the short leg. It is not the case. Nonetheless, this is normal
because if the risk is perfectly split between the long leg and the short
leg, the risk exposure to this factor vanishes!
\end{enumerate}
\end{enumerate}

\section{Risk allocation with the expected shortfall risk measure}

\begin{enumerate}
\item

\begin{enumerate}
\item We have:%
\begin{eqnarray*}
L\left( x\right) &=&-R\left( x\right) \\
&=&-x^{\top }R
\end{eqnarray*}%
It follows that:%
\begin{equation*}
L\left( x\right) \sim \mathcal{N}\left( -\mu \left( x\right) ,\sigma \left(
x\right) \right)
\end{equation*}%
with $\mu \left( x\right) =x^{\top }\mu $ and $\sigma \left( x\right) =\sqrt{%
x^{\top }\Sigma x}$.

\item The expected shortfall $\limfunc{ES}\nolimits_{\alpha }\left( L\right)
$ is the average of value-at-risks at level $\alpha $ and higher:%
\begin{equation*}
\limfunc{ES}\nolimits_{\alpha }\left( L\right) =\mathbb{E}\left[ L\mid L\geq
\limfunc{VaR}\nolimits_{\alpha }\left( L\right) \right]
\end{equation*}%
We know that the value-at-risk is (TR-RPB, page 74):%
\begin{equation*}
\limfunc{VaR}\nolimits_{\alpha }\left( x\right) =-x^{\top }\mu +\Phi
^{-1}\left( \alpha \right) \sqrt{x^{\top }\Sigma x}
\end{equation*}%
We deduce that:%
\begin{equation*}
\limfunc{ES}\nolimits_{\alpha }\left( x\right) =\frac{1}{1-\alpha }%
\int_{u^{-}}^{\infty }\frac{u}{\sigma \left( x\right) \sqrt{2\pi }}\exp \left( -%
\frac{1}{2}\left( \frac{u+\mu \left( x\right) }{\sigma \left( x\right) }%
\right) ^{2}\right) \,\mathrm{d}u
\end{equation*}%
where $u^{-}=-\mu \left( x\right) +\sigma \left( x\right) \Phi ^{-1}\left(
\alpha \right) $. With the change of variable $t=\sigma \left( x\right)
^{-1}\left( u+\mu \left( x\right) \right) $, we obtain:%
\begin{eqnarray*}
\limfunc{ES}\nolimits_{\alpha }\left( x\right)  &=&\frac{1}{1-\alpha }%
\int_{\Phi ^{-1}\left( \alpha \right) }^{\infty }\left( -\mu \left( x\right)
+\sigma \left( x\right) t\right) \frac{1}{\sqrt{2\pi }}\exp \left( -\frac{1}{%
2}t^{2}\right) \,\mathrm{d}t \\
&=&-\frac{\mu \left( x\right) }{1-\alpha }\left[ \Phi \left( t\right) \right]
_{\Phi ^{-1}\left( \alpha \right) }^{\infty }+ \\
&&\frac{\sigma \left( x\right) }{\left( 1-\alpha \right) \sqrt{2\pi }}%
\int_{\Phi ^{-1}\left( \alpha \right) }^{\infty }t\exp \left( -\frac{1}{2}%
t^{2}\right) \,\mathrm{d}t \\
&=&-\mu \left( x\right) +\frac{\sigma \left( x\right) }{\left( 1-\alpha
\right) \sqrt{2\pi }}\left[ -\exp \left( -\frac{1}{2}t^{2}\right) \right]
_{\Phi ^{-1}\left( \alpha \right) }^{\infty } \\
&=&-\mu \left( x\right) +\frac{\sigma \left( x\right) }{\left( 1-\alpha
\right) \sqrt{2\pi }}\exp \left( -\frac{1}{2}\left[ \Phi ^{-1}\left( \alpha
\right) \right] ^{2}\right)
\end{eqnarray*}%
The expected shortfall of portfolio $x$ is then (TR-RPB, page 75):%
\begin{equation*}
\limfunc{ES}\nolimits_{\alpha }\left( x\right) =-x^{\top }\mu +\frac{\phi
\left( \Phi ^{-1}\left( \alpha \right) \right) }{\left( 1-\alpha \right) }%
\sqrt{x^{\top }\Sigma x}
\end{equation*}

\item The vector of marginal risk is defined as follows (TR-RPB, page 80):%
\begin{eqnarray*}
\mathcal{MR} &=&\frac{\partial \,\limfunc{ES}\nolimits_{\alpha }\left(
x\right) }{\partial \,x} \\
&=&-\mu +\frac{\phi \left( \Phi ^{-1}\left( \alpha \right) \right) }{\left(
1-\alpha \right) }\frac{\Sigma x}{\sqrt{x^{\top }\Sigma x}}
\end{eqnarray*}%
We deduce that the risk contribution $\mathcal{RC}_{i}$ of the asset $i$ is:%
\begin{eqnarray*}
\mathcal{RC}_{i} &=&x_{i}\cdot \mathcal{MR}_{i} \\
&=&-x_{i}\mu _{i}+\frac{\phi \left( \Phi ^{-1}\left( \alpha \right) \right)
}{\left( 1-\alpha \right) }\frac{x_{i}\cdot \left( \Sigma x\right) _{i}}{%
\sqrt{x^{\top }\Sigma x}}
\end{eqnarray*}
It follows that:%
\begin{eqnarray*}
\sum_{i=1}^{n}\mathcal{RC}_{i} &=&\sum_{i=1}^{n}-x_{i}\mu _{i}+\frac{\phi
\left( \Phi ^{-1}\left( \alpha \right) \right) }{\left( 1-\alpha \right) }%
\frac{x_{i}\cdot \left( \Sigma x\right) _{i}}{\sqrt{x^{\top }\Sigma x}} \\
&=&-x^{\top }\mu +\frac{\phi \left( \Phi ^{-1}\left( \alpha \right) \right)
}{\left( 1-\alpha \right) }\frac{x^{\top }\left( \Sigma x\right) }{\sqrt{%
x^{\top }\Sigma x}} \\
&=&\limfunc{ES}\nolimits_{\alpha }\left( x\right)
\end{eqnarray*}%
The expected shortfall then verifies the Euler allocation principle (TR-RPB,
page 78).
\end{enumerate}

\item

\begin{enumerate}
\item We have:%
\begin{equation*}
\begin{tabular}{|c|rrrr|}
\hline
Asset & \multicolumn{1}{c}{$x_{i}$} & \multicolumn{1}{c}{$\mathcal{MR}_{i}$} &
\multicolumn{1}{c}{$\mathcal{RC}_{i}$} & \multicolumn{1}{c|}{$\mathcal{RC}_{i}^{\star }$} \\
\hline
$1$                          & $30.00\%$ & $18.40\%$ & $ 5.52\%$ & $45.57\%$  \\
$2$                          & $30.00\%$ & $22.95\%$ & $ 6.89\%$ & $56.84\%$  \\
$3$                          & $40.00\%$ & $-0.73\%$ & $-0.29\%$ & $-2.41\%$  \\ \hdashline
$\limfunc{ES}\nolimits_{\alpha }\left( x\right)$ & & & $12.11\%$ &             \\
\hline
\end{tabular}%
\end{equation*}

\item The ERC portfolio is:%
\begin{equation*}
\begin{tabular}{|c|rrrr|}
\hline
Asset & \multicolumn{1}{c}{$x_{i}$} & \multicolumn{1}{c}{$\mathcal{MR}_{i}$} &
\multicolumn{1}{c}{$\mathcal{RC}_{i}$} & \multicolumn{1}{c|}{$\mathcal{RC}_{i}^{\star }$} \\
\hline
$1$                          & $18.53\%$ & $14.83\%$ & $2.75\%$ & $33.33\%$  \\
$2$                          & $18.45\%$ & $14.89\%$ & $2.75\%$ & $33.33\%$  \\
$3$                          & $63.02\%$ & $ 4.36\%$ & $2.75\%$ & $33.33\%$  \\ \hdashline
$\limfunc{ES}\nolimits_{\alpha }\left( x\right)$ & & & $8.24\%$ &             \\
\hline
\end{tabular}%
\end{equation*}

\item If the risk budgets are equal to $b=\left( 70\%,20\%,10\%\right) $, we obtain:
\begin{equation*}
\begin{tabular}{|c|rrrr|}
\hline
Asset & \multicolumn{1}{c}{$x_{i}$} & \multicolumn{1}{c}{$\mathcal{MR}_{i}$} &
\multicolumn{1}{c}{$\mathcal{RC}_{i}$} & \multicolumn{1}{c|}{$\mathcal{RC}_{i}^{\star }$} \\
\hline
$1$                          & $33.16\%$ & $21.57\%$ & $ 7.15\%$ & $70.00\%$  \\
$2$                          & $15.91\%$ & $12.85\%$ & $ 2.04\%$ & $20.00\%$  \\
$3$                          & $50.93\%$ & $ 2.01\%$ & $ 1.02\%$ & $10.00\%$  \\ \hdashline
$\limfunc{ES}\nolimits_{\alpha }\left( x\right)$ & & & $10.22\%$ &             \\
\hline
\end{tabular}%
\end{equation*}

\item We have:%
\begin{equation*}
\begin{tabular}{|c|rrrr|}
\hline
Asset & \multicolumn{1}{c}{$x_{i}$} & \multicolumn{1}{c}{$\mathcal{MR}_{i}$} &
\multicolumn{1}{c}{$\mathcal{RC}_{i}$} & \multicolumn{1}{c|}{$\mathcal{RC}_{i}^{\star }$} \\
\hline
$1$                          & $ 80.00\%$ & $21.54\%$ & $17.24\%$ & $57.93\%$  \\
$2$                          & $ 50.00\%$ & $21.49\%$ & $10.75\%$ & $36.12\%$  \\
$3$                          & $-30.00\%$ & $-5.89\%$ & $ 1.77\%$ & $ 5.94\%$  \\ \hdashline
$\limfunc{ES}\nolimits_{\alpha }\left( x\right)$ & &  & $29.75\%$ &             \\
\hline
\end{tabular}%
\end{equation*}
We notice that the risk contributions are all positive even if we consider a long-short portfolio.
We can therefore think that there may be several solutions to the risk budgeting problem if we consider long-short
portfolios.

\item Here is a long-short ERC portfolio:%
\begin{equation*}
\begin{tabular}{|c|rrrr|}
\hline
Asset & \multicolumn{1}{c}{$x_{i}$} & \multicolumn{1}{c}{$\mathcal{MR}_{i}$} &
\multicolumn{1}{c}{$\mathcal{RC}_{i}$} & \multicolumn{1}{c|}{$\mathcal{RC}_{i}^{\star }$} \\
\hline
$1$                          & $ 38.91\%$ & $ 13.22\%$ & $ 5.14\%$ & $33.33\%$  \\
$2$                          & $-21.02\%$ & $-24.47\%$ & $ 5.14\%$ & $33.33\%$  \\
$3$                          & $ 82.11\%$ & $  6.26\%$ & $ 5.14\%$ & $33.33\%$  \\ \hdashline
$\limfunc{ES}\nolimits_{\alpha }\left( x\right)$ &  &  & $15.43\%$ &             \\
\hline
\end{tabular}%
\end{equation*}
Nevertheless, this solution is not unique. For instance, here is another long-short ERC portfolio:%
\begin{equation*}
\begin{tabular}{|c|rrrr|}
\hline
Asset & \multicolumn{1}{c}{$x_{i}$} & \multicolumn{1}{c}{$\mathcal{MR}_{i}$} &
\multicolumn{1}{c}{$\mathcal{RC}_{i}$} & \multicolumn{1}{c|}{$\mathcal{RC}_{i}^{\star }$} \\
\hline
$1$                          & $-58.65\%$ & $-23.04\%$ & $13.51\%$ & $33.33\%$  \\
$2$                          & $-40.03\%$ & $-33.75\%$ & $13.51\%$ & $33.33\%$  \\
$3$                          & $198.68\%$ & $  6.80\%$ & $13.51\%$ & $33.33\%$  \\ \hdashline
$\limfunc{ES}\nolimits_{\alpha }\left( x\right)$ &  &  & $40.53\%$ &             \\
\hline
\end{tabular}%
\end{equation*}

\item Here are three long-short portfolios that satisfy the risk budgets $%
b=\left( 70\%,20\%,10\%\right) $:%
\begin{equation*}
\begin{tabular}{|c|c|rrrr|}
\hline
Solution & Asset & \multicolumn{1}{c}{$x_{i}$} & \multicolumn{1}{c}{$\mathcal{MR}_{i}$} &
  \multicolumn{1}{c}{$\mathcal{RC}_{i}$} & \multicolumn{1}{c|}{$\mathcal{RC}_{i}^{\star }$} \\
  \hline
\multirow{4}{*}{$\mathcal{S}_{1}$}
& $1$                          & $115.31\%$ & $23.56\%$ & $27.17\%$ & $70.00\%$  \\
& $2$                          & $ 45.56\%$ & $17.04\%$ & $ 7.76\%$ & $20.00\%$  \\
& $3$                          & $-60.87\%$ & $-6.38\%$ & $ 3.88\%$ & $10.00\%$  \\
& $\limfunc{ES}\nolimits_{\alpha }\left( x\right)$ &  &  & $38.81\%$ &             \\
\hline
\multirow{4}{*}{$\mathcal{S}_{2}$}
& $1$                          & $ 60.76\%$ & $ 20.97\%$ & $12.74\%$ & $70.00\%$  \\
& $2$                          & $-20.82\%$ & $-17.48\%$ & $ 3.64\%$ & $20.00\%$  \\
& $3$                          & $ 60.07\%$ & $  3.03\%$ & $ 1.82\%$ & $10.00\%$  \\
& $\limfunc{ES}\nolimits_{\alpha }\left( x\right)$ &  &  & $18.20\%$ &             \\
\hline
\multirow{4}{*}{$\mathcal{S}_{3}$}
& $1$                          & $-72.96\%$ & $-24.32\%$ & $17.74\%$ & $70.00\%$  \\
& $2$                          & $ 54.53\%$ & $  9.30\%$ & $ 5.07\%$ & $20.00\%$  \\
& $3$                          & $118.43\%$ & $  2.14\%$ & $ 2.53\%$ & $10.00\%$  \\
& $\limfunc{ES}\nolimits_{\alpha }\left( x\right)$ &  &  & $25.34\%$ &             \\
\hline
\end{tabular}%
\end{equation*}

\item Contrary to the long-only case, the RB portfolio may not be unique when the portfolio is long-short.
\end{enumerate}

\item

\begin{enumerate}
\item We have:%
\begin{eqnarray*}
L\left( x\right)  &=&-\sum_{i=1}^{n}x_{i}R_{i} \\
&=&\sum_{i=1}^{n}L_{i}
\end{eqnarray*}%
with $L_{i}=-x_{i}R_{i}$. We know that (TR-RPB, page 85):%
\begin{eqnarray*}
\mathcal{RC}_{i} &=&\mathbb{E}\left[ L_{i}\mid L\geq \limfunc{VaR}%
\nolimits_{\alpha }\left( L\right) \right]  \\
&=&\frac{\mathbb{E}\left[ L_{i}\cdot \mathds{1}\left\{ L\geq \limfunc{VaR}%
\nolimits_{\alpha }\left( L\right) \right\} \right] }{\mathbb{E}\left[
\mathds{1}\left\{ L\geq \limfunc{VaR}\nolimits_{\alpha }\left( L\right)
\right\} \right] } \\
&=&\frac{\mathbb{E}\left[ L_{i}\cdot \mathds{1}\left\{ L\geq \limfunc{VaR}%
\nolimits_{\alpha }\left( L\right) \right\} \right] }{1-\alpha }
\end{eqnarray*}%
We deduce that:%
\begin{equation*}
\mathcal{RC}_{i}=-\frac{x_{i}}{1-\alpha }\mathbb{E}\left[ R_{i}\cdot \mathbf{%
1}\left\{ R\left( x\right) \leq -\limfunc{VaR}\nolimits_{\alpha }\left(
L\right) \right\} \right]
\end{equation*}

\item We know that the random vector $\left( R,R\left( x\right) \right) $
has a multivariate normal distribution:%
\begin{equation*}
\left(
\begin{array}{c}
R \\
R\left( x\right)
\end{array}%
\right) \sim \mathcal{N}\left( \left(
\begin{array}{c}
\mu  \\
x^{\top }\mu
\end{array}%
\right) ,\left(
\begin{array}{ll}
\Sigma  & \Sigma x \\
x^{\top }\Sigma  & x^{\top }\Sigma x%
\end{array}%
\right) \right)
\end{equation*}%
We deduce that:%
\begin{equation*}
\left(
\begin{array}{c}
R_{i} \\
R\left( x\right)
\end{array}%
\right) \sim \mathcal{N}\left( \left(
\begin{array}{c}
\mu _{i} \\
x^{\top }\mu
\end{array}%
\right) ,\left(
\begin{array}{ll}
\Sigma _{i,i} & \left( \Sigma x\right) _{i} \\
\left( \Sigma x\right) _{i} & x^{\top }\Sigma x%
\end{array}%
\right) \right)
\end{equation*}%
Let $I=\mathbb{E}\left[ R_{i}\cdot \mathds{1}\left\{ R\left( x\right) \leq -%
\limfunc{VaR}\nolimits_{\alpha }\left( L\right) \right\} \right] $. We note $%
f$ the density function of the random vector $\left( R_{i},R\left( x\right)
\right) $ and $\rho =\Sigma _{i,i}^{-1/2}\left( x^{\top }\Sigma x\right)
^{-1/2}\left( \Sigma x\right) _{i}$ the correlation between $R_{i}$ and $%
R\left( x\right) $. It follows that:%
\begin{eqnarray*}
I &=&\int_{-\infty }^{+\infty }\int_{-\infty }^{+\infty }r\cdot \mathbf{1}%
\left\{ s\leq -\limfunc{VaR}\nolimits_{\alpha }\left( L\right) \right\}
f\left( r,s\right) \,\mathrm{d}r\,\mathrm{d}s \\
&=&\int_{-\infty }^{+\infty }\int_{-\infty }^{-\limfunc{VaR}%
\nolimits_{\alpha }\left( L\right) }rf\left( r,s\right) \,\mathrm{d}r\,%
\mathrm{d}s
\end{eqnarray*}%
Let $t=\left( r-\mu _{i}\right) /\sqrt{\Sigma _{i,i}}$ and $u=\left(
s-x^{\top }\mu \right) /\sqrt{x^{\top }\Sigma x}$. We deduce that\footnote{%
Because we have $\Phi ^{-1}\left( 1-\alpha \right) =-\Phi ^{-1}\left( \alpha
\right) $.}:%
\begin{equation*}
I=\int_{-\infty }^{+\infty }\int_{-\infty }^{\Phi ^{-1}\left( 1-\alpha
\right) }\frac{\mu _{i}+\sqrt{\Sigma _{i,i}}t}{2\pi \sqrt{1-\rho ^{2}}}\exp
\left( -\frac{t^{2}+u^{2}-2\rho tu}{2\left( 1-\rho ^{2}\right) }\right) \,%
\mathrm{d}t\,\mathrm{d}u
\end{equation*}%
By considering the change of variables $\left( t,u\right) =\varphi \left(
t,v\right) $ such that $u=\rho t+\sqrt{1-\rho ^{2}}v$, we obtain\footnote{%
We use the fact that $\mathrm{d}t\,\mathrm{d}v=\sqrt{1-\rho ^{2}}\,\mathrm{d}%
t\,\mathrm{d}u$ because the determinant of the Jacobian matrix containing
the partial derivatives $D\varphi $ is $\sqrt{1-\rho ^{2}}$.}:
\begin{eqnarray*}
I &=&\int_{-\infty }^{+\infty }\int_{-\infty }^{g\left( t\right) }\frac{\mu
_{i}+\sqrt{\Sigma _{i,i}}t}{2\pi }\exp \left( -\frac{t^{2}+v^{2}}{2}\right)
\,\mathrm{d}t\,\mathrm{d}v \\
&=&\mu _{i}\int_{-\infty }^{+\infty }\int_{-\infty }^{g\left( t\right) }%
\frac{1}{2\pi }\exp \left( -\frac{t^{2}+v^{2}}{2}\right) \,\mathrm{d}t\,%
\mathrm{d}v+ \\
&&\sqrt{\Sigma _{i,i}}\int_{-\infty }^{+\infty }\int_{-\infty }^{g\left(
t\right) }\frac{t}{2\pi }\exp \left( -\frac{t^{2}+v^{2}}{2}\right) \,\mathrm{%
d}t\,\mathrm{d}v+ \\
&=&\mu _{i}I_{1}+\sqrt{\Sigma _{i,i}}I_{2}
\end{eqnarray*}%
where the bound $g\left( t\right) $ is defined as follows:%
\begin{equation*}
g\left( t\right) =\frac{\Phi ^{-1}\left( 1-\alpha \right) -\rho t}{\sqrt{%
1-\rho ^{2}}}
\end{equation*}%
For the first integral, we have\footnote{%
We use the fact that:%
\begin{equation*}
\mathbb{E}\left[ \Phi \left( \frac{\Phi ^{-1}\left( p\right) -\rho T}{\sqrt{%
1-\rho ^{2}}}\right) \right] =p
\end{equation*}%
where $T\sim \mathcal{N}\left( 0,1\right) $.}:%
\begin{eqnarray*}
I_{1} &=&\int_{-\infty }^{+\infty }%
\frac{1}{\sqrt{2\pi }}\exp \left( -\frac{t^{2}}{2}\right) \left(
\int_{-\infty }^{g\left( t\right) }\frac{1}{\sqrt{2\pi }}\exp \left( -\frac{%
v^{2}}{2}\right) \,\mathrm{d}v\right) \,\mathrm{d}t \\
&=&\int_{-\infty }^{+\infty }\Phi \left( \frac{\Phi ^{-1}\left( 1-\alpha
\right) -\rho t}{\sqrt{1-\rho ^{2}}}\right) \phi \left( t\right) \,\mathrm{d}%
t \\
&=&1-\alpha
\end{eqnarray*}%
The computation of the second integral $I_{2}$ is a little bit more tedious.
Integration by parts with the derivative function $t\phi \left( t\right) $
gives:%
\begin{eqnarray*}
I_{2} &=&\int_{-\infty }^{+\infty }\Phi \left( \frac{\Phi ^{-1}\left(
1-\alpha \right) -\rho t}{\sqrt{1-\rho ^{2}}}\right) t\phi \left( t\right) \,%
\mathrm{d}t \\
&=&-\frac{\rho }{\sqrt{1-\rho ^{2}}}\int_{-\infty }^{+\infty }\phi \left(
\frac{\Phi ^{-1}\left( 1-\alpha \right) -\rho t}{\sqrt{1-\rho ^{2}}}\right)
\phi \left( t\right) \,\mathrm{d}t \\
&=&-\frac{\rho }{\sqrt{1-\rho ^{2}}}\phi \left( \Phi ^{-1}\left( 1-\alpha
\right) \right) \int_{-\infty }^{+\infty }\phi \left( \frac{t-\rho \Phi
^{-1}\left( 1-\alpha \right) }{\sqrt{1-\rho ^{2}}}\right) \,\mathrm{d}t \\
&=&-\rho \phi \left( \Phi ^{-1}\left( 1-\alpha \right) \right)
\end{eqnarray*}%
We could then deduce the value of $I$:%
\begin{eqnarray*}
I &=&\mu _{i}\left( 1-\alpha \right) -\rho \sqrt{\Sigma _{i,i}}\phi \left(
\Phi ^{-1}\left( 1-\alpha \right) \right)  \\
&=&\mu _{i}\left( 1-\alpha \right) -\frac{\left( \Sigma x\right) _{i}}{\sqrt{%
x^{\top }\Sigma x}}\phi \left( \Phi ^{-1}\left( \alpha \right) \right)
\end{eqnarray*}%
We finally obtain that:%
\begin{equation*}
\mathcal{RC}_{i}=-x_{i}\mu _{i}+\frac{\phi \left( \Phi ^{-1}\left( \alpha
\right) \right) }{\left( 1-\alpha \right) }\frac{x_{i}\cdot \left( \Sigma
x\right) _{i}}{\sqrt{x^{\top }\Sigma x}}
\end{equation*}%
We obtain the same expression as found in Question 1(c).

\item We have%
\begin{equation*}
\mathcal{RC}_{i}=-\frac{x_{i}}{1-\alpha }\mathbb{E}\left[ R_{i}\cdot \mathbf{%
1}\left\{ R\left( x\right) \leq -\limfunc{VaR}\nolimits_{\alpha }\left(
L\right) \right\} \right]
\end{equation*}%
Let $R_{i,t}$ be the asset return for the observation $t$. The portfolio
return is then $R_{t}\left( x\right) =\sum_{i=1}^{n}x_{i}R_{i,t}$ at time $t$%
. We note $R_{j:T}\left( x\right) $ the $j^{\mathrm{th}}$ order
statistic. The estimated value-at-risk is then:%
\begin{equation*}
\limfunc{VaR}\nolimits_{\alpha }=-R_{\left( 1-\alpha \right) T:T}\left(
x\right)
\end{equation*}%
We deduce that the estimated risk contribution is:%
\begin{equation*}
\mathcal{RC}_{i}=-\frac{x_{i}}{\left( 1-\alpha \right) T}%
\sum_{t=1}^{T}R_{i,t}\cdot \mathds{1}\left\{ R_{t}\left( x\right) \leq
R_{\left( 1-\alpha \right) T:T}\left( x\right) \right\}
\end{equation*}

\item We note $\mathcal{RC}_{i}^{\left( j\right) }$ the estimated risk
contribution of the asset $i$ for the simulation $j$:%
\begin{equation*}
\mathcal{RC}_{i}^{\left( j\right) }=-\frac{x_{i}}{\left( 1-\alpha \right) T}%
\sum_{t=1}^{T}R_{i,t}^{\left( j\right) }\cdot \mathds{1}\left\{
R_{t}^{\left( j\right) }\left( x\right) \leq R_{\left( 1-\alpha \right)
T:T}^{\left( j\right) }\left( x\right) \right\}
\end{equation*}%
where $R_{i,t}^{\left( j\right) }$ is the simulated value of $R_{i,t}$ for the
simulation $j$. We consider one million of simulated observations\footnote{%
It means that $T=1\,000\,000$.} and 100 Monte Carlo replications. We estimate
the risk contribution as the average of $\mathcal{RC}_{i}^{\left( j\right) }$%
:%
\begin{equation*}
\mathcal{RC}_{i}=\frac{1}{100}\sum_{j=1}^{100}\mathcal{RC}_{i}^{\left(
j\right) }
\end{equation*}%
Using the numerical values of the parameters, we obtain the following
results:%
\begin{equation*}
\begin{tabular}{|c|rrrr|}
\hline
Asset & \multicolumn{1}{c}{$x_{i}$} & \multicolumn{1}{c}{$\mathcal{MR}_{i}$} &
\multicolumn{1}{c}{$\mathcal{RC}_{i}$} & \multicolumn{1}{c|}{$\mathcal{RC}_{i}^{\star }$} \\
\hline
$1$                          & $30.00\%$ & $29.12\%$ & $ 8.74\%$ & $25.91\%$  \\
$2$                          & $30.00\%$ & $36.48\%$ & $10.94\%$ & $32.46\%$  \\
$3$                          & $40.00\%$ & $35.11\%$ & $14.04\%$ & $41.63\%$  \\ \hdashline
$\limfunc{ES}\nolimits_{\alpha }\left( x\right)$ & & & $33.73\%$ &             \\
\hline
\end{tabular}%
\end{equation*}
If $\nu _{i}\rightarrow \infty $, we have:%
\begin{equation*}
\frac{R_{i}-\mu _{i}}{\sigma _{i}}\sim \mathcal{N}\left( 0,1\right)
\end{equation*}%
One would think that this numerical application is very close to the one
given in Question 2(a). However, we obtain very different risk
contributions. When we assume that asset returns are Gaussian, the risk
contribution of the third asset is negative ($\mathcal{RC}_{i}^{\star
}=-2.41\%$) whereas the third asset is the main contributor here ($\mathcal{%
RC}_{i}^{\star }=41.63\%$). It is due to the fat tail effect of the Student $%
t$ distribution. This effect also explains why the expected shortfall has been multiplied by a factor greater than $2$.
\end{enumerate}
\end{enumerate}

\section{ERC optimization problem}

\begin{enumerate}
\item

\begin{enumerate}
\item The weights of the three portfolios are:
\begin{equation*}
\begin{tabular}{|c|rrr|}
\hline
Asset & \multicolumn{1}{c}{MV} & \multicolumn{1}{c}{ERC} & \multicolumn{1}{c|}{EW} \\
\hline
$1$                          & $87.51\%$ & $37.01\%$ & $25.00\%$  \\
$2$                          & $ 4.05\%$ & $24.68\%$ & $25.00\%$  \\
$3$                          & $ 4.81\%$ & $20.65\%$ & $25.00\%$  \\
$4$                          & $ 3.64\%$ & $17.66\%$ & $25.00\%$  \\
\hline
\end{tabular}%
\end{equation*}

\item The Lagrange function is:%
\begin{eqnarray*}
\mathcal{L}\left( x;\lambda ,\lambda _{0},\lambda _{c}\right)  &=&\sqrt{%
x^{\top }\Sigma x}-\lambda ^{\top }x-\lambda _{0}\left( \mathbf{1}^{\top
}x-1\right) - \\
& & \lambda _{c}\left( \sum_{i=1}^{n}\ln x_{i}-c\right)  \\
&=&\left( \sqrt{x^{\top }\Sigma x}-\lambda _{c}\sum_{i=1}^{n}\ln
x_{i}\right) -\lambda ^{\top }x- \\
& & \lambda _{0}\left( \mathbf{1}^{\top
}x-1\right) +\lambda _{c}c
\end{eqnarray*}%
We deduce that an equivalent optimization problem is:%
\begin{eqnarray*}
\tilde{x}^{\star }\left( \lambda _{c}\right)  &=&\arg \min \sqrt{\tilde{x}%
^{\top }\Sigma \tilde{x}}-\lambda _{c}\sum_{i=1}^{n}\ln \tilde{x}_{i} \\
&\text{u.c.}&\left\{
\begin{array}{l}
\mathbf{1}^{\top }\tilde{x}=1 \\
\tilde{x}\geq \mathbf{0}%
\end{array}%
\right.
\end{eqnarray*}%
We notice a strong difference between the two problems because they don't
use the same control variable. However, the control variable $c$ of the
first problem may be deduced from the solution of the second problem:%
\begin{equation*}
c=\sum_{i=1}^{n}\ln \tilde{x}_{i}^{\star }\left( \lambda _{c}\right)
\end{equation*}%
We also know that (TR-RPB, page 131):%
\begin{equation*}
c_{-}\leq \sum_{i=1}^{n}\ln x_{i}\leq c_{+}
\end{equation*}%
where $c_{-}=\sum_{i=1}^{n}\ln \left( x_{\mathrm{mv}}\right) _{i}$ and $%
c_{+}=-n\ln n$. It follows that:%
\begin{equation*}
\left\{
\begin{array}{ll}
x^{\star }\left( c\right) =\tilde{x}^{\star }\left( 0\right)  & \text{if}\quad%
c\leq c_{-} \\
x^{\star }\left( c\right) =\tilde{x}^{\star }\left( \infty \right)  & \text{if}\quad%
c\geq c_{+}%
\end{array}%
\right.
\end{equation*}%
If $c\in \left] c_{-},c_{+}\right[ $, there exists a scalar $\lambda _{c}>0$
such that:%
\begin{equation*}
x^{\star }\left( c\right) =\tilde{x}^{\star }\left( \lambda _{c}\right)
\end{equation*}

\item For a given value $\lambda _{c}\in \left[ 0,+\infty \right[ $, we
solve numerically the second problem and find the optimized portfolio $%
\tilde{x}^{\star }\left( \lambda _{c}\right) $. Then, we calculate $%
c=\sum_{i=1}^{n}\ln \tilde{x}_{i}^{\star }\left( \lambda _{c}\right) $ and
deduce that $x^{\star }\left( c\right) =\tilde{x}^{\star }\left( \lambda
_{c}\right) $. We finally obtain $\sigma \left( x^{\star }\left( c\right)
\right) =\sigma \left( \tilde{x}^{\star }\left( \lambda _{c}\right) \right) $
and $\mathcal{I}^{\star }\left( x^{\star }\left( c\right) \right) =\mathcal{I%
}^{\star }\left( \tilde{x}^{\star }\left( \lambda _{c}\right) \right) $. The
relationships between $\lambda _{c}$, $c$, $\mathcal{I}^{\star }\left(
x^{\star }\left( c\right) \right) $ and $\sigma \left( x^{\star }\left(
c\right) \right) $ are reported in Figure \ref{fig:app2-2-8-1}.

\begin{figure}[tbph]
\centering
\includegraphics[width = \figurewidth, height = \figureheight]{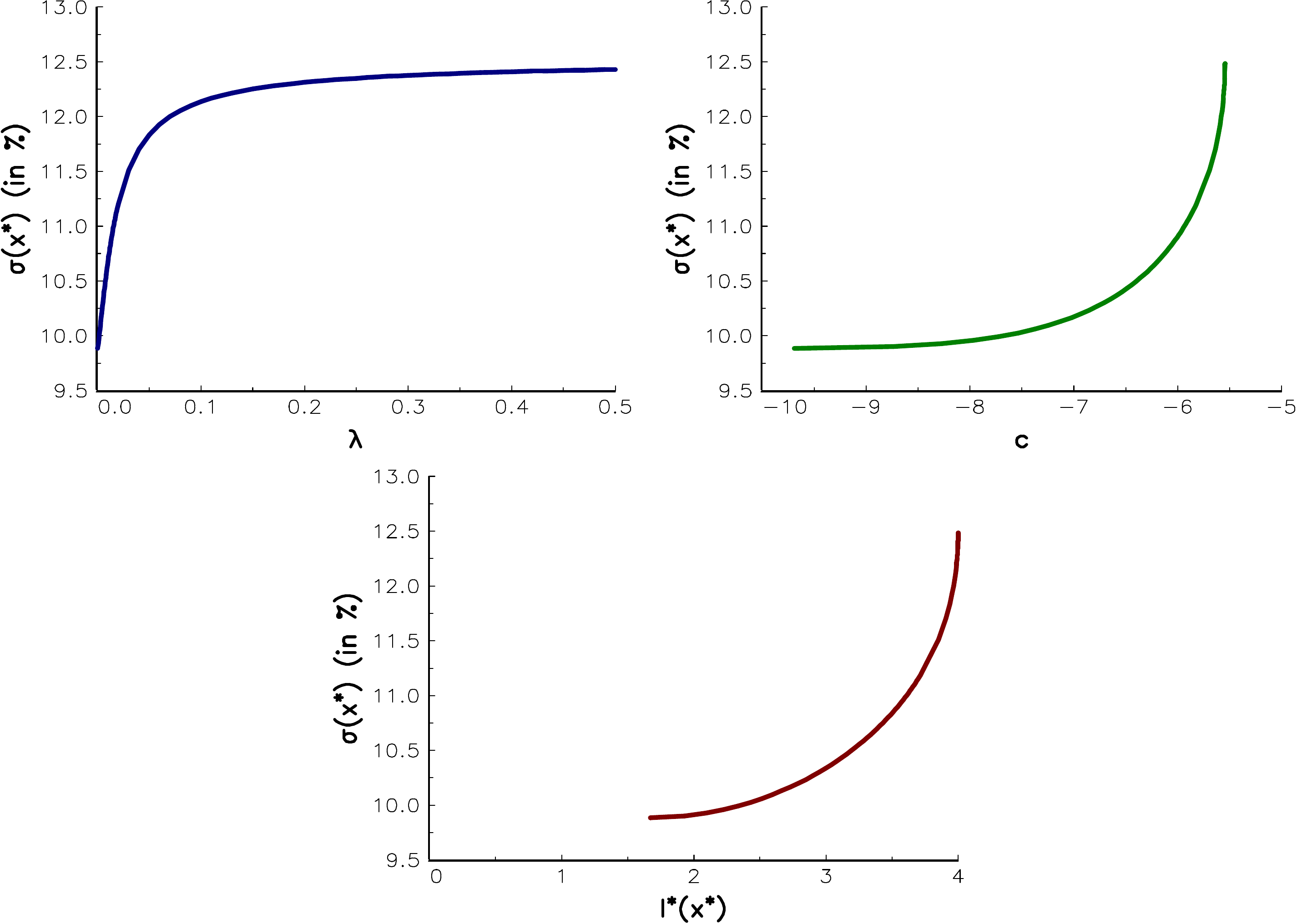}
\caption{Relationship between $\lambda _{c}$, $c$, $\mathcal{I}^{\star }\left(
x^{\star }\left( c\right) \right) $ and $\sigma \left( x^{\star }\left(
c\right) \right) $}
\label{fig:app2-2-8-1}
\end{figure}

\item If we consider $\mathcal{I}^{\star }\left( \mathcal{RC}\right) $ in place of
$\sigma \left( x^{\star }\left(c\right) \right) $, we obtain Figure \ref{fig:app2-2-8-2}.

\begin{figure}[tbph]
\centering
\includegraphics[width = \figurewidth, height = \figureheight]{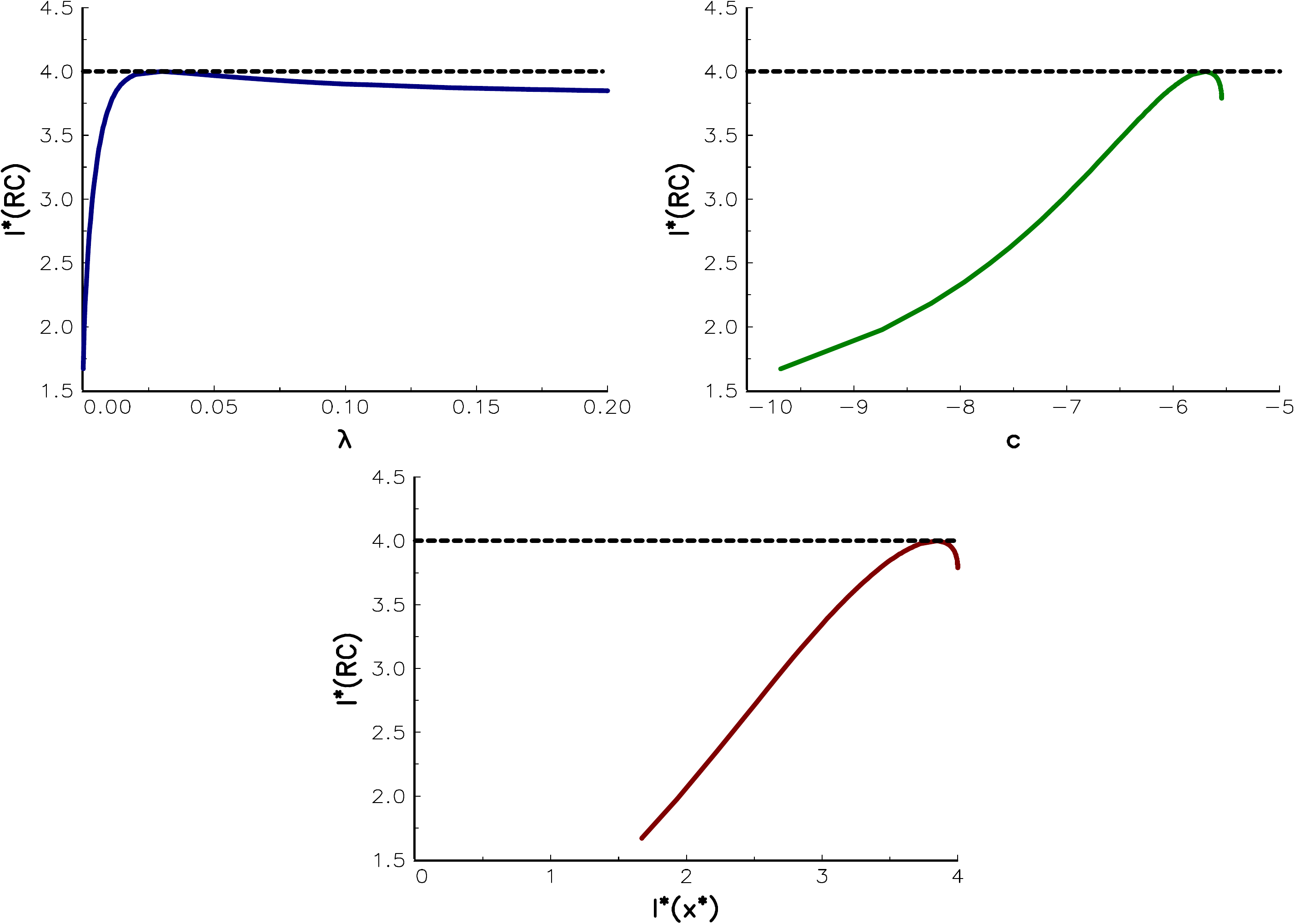}
\caption{Relationship between $\lambda _{c}$, $c$, $\mathcal{I}^{\star }\left(
x^{\star }\left( c\right) \right) $ and $\mathcal{I}^{\star }\left( \mathcal{RC}\right) $}
\label{fig:app2-2-8-2}
\end{figure}

\item In Figure \ref{fig:app2-2-8-3}, we have reported the relationship between $\sigma \left(
x^{\star }\left( c\right) \right) $ and $\mathcal{I}^{\star }\left( \mathcal{%
RC}\right) $. The ERC portfolio satisfies the equation $\mathcal{I}^{\star
}\left( \mathcal{RC}\right) =n$.
\end{enumerate}

\begin{figure}[tbph]
\centering
\includegraphics[width = \figurewidth, height = \figureheight]{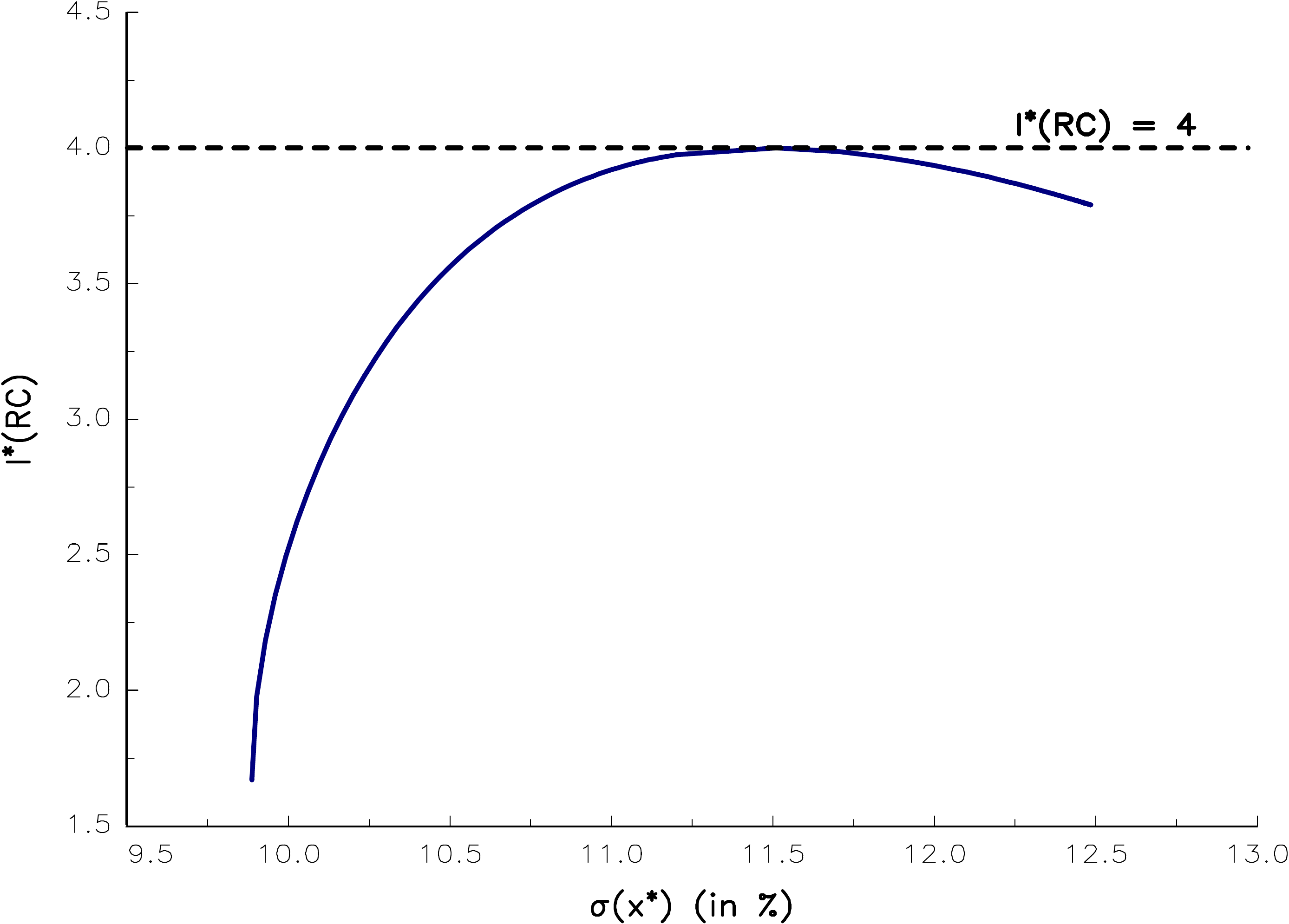}
\caption{Relationship between $\sigma \left( x^{\star }\left(c\right) \right) $
and $\mathcal{I}^{\star }\left( \mathcal{RC}\right) $}
\label{fig:app2-2-8-3}
\end{figure}

\item

\begin{enumerate}
\item Let us consider the optimization problem when we impose the constraint
$\mathbf{1}^{\top }x=1$. The first-order condition is:%
\begin{equation*}
\frac{\partial \,\sigma \left( x\right) }{\partial \,x_{i}}-\lambda
_{i}-\lambda _{0}-\frac{\lambda _{c}}{x_{i}}=0
\end{equation*}%
Because $x_{i}>0$, we deduce that $\lambda _{i}=0$ and:%
\begin{equation*}
x_{i}\frac{\partial \,\sigma \left( x\right) }{\partial \,x_{i}}=\lambda
_{0}x_{i}+\lambda _{c}
\end{equation*}%
If this solution corresponds to the ERC portfolio, we obtain:%
\begin{equation*}
\mathcal{RC}_{i}=\mathcal{RC}_{j}\Leftrightarrow \lambda _{0}x_{i}+\lambda
_{c}=\lambda _{0}x_{j}+\lambda _{c}
\end{equation*}%
If $\lambda _{0}\neq 0$, we deduce that:%
\begin{equation*}
x_{i}=x_{j}
\end{equation*}%
It corresponds to the EW portfolio meaning that the assumption $\mathcal{RC}%
_{i}=\mathcal{RC}_{j}$ is false.

\item If $c$ is equal to $-10$, we obtain the following results:%
\begin{equation*}
\begin{tabular}{|c|rrrr|}
\hline
Asset & \multicolumn{1}{c}{$x_{i}$} & \multicolumn{1}{c}{$\mathcal{MR}_{i}$} &
\multicolumn{1}{c}{$\mathcal{RC}_{i}$} & \multicolumn{1}{c|}{$\mathcal{RC}_{i}^{\star }$} \\
\hline
$1$                          & $12.65\%$ & $ 7.75\%$ & $0.98\%$ & $25.00\%$  \\
$2$                          & $ 8.43\%$ & $11.63\%$ & $0.98\%$ & $25.00\%$  \\
$3$                          & $ 7.06\%$ & $13.89\%$ & $0.98\%$ & $25.00\%$  \\
$4$                          & $ 6.03\%$ & $16.25\%$ & $0.98\%$ & $25.00\%$  \\ \hdashline
$\sigma\left( x\right)$                          & & & $3.92\%$ &             \\
\hline
\end{tabular}%
\end{equation*}

\item If $c$ is equal to $0$, we obtain the following results:%
\begin{equation*}
\begin{tabular}{|c|rrrr|}
\hline
Asset & \multicolumn{1}{c}{$x_{i}$} & \multicolumn{1}{c}{$\mathcal{MR}_{i}$} &
\multicolumn{1}{c}{$\mathcal{RC}_{i}$} & \multicolumn{1}{c|}{$\mathcal{RC}_{i}^{\star }$} \\
\hline
$1$                          & $154.07\%$ & $ 7.75\%$ & $11.94\%$ & $25.00\%$  \\
$2$                          & $102.72\%$ & $11.63\%$ & $11.94\%$ & $25.00\%$  \\
$3$                          & $ 85.97\%$ & $13.89\%$ & $11.94\%$ & $25.00\%$  \\
$4$                          & $ 73.50\%$ & $16.25\%$ & $11.94\%$ & $25.00\%$  \\ \hdashline
$\sigma\left( x\right)$                           & & &  $47.78\%$ &             \\
\hline
\end{tabular}%
\end{equation*}

\item In this case, the first-order condition is:%
\begin{equation*}
\frac{\partial \,\sigma \left( x\right) }{\partial \,x_{i}}-\lambda _{i}-%
\frac{\lambda _{c}}{x_{i}}=0
\end{equation*}%
As previously, $\lambda _{i}=0$ because $x_{i}>0$ and we obtain:%
\begin{equation*}
x_{i}\frac{\partial \,\sigma \left( x\right) }{\partial \,x_{i}}=\lambda _{c}
\end{equation*}%
The solution of the second optimization problem is then a non-normalized ERC
portfolio because $\sum_{i=1}^{n} x_i$ is not necessarily equal to 1. If we note
$c_{\mathrm{erc}}=\sum_{i=1}^{n}\ln \left( x_{\mathrm{erc}}\right) _{i}$, we
deduce that:%
\begin{eqnarray*}
x_{\mathrm{erc}} &=&\arg \min \sqrt{x^{\top }\Sigma x} \\
&\text{u.c.}&\left\{
\begin{array}{l}
\sum_{i=1}^{n}\ln x_{i}\geq c_{\mathrm{erc}} \\
x\geq \mathbf{0}%
\end{array}%
\right.
\end{eqnarray*}%
Let $x^{\star }\left( c\right) $ be the portfolio defined by:%
\begin{equation*}
x^{\star }\left( c\right) =\exp \left( \frac{c-c_{\mathrm{erc}}}{n}\right)
x_{\mathrm{erc}}
\end{equation*}%
We have $x^{\star }\left( c\right) >\mathbf{0}$,
\begin{equation*}
\sqrt{x^{\star }\left( c\right) ^{\top }\Sigma x^{\star }\left( c\right) }%
=\exp \left( \frac{c-c_{\mathrm{erc}}}{n}\right) \sqrt{x_{\mathrm{erc}%
}^{\top }\Sigma x_{\mathrm{erc}}}
\end{equation*}%
and:%
\begin{eqnarray*}
\sum_{i=1}^{n}\ln x_{i}^{\star }\left( c\right)  &=&\sum_{i=1}^{n}\ln \left(
\exp \left( \frac{c-c_{\mathrm{erc}}}{n}\right) x_{\mathrm{erc}}\right) _{i}
\\
&=&c-c_{\mathrm{erc}}+\sum_{i=1}^{n}\ln \left( x_{\mathrm{erc}}\right) _{i} \\
&=& c
\end{eqnarray*}%
We conclude that $x^{\star }\left( c\right) $ is the solution of the
optimization problem. $x^{\star }\left( c\right) $ is then a leveraged ERC
portfolio if $c>c_{\mathrm{erc}}$ and a deleveraged ERC portfolio if $c<c_{%
\mathrm{erc}}$. In our example, $c_{\mathrm{erc}}$ is equal to $-5.7046$. If
$c=-10$, we have:%
\begin{equation*}
\exp \left( \frac{c-c_{\mathrm{erc}}}{n}\right) =34.17\%
\end{equation*}%
We verify that the solution of Question 2(b) is such that $%
\sum_{i=1}^{n}x_{i}=34.17\%$ and $RC_{i}^{\star }=RC_{j}^{\star }$. If $c=0$%
, we obtain:%
\begin{equation*}
\exp \left( \frac{c-c_{\mathrm{erc}}}{n}\right) =416.26\%
\end{equation*}%
In this case, the solution is a leveraged ERC portfolio.

\item From the previous question, we know that the ERC optimization
portfolio is the solution of the second optimization problem if we use $c_{%
\mathrm{erc}}$ for the control variable. In this case, we have $%
\sum_{i=1}^{n}x_{i}^{\star }\left( c_{\mathrm{erc}}\right) =1$ meaning that $%
x_{\mathrm{erc}}$ is also the solution of the first optimization problem. We
deduce that $\lambda _{0}=0$ if $c=c_{\mathrm{erc}}$. The first optimization
problem is a convex problem with a convex inequality constraint.
The objective function is then an increasing function of the control
variable $c$:%
\begin{equation*}
c_{1}\leq c_{2}\Rightarrow \sigma \left( x^{\star }\left( c_{1}\right)
\right) \geq \sigma \left( x^{\star }\left( c_{2}\right) \right)
\end{equation*}%
We have seen that the minimum variance portfolio corresponds to $c=-\infty $,
that the EW portfolio is obtained with $c=-n\ln n$ and that the ERC portfolio is the
solution of the optimization problem when $c$ is equal to $c_{\mathrm{erc}}$.
Moreover, we have $-\infty \leq c_{\mathrm{erc}}\leq -n\ln n$. We deduce
that the volatility of the ERC portfolio is between the volatility of the
long-only minimum variance portfolio and the volatility of the
equally weighted portfolio:%
\begin{equation*}
\sigma \left( x_{\mathrm{mv}}\right) \leq \sigma \left( x_{\mathrm{erc}%
}\right) \leq \sigma \left( x_{\mathrm{ew}}\right)
\end{equation*}
\end{enumerate}
\end{enumerate}

\section{Risk parity portfolios with skewness and kurtosis}

\begin{enumerate}
\item

\begin{enumerate}
\item We use the formulas given in TR-RPB on page 94. The mean $\mu $
corresponds to $M_{1}$:%
\begin{equation*}
\mu =\left(
\begin{array}{r}
0.225\% \\
0.099\% \\
0.087\%%
\end{array}%
\right)
\end{equation*}%
Let $\mu _{r}$ be the centered $r$-order moment. We have $\sigma =\sqrt{\mu
_{2}}$, $\gamma _{1}=\mu _{3}/\mu _{2}^{3/2}$ and $\gamma _{2}=\mu _{4}/\mu
_{2}^{2}-3$. The difficulty is to read the good value of $\mu _{r}^{\left(
i\right) }$ for the asset $i$ from the matrices $M_{2}$, $M_{3}$ and $M_{4}$%
. We have $\mu _{r}^{\left( i\right) }=\left( M_{r}\right) _{i,j}$ where the
index $j$ is given by the following table:
\begin{equation*}
\begin{tabular}{|c|ccc|}
\hline
& \multicolumn{3}{c|}{$r$} \\
$i$ & $2$ & $3$ & $4$ \\ \hline
$1$ & $1$ & $1$ & $1$ \\
$2$ & $2$ & $5$ & $14$ \\
$3$ & $3$ & $9$ & $27$ \\ \hline
\end{tabular}%
\end{equation*}%
The volatility is then:%
\begin{equation*}
\sigma =\left(
\begin{array}{r}
2.652\% \\
0.874\% \\
2.498\%%
\end{array}%
\right)
\end{equation*}%
For the skewness, we obtain:%
\begin{equation*}
\gamma _{1}=\left(
\begin{array}{r}
-0.351 \\
-0.027 \\
-0.248%
\end{array}%
\right)
\end{equation*}%
whereas the excess kurtosis is:%
\begin{equation*}
\gamma _{2}=\left(
\begin{array}{r}
1.242 \\
-0.088 \\
0.930%
\end{array}%
\right)
\end{equation*}

\item Let $x$ be the portfolio. We know that (TR-RPB, page 94):%
\begin{equation*}
\mu _{r}\left( \Pi \right) =xM_{r}\left( \overset{r}{\underset{j=1}{\otimes }%
}x\right)
\end{equation*}%
We obtain $\mu _{1}\left( \Pi \right) =13.732\times 10^{-4}$, $\mu
_{2}\left( \Pi \right) =2.706\times 10^{-4}$, $\mu _{3}\left( \Pi \right)
=-1.117\times 10^{-6}$ and $\mu _{4}\left( \Pi \right) =0.252\times 10^{-6}$.

\item We have:%
\begin{eqnarray*}
\mu \left( L\right)  &=&-\mu _{1}\left( \Pi \right) =-0.137\% \\
\sigma \left( L\right)  &=&\sqrt{\mu _{2}\left( \Pi \right) }=1.645\% \\
\gamma _{1}\left( L\right)  &=&-\frac{\mu _{3}\left( \Pi \right) }{\sigma
^{3}\left( L\right) }=0.251 \\
\gamma _{2}\left( L\right)  &=&\frac{\mu _{4}\left( \Pi \right) }{\sigma
^{4}\left( L\right) }=0.438
\end{eqnarray*}

\item The risk allocation of the EW portfolio is:
\begin{equation*}
\begin{tabular}{|c|rrrr|}
\hline Asset & \multicolumn{1}{c}{$x_{i}$} &
\multicolumn{1}{c}{$\mathcal{MR}_{i}$} &
\multicolumn{1}{c}{$\mathcal{RC}_{i}$} & \multicolumn{1}{c|}{$\mathcal{RC}_{i}^{\star }$} \\
\hline
$1$ & $33.333\%$ & $3.719\%$ & $1.240\%$ & $48.272\%$ \\
$2$ & $33.333\%$ & $0.372\%$ & $0.124\%$ & $ 4.825\%$ \\
$3$ & $33.333\%$ & $3.614\%$ & $1.205\%$ & $46.903\%$ \\ \hdashline
$\limfunc{VaR}\nolimits_{\alpha }\left( x\right)$ & &  & $2.568\%$ &             \\
\hline
\end{tabular}%
\end{equation*}

\item If we consider the Cornish-Fisher value-at-risk, we get:
\begin{equation*}
\begin{tabular}{|c|rrrr|}
\hline Asset & \multicolumn{1}{c}{$x_{i}$} &
\multicolumn{1}{c}{$\mathcal{MR}_{i}$} &
\multicolumn{1}{c}{$\mathcal{RC}_{i}$} & \multicolumn{1}{c|}{$\mathcal{RC}_{i}^{\star }$} \\
\hline
$1$ & $33.333\%$ & $3.919\%$ & $1.306\%$ & $48.938\%$ \\
$2$ & $33.333\%$ & $0.319\%$ & $0.106\%$ & $ 3.977\%$ \\
$3$ & $33.333\%$ & $3.770\%$ & $1.257\%$ & $47.085\%$ \\ \hdashline
$\limfunc{VaR}\nolimits_{\alpha }\left( x\right)$ & &  & $2.669\%$ &             \\
\hline
\end{tabular}%
\end{equation*}

\end{enumerate}

\item

\begin{enumerate}
\item We obtain the following results:
\begin{equation*}
\begin{tabular}{|c|rrrr|}
\hline Asset & \multicolumn{1}{c}{$x_{i}$} &
\multicolumn{1}{c}{$\mathcal{MR}_{i}$} &
\multicolumn{1}{c}{$\mathcal{RC}_{i}$} & \multicolumn{1}{c|}{$\mathcal{RC}_{i}^{\star }$} \\
\hline
$1$ & $17.371\%$ & $3.151\%$ & $0.547\%$ & $33.333\%$ \\
$2$ & $65.208\%$ & $0.840\%$ & $0.547\%$ & $33.333\%$ \\
$3$ & $17.421\%$ & $3.142\%$ & $0.547\%$ & $33.333\%$ \\ \hdashline
$\limfunc{VaR}\nolimits_{\alpha }\left( x\right)$ & &  & $1.642\%$ &             \\
\hline
\end{tabular}%
\end{equation*}

\item If we consider the Cornish-Fisher value-at-risk, the results become:
\begin{equation*}
\begin{tabular}{|c|rrrr|}
\hline Asset & \multicolumn{1}{c}{$x_{i}$} &
\multicolumn{1}{c}{$\mathcal{MR}_{i}$} &
\multicolumn{1}{c}{$\mathcal{RC}_{i}$} & \multicolumn{1}{c|}{$\mathcal{RC}_{i}^{\star }$} \\
\hline
$1$ & $17.139\%$ & $3.253\%$ & $0.558\%$ & $33.333\%$ \\
$2$ & $65.659\%$ & $0.849\%$ & $0.558\%$ & $33.333\%$ \\
$3$ & $17.202\%$ & $3.241\%$ & $0.558\%$ & $33.333\%$ \\ \hdashline
$\limfunc{VaR}\nolimits_{\alpha }\left( x\right)$ & &  & $1.673\%$ &             \\
\hline
\end{tabular}%
\end{equation*}
We notice that the weights of the portfolio are very close to the weights obtained with
the Gaussian value-at-risk. The impact of the skewness and kurtosis is thus limited.

\item

If $\alpha$ is equal to $99\%$, the ERC portfolio with the Gaussian value-at-risk is:
\begin{equation*}
\begin{tabular}{|c|rrrr|}
\hline Asset & \multicolumn{1}{c}{$x_{i}$} &
\multicolumn{1}{c}{$\mathcal{MR}_{i}$} &
\multicolumn{1}{c}{$\mathcal{RC}_{i}$} & \multicolumn{1}{c|}{$\mathcal{RC}_{i}^{\star }$} \\
\hline
$1$ & $17.403\%$ & $4.559\%$ & $0.793\%$ & $33.333\%$ \\
$2$ & $64.950\%$ & $1.222\%$ & $0.793\%$ & $33.333\%$ \\
$3$ & $17.647\%$ & $4.497\%$ & $0.793\%$ & $33.333\%$ \\ \hdashline
$\limfunc{VaR}\nolimits_{\alpha }\left( x\right)$ & &  & $2.380\%$ &             \\
\hline
\end{tabular}%
\end{equation*}
whereas the ERC portfolio with the Cornish-Fisher value-at-risk is:
\begin{equation*}
\begin{tabular}{|c|rrrr|}
\hline Asset & \multicolumn{1}{c}{$x_{i}$} &
\multicolumn{1}{c}{$\mathcal{MR}_{i}$} &
\multicolumn{1}{c}{$\mathcal{RC}_{i}$} & \multicolumn{1}{c|}{$\mathcal{RC}_{i}^{\star }$} \\
\hline
$1$ & $16.467\%$ & $4.770\%$ & $0.785\%$ & $33.333\%$ \\
$2$ & $67.860\%$ & $1.157\%$ & $0.785\%$ & $33.333\%$ \\
$3$ & $15.672\%$ & $5.012\%$ & $0.785\%$ & $33.333\%$ \\ \hdashline
$\limfunc{VaR}\nolimits_{\alpha }\left( x\right)$ & &  & $2.356\%$ &             \\
\hline
\end{tabular}%
\end{equation*}
The impact is higher. In particular, we see that the weight of bonds increases if we take
into account skewness and kurtosis.
\end{enumerate}
\end{enumerate}

\chapter{Exercises related to risk parity applications}

\section{Computation of heuristic portfolios}

\begin{enumerate}
\item All the results are expressed in \%.
\begin{enumerate}
\item To compute the unconstrained tangency portfolio, we use the analytical
formula (TR-RPB, page 14):%
\begin{equation*}
x^{\star }=\frac{\Sigma ^{-1}\left( \mu -r\mathbf{1}\right) }{\mathbf{1}%
^{\top }\Sigma ^{-1}\left( \mu -r\mathbf{1}\right) }
\end{equation*}%
We obtain the following results:%
\begin{equation*}
\begin{tabular}{|c|rrrr|}
\hline
Asset & \multicolumn{1}{c}{$x_{i}$}  & \multicolumn{1}{c}{$\mathcal{MR}_{i}$} &
        \multicolumn{1}{c}{$\mathcal{RC}_{i}$} & \multicolumn{1}{c|}{$\mathcal{RC}^{\star}_{i}$} \\ \hline
$1$  & $11.11\%$ & $ 6.56\%$ & $0.73\%$ & $ 5.96\%$ \\
$2$  & $17.98\%$ & $13.12\%$ & $2.36\%$ & $19.27\%$ \\
$3$  & $ 2.55\%$ & $ 6.56\%$ & $0.17\%$ & $ 1.37\%$ \\
$4$  & $33.96\%$ & $ 9.84\%$ & $3.34\%$ & $27.31\%$ \\
$5$  & $34.40\%$ & $16.40\%$ & $5.64\%$ & $46.09\%$ \\
\hline
\end{tabular}
\end{equation*}

\item We obtain the following results for the equally weighted portfolio:%
\begin{equation*}
\begin{tabular}{|c|rrrr|}
\hline
Asset & \multicolumn{1}{c}{$x_{i}$}  & \multicolumn{1}{c}{$\mathcal{MR}_{i}$} &
        \multicolumn{1}{c}{$\mathcal{RC}_{i}$} & \multicolumn{1}{c|}{$\mathcal{RC}^{\star}_{i}$} \\ \hline
$1$  & $20.00\%$ & $ 7.47\%$ & $1.49\%$ & $13.43\%$ \\
$2$  & $20.00\%$ & $15.83\%$ & $3.17\%$ & $28.48\%$ \\
$3$  & $20.00\%$ & $ 9.98\%$ & $2.00\%$ & $17.96\%$ \\
$4$  & $20.00\%$ & $ 9.89\%$ & $1.98\%$ & $17.80\%$ \\
$5$  & $20.00\%$ & $12.41\%$ & $2.48\%$ & $22.33\%$ \\
\hline
\end{tabular}
\end{equation*}

\item For the minimum variance portfolio, we have:%
\begin{equation*}
\begin{tabular}{|c|rrrr|}
\hline
Asset & \multicolumn{1}{c}{$x_{i}$}  & \multicolumn{1}{c}{$\mathcal{MR}_{i}$} &
        \multicolumn{1}{c}{$\mathcal{RC}_{i}$} & \multicolumn{1}{c|}{$\mathcal{RC}^{\star}_{i}$} \\ \hline
$1$  & $ 74.80\%$ & $9.08\%$ & $ 6.79\%$ & $ 74.80\%$ \\
$2$  & $-15.04\%$ & $9.08\%$ & $-1.37\%$ & $-15.04\%$ \\
$3$  & $ 21.63\%$ & $9.08\%$ & $ 1.96\%$ & $ 21.63\%$ \\
$4$  & $ 10.24\%$ & $9.08\%$ & $ 0.93\%$ & $ 10.24\%$ \\
$5$  & $  8.36\%$ & $9.08\%$ & $ 0.76\%$ & $  8.36\%$ \\
\hline
\end{tabular}
\end{equation*}

\item For the most diversified portfolio, we have:%
\begin{equation*}
\begin{tabular}{|c|rrrr|}
\hline
Asset & \multicolumn{1}{c}{$x_{i}$}  & \multicolumn{1}{c}{$\mathcal{MR}_{i}$} &
        \multicolumn{1}{c}{$\mathcal{RC}_{i}$} & \multicolumn{1}{c|}{$\mathcal{RC}^{\star}_{i}$} \\ \hline
$1$  & $-14.47\%$ & $ 4.88\%$ & $-0.71\%$ & $-5.34\%$ \\
$2$  & $  4.83\%$ & $ 9.75\%$ & $ 0.47\%$ & $ 3.56\%$ \\
$3$  & $ 18.94\%$ & $ 7.31\%$ & $ 1.38\%$ & $10.47\%$ \\
$4$  & $ 49.07\%$ & $12.19\%$ & $ 5.98\%$ & $45.24\%$ \\
$5$  & $ 41.63\%$ & $14.63\%$ & $ 6.09\%$ & $46.06\%$ \\
\hline
\end{tabular}
\end{equation*}

\item For the ERC portfolio, we have:%
\begin{equation*}
\begin{tabular}{|c|rrrr|}
\hline
Asset & \multicolumn{1}{c}{$x_{i}$}  & \multicolumn{1}{c}{$\mathcal{MR}_{i}$} &
        \multicolumn{1}{c}{$\mathcal{RC}_{i}$} & \multicolumn{1}{c|}{$\mathcal{RC}^{\star}_{i}$} \\ \hline
$1$  & $27.20\%$ & $ 7.78\%$ & $2.12\%$ & $20.00$ \\
$2$  & $13.95\%$ & $15.16\%$ & $2.12\%$ & $20.00$ \\
$3$  & $20.86\%$ & $10.14\%$ & $2.12\%$ & $20.00$ \\
$4$  & $19.83\%$ & $10.67\%$ & $2.12\%$ & $20.00$ \\
$5$  & $18.16\%$ & $11.65\%$ & $2.12\%$ & $20.00$ \\
\hline
\end{tabular}
\end{equation*}

\item We recall the definition of the statistics:%
\begin{eqnarray*}
\mu \left( x\right)  &=&\mu ^{\top }x \\
\sigma \left( x\right)  &=&\sqrt{x^{\top }\Sigma x} \\
\limfunc{SR}\left( x\mid r\right)  &=&\frac{\mu \left( x\right) -r}{\sigma
\left( x\right) } \\
\sigma \left( x\mid b\right)  &=&\sqrt{\left( x-b\right) ^{\top }\Sigma
\left( x-b\right) } \\
\beta \left( x\mid b\right)  &=&\frac{x^{\top }\Sigma b}{b^{\top }\Sigma b}
\\
\rho \left( x\mid b\right)  &=&\frac{x^{\top }\Sigma b}{\sqrt{x^{\top
}\Sigma x}\sqrt{b^{\top }\Sigma b}}
\end{eqnarray*}%
We obtain the following results:%
\begin{equation*}
\begin{tabular}{|c|rrrrr|}
\hline
Statistic & \multicolumn{1}{c}{$x^{\star}$}  & \multicolumn{1}{c}{$x_{\mathrm{ew}}$} &
        \multicolumn{1}{c}{$x_{\mathrm{mv}}$} & \multicolumn{1}{c}{$x_{\mathrm{mdp}}$} &
        \multicolumn{1}{c|}{$x_{\mathrm{erc}}$} \\ \hline
$\mu \left( x\right)$                & $  9.46\%$ & $ 8.40\%$ & $ 6.11\%$ & $  9.67\%$ & $ 8.04\%$ \\
$\sigma \left( x\right)$             & $ 12.24\%$ & $11.12\%$ & $ 9.08\%$ & $ 13.22\%$ & $10.58\%$ \\
$\limfunc{SR}\left( x\mid r\right)$  & $ 60.96\%$ & $57.57\%$ & $45.21\%$ & $ 58.03\%$ & $57.15\%$ \\
$\sigma \left( x\mid b\right)$       & $  0.00\%$ & $ 4.05\%$ & $ 8.21\%$ & $  4.06\%$ & $ 4.35\%$ \\
$\beta \left( x\mid b\right)$        & $100.00\%$ & $85.77\%$ & $55.01\%$ & $102.82\%$ & $81.00\%$ \\
$\rho \left( x\mid b\right)$         & $100.00\%$ & $94.44\%$ & $74.17\%$ & $ 95.19\%$ & $93.76\%$ \\
\hline
\end{tabular}
\end{equation*}
We notice that all the portfolios present similar performance in terms of Sharpe Ratio.
The minimum variance portfolio shows the smallest Sharpe ratio, but it also shows
the lowest correlation with the tangency portfolio.
\end{enumerate}

\item The tangency portfolio, the equally weighted portfolio and the ERC
portfolio are already long-only. For the minimum variance portfolio, we
obtain:%
\begin{equation*}
\begin{tabular}{|c|rrrr|}
\hline
Asset & \multicolumn{1}{c}{$x_{i}$}  & \multicolumn{1}{c}{$\mathcal{MR}_{i}$} &
        \multicolumn{1}{c}{$\mathcal{RC}_{i}$} & \multicolumn{1}{c|}{$\mathcal{RC}^{\star}_{i}$} \\ \hline
$1$  & $65.85\%$ & $ 9.37\%$ & $6.17\%$ & $65.85\%$ \\
$2$  & $ 0.00\%$ & $13.11\%$ & $0.00\%$ & $ 0.00\%$ \\
$3$  & $16.72\%$ & $ 9.37\%$ & $1.57\%$ & $16.72\%$ \\
$4$  & $ 9.12\%$ & $ 9.37\%$ & $0.85\%$ & $ 9.12\%$ \\
$5$  & $ 8.32\%$ & $ 9.37\%$ & $0.78\%$ & $ 8.32\%$ \\
\hline
\end{tabular}
\end{equation*}
whereas we have for the most diversified portfolio:%
\begin{equation*}
\begin{tabular}{|c|rrrr|}
\hline
Asset & \multicolumn{1}{c}{$x_{i}$}  & \multicolumn{1}{c}{$\mathcal{MR}_{i}$} &
        \multicolumn{1}{c}{$\mathcal{RC}_{i}$} & \multicolumn{1}{c|}{$\mathcal{RC}^{\star}_{i}$} \\ \hline
$1$  & $ 0.00\%$ & $ 5.50\%$ & $0.00\%$ & $ 0.00\%$ \\
$2$  & $ 1.58\%$ & $ 9.78\%$ & $0.15\%$ & $ 1.26\%$ \\
$3$  & $16.81\%$ & $ 7.34\%$ & $1.23\%$ & $10.04\%$ \\
$4$  & $44.13\%$ & $12.23\%$ & $5.40\%$ & $43.93\%$ \\
$5$  & $37.48\%$ & $14.68\%$ & $5.50\%$ & $44.77\%$ \\
\hline
\end{tabular}
\end{equation*}%
The results become:%
\begin{equation*}
\begin{tabular}{|c|rrrrr|}
\hline
Statistic & \multicolumn{1}{c}{$x^{\star}$}  & \multicolumn{1}{c}{$x_{\mathrm{ew}}$} &
        \multicolumn{1}{c}{$x_{\mathrm{mv}}$} & \multicolumn{1}{c}{$x_{\mathrm{mdp}}$} &
        \multicolumn{1}{c|}{$x_{\mathrm{erc}}$} \\ \hline
$\mu \left( x\right)$                & $  9.46\%$ & $ 8.40\%$ & $ 6.68\%$ & $ 9.19\%$ & $ 8.04\%$ \\
$\sigma \left( x\right)$             & $ 12.24\%$ & $11.12\%$ & $ 9.37\%$ & $12.29\%$ & $10.58\%$ \\
$\limfunc{SR}\left( x\mid r\right)$  & $ 60.96\%$ & $57.57\%$ & $49.99\%$ & $58.56\%$ & $57.15\%$ \\
$\sigma \left( x\mid b\right)$       & $  0.00\%$ & $ 4.05\%$ & $ 7.04\%$ & $ 3.44\%$ & $ 4.35\%$ \\
$\beta \left( x\mid b\right)$        & $100.00\%$ & $85.77\%$ & $62.74\%$ & $96.41\%$ & $81.00\%$ \\
$\rho \left( x\mid b\right)$         & $100.00\%$ & $94.44\%$ & $82.00\%$ & $96.06\%$ & $93.76\%$ \\
\hline
\end{tabular}
\end{equation*}
\end{enumerate}

\section{Equally weighted portfolio}

\begin{enumerate}
\item
\begin{enumerate}
\item The elements of the covariance matrix are $\Sigma _{i,j}=\rho
_{i,j}\sigma _{i}\sigma _{j}$. If we consider a portfolio $x=\left(
x_{1},\ldots ,x_{n}\right) $, its volatility is:%
\begin{eqnarray*}
\sigma \left( x\right) &=&\sqrt{x^{\top }\Sigma x} \\
&=&\sqrt{\sum_{i=1}^{n}x_{i}^{2}\sigma _{i}^{2}+2\sum_{i>j}x_{i}x_{j}\rho
_{i,j}\sigma _{i}\sigma _{j}}
\end{eqnarray*}%
For the equally weighted portfolio, we have $x_{i}=n^{-1}$ and:%
\begin{equation*}
\sigma \left( x\right) =\frac{1}{n}\sqrt{\sum_{i=1}^{n}\sigma
_{i}^{2}+2\sum_{i>j}\rho _{i,j}\sigma _{i}\sigma _{j}}
\end{equation*}

\item We have:%
\begin{equation*}
\sigma _{0}\left( x\right) =\frac{1}{n}\sqrt{\sum_{i=1}^{n}\sigma _{i}^{2}}
\end{equation*}%
and:%
\begin{eqnarray*}
\sigma _{1}\left( x\right) &=&\frac{1}{n}\sqrt{\sum_{i=1}^{n}\sum_{j=1}^{n}%
\sigma _{i}\sigma _{j}} \\
&=&\frac{1}{n}\sqrt{\sum_{i=1}^{n}\sigma _{i}\sum_{j=1}^{n}\sigma _{j}} \\
&=&\frac{1}{n}\sqrt{\left( \sum_{i=1}^{n}\sigma _{i}\right) ^{2}} \\
&=&\frac{\sum_{i=1}^{n}\sigma _{i}}{n} \\
&=&\bar{\sigma}
\end{eqnarray*}

\item If $\sigma _{i}=\sigma _{j}=\sigma $, we obtain:%
\begin{equation*}
\sigma \left( x\right) =\frac{\sigma }{n}\sqrt{n+2\sum_{i>j}\rho _{i,j}}
\end{equation*}%
Let $\bar{\rho}$ be the mean correlation. We have:%
\begin{equation*}
\bar{\rho}=\frac{2}{n^{2}-n}\sum_{i>j}\rho _{i,j}
\end{equation*}%
We deduce that:%
\begin{equation*}
\sum_{i>j}\rho _{i,j}=\frac{n\left( n-1\right) }{2}\bar{\rho}
\end{equation*}%
We finally obtain:%
\begin{eqnarray*}
\sigma \left( x\right) &=&\frac{\sigma }{n}\sqrt{n+n\left( n-1\right) \bar{%
\rho}} \\
&=&\sigma \sqrt{\frac{1+\left( n-1\right) \bar{\rho}}{n}}
\end{eqnarray*}%
When $\bar{\rho}$ is equal to zero, the volatility $\sigma \left( x\right) $
is equal to $\sigma /\sqrt{n}$. When the number of assets tends to $+\infty $%
, it follows that:%
\begin{equation*}
\lim_{n\rightarrow \infty }\sigma \left( x\right) =\sigma \sqrt{\bar{\rho}}
\end{equation*}

\item If $\rho _{i,j}=\rho $, we obtain:%
\begin{eqnarray*}
\sigma \left( x\right) &=&\frac{1}{n}\sqrt{\sum_{i=1}^{n}\sum_{j=1}^{n}\rho
_{i,j}\sigma _{i}\sigma _{j}} \\
&=&\frac{1}{n}\sqrt{\sum_{i=1}^{n}\sigma _{i}^{2}+\rho
\sum_{i=1}^{n}\sum_{j=1}^{n}\sigma _{i}\sigma _{j}-\rho \sum_{i=1}^{n}\sigma
_{i}^{2}} \\
&=&\frac{1}{n}\sqrt{\left( 1-\rho \right) \sum_{i=1}^{n}\sigma _{i}^{2}+\rho
\sum_{i=1}^{n}\sum_{j=1}^{n}\sigma _{i}\sigma _{j}}
\end{eqnarray*}%
We have:%
\begin{equation*}
\sum_{i=1}^{n}\sigma _{i}^{2}=n^{2}\sigma _{0}^{2}\left( x\right)
\end{equation*}%
and:%
\begin{equation*}
\sum_{i=1}^{n}\sum_{j=1}^{n}\sigma _{i}\sigma _{j}=n^{2}\sigma
_{1}^{2}\left( x\right)
\end{equation*}%
It follows that:%
\begin{equation*}
\sigma \left( x\right) =\sqrt{\left( 1-\rho \right) \sigma _{0}^{2}\left(
x\right) +\rho \sigma _{1}^{2}\left( x\right) }
\end{equation*}%
When the correlation is uniform, the variance $\sigma ^{2}\left( x\right) $
is the weighted average between $\sigma _{0}^{2}\left( x\right) $ and $%
\sigma _{1}^{2}\left( x\right) $.
\end{enumerate}

\item

\begin{enumerate}
\item The risk contributions are equal to:%
\begin{equation*}
\mathcal{RC}_{i}^{\star }=\frac{x_{i}\cdot \left( \Sigma x\right) _{i}}{%
\sigma ^{2}\left( x\right) }
\end{equation*}%
In the case of the EW portfolio, we obtain:%
\begin{eqnarray*}
\mathcal{RC}_{i}^{\star } &=&\frac{\sum_{j=1}^{n}\rho _{i,j}\sigma
_{i}\sigma _{j}}{n^{2}\sigma ^{2}\left( x\right) } \\
&=&\frac{\sigma _{i}^{2}+\sigma _{i}\sum_{j\neq i}\rho _{i,j}\sigma _{j}}{%
n^{2}\sigma ^{2}\left( x\right) }
\end{eqnarray*}

\item If asset returns are independent, we have:%
\begin{equation*}
\mathcal{RC}_{i}^{\star }=\frac{\sigma _{i}^{2}}{\sum_{i=1}^{n}\sigma
_{i}^{2}}
\end{equation*}%
In the case of perfect correlation, we obtain:%
\begin{eqnarray*}
\mathcal{RC}_{i}^{\star } &=&\frac{\sigma _{i}^{2}+\sigma _{i}\sum_{j\neq
i}\sigma _{j}}{n^{2}\bar{\sigma}^{2}} \\
&=&\frac{\sigma _{i}\sum_{j}\sigma _{j}}{n^{2}\bar{\sigma}^{2}} \\
&=&\frac{\sigma _{i}}{n\bar{\sigma}}
\end{eqnarray*}

\item If $\sigma _{i}=\sigma _{j}=\sigma $, we obtain:%
\begin{eqnarray*}
\mathcal{RC}_{i}^{\star } &=&\frac{\sigma ^{2}+\sigma ^{2}\sum_{j\neq i}\rho
_{i,j}}{n^{2}\sigma ^{2}\left( x\right) } \\
&=&\frac{\sigma ^{2}+\left( n-1\right) \sigma ^{2}\bar{\rho}_{i}}{%
n^{2}\sigma ^{2}\left( x\right) } \\
&=&\frac{1+\left( n-1\right) \bar{\rho}_{i}}{n\left( 1+\left( n-1\right)
\bar{\rho}\right) }
\end{eqnarray*}%
It follows that:%
\begin{equation*}
\lim_{n\rightarrow \infty }\frac{1+\left( n-1\right) \bar{\rho}_{i}}{%
1+\left( n-1\right) \bar{\rho}}=\frac{\bar{\rho}_{i}}{\bar{\rho}}
\end{equation*}%
We deduce that the risk contributions are proportional to the ratio between
the mean correlation of asset $i$ and the mean correlation of the asset
universe.

\item We recall that we have:%
\begin{equation*}
\sigma \left( x\right) =\sqrt{\left( 1-\rho \right) \sigma _{0}^{2}\left(
x\right) +\rho \sigma _{1}^{2}\left( x\right) }
\end{equation*}%
It follows that:%
\begin{eqnarray*}
\mathcal{RC}_{i} &=&x_{i}\cdot \frac{\left( 1-\rho \right) \sigma _{0}\left(
x\right) \partial _{x_{i}}\sigma _{0}\left( x\right) +\rho \sigma _{1}\left(
x\right) \partial _{x_{i}}\sigma _{1}\left( x\right) }{\sqrt{\left( 1-\rho
\right) \sigma _{0}^{2}\left( x\right) +\rho \sigma _{1}^{2}\left( x\right) }%
} \\
&=&\frac{\left( 1-\rho \right) \sigma _{0}\left( x\right) \mathcal{RC}%
_{0,i}+\rho \sigma _{1}\left( x\right) \mathcal{RC}_{1,i}}{\sqrt{\left(
1-\rho \right) \sigma _{0}^{2}\left( x\right) +\rho \sigma _{1}^{2}\left(
x\right) }}
\end{eqnarray*}%
We then obtain:%
\begin{equation*}
\mathcal{RC}_{i}^{\star }=\frac{\left( 1-\rho \right) \sigma _{0}^{2}\left(
x\right) }{\sigma ^{2}\left( x\right) }\mathcal{RC}_{0,i}^{\star }+\frac{%
\rho \sigma _{1}\left( x\right) }{\sigma ^{2}\left( x\right) }\mathcal{RC}%
_{1,i}^{\star }
\end{equation*}%
We verify that the risk contribution $\mathcal{RC}_{i}$ is a weighted
average of $\mathcal{RC}_{0,i}^{\star }$ and $\mathcal{RC}_{1,i}^{\star }$.
\end{enumerate}

\item

\begin{enumerate}
\item We have:%
\begin{equation*}
\Sigma =\beta \beta ^{\top }\sigma _{m}^{2}+D
\end{equation*}%
We deduce that:%
\begin{equation*}
\sigma \left( x\right) =\frac{1}{n}\sqrt{\sigma
_{m}^{2}\sum_{i=1}^{n}\sum_{j=1}^{n}\beta _{i}\beta _{j}+\sum_{i=1}^{n}%
\tilde{\sigma}_{i}^{2}}
\end{equation*}

\item The risk contributions are equal to:%
\begin{equation*}
\mathcal{RC}_{i}=\frac{x_{i}\cdot \left( \Sigma x\right) _{i}}{\sigma \left(
x\right) }
\end{equation*}%
In the case of the EW portfolio, we obtain:%
\begin{eqnarray*}
\mathcal{RC}_{i} &=&\frac{x_{i}\cdot \left( \sigma _{m}^{2}\beta
_{i}\sum_{j=1}^{n}x_{j}\beta _{j}+x_{i}\tilde{\sigma}_{i}^{2}\right) }{%
n^{2}\sigma \left( x\right) } \\
&=&\frac{\sigma _{m}^{2}\beta _{i}\sum_{j=1}^{n}\beta _{j}+\tilde{\sigma}%
_{i}^{2}}{n^{2}\sigma \left( x\right) } \\
&=&\frac{n\sigma _{m}^{2}\beta _{i}\bar{\beta}+\tilde{\sigma}_{i}^{2}}{%
n^{2}\sigma \left( x\right) }
\end{eqnarray*}

\item When the number of assets is large and $\beta _{i}>0$, we obtain:%
\begin{equation*}
\mathcal{RC}_{i}\simeq \frac{\sigma _{m}^{2}\beta _{i}\bar{\beta}}{n\sigma
\left( x\right) }
\end{equation*}%
because $\bar{\beta}>0$. We deduce that the risk contributions are
approximately proportional to the beta coefficients:%
\begin{equation*}
\mathcal{RC}_{i}^{\star }\simeq \frac{\beta _{i}}{\sum_{j=1}^{n}\beta _{j}}
\end{equation*}%
In Figure \ref{fig:app2-3-2-1}, we compare the exact and approximated values of $\mathcal{RC}%
_{i}^{\star }$. For that, we simulate $\beta _{i}$ and $\tilde{\sigma}_{i}$
with $\beta _{i}\sim \mathcal{U}_{\left[ 0.5,1.5\right] }$ and $\tilde{\sigma%
}_{i}\sim \mathcal{U}_{\left[ 0,20\%\right] }$ whereas $\sigma _{m}$ is set
to 25\%. We notice that the approximated value is very close to the exact
value when $n$ increases.
\end{enumerate}
\end{enumerate}

\begin{figure}[tbph]
\centering
\includegraphics[width = \figurewidth, height = \figureheight]{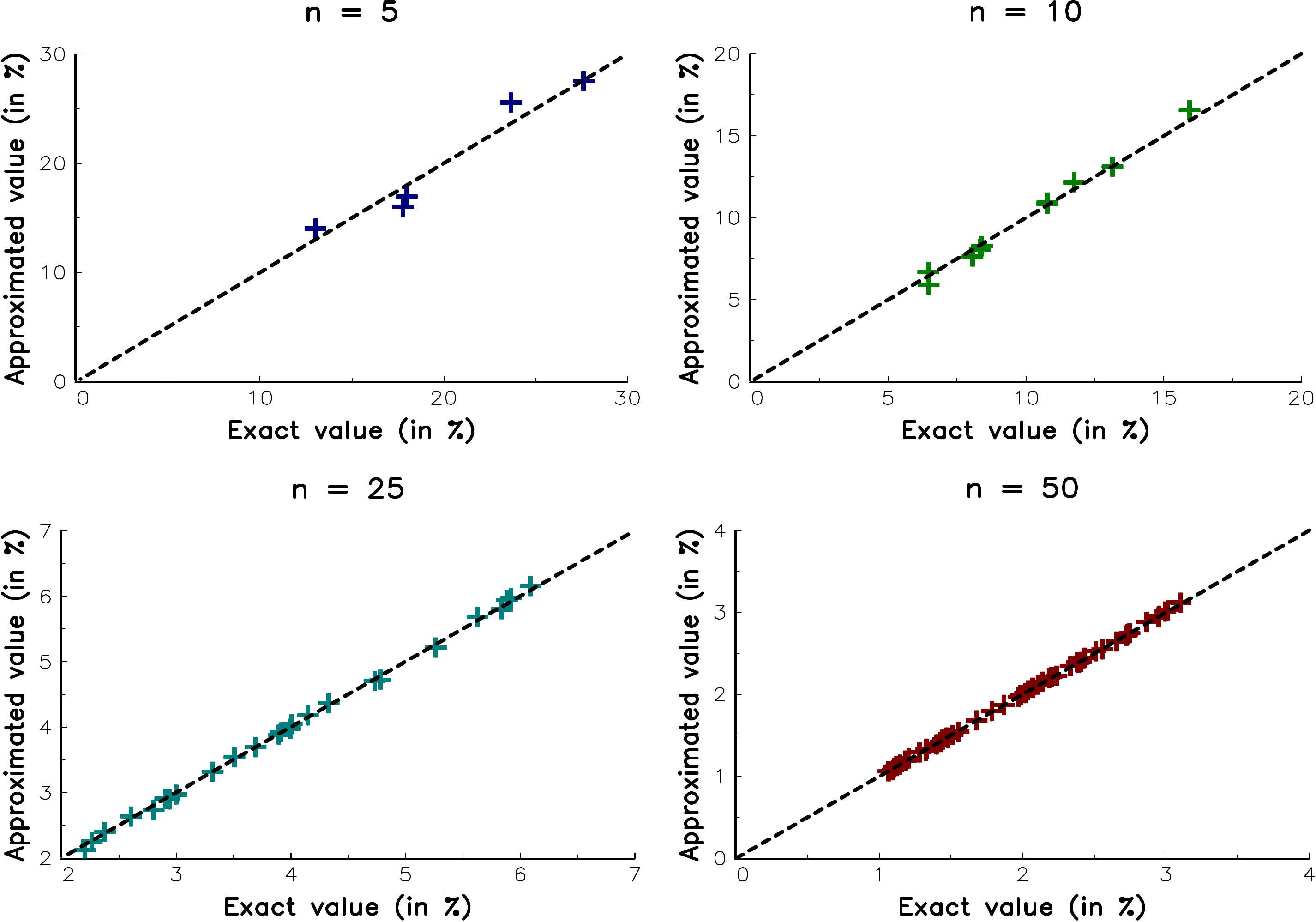}
\caption{Comparing the exact and approximated values of $\mathcal{RC}_{i}^{\star }$}
\label{fig:app2-3-2-1}
\end{figure}

\section{Minimum variance portfolio}

\begin{enumerate}
\item
\begin{enumerate}
\item The optimization program is:%
\begin{eqnarray*}
x^{\star } &=&\arg \min \frac{1}{2}x^{\top }\Sigma x \\
&\text{u.c.}&\mathbf{1}^{\top }x=1
\end{eqnarray*}%
We can show that the optimal portfolio is then equal to (TR-RPB, page 11):%
\begin{equation*}
x^{\star }=\frac{\Sigma ^{-1}\mathbf{1}}{\mathbf{1}^{\top }\Sigma ^{-1}%
\mathbf{1}}
\end{equation*}%
Let $C=C_{n}\left( \rho \right) $ be the constant correlation matrix. We
have $\Sigma =\sigma \sigma ^{\top }\circ\, C$ and $\Sigma ^{-1}=\Gamma \circ\,
C^{-1}$ with $\Gamma _{i,j}=\left( \sigma _{i}\sigma _{j}\right) ^{-1}$. The
computation of $C^{-1}$ gives\footnote{%
We use the relationship:%
\begin{equation*}
C^{-1}=\frac{1}{\det C}\tilde{C}^{\top }
\end{equation*}%
with $\tilde{C}$ the cofactor matrix of $C$.}:%
\begin{equation*}
C^{-1}=\frac{\rho \mathbf{11}^{\top }-\left( \left( n-1\right) \rho
+1\right) I_{n}}{\left( n-1\right) \rho ^{2}-\left( n-2\right) \rho -1}
\end{equation*}%
Because $\limfunc{tr}\left( AB\right) =\limfunc{tr}\left( BA\right) $, we
obtain:%
\begin{equation*}
\left( \Sigma ^{-1}\right) _{i,j}\mathbf{=}\frac{\rho }{\left( n-1\right)
\rho ^{2}-\left( n-2\right) \rho -1}\left( \sigma _{i}\sigma _{j}\right)
^{-1}
\end{equation*}%
if $i\neq j$ and:%
\begin{equation*}
\left( \Sigma ^{-1}\right) _{i,j}\mathbf{=-}\frac{\left( n-2\right) \rho +1}{%
\left( n-1\right) \rho ^{2}-\left( n-2\right) \rho -1}\sigma _{i}^{-2}
\end{equation*}%
It follows that:%
\begin{eqnarray*}
\left( \Sigma ^{-1}\mathbf{1}\right) _{i} &\mathbf{=}&\mathbf{-}\frac{\left(
n-2\right) \rho +1}{\left( n-1\right) \rho ^{2}-\left( n-2\right) \rho -1}%
\sigma _{i}^{-2}+ \\
&&\sum_{j\neq i}\frac{\rho }{\left( n-1\right) \rho ^{2}-\left( n-2\right)
\rho -1}\left( \sigma _{i}\sigma _{j}\right) ^{-1} \\
&=&\frac{-\left( \left( n-1\right) \rho +1\right) \sigma _{i}^{-2}+\rho
\sum_{j=1}^{n}\left( \sigma _{i}\sigma _{j}\right) ^{-1}}{\left( n-1\right)
\rho ^{2}-\left( n-2\right) \rho -1}
\end{eqnarray*}%
We deduce that:%
\begin{eqnarray*}
x_{i}^{\star } &=&\frac{-\left( \left( n-1\right) \rho +1\right) \sigma
_{i}^{-2}+\rho \sum_{j=1}^{n}\left( \sigma _{i}\sigma _{j}\right) ^{-1}}{%
\sum_{k=1}^{n}\left( -\left( \left( n-1\right) \rho +1\right) \sigma
_{k}^{-2}+\rho \sum_{j=1}^{n}\left( \sigma _{k}\sigma _{j}\right)
^{-1}\right) } \\
&\propto &-\left( \left( n-1\right) \rho +1\right) \sigma _{i}^{-2}+\rho
\sum_{j=1}^{n}\left( \sigma _{i}\sigma _{j}\right) ^{-1}
\end{eqnarray*}

\item When $\rho =1$, we obtain:%
\begin{eqnarray*}
x_{i}^{\star } &\propto &-n\sigma _{i}^{-2}+\sum_{j=1}^{n}\left( \sigma
_{i}\sigma _{j}\right) ^{-1} \\
&=&\sigma _{i}^{-1}\left( \sum_{j=1}^{n}\sigma _{j}^{-1}-n\sigma
_{i}^{-1}\right)  \\
&=&n\sigma _{i}^{-1}\left( H^{-1}-\sigma _{i}^{-1}\right)  \\
&=&\frac{n\sigma _{i}^{-1}}{H\sigma _{i}}\left( \sigma _{i}-H\right)
\end{eqnarray*}%
where $H$ is the harmonic mean:%
\begin{equation*}
H=\left( \frac{1}{n}\sum_{j=1}^{n}\sigma _{j}^{-1}\right) ^{-1}
\end{equation*}%
The weights are all positive if the sign of $\sigma _{i}-H$ is the same for
all the assets. By definition of the harmonic mean, the only solution is
when the volatilities are the same for all the assets.

\item When $\rho =0$, we obtain $x_{i}^{\star }\propto -\sigma _{i}^{-2}$.
Because $\mathbf{1}^{\top }x^{\star }=1$, the solution is:%
\begin{equation*}
x_{i}^{\star }=\frac{\sigma _{i}^{-2}}{\sum_{j=1}^{n}\sigma _{i}^{-2}}
\end{equation*}%
The weight of asset $i$ is inversely proportional to its variance.

\item We have:%
\begin{eqnarray*}
C &=&\rho \mathbf{11}^{\top }-\left( \rho -1\right) I_{n} \\
&=&\rho \left( \mathbf{11}^{\top }-\frac{\left( \rho -1\right) }{\rho }%
I_{n}\right)
\end{eqnarray*}%
Let $\mathds{P}_{A}\left( \lambda \right) $ be the characteristic polynomial
of $A$. It follows that:%
\begin{eqnarray*}
\det C &=&\rho ^{n}\det \left( \mathbf{11}^{\top }-\frac{\left( \rho
-1\right) }{\rho }I_{n}\right)  \\
&=&\rho ^{n}\mathds{P}_{\mathbf{11}^{\top }}\left( \frac{\rho -1}{\rho }%
\right)  \\
&=&\rho ^{n}\left( -1\right) ^{n}\left( \frac{\rho -1}{\rho }\right)
^{n-1}\left( \frac{\rho -1}{\rho }-n\right)  \\
&=&\left( 1-\rho \right) ^{n-1}\left( \left( n-1\right) \rho +1\right)
\end{eqnarray*}%
A necessary condition for $C$ to be definite positive is that the
determinant is positive:%
\begin{equation*}
\det C>0\Leftrightarrow \rho >-\frac{1}{n-1}
\end{equation*}%
The lower bound is then $\rho ^{-}=-\left( n-1\right) ^{-1}$. In this case,
we obtain:%
\begin{equation*}
x_{i}^{\star }\propto -\frac{1}{n-1}\sum_{j=1}^{n}\left( \sigma _{i}\sigma
_{j}\right) ^{-1}
\end{equation*}%
We deduce that:%
\begin{eqnarray*}
x_{i}^{\star } &=&\frac{\sum_{j=1}^{n}\left( \sigma _{i}\sigma _{j}\right)
^{-1}}{\sum_{k=1}^{n}\sum_{j=1}^{n}\left( \sigma _{i}\sigma _{j}\right) ^{-1}%
} \\
&=&\frac{\sigma _{i}^{-1}}{\sum_{j=1}^{n}\sigma _{i}^{-1}}
\end{eqnarray*}%
When $\rho =\rho ^{-}$, the MV portfolio coincides with the ERC portfolio.

\item We have%
\begin{equation*}
x_{i}^{\star }=\frac{a_{i}+b_{i}}{\sum_{j=1}^{n}a_{j}+b_{j}}
\end{equation*}%
with:%
\begin{eqnarray*}
a_{i} &=&\left( n-1\right) \rho \sigma _{i}^{-2} \\
b_{i} &=&\sigma _{i}^{-2}-\rho \sum_{j=1}^{n}\left( \sigma _{i}\sigma
_{j}\right) ^{-1}
\end{eqnarray*}%
If $\rho ^{-}\leq \rho \leq 0$, we have $a_{i}\leq 0$ and $b_{i}\geq 0$. We
would like to show that $x_{i}^{\star }$ or equivalently $a_{i}+b_{i}$ is
positive. We then have to show that:%
\begin{equation*}
\rho n\left( \sigma _{i}^{-1}-H^{-1}\right) +\left( 1-\rho \right) \sigma
_{i}^{-1}\geq 0
\end{equation*}%
for every asset. If $\sigma _{i}\geq H$, this inequality is satisfied. If $%
\sigma _{i}<H$, it means that:%
\begin{equation*}
\sigma _{i}<\left( \frac{1}{n}\sum_{j=1}^{n}\sigma _{j}^{-1}\right) ^{-1}
\end{equation*}%
If $\rho ^{-}\leq \rho \leq 0$, we conclude that $x_{i}^{\star }\geq 0$.
Moreover, we know that if $\rho =1$, at least one asset has a negative
weight. It implies that there exists a correlation $\rho ^{\star }>0$ such
that the weights are all positive if $\rho \leq \rho ^{\star }$. It is
obvious that $\rho ^{\star }$ must satisfy this equation:%
\begin{equation*}
-\left( \left( n-1\right) \rho +1\right) \sigma _{i}^{-1}+\rho
\sum_{j=1}^{n}\sigma _{j}^{-1}=0
\end{equation*}%
for one asset. We finally obtain:%
\begin{eqnarray*}
\rho ^{\star } &=&\inf\nolimits_{\left[ 0,1\right] } \frac{\sigma _{i}^{-1}}{\sum_{j=1}^{n}\sigma
_{j}^{-1}-\left( n-1\right) \sigma _{i}^{-1}} \\
&=&\frac{\sigma _{+}^{-1}}{\sum_{j=1}^{n}\sigma _{j}^{-1}-\left( n-1\right)
\sigma _{+}^{-1}}
\end{eqnarray*}%
with $\sigma _{+}=\sup \sigma _{i}$.

\item We have reported the relationship between $\inf x_{i}^{\star }$ and $\rho $
in Figure \ref{fig:app2-3-3-1}. We notice that $\inf x_{i}^{\star }$ is close to one for the
second set of parameters (i.e. when the dispersion across $\sigma _{i}$
is high) and $\inf x_{i}^{\star }$ is close to zero for the third set of
parameters (i.e. when the dispersion across $\sigma _{i}$ is low). We may
then postulate these two rules\footnote{%
If you are interested to prove these two rules, you have to use the
following inequality:%
\begin{equation*}
A\left( \sigma _{i}\right) -H\left( \sigma _{i}\right) \geq \frac{\limfunc{%
var}\left( \sigma _{i}\right) }{2\sup \sigma _{i}}
\end{equation*}%
where $A\left( \sigma _{i}\right) $ and $H\left( \sigma _{i}\right) $ are the
arithmetic and harmonic means of $\left\{ \sigma _{1},\ldots ,\sigma
_{n}\right\} $. }:%
\begin{equation*}
\lim_{\limfunc{var}\left( \sigma _{i}\right) \rightarrow 0}\rho ^{\star }=0
\end{equation*}%
and:%
\begin{equation*}
\lim_{\limfunc{var}\left( \sigma _{i}\right) \rightarrow \infty }\rho
^{\star }=1
\end{equation*}%
\begin{figure}[tbp]
\centering
\includegraphics[width = \figurewidth, height = \figureheight]{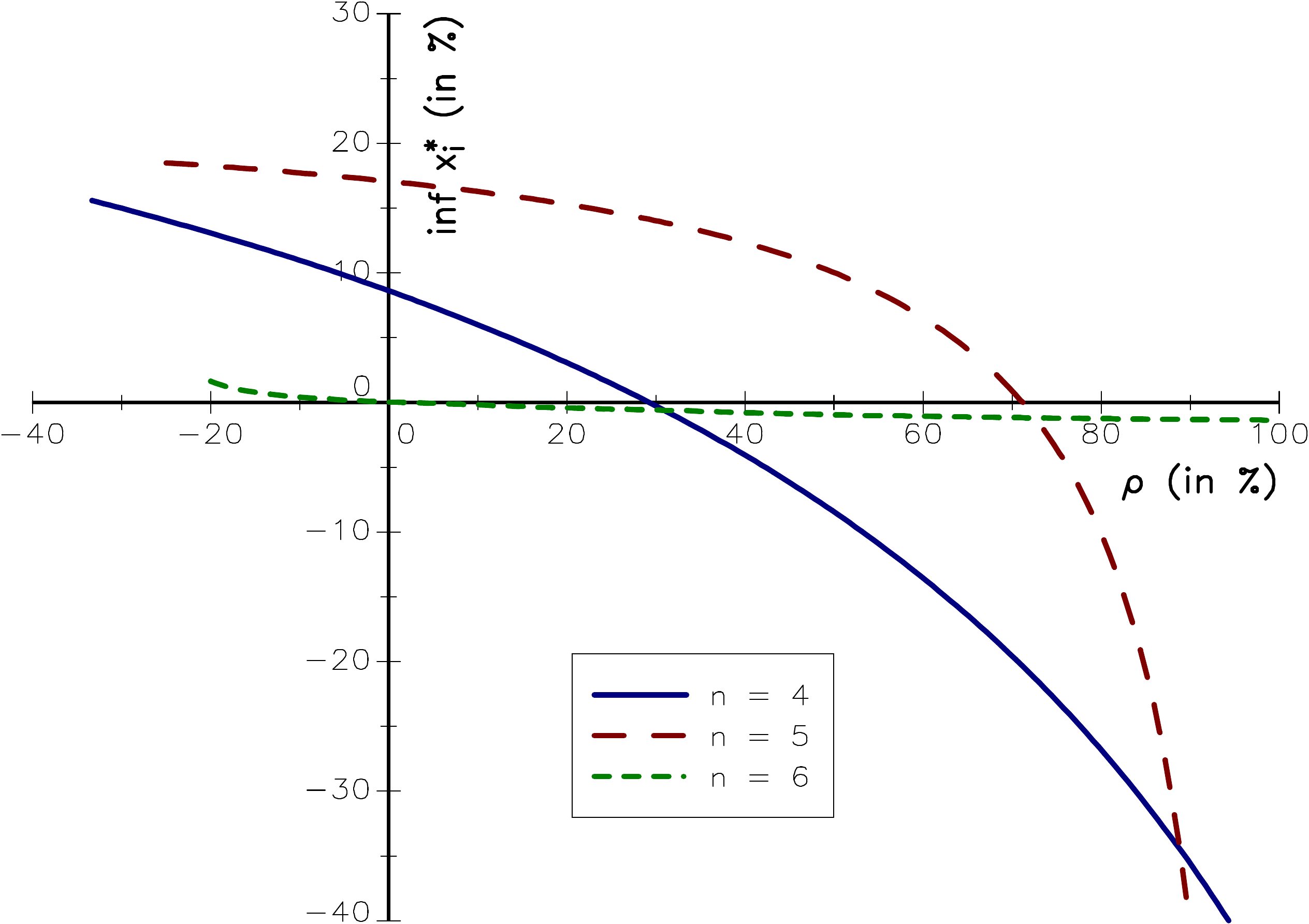}
\caption{Relationship between $\rho$ and $\inf x_{i}^{\star }$}
\label{fig:app2-3-3-1}
\end{figure}

\item
For the first set of parameters, $\rho^{\star}$ is equal to $29.27\%$ and
the optimal weights are:%
\begin{equation*}
\begin{tabular}{|c|rrrrr|}
\hline
$\rho$ & \multicolumn{1}{c}{$\rho ^{-}$} & \multicolumn{1}{c}{$0$} &
\multicolumn{1}{c}{$\rho ^{\star }$} & \multicolumn{1}{c}{$0.9$} & \multicolumn{1}{c|}{$1$} \\
\hline
$x_{1}^{\star}$ & $38.96\%$ & $53.92\%$ & $72.48\%$ & $149.07\%$ & $172.69\%$ \\
$x_{2}^{\star}$ & $25.97\%$ & $23.97\%$ & $21.48\%$ & $ 11.20\%$ & $  8.03\%$ \\
$x_{3}^{\star}$ & $19.48\%$ & $13.48\%$ & $ 6.04\%$ & $-24.67\%$ & $-34.14\%$ \\
$x_{4}^{\star}$ & $15.58\%$ & $ 8.63\%$ & $ 0.00\%$ & $-35.61\%$ & $-46.59\%$ \\
\hline
\end{tabular}%
\end{equation*}
For the second set of parameters, $\rho^{\star}$ is equal to $71.15\%$ and
the optimal weights are:%
\begin{equation*}
\begin{tabular}{|c|rrrrr|}
\hline
$\rho$ & \multicolumn{1}{c}{$\rho ^{-}$} & \multicolumn{1}{c}{$0$} &
\multicolumn{1}{c}{$\rho ^{\star }$} & \multicolumn{1}{c}{$0.9$} & \multicolumn{1}{c|}{$1$} \\
\hline
$x_{1}^{\star}$ & $21.42\%$ & $22.89\%$ & $40.38\%$ & $ 82.16\%$ & $ 621.36\%$ \\
$x_{2}^{\star}$ & $20.35\%$ & $20.65\%$ & $24.30\%$ & $ 33.00\%$ & $ 145.26\%$ \\
$x_{3}^{\star}$ & $19.38\%$ & $18.73\%$ & $11.02\%$ & $ -7.40\%$ & $-245.13\%$ \\
$x_{4}^{\star}$ & $18.50\%$ & $17.07\%$ & $ 0.00\%$ & $-40.75\%$ & $-566.75\%$ \\
$x_{5}^{\star}$ & $20.35\%$ & $20.65\%$ & $24.30\%$ & $ 33.00\%$ & $ 145.26\%$ \\
\hline
\end{tabular}%
\end{equation*}
For the third set of parameters, $\rho^{\star}$ is equal to $1.77\%$ and
the optimal weights are:%
\begin{equation*}
\begin{tabular}{|c|rrrrr|}
\hline
$\rho$ & \multicolumn{1}{c}{$\rho ^{-}$} & \multicolumn{1}{c}{$0$} &
\multicolumn{1}{c}{$\rho ^{\star }$} & \multicolumn{1}{c}{$0.9$} & \multicolumn{1}{c|}{$1$} \\
\hline
$x_{1}^{\star}$ & $81.41\%$ & $98.56\%$ & $98.98\%$ & $104.07\%$ & $104.21\%$ \\
$x_{2}^{\star}$ & $ 8.14\%$ & $ 0.99\%$ & $ 0.81\%$ & $ -1.32\%$ & $ -1.37\%$ \\
$x_{3}^{\star}$ & $ 4.07\%$ & $ 0.25\%$ & $ 0.15\%$ & $ -0.98\%$ & $ -1.01\%$ \\
$x_{4}^{\star}$ & $ 2.71\%$ & $ 0.11\%$ & $ 0.04\%$ & $ -0.73\%$ & $ -0.75\%$ \\
$x_{5}^{\star}$ & $ 2.04\%$ & $ 0.06\%$ & $ 0.01\%$ & $ -0.57\%$ & $ -0.59\%$ \\
$x_{6}^{\star}$ & $ 1.63\%$ & $ 0.04\%$ & $ 0.00\%$ & $ -0.47\%$ & $ -0.48\%$ \\
\hline
\end{tabular}%
\end{equation*}
\end{enumerate}

\item

\begin{enumerate}
\item We have:%
\begin{equation*}
\Sigma =\beta \beta ^{\top }\sigma _{m}^{2}+D
\end{equation*}%
The Sherman-Morrison-Woodbury formula is (TR-RPB, page 167):%
\begin{equation*}
\left( A+uv^{\top }\right) ^{-1}=A^{-1}-\frac{1}{1+v^{\top }A^{-1}u}%
A^{-1}uv^{\top }A^{-1}
\end{equation*}%
where $u$ and $v$ are two vectors and $A$ is an invertible square matrix. By
setting $A=D$ and $u=v=\sigma _{m}\beta $, we obtain:%
\begin{equation*}
\Sigma ^{-1}=D^{-1}-\frac{\sigma _{m}^{2}}{1+\sigma _{m}^{2}\beta ^{\top
}D^{-1}\beta }D^{-1}\beta \beta ^{\top }D^{-1}
\end{equation*}%
We note $\tilde{\beta}_{i}=\beta _{i}/\tilde{\sigma}_{i}^{2}$ and $\kappa =%
\tilde{\beta}^{\top }\beta $. We have $\tilde{\beta}=D^{-1}\beta $ and:%
\begin{equation*}
\Sigma ^{-1}=D^{-1}-\frac{\sigma _{m}^{2}}{1+\sigma _{m}^{2}\kappa }\tilde{%
\beta}\tilde{\beta}^{\top }
\end{equation*}

\item The analytical expression of the minimum variance portfolio is:%
\begin{equation*}
x^{\star }=\frac{\Sigma ^{-1}\mathbf{1}}{\mathbf{1}^{\top }\Sigma ^{-1}%
\mathbf{1}}
\end{equation*}%
We have:%
\begin{eqnarray*}
\sigma ^{2}\left( x^{\star }\right)  &=&x^{\star ^{\top }}\Sigma x^{\star }
\\
&=&\frac{\mathbf{1}^{\top }\Sigma ^{-1}}{\mathbf{1}^{\top }\Sigma ^{-1}%
\mathbf{1}}\Sigma \frac{\Sigma ^{-1}\mathbf{1}}{\mathbf{1}^{\top }\Sigma
^{-1}\mathbf{1}} \\
&=&\frac{1}{\mathbf{1}^{\top }\Sigma ^{-1}\mathbf{1}}
\end{eqnarray*}%
We deduce that:%
\begin{equation*}
x^{\star }=\sigma ^{2}\left( x^{\star }\right) \left( D^{-1}\mathbf{1}-\frac{%
\sigma _{m}^{2}}{1+\sigma _{m}^{2}\kappa }\tilde{\beta}\tilde{\beta}^{\top }%
\mathbf{1}\right)
\end{equation*}

\item We have:%
\begin{eqnarray*}
x_{i}^{\star } &=&\sigma ^{2}\left( x^{\star }\right) \left( \frac{1}{\tilde{%
\sigma}_{i}^{2}}-\frac{\sigma _{m}^{2}\tilde{\beta}^{\top }\mathbf{1}}{%
1+\sigma _{m}^{2}\kappa }\tilde{\beta}_{i}\right)  \\
&=&\sigma ^{2}\left( x^{\star }\right) \left( \frac{1}{\tilde{\sigma}_{i}^{2}%
}-\frac{\sigma _{m}^{2}\tilde{\beta}^{\top }\mathbf{1}}{1+\sigma
_{m}^{2}\kappa }\frac{\beta _{i}}{\tilde{\sigma}_{i}^{2}}\right)  \\
&=&\frac{\sigma ^{2}\left( x^{\star }\right) }{\tilde{\sigma}_{i}^{2}}\left(
1-\frac{\beta _{i}}{\beta ^{\star }}\right)
\end{eqnarray*}%
with:%
\begin{eqnarray*}
\beta ^{\star } &=&\frac{1+\sigma _{m}^{2}\kappa }{\sigma _{m}^{2}\tilde{%
\beta}^{\top }\mathbf{1}} \\
&=&\frac{1+\sigma _{m}^{2}\sum_{j=1}^{n}\beta _{j}^{2}/\tilde{\sigma}_{j}^{2}%
}{\sigma _{m}^{2}\sum_{j=1}^{n}\beta _{j}/\tilde{\sigma}_{j}^{2}}
\end{eqnarray*}

\item The optimal weight $x_{i}^{\star }$ is positive if:%
\begin{equation*}
1-\frac{\beta _{i}}{\beta ^{\star }}\geq 0
\end{equation*}%
or equivalently:%
\begin{equation*}
\beta ^{\star }\geq \beta _{i}
\end{equation*}%
If $\beta _{i}=\beta _{j}=\beta $, we obtain:%
\begin{eqnarray*}
\beta ^{\star } &=&\frac{1+\sigma _{m}^{2}\beta ^{2}\sum_{j=1}^{n}1/\tilde{%
\sigma}_{j}^{2}}{\sigma _{m}^{2}\beta \sum_{j=1}^{n}1/\tilde{\sigma}_{j}^{2}}
\\
&=&\frac{1}{\sigma _{m}^{2}\beta \sum_{j=1}^{n}1/\tilde{\sigma}_{j}^{2}}%
+\beta  \\
&\geq &\beta
\end{eqnarray*}%
We deduce that the weights are positive for all the assets if the betas are
the same. If $\tilde{\sigma}_{i}=\tilde{\sigma}_{j}=\tilde{\sigma}$, we have:%
\begin{eqnarray*}
\beta ^{\star }-\beta _{i} &=&\frac{1+\sigma _{m}^{2}/\tilde{\sigma}%
^{2}\sum_{j=1}^{n}\beta _{j}^{2}}{\sigma _{m}^{2}/\tilde{\sigma}%
^{2}\sum_{j=1}^{n}\beta _{j}}-\beta _{i} \\
&=&\frac{1}{\sum_{j=1}^{n}\beta _{j}}\left( \frac{\tilde{\sigma}^{2}}{\sigma
_{m}^{2}}+\sum_{j=1}^{n}\left( \beta _{j}-\beta _{i}\right) \beta
_{j}\right)
\end{eqnarray*}%
The weights are all positive if and only if:%
\begin{equation*}
\frac{\tilde{\sigma}^{2}}{\sigma _{m}^{2}}\geq \sum_{j=1}^{n}\left( \sup
\beta _{j}-\beta _{j}\right) \beta _{j}
\end{equation*}%
If $\tilde{\sigma}\gg \sigma _{m}$, the previous inequality holds. Except
in this case, the inequality cannot be verified, i.e.
the weights cannot be all positive.

\item We obtain the following results:

\begin{equation*}
\begin{tabular}{|c|rrrrr|}
\hline
$\sigma_m$      & $ 5.00\%$     & $ 10.00\%$     & $ 15.00\%$     & $ 20.00\%$     & $  25.00\%$     \\ \hline
$x_{1}^{\star}$ & $49.41\%$     & $ 90.71\%$     & $125.69\%$     & $149.41\%$     & $ 164.78\%$     \\
$x_{2}^{\star}$ & $10.26\%$     & $ 16.52\%$     & $ 21.83\%$     & $ 25.43\%$     & $  27.76\%$     \\
$x_{3}^{\star}$ & $ 8.16\%$     & $ 10.37\%$     & $ 12.24\%$     & $ 13.50\%$     & $  14.32\%$     \\
$x_{4}^{\star}$ & $24.24\%$     & $ 16.84\%$     & $ 10.57\%$     & $  6.31\%$     & $   3.56\%$     \\
$x_{5}^{\star}$ & $ 7.46\%$     & $-32.41\%$     & $-66.18\%$     & $-89.08\%$     & $-103.93\%$     \\
$x_{6}^{\star}$ & $ 0.47\%$     & $ -2.03\%$     & $ -4.14\%$     & $ -5.57\%$     & $  -6.50\%$     \\ \hdashline
$\beta^{\star}$ & $ 1.29{\bPp}$ & $  1.07{\bPp}$ & $  1.03{\bPp}$ & $  1.01{\bPp}$ & $   1.01{\bPp}$ \\
\hline
\end{tabular}%
\end{equation*}

\end{enumerate}
\end{enumerate}

\section{Most diversified portfolio}

\begin{enumerate}
\item
\begin{enumerate}
\item Let $\mathcal{R}\left( x\right) $ be the risk measure of the portfolio
$x$. We note $\mathcal{R}_{i}=\mathcal{R}\left( \mathbf{e}_{i}\right) $ the
risk associated to the $i^{\mathrm{th}}$ asset. The diversification ratio is
the ratio between the weighted mean of the individual risks and the
portfolio risk (TR-RPB, page 168):%
\begin{equation*}
\mathcal{DR}\left( x\right) =\frac{\sum_{i=1}^{n}x_{i}\mathcal{R}_{i}}{%
\mathcal{R}\left( x\right) }
\end{equation*}%
If we assume that the risk measure satisfies the Euler allocation principle,
we have:%
\begin{equation*}
\mathcal{DR}\left( x\right) =\frac{\sum_{i=1}^{n}x_{i}\mathcal{R}_{i}}{%
\sum_{i=1}^{n}\mathcal{RC}_{i}}
\end{equation*}

\item If $\mathcal{R}\left( x\right) $ satisfies the Euler allocation
principle, we know that $\mathcal{R}_{i}\geq \mathcal{MR}_{i}$ (TR-RPB, page
78). We deduce that:%
\begin{eqnarray*}
\mathcal{DR}\left( x\right)  &\geq &\frac{\sum_{i=1}^{n}x_{i}\mathcal{R}_{i}%
}{\sum_{i=1}^{n}x_{i}\mathcal{R}_{i}} \\
&\geq &1
\end{eqnarray*}%
Let $x_{\mathrm{mr}}$ be the portfolio that minimizes the risk measure. We
have:%
\begin{equation*}
\mathcal{DR}\left( x\right) \leq \frac{\sup \mathcal{R}_{i}}{\mathcal{R}%
\left( x_{\mathrm{mr}}\right) }
\end{equation*}

\item If we consider the volatility risk measure, the minimum risk portfolio
is the minimum variance portfolio. We have (TR-RPB, page 164):%
\begin{equation*}
\sigma \left( x_{\mathrm{mv}}\right) =\frac{1}{\sqrt{\mathbf{1}^{\top
}\Sigma \mathbf{1}}}
\end{equation*}%
We deduce that:%
\begin{equation*}
\mathcal{DR}\left( x\right) \leq \sqrt{\mathbf{1}^{\top }\Sigma ^{-1}\mathbf{%
1}}\cdot \sup \sigma _{i}
\end{equation*}

\item The MDP is the portfolio which maximizes the diversification ratio
when the risk measure is the volatility (TR-RPB, page 168). We have:%
\begin{eqnarray*}
x^{\star } &=&\arg \max \mathcal{DR}\left( x\right)  \\
&\text{u.c.}&\mathbf{1}^{\top }x=1
\end{eqnarray*}%
If we consider that the risk premium $\pi _{i}=\mu _{i}-r$ of the asset $i$ is
proportional to its volatility $\sigma _{i}$, we obtain:%
\begin{eqnarray*}
\limfunc{SR}\left( x\mid r\right)  &=&\frac{\mu \left( x\right) -r}{\sigma
\left( x\right) } \\
&=&\frac{\sum_{i=1}^{n}x_{i}\left( \mu _{i}-r\right) }{\sigma \left(
x\right) } \\
&=&s\frac{\sum_{i=1}^{n}x_{i}\sigma _{i}}{\sigma \left( x\right) } \\
&=&s\cdot \mathcal{DR}\left( x\right)
\end{eqnarray*}%
where $s$ is the coefficient of proportionality. Maximizing the
diversification ratio is equivalent to maximizing the Sharpe ratio. We recall
that the expression of the tangency portfolio is:%
\begin{equation*}
x^{\star }=\frac{\Sigma ^{-1}\left( \mu -r\mathbf{1}\right) }{\mathbf{1}%
^{\top }\Sigma ^{-1}\left( \mu -r\mathbf{1}\right) }
\end{equation*}%
We deduce that the weights of the MDP are:%
\begin{equation*}
x^{\star }=\frac{\Sigma ^{-1}\sigma }{\mathbf{1}^{\top }\Sigma ^{-1}\sigma }
\end{equation*}%
The volatility of the MDP is then:%
\begin{eqnarray*}
\sigma \left( x^{\star }\right)  &=&\sqrt{\frac{\sigma ^{\top }\Sigma ^{-1}}{%
\mathbf{1}^{\top }\Sigma ^{-1}\sigma }\Sigma \frac{\Sigma ^{-1}\sigma }{%
\mathbf{1}^{\top }\Sigma ^{-1}\sigma }} \\
&=&\frac{\sqrt{\sigma ^{\top }\Sigma ^{-1}\sigma }}{\mathbf{1}^{\top }\Sigma
^{-1}\sigma }
\end{eqnarray*}

\item We have seen in Exercise 1.11 that another expression of the
unconstrained tangency portfolio is:%
\begin{equation*}
x^{\star }=\frac{\sigma ^{2}\left( x^{\star }\right) }{\left( \mu \left(
x^{\star }\right) -r\right) }\Sigma ^{-1}\left( \mu -r\mathbf{1}\right)
\end{equation*}%
We deduce that the MDP is also:%
\begin{equation*}
x^{\star }=\frac{\sigma ^{2}\left( x^{\star }\right) }{\bar{\sigma}\left(
x^{\star }\right) }\Sigma ^{-1}\sigma
\end{equation*}%
where $\bar{\sigma}\left( x^{\star }\right) =x^{\star \top }\sigma $.
Nevertheless, this solution is endogenous.
\end{enumerate}

\item

\begin{enumerate}
\item We have:%
\begin{equation*}
\limfunc{cov}\left( R_{i},R_{m}\right) =\beta _{i}\sigma _{m}^{2}
\end{equation*}%
We deduce that:%
\begin{eqnarray}
\rho _{i,m} &=&\frac{\limfunc{cov}\left( R_{i},R_{m}\right) }{\sigma
_{i}\sigma _{m}}  \notag \\
&=&\beta _{i}\frac{\sigma _{m}}{\sigma _{i}}  \label{eq:app2-mdp2}
\end{eqnarray}%
and:%
\begin{eqnarray}
\tilde{\sigma}_{i} &=&\sqrt{\sigma _{i}^{2}-\beta _{i}^{2}\sigma _{m}^{2}}
\notag \\
&=&\sigma _{i}\sqrt{1-\rho _{i,m}^{2}}  \label{eq:app2-mdp3}
\end{eqnarray}

\item We know that (TR-RPB, page 167):
\begin{equation*}
\Sigma ^{-1}=D^{-1}-\frac{1}{\sigma _{m}^{-2}+\tilde{\beta}^{\top }\beta }%
\tilde{\beta}\tilde{\beta}^{\top }
\end{equation*}%
where $\tilde{\beta}_{i}=\beta _{i}/\tilde{\sigma}_{i}^{2}$. We deduce that:%
\begin{equation*}
x^{\star }=\frac{\sigma ^{2}\left( x^{\star }\right) }{\bar{\sigma}\left(
x^{\star }\right) }\left( D^{-1}\sigma -\frac{1}{\sigma _{m}^{-2}+\tilde{%
\beta}^{\top }\beta }\tilde{\beta}\tilde{\beta}^{\top }\sigma \right)
\end{equation*}%
and:%
\begin{eqnarray*}
x_{i}^{\star } &=&\frac{\sigma ^{2}\left( x^{\star }\right) }{\bar{\sigma}%
\left( x^{\star }\right) }\left( \frac{\sigma _{i}}{\tilde{\sigma}_{i}^{2}}-%
\frac{\tilde{\beta}^{\top }\sigma }{\sigma _{m}^{-2}+\tilde{\beta}^{\top
}\beta }\tilde{\beta}_{i}\right)  \\
&=&\frac{\sigma _{i}\sigma ^{2}\left( x^{\star }\right) }{\bar{\sigma}\left(
x^{\star }\right) \tilde{\sigma}_{i}^{2}}\left( 1-\frac{\tilde{\beta}^{\top
}\sigma }{\sigma _{m}^{-1}+\sigma _{m}\tilde{\beta}^{\top }\beta }\frac{%
\sigma _{m}\tilde{\sigma}_{i}^{2}\tilde{\beta}_{i}}{\sigma _{i}}\right)  \\
&=&\frac{\sigma _{i}\sigma ^{2}\left( x^{\star }\right) }{\bar{\sigma}\left(
x^{\star }\right) \tilde{\sigma}_{i}^{2}}\left( 1-\frac{\tilde{\beta}^{\top
}\sigma }{\sigma _{m}^{-1}+\sigma _{m}\tilde{\beta}^{\top }\beta }\rho
_{i,m}\right)  \\
&=&\mathcal{DR}\left( x^{\star }\right) \frac{\sigma _{i}\sigma \left(
x^{\star }\right) }{\tilde{\sigma}_{i}^{2}}\left( 1-\frac{\rho _{i,m}}{\rho
^{\star }}\right)
\end{eqnarray*}%
Using Equations (\ref{eq:app2-mdp2}) and (\ref{eq:app2-mdp3}), $\rho ^{\star
}$ is defined as follows:%
\begin{eqnarray*}
\rho ^{\star } &=&\frac{\sigma _{m}^{-1}+\sigma _{m}\tilde{\beta}^{\top
}\beta }{\tilde{\beta}^{\top }\sigma } \\
&=&\left. \left( 1+\sum_{j=1}^{n}\frac{\sigma _{m}^{2}\beta _{j}^{2}}{\tilde{%
\sigma}_{j}^{2}}\right) \right/ \left( \sum_{j=1}^{n}\frac{\sigma _{m}\beta
_{j}\sigma _{j}}{\tilde{\sigma}_{j}^{2}}\right)  \\
&=&\left. \left( 1+\sum_{j=1}^{n}\frac{\rho _{j,m}^{2}}{1-\rho _{j,m}^{2}}%
\right) \right/ \left( \sum_{j=1}^{n}\frac{\rho _{j,m}}{1-\rho _{j,m}^{2}}%
\right)
\end{eqnarray*}

\item The optimal weight $x_{i}^{\star }$ is positive if:%
\begin{equation*}
1-\frac{\rho _{i,m}}{\rho ^{\star }}\geq 0
\end{equation*}%
or equivalently:%
\begin{equation*}
\rho _{i,m}\leq \rho ^{\star }
\end{equation*}

\item We recall that:%
\begin{eqnarray*}
\rho _{i,m} &=&\beta _{i}\frac{\sigma _{m}}{\sigma _{i}} \\
&=&\frac{\beta _{i}\sigma _{m}}{\sqrt{\beta _{i}^{2}\sigma _{m}^{2}+\tilde{%
\sigma}_{i}^{2}}}
\end{eqnarray*}%
If $\beta _{i}<0$, an increase of the idiosyncratic volatility $\tilde{\sigma%
}_{i}$ increases $\rho _{i,m}$ and decreases the ratio $\sigma _{i}/\tilde{%
\sigma}_{i}^{2}$. We deduce that the weight is a decreasing function of the
specific volatility $\tilde{\sigma}_{i}$. If $\beta _{i}>0$, an increase of
the idiosyncratic volatility $\tilde{\sigma}_{i}$ decreases $\rho _{i,m}$
and decreases the ratio $\sigma _{i}/\tilde{\sigma}_{i}^{2}$. We cannot
conclude in this case.
\end{enumerate}

\item

\begin{enumerate}

\item The MDP coincide with the MV portfolio when the volatility is the same for all the assets.

\item The formula cannot be used directly, because it depends on $\sigma
\left( x^{\star }\right) $ and $\mathcal{DR}\left( x^{\star }\right) $.
However, we notice that:
\begin{equation*}
x_{i}^{\star }\propto \frac{\sigma _{i}}{\tilde{\sigma}_{i}^{2}}\left( 1-%
\frac{\rho _{i,m}}{\rho ^{\star }}\right)
\end{equation*}%
It suffices then to rescale these weights to obtain the solution. Using the
numerical values of the parameters, $\rho ^{\star }=98.92\%$ and we obtain
the following results:%
\begin{equation*}
\begin{tabular}{|c|rr:rr:rr|}
\hline
                                & \multirow{2}{*}{$\beta_i{\bN}$} & \multirow{2}{*}{$\rho_{i,m}{\bN}$} &
                                \multicolumn{2}{c:}{$x_i \in \mathbb{R}$} & \multicolumn{2}{c|}{$x_i \geq 0$} \\
                                &        &           & \multicolumn{1}{c}{MDP} & \multicolumn{1}{c:}{MV} &
                                \multicolumn{1}{c}{MDP} & \multicolumn{1}{c|}{MV} \\ \hline
$x_{1}^{\star}$                 & $0.80$ & $99.23\%$ & $-27.94\%$ & $211.18\%$ & $ 0.00\%$ & $100.00\%$ \\
$x_{2}^{\star}$                 & $0.90$ & $96.35\%$ & $ 43.69\%$ & $-51.98\%$ & $25.00\%$ & $  0.00\%$ \\
$x_{3}^{\star}$                 & $1.10$ & $82.62\%$ & $ 43.86\%$ & $-24.84\%$ & $39.24\%$ & $  0.00\%$ \\
$x_{4}^{\star}$                 & $1.20$ & $84.80\%$ & $ 40.39\%$ & $-34.37\%$ & $35.76\%$ & $  0.00\%$ \\ \hdashline
$\sigma\left( x^{\star}\right)$ &        &           & $ 24.54\%$ & $ 13.42\%$ & $23.16\%$ & $ 16.12\%$ \\
\hline
\end{tabular}%
\end{equation*}
\item The results are:%
\begin{equation*}
\begin{tabular}{|c|rr:rr|}
\hline
                                & \multicolumn{2}{c:}{$x_i \in \mathbb{R}$} & \multicolumn{2}{c|}{$x_i \geq 0$} \\
                                & \multicolumn{1}{c}{MDP} & \multicolumn{1}{c:}{MV} &
                                \multicolumn{1}{c}{MDP} & \multicolumn{1}{c|}{MV} \\ \hline
$x_{1}^{\star}$                 & $-36.98\%$ & $ 60.76\%$ & $ 0.00\%$ & $48.17\%$ \\
$x_{2}^{\star}$                 & $-36.98\%$ & $ 60.76\%$ & $ 0.00\%$ & $48.17\%$ \\
$x_{3}^{\star}$                 & $ 91.72\%$ & $-18.54\%$ & $50.00\%$ & $ 0.00\%$ \\
$x_{4}^{\star}$                 & $ 82.25\%$ & $ -2.98\%$ & $50.00\%$ & $ 3.66\%$ \\  \hdashline
$\sigma\left( x^{\star}\right)$ & $ 48.59\%$ & $  6.43\%$ & $30.62\%$ & $ 9.57\%$ \\
\hline
\end{tabular}%
\end{equation*}

\item These two examples show that the MDP may have a different behavior
than the minimum variance portfolio. Contrary to the latter, the most
diversified portfolio is not necessarily a low-beta or a low-volatility
portfolio.
\end{enumerate}
\end{enumerate}

\section{Risk allocation with yield curve factors}

\begin{enumerate}
\item
\begin{enumerate}
\item Let $v_{i}$ be the $i^{\mathrm{th}}$ eigenvector. We have:%
\begin{equation*}
Av_{i}=\lambda _{i}v_{i}
\end{equation*}%
Generally, we assume that the eigenvector is normalized, that is $%
v_{i}^{\top }v_{i}=1$. In a matrix form, the previous relationship becomes:%
\begin{equation*}
AV=V\Lambda
\end{equation*}%
with $\Lambda =\limfunc{diag}\left( \lambda _{1},\ldots ,\lambda _{n}\right)
$ and $V=\left(
\begin{array}{ccc}
v_{1} & \cdots & v_{n}%
\end{array}%
\right) $. We deduce that:%
\begin{equation*}
A=V\Lambda V^{-1}
\end{equation*}%
We have:%
\begin{eqnarray*}
\limfunc{tr}\left( A\right) &=&\limfunc{tr}\left( V\Lambda
V^{-1}\right) \\
&=&\limfunc{tr}\left( \Lambda V^{-1}V\right) \\
&=&\limfunc{tr}\left( \Lambda \right) \\
&=&\sum_{i=1}^{n}\lambda _{i}
\end{eqnarray*}%
and\footnote{%
Because $\det \left( V\right) \cdot \det \left( V^{-1}\right) =\det \left(
I\right) =1$.}:%
\begin{eqnarray*}
\det \left( A\right) &=&\det \left( V\Lambda V^{-1}\right) \\
&=&\det \left( V\right) \cdot \det \left( \Lambda \right) \cdot \det \left(
V^{-1}\right) \\
&=&\det \left( \Lambda \right) \\
&=&\prod_{i=1}^{n}\lambda _{i}
\end{eqnarray*}

\item We have $\Sigma v_{i}=\lambda _{i}v_{i}$ or $v_{i}^{\top }\Sigma
=\lambda _{i}v_{i}^{\top }$. We deduce that:%
\begin{equation*}
v_{i}^{\top }\Sigma v_{j}=\lambda _{i}v_{i}^{\top }v_{j}
\end{equation*}%
and:%
\begin{equation*}
v_{j}^{\top }\Sigma v_{i}=\lambda _{j}v_{j}^{\top }v_{i}=\lambda
_{j}v_{i}^{\top }v_{j}
\end{equation*}%
Moreover, we have:%
\begin{equation*}
v_{i}^{\top }\Sigma v_{j}=\left( v_{i}^{\top }\Sigma v_{j}\right) ^{\top
}=v_{j}^{\top }\Sigma v_{i}
\end{equation*}%
We finally obtain $\lambda _{i}v_{i}^{\top }v_{j}=\lambda _{j}v_{i}^{\top
}v_{j}$ or:%
\begin{equation*}
\left( \lambda _{i}-\lambda _{j}\right) v_{i}^{\top }v_{j}=0
\end{equation*}%
Because $\lambda _{i}\neq \lambda _{j}$, we conclude that $v_{i}^{\top
}v_{j} $ meaning that eigenvectors are orthogonal. We then have $V^{\top
}V=I$ and:%
\begin{eqnarray*}
\Sigma &=&V\Lambda V^{-1} \\
&=&V\Lambda V^{\top }
\end{eqnarray*}

\item If $C=C_{n}\left( \rho \right) $, we know that $\lambda _{2}=\ldots
=\lambda _{n}=\lambda $. It follows that:%
\begin{equation*}
\sum_{i=1}^{n}\lambda _{i}=\lambda _{1}+\left( n-1\right) \lambda =n
\end{equation*}%
and\footnote{%
See Exercise 3.2.}:%
\begin{equation*}
\prod_{i=1}^{n}\lambda _{i}=\lambda _{1}\lambda ^{n-1}=\left( 1-\rho \right)
^{n-1}\left( \left( n-1\right) \rho +1\right)
\end{equation*}%
We deduce that:%
\begin{equation*}
\lambda _{1}=1+\left( n-1\right) \rho
\end{equation*}%
and:%
\begin{equation*}
\lambda _{i}=1-\rho \text{\quad if\quad }i>1
\end{equation*}%
It proves the result because $\bar{\rho}=\rho $. We note $\pi _{1}$ the
percentage of variance explained by the first eigenvalue. We have:%
\begin{eqnarray*}
\pi _{1} &=&\frac{\lambda _{1}}{\limfunc{tr}\left( C\right) } \\
&=&\frac{\lambda _{1}}{n}
\end{eqnarray*}%
We get $\pi _{1}\geq \pi _{1}^{-}$ with:%
\begin{eqnarray*}
\pi _{1}^{-} &=&\frac{1+\left( n-1\right) \bar{\rho}}{n} \\
&=&\bar{\rho}+\frac{\left( 1-\bar{\rho}\right) }{n}
\end{eqnarray*}%
$\pi _{1}^{-}$ takes the following values:%
\begin{equation*}
\begin{tabular}{|c|ccccc|}
\hline
$n/\bar{\rho}$ & 10\% & 20\% & 50\% & 70\% & 90\% \\ \hline
2  & $55.0\%$ & $60.0\%$ & $75.0\%$ & $85.0\%$ & $95.0\%$ \\
3  & $40.0\%$ & $46.7\%$ & $66.7\%$ & $80.0\%$ & $93.3\%$ \\
5  & $28.0\%$ & $36.0\%$ & $60.0\%$ & $76.0\%$ & $92.0\%$ \\
10 & $19.0\%$ & $28.0\%$ & $55.0\%$ & $73.0\%$ & $91.0\%$ \\ \hline
\end{tabular}%
\end{equation*}%
We notice that $\pi _{1}^{-}\simeq \bar{\rho}$ when $n$ is large.

\item We obtain the following results:%
\begin{equation*}
\begin{tabular}{|c|rrrr|}
\hline
Asset & \multicolumn{1}{c}{$v_{1}$} & \multicolumn{1}{c}{$v_{2}$} & \multicolumn{1}{c}{$v_{3}$} & \multicolumn{1}{c|}{$v_{4}$} \\
\hline
1              & $0.2704$ & $ 0.5900$ & $ 0.3351$ & $ 0.6830$ \\
2              & $0.4913$ & $ 0.3556$ & $ 0.3899$ & $-0.6929$ \\
3              & $0.4736$ & $ 0.2629$ & $-0.8406$ & $-0.0022$ \\
4              & $0.6791$ & $-0.6756$ & $ 0.1707$ & $ 0.2309$ \\ \hdashline
$\lambda _{i}$ & $0.0903$ & $ 0.0416$ & $ 0.0358$ & $ 0.0156$ \\
\hline
\end{tabular}%
\end{equation*}%
Let us consider the following optimization problem:%
\begin{eqnarray*}
x^{\star } &=&\arg \max \frac{1}{2}x^{\top }\Sigma x \\
&\text{u.c.}&x^{\top }x=1
\end{eqnarray*}%
The first-order condition is $\Sigma x-\lambda x=0$ where $\lambda $ is the
Lagrange coefficient associated to the normalization constraint $x^{\top }x=1
$. It corresponds precisely to the definition of the eigendecomposition. We
deduce that $x^{\star }$ is the first eigenvector:%
\begin{equation*}
x^{\star }=v_{1}
\end{equation*}%
It means that the first eigenvector is the maximum variance portfolio under
the normalization constraint. In the same way, we can show that the last
eigenvector is the minimum variance portfolio under the normalization
constraint:%
\begin{eqnarray*}
v_{n} &=&\arg \min \frac{1}{2}x^{\top }\Sigma x \\
&\text{u.c.}&x^{\top }x=1
\end{eqnarray*}

\item In the case of the correlation matrix, the eigendecomposition is:%
\begin{equation*}
\begin{tabular}{|c|rrrr|}
\hline
Asset & \multicolumn{1}{c}{$v_{1}$} & \multicolumn{1}{c}{$v_{2}$} & \multicolumn{1}{c}{$v_{3}$} & \multicolumn{1}{c|}{$v_{4}$} \\
\hline
1              & $0.4742$ & $ 0.6814$ & $ 0.0839$ & $-0.5512$ \\
2              & $0.6026$ & $ 0.2007$ & $-0.2617$ & $ 0.7267$ \\
3              & $0.4486$ & $-0.3906$ & $ 0.8035$ & $ 0.0253$ \\
4              & $0.4591$ & $-0.5855$ & $-0.5281$ & $-0.4092$ \\ \hdashline
$\lambda _{i}$ & $1.9215$ & $ 0.9467$ & $ 0.7260$ & $ 0.4059$ \\
\hline
\end{tabular}%
\end{equation*}%
We obtain $\pi _{1}=48.04\%$ and $\pi _{1}^{-}=47.50\%$ because $\bar{\rho}%
=30\%$. We notice that the lower bound is close to the true value.

\item Let us specify the risk model as follows (TR-RPB, page 38):%
\begin{equation*}
R_{t}=A\mathcal{F}_{t}+\varepsilon _{t}
\end{equation*}%
where $R_{t}$ is the vector of asset returns, $\mathcal{F}_{t}$ is the vector of
risk factors and $\varepsilon _{t}$ is the vector of idiosyncratic factors. We
assume that $\limfunc{cov}\left( R_{t}\right) =\Sigma $, $\limfunc{cov}%
\left( \mathcal{F}_{t}\right) =\Omega $ and $\limfunc{cov}\left( \varepsilon
_{t}\right) =D$. Moreover, we suppose that the risk factors and the
idiosyncratic factors are independent. We know that (TR-RPB, page 38):%
\begin{equation*}
\Sigma =A\Omega A^{\top }+D
\end{equation*}%
If we consider a principal component analysis, we have (TR-RPB, page 216):%
\begin{equation*}
\Sigma =V\Lambda V^{\top }
\end{equation*}%
We could then consider that the risk factors correspond to the
eigenfactors and we have $A=V$, $\Omega =\Lambda $ and $D=\mathbf{0}$. In the
case of two factors, the risk factors are the first two eigenfactors and we
have $A=\left(
\begin{array}{cc}
v_{1} & v_{2}%
\end{array}%
\right) $, $\Omega =\limfunc{diag}\left( \lambda _{1},\lambda _{2}\right) $
and $D=\Sigma -A\Omega A^{\top }$\textbf{.} We notice that:%
\begin{eqnarray*}
R_{t} &=&\sum_{j=1}^{4}v_{j}\mathcal{F}_{j,t} \\
&=&\sum_{j=1}^{2}v_{j}\mathcal{F}_{j,t}+\sum_{j=3}^{4}v_{j}\mathcal{F}_{j,t}
\\
&=&\sum_{j=1}^{2}v_{j}\mathcal{F}_{j,t}+\varepsilon _{t}
\end{eqnarray*}%
Another expression of $D$ is then $D=B\Phi B^{\top }$ with $B=\left(
\begin{array}{cc}
v_{3} & v_{4}%
\end{array}%
\right) $ and $\Phi =\limfunc{diag}\left( \lambda _{3},\lambda _{4}\right) $%
. We also verify that $\limfunc{cov}\left( \mathcal{F}_{t},\varepsilon
_{t}\right) =\mathbf{0}$ by definition of the eigendecomposition. If we
would like to impose that $D$ is diagonal, a simple way is to consider only
the diagonal elements of the matrix $\Sigma -A\Omega A^{\top }$. We have:%
\begin{eqnarray*}
\mathcal{F}_{1,t} &=&0.47\cdot R_{1,t}+0.60\cdot R_{2,t}+0.45\cdot
R_{3,t}+0.46\cdot R_{4,t} \\
\mathcal{F}_{1,t} &=&0.68\cdot R_{1,t}+0.20\cdot R_{2,t}-0.39\cdot
R_{3,t}-0.59\cdot R_{4,t}
\end{eqnarray*}%
The first factor is a long-only portfolio. It represents the market factor.
The second factor is a long-short portfolio. It represents an arbitrage
factor between the first two assets and the last two assets.
\end{enumerate}

\item

\begin{enumerate}
\item We consider that the risk factors are the zero-coupon rates.
For each maturity, we compute the sensitivity  (TR-RPB,
page 204):%
\begin{equation*}
\delta \left( T_{i}\right) =-D_{t}\left( T_{i}\right) \cdot B_{t}\left(
T_{i}\right) \cdot C\left( T_{i}\right)
\end{equation*}%
where $D_{t}\left( T_{i}\right) $, $B_{t}\left( T_{i}\right) $ and $C\left(
T_{i}\right) $ are the duration, the price and the coupon of the zero-coupon
bond with maturity $T_{i}$. Using Equation (4.2) of TR-RPB on page 204, we
obtain the following results:%
\begin{equation*}
\begin{tabular}{|c|rrrr|}
\hline
$T_i$ & \multicolumn{1}{c}{$\delta \left( T_{i}\right)$} & \multicolumn{1}{c}{$\mathcal{MR}_{i}$} &
\multicolumn{1}{c}{$\mathcal{RC}_{i}$} & \multicolumn{1}{c|}{$\mathcal{RC}_{i}^{\star }$} \\
\hline
1Y     &      $-1.00$ & $-0.21\%$ & $0.21\%$ & $ 1.98\%$ \\
3Y     &      $-2.94$ & $-0.41\%$ & $1.20\%$ & $11.20\%$ \\
5Y     &      $-4.77$ & $-0.47\%$ & $2.25\%$ & $21.07\%$ \\
7Y     &      $-6.37$ & $-0.48\%$ & $3.08\%$ & $28.75\%$ \\
10Y    &      $-8.36$ & $-0.47\%$ & $3.96\%$ & $37.01\%$ \\ \hdashline
$\limfunc{VaR}\nolimits_{\alpha }\left( x\right)$
       &              &         & $10.70\%$ &           \\
\hline
\end{tabular}%
\end{equation*}
If we prefer to consider the number $n_{i}$ of the zero-coupon bond with
maturity $T_{i}$, we have:%
\begin{equation*}
\begin{tabular}{|c|rrrr|}
\hline
$T_i$ & \multicolumn{1}{c}{$n_{i}$} & \multicolumn{1}{c}{$\mathcal{MR}_{i}$} &
\multicolumn{1}{c}{$\mathcal{RC}_{i}$} & \multicolumn{1}{c|}{$\mathcal{RC}_{i}^{\star }$} \\
\hline
1Y     &      $1.00$ & $0.21\%$ & $0.21\%$ & $ 1.98\%$ \\
3Y     &      $1.00$ & $1.20\%$ & $1.20\%$ & $11.20\%$ \\
5Y     &      $1.00$ & $2.25\%$ & $2.25\%$ & $21.07\%$ \\
7Y     &      $1.00$ & $3.08\%$ & $3.08\%$ & $28.75\%$ \\
10Y    &      $1.00$ & $3.96\%$ & $3.96\%$ & $37.01\%$ \\ \hdashline
$\limfunc{VaR}\nolimits_{\alpha }\left( x\right)$
       &              &         & $10.70\%$ &           \\
\hline
\end{tabular}%
\end{equation*}
Finally, if we measure the exposures in terms of relative weights, we obtain
(TR-RPB, Equation (4.3), page 205):%
\begin{equation*}
\begin{tabular}{|c|rrrr|}
\hline
$T_i$ & \multicolumn{1}{c}{$\varpi _{i}$} & \multicolumn{1}{c}{$\mathcal{MR}_{i}$} &
\multicolumn{1}{c}{$\mathcal{RC}_{i}$} & \multicolumn{1}{c|}{$\mathcal{RC}_{i}^{\star }$} \\
\hline
1Y     &      $21.28\%$ & $ 1.00\%$ & $0.21\%$ & $ 1.98\%$ \\
3Y     &      $20.99\%$ & $ 5.71\%$ & $1.20\%$ & $11.20\%$ \\
5Y     &      $20.39\%$ & $11.06\%$ & $2.25\%$ & $21.07\%$ \\
7Y     &      $19.46\%$ & $15.81\%$ & $3.08\%$ & $28.75\%$ \\
10Y    &      $17.88\%$ & $22.16\%$ & $3.96\%$ & $37.01\%$ \\ \hdashline
$\limfunc{VaR}\nolimits_{\alpha }\left( x\right)$
       &              &         & $10.70\%$ &           \\
\hline
\end{tabular}%
\end{equation*}
\item The eigendecomposition of $\Sigma $ is:%
\begin{equation*}
\begin{tabular}{|c|rrrrr|}
\hline
$T_i$ & \multicolumn{1}{c}{$v_{1}$} & \multicolumn{1}{c}{$v_{2}$} &
        \multicolumn{1}{c}{$v_{3}$} & \multicolumn{1}{c}{$v_{4}$} & \multicolumn{1}{c|}{$v_{5}$} \\
\hline
1Y     &                        $ 0.2475$ & $-0.7909$ & $-0.5471$ & $ 0.1181$ & $  0.0016$ \\
3Y     &                        $ 0.4427$ & $-0.3469$ & $ 0.5747$ & $-0.5898$ & $  0.0739$ \\
5Y     &                        $ 0.5012$ & $ 0.0209$ & $ 0.3292$ & $ 0.6215$ & $ -0.5038$ \\
7Y     &                        $ 0.5040$ & $ 0.2487$ & $-0.0736$ & $ 0.2583$ & $  0.7823$ \\
10Y    &                        $ 0.4874$ & $ 0.4380$ & $-0.5066$ & $-0.4304$ & $ -0.3588$ \\ \hdashline
$\lambda _{i}$ ($\times 10^{-6}$) &       $16.6446$ & $ 1.5685$ & $ 0.2421$ & $ 0.0273$ & $  0.0045$ \\
$\pi_{i}$ (in \%)       &       $90.0340$ & $ 8.4846$ & $ 1.3095$ & $ 0.1478$ & $  0.0241$ \\
$\pi _{i}^{\star }$ (in \%) &    $90.0340$ & $98.5186$ & $99.8281$ & $99.9759$ & $100.0000$ \\
\hline
\end{tabular}%
\end{equation*}
where $\pi _{i}^{\star }=\sum_{j=1}^{i}\pi _{i}$ is the percentage of
variance explained by the top $i$ eigenvectors. The first three factors are
the level, slope and convexity factors. We have represented them in Figure
\ref{fig:app2-3-5-2}. They differ slightly from those reported in TR-RPB on page 197, because
the set of maturities is different. Note also that the convexity factor is
the opposite of the one obtained in TR-RPB because eigenvectors are not
signed.

\begin{figure}[tbph]
\centering
\includegraphics[width = \figurewidth, height = \figureheight]{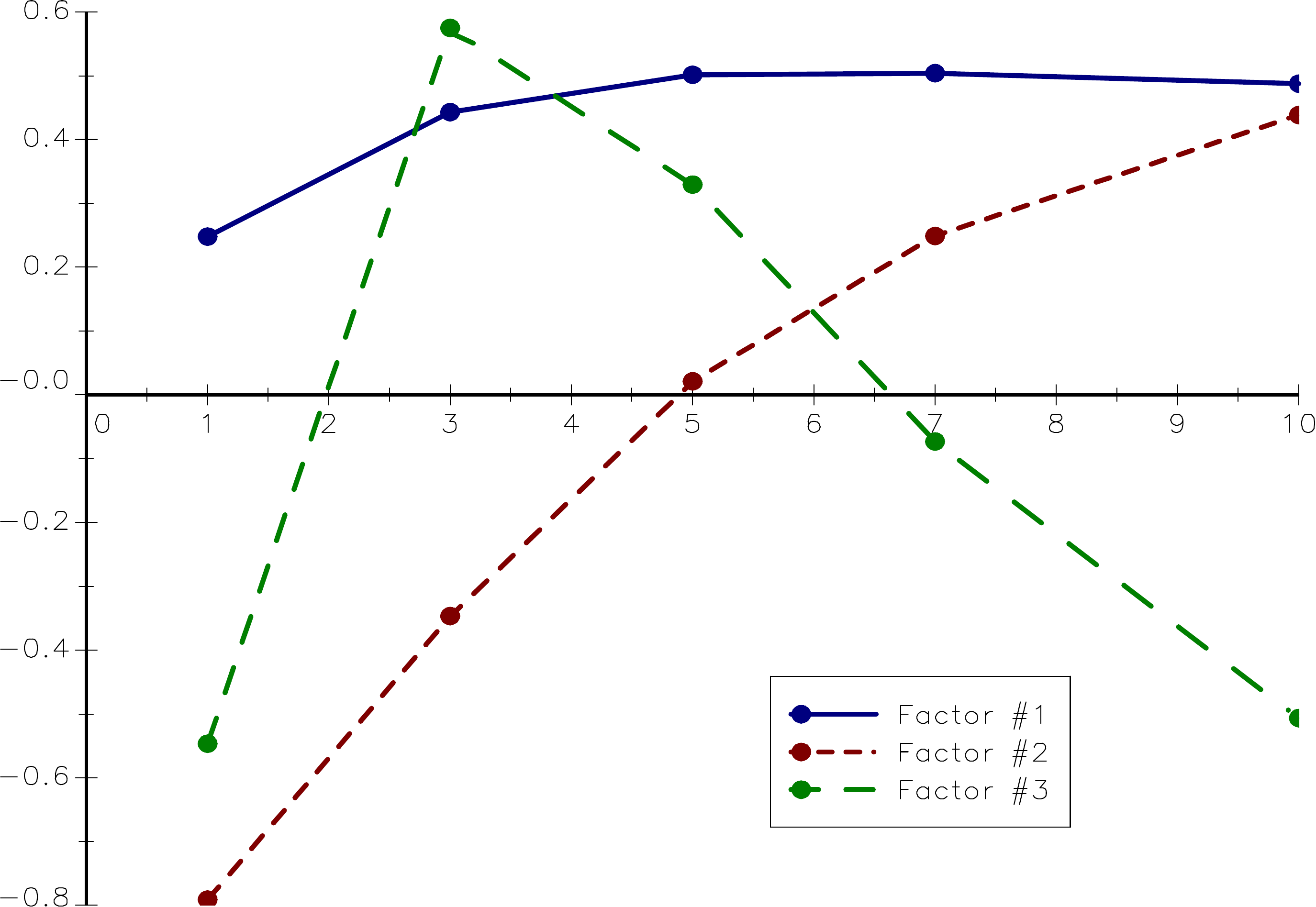}
\caption{Representation of the first three PCA factors}
\label{fig:app2-3-5-2}
\end{figure}

\item Let $V\Lambda V^{\top }$ be the eigendecomposition of $\Sigma $. We
have $A=V$, $\Omega =\Lambda $ and $D=\mathbf{0}$. The risk contribution of
the $j^{\mathrm{th}}$ factor is then (TR-RPB, page 142):%
\begin{equation*}
\mathcal{RC}\left( \mathcal{F}_{j}\right) =\Phi ^{-1}\left( \alpha \right)
\cdot \left( A^{\top }\delta \right) _{j}\cdot \left( \frac{A^{+}\Sigma
\delta }{\sqrt{\delta ^{\top }\Sigma \delta }}\right) _{j}
\end{equation*}%
We obtain the following risk decomposition:%
\begin{equation*}
\begin{tabular}{|c|rrrr|}
\hline
PCA factor & \multicolumn{1}{c}{$y_{j}$} & \multicolumn{1}{c}{$\mathcal{MR}_{j}$} &
\multicolumn{1}{c}{$\mathcal{RC}_{j}$} & \multicolumn{1}{c|}{$\mathcal{RC}_{j}^{\star }$} \\
\hline
$\mathcal{F}_{1}$ &      $-11.22$ & $-0.94\%$ & $10.60\%$ & $99.03\%$ \\
$\mathcal{F}_{2}$ &      $ -3.54$ & $-0.03\%$ & $ 0.10\%$ & $ 0.93\%$ \\
$\mathcal{F}_{3}$ &      $  1.99$ & $ 0.00\%$ & $ 0.00\%$ & $ 0.05\%$ \\
$\mathcal{F}_{4}$ &      $  0.61$ & $ 0.00\%$ & $ 0.00\%$ & $ 0.00\%$ \\
$\mathcal{F}_{5}$ &      $  0.20$ & $ 0.00\%$ & $ 0.00\%$ & $ 0.00\%$ \\ \hdashline
$\limfunc{VaR}\nolimits_{\alpha }\left( x\right)$
       &              &         & $10.70\%$ &           \\
\hline
\end{tabular}%
\end{equation*}
We notice that most of the risk is explained by the level factor (about $99\%$).

\item The ERC portfolio is the following:%
\begin{equation*}
\begin{tabular}{|c|rrrr|}
\hline
$T_i$ & \multicolumn{1}{c}{$n_{i}$} & \multicolumn{1}{c}{$\mathcal{MR}_{i}$} &
\multicolumn{1}{c}{$\mathcal{RC}_{i}$} & \multicolumn{1}{c|}{$\mathcal{RC}_{i}^{\star }$} \\
\hline
1Y     &      $14.08$ & $0.26\%$ & $3.71\%$ & $20.00\%$ \\
3Y     &      $ 2.95$ & $1.26\%$ & $3.71\%$ & $20.00\%$ \\
5Y     &      $ 1.66$ & $2.24\%$ & $3.71\%$ & $20.00\%$ \\
7Y     &      $ 1.25$ & $2.96\%$ & $3.71\%$ & $20.00\%$ \\
10Y    &      $ 1.00$ & $3.71\%$ & $3.71\%$ & $20.00\%$ \\ \hdashline
$\limfunc{VaR}\nolimits_{\alpha }\left( x\right)$
       &              &         & $18.54\%$ &           \\
\hline
\end{tabular}%
\end{equation*}
whereas its risk decomposition with respect to PCA factors is:%
\begin{equation*}
\begin{tabular}{|c|rrrr|}
\hline
PCA factor & \multicolumn{1}{c}{$y_{j}$} & \multicolumn{1}{c}{$\mathcal{MR}_{j}$} &
\multicolumn{1}{c}{$\mathcal{RC}_{j}$} & \multicolumn{1}{c|}{$\mathcal{RC}_{j}^{\star }$} \\
\hline
$\mathcal{F}_{1}$ &      $-19.36$ & $-0.94\%$ & $18.21\%$ & $98.22\%$ \\
$\mathcal{F}_{2}$ &      $  8.28$ & $ 0.04\%$ & $ 0.31\%$ & $ 1.69\%$ \\
$\mathcal{F}_{3}$ &      $  4.90$ & $ 0.00\%$ & $ 0.02\%$ & $ 0.09\%$ \\
$\mathcal{F}_{4}$ &      $  0.09$ & $ 0.00\%$ & $ 0.00\%$ & $ 0.00\%$ \\
$\mathcal{F}_{5}$ &      $  0.07$ & $ 0.00\%$ & $ 0.00\%$ & $ 0.00\%$ \\ \hdashline
$\limfunc{VaR}\nolimits_{\alpha }\left( x\right)$
       &              &         & $18.54\%$ &           \\
\hline
\end{tabular}%
\end{equation*}
We notice that the ERC portfolio is more exposed to the slope factor than the previous EW portfolio.

\item We suppose now that portfolio $x$ is equal to the eigenvector $v_i$.
In the following table, we have reported the normalized risk contributions $\mathcal{RC}_{j}^{\star }$:
\begin{equation*}
\begin{tabular}{|c|rrrrr|}
\hline
PCA factor & \multicolumn{1}{c}{$v_{1}$} & \multicolumn{1}{c}{$v_{2}$} &
\multicolumn{1}{c}{$v_{3}$} & \multicolumn{1}{c}{$v_{4}$} & \multicolumn{1}{c|}{$v_{5}$} \\
\hline
$\mathcal{F}_{1}$ &  $98.72\%$ & $81.43\%$ & $55.57\%$ & $22.62\%$ & $14.62\%$ \\
$\mathcal{F}_{2}$ &  $ 1.24\%$ & $17.24\%$ & $28.41\%$ & $23.43\%$ & $17.07\%$ \\
$\mathcal{F}_{3}$ &  $ 0.04\%$ & $ 1.32\%$ & $15.72\%$ & $26.63\%$ & $15.44\%$ \\
$\mathcal{F}_{4}$ &  $ 0.00\%$ & $ 0.01\%$ & $ 0.29\%$ & $27.16\%$ & $ 6.75\%$ \\
$\mathcal{F}_{5}$ &  $ 0.00\%$ & $ 0.00\%$ & $ 0.00\%$ & $ 0.17\%$ & $46.13\%$ \\
\hline
\end{tabular}%
\end{equation*}
We notice that the risk contribution of the level factor decreases when we
consider higher orders of eigenvectors. For instance, the risk contribution is equal
to $14.62\%$ if the portfolio is the fifth eigenvector whereas it is equal to $98.72\%$
if the portfolio is the first eigenvector. We also observe that the risk contribution
$\mathcal{RC}_{j}^{\star }$ of the $j^\mathrm{th}$ factor is generally high when the portfolio
corresponds to the $j^\mathrm{th}$ eigenvector.
\end{enumerate}
\end{enumerate}

\section{Credit risk analysis of sovereign bond portfolios}

\begin{enumerate}
\item We recall that the credit risk measure of a bond portfolio is (TR-RPB,
page 227):%
\begin{equation*}
\mathcal{R}\left( x\right) =\sqrt{x^{\top }\Sigma x}
\end{equation*}%
where $\Sigma =\left( \Sigma _{i,j}\right) $ and $\Sigma _{i,j}$ is the
credit covariance between the bond $i$ and the bond $j$. We have:%
\begin{equation*}
\Sigma _{i,j}=\rho _{i,j}\sigma _{i}^{\mathfrak{c}}\sigma _{i}^{\mathfrak{c}}
\end{equation*}%
and:%
\begin{equation*}
\sigma _{i}^{\mathfrak{c}}=D_{i}\sigma _{i}^{\mathfrak{s}}\mathfrak{s}%
_{i}\left( t\right)
\end{equation*}%
where $D_{i}$ is the duration of bond $i$.

\begin{enumerate}
\item We obtain the following results:%
\begin{equation*}
\begin{tabular}{|c|cccc|}
\hline
Bond & $x^{\left(1\right)}_{i}$  & $\mathcal{MR}_{i}$ & $\mathcal{RC}_{i}$ & $\mathcal{RC}^{\star}_{i}$ \\ \hline
$1$  &     $10$ & $0.022$ & $0.221$ &     $14.8\%$ \\
$2$  &     $12$ & $0.012$ & $0.147$ & ${\bP}9.8\%$ \\
$3$  & ${\bP}8$ & $0.066$ & $0.526$ &     $35.2\%$ \\
$4$  & ${\bP}7$ & $0.086$ & $0.602$ &     $40.2\%$ \\ \hdashline
$\mathcal{R}\left(x^{\left(1\right)}\right) $ &  &  & $1.495$ &  \\ \hline
\end{tabular}
\end{equation*}

\item We obtain the following results:%
\begin{equation*}
\begin{tabular}{|c|cccc|}
\hline
Bond & $x^{\left(2\right)}_{i}$  & $\mathcal{MR}_{i}$ & $\mathcal{RC}_{i}$ & $\mathcal{RC}^{\star}_{i}$ \\ \hline
$1$  & ${\bP}8$ & $0.023$ & $0.186$ & ${\bP}6.6\%$ \\
$2$  & ${\bP}8$ & $0.010$ & $0.084$ & ${\bP}3.0\%$ \\
$3$  &     $12$ & $0.065$ & $0.785$ &     $27.6\%$ \\
$4$  &     $14$ & $0.128$ & $1.788$ &     $62.9\%$ \\ \hdashline
$\mathcal{R}\left(x^{\left(2\right)}\right) $ &  &  & $2.843$ &  \\
\hline
\end{tabular}
\end{equation*}

\item The portfolio is now composed by two bonds per country, i.e. eight
bonds. The correlation matrix becomes:%
\begin{equation*}
\rho =\left(
\begin{array}{cccc:cccc}
$1.00$ & $0.65$ & $0.67$ & $0.64$ & $1.00$ & $0.65$ & $0.67$ & $0.64$ \\
$0.65$ & $1.00$ & $0.70$ & $0.67$ & $0.65$ & $1.00$ & $0.70$ & $0.67$ \\
$0.67$ & $0.70$ & $1.00$ & $0.83$ & $0.67$ & $0.70$ & $1.00$ & $0.83$ \\
$0.64$ & $0.67$ & $0.83$ & $1.00$ & $0.64$ & $0.67$ & $0.83$ & $1.00$ \\ \hdashline
$1.00$ & $0.65$ & $0.67$ & $0.64$ & $1.00$ & $0.65$ & $0.67$ & $0.64$ \\
$0.65$ & $1.00$ & $0.70$ & $0.67$ & $0.65$ & $1.00$ & $0.70$ & $0.67$ \\
$0.67$ & $0.70$ & $1.00$ & $0.83$ & $0.67$ & $0.70$ & $1.00$ & $0.83$ \\
$0.64$ & $0.67$ & $0.83$ & $1.00$ & $0.64$ & $0.67$ & $0.83$ & $1.00$
\end{array}%
\right)
\end{equation*}%
We obtain the following results:%
\begin{equation*}
\begin{tabular}{|c|cccc|}
\hline Bond & $x^{\left(1+2\right)}_{i}$  & $\mathcal{MR}_{i}$ &
$\mathcal{RC}_{i}$ & $\mathcal{RC}^{\star}_{i}$ \\ \hline
$1$ (\#1) &     $10$ & $0.021$ & $0.210$ & ${\bP}4.8\%$ \\
$2$ (\#1) &     $12$ & $0.012$ & $0.141$ & ${\bP}3.3\%$ \\
$3$ (\#1) & ${\bP}8$ & $0.065$ & $0.520$ &     $12.0\%$ \\
$4$ (\#1) & ${\bP}7$ & $0.088$ & $0.618$ &     $14.3\%$ \\
$1$ (\#2) & ${\bP}8$ & $0.024$ & $0.192$ & ${\bP}4.4\%$ \\
$2$ (\#2) & ${\bP}8$ & $0.011$ & $0.086$ & ${\bP}2.0\%$ \\
$3$ (\#2) &     $12$ & $0.066$ & $0.793$ &     $18.3\%$ \\
$4$ (\#2) &     $14$ & $0.126$ & $1.769$ &     $40.9\%$ \\ \hdashline
$\mathcal{R}\left(x^{\left(1+2\right)}\right) $ &  &  & $4.329$ &  \\
\hline
\end{tabular}
\end{equation*}%
We notice that $\mathcal{R}\left( x^{\left( 1+2\right) }\right) \simeq
\mathcal{R}\left( x^{\left( 1\right) }\right) +\mathcal{R}\left( x^{\left(
2\right) }\right) $. The diversification effect is limited because the two
portfolios $x^{\left( 1\right) }$ and $x^{\left( 2\right) }$ are highly
correlated:%
\begin{equation*}
\rho \left( x^{\left( 1\right) },x^{\left( 2\right) }\right) = 99.01\%
\end{equation*}

\item The notional of the meta-bond for the country $i$ is the sum of
notional of the two bonds, which belong to this country:%
\begin{equation*}
x_{i}^{\left( 3\right) }=x_{i}^{\left( 1\right) }+x_{i}^{\left( 2\right) }
\end{equation*}%
Its duration is the weighted average:%
\begin{equation*}
D_{i}^{\left( 3\right) }=\frac{x_{i}^{\left( 1\right) }}{x_{i}^{\left(
1\right) }+x_{i}^{\left( 2\right) }}D_{i}^{\left( 1\right) }+\frac{%
x_{i}^{\left( 2\right) }}{x_{i}^{\left( 1\right) }+x_{i}^{\left( 2\right) }}%
D_{i}^{\left( 2\right) }
\end{equation*}%
We obtain the following results:%
\begin{equation*}
\begin{tabular}{|c|ccccc|}
\hline
Bond & $x^{\left(3\right)}_{i}$  & $D_{i}^{\left( 3\right) }$ & $\mathcal{MR}_{i}$ & $\mathcal{RC}_{i}$ & $\mathcal{RC}^{\star}_{i}$ \\ \hline
$1$  & $18$ & $6.600$ & $0.022$ & $0.402$ & ${\bP}9.3\%$ \\
$2$  & $20$ & $7.260$ & $0.011$ & $0.227$ & ${\bP}5.2\%$ \\
$3$  & $20$ & $6.460$ & $0.066$ & $1.313$ &     $30.3\%$ \\
$4$  & $21$ & $7.467$ & $0.114$ & $2.387$ &     $55.1\%$ \\ \hdashline
$\mathcal{R}\left(x^{\left(3\right)}\right) $ &  &  & $4.329$ &  & \\
\hline
\end{tabular}
\end{equation*}%
Let us consider the results of the previous question. If we aggregate the
risk contributions by countries, we obtain:%
\begin{equation*}
\begin{tabular}{|c|ccc|}
\hline
Country & $x^{\left(1+2\right)}_{j}$  & $\mathcal{RC}_{j}$ & $\mathcal{RC}^{\star}_{j}$ \\ \hline
France  & $18$ & $0.402$ & ${\bP}9.3\%$ \\
Germany  & $20$ & $0.227$ & ${\bP}5.2\%$ \\
Italy  & $20$ & $1.313$ &     $30.3\%$ \\
Spain  & $21$ & $2.387$ &     $55.1\%$ \\
\hline
\end{tabular}
\end{equation*}%
We notice that we have exactly the same results. We can postulate the
hypothesis that computing the risk contributions based on the portfolio in
its entirety is equivalent to compute the risk contributions by considering
one meta-bond by country.
\item The notional invested in each country is very close to $20$ billions of dollars. However,
most of the credit risk in concentrated in Italian and Spanish bonds. Indeed, these two countries
represent about $85\%$ of the credit risk of the portfolio.
\end{enumerate}

\item
\begin{enumerate}
\item The last two assets are perfectly correlated. It means that:%
\begin{equation*}
\rho _{i,n}=\rho _{i,n+1}
\end{equation*}%
If $i<n$, we have:%
\begin{eqnarray*}
\mathcal{RC}_{i}\left( x\right)  &=&\frac{x_{i}\sigma _{i}}{\sigma \left(
x\right) }\left( \sum_{j=1}^{n-1}x_{j}\rho _{i,j}\sigma _{j}+x_{n}\rho
_{i,n}\sigma _{n}+x_{n+1}\rho _{i,n+1}\sigma _{n+1}\right)  \\
&=&\frac{x_{i}\sigma _{i}}{\sigma \left( x\right) }\left(
\sum_{j=1}^{n-1}x_{j}\rho _{i,j}\sigma _{j}+\rho _{i,n}\left( x_{n}\sigma
_{n}+x_{n+1}\sigma _{n+1}\right) \right)
\end{eqnarray*}%
For the last two assets, we have:%
\begin{equation*}
\mathcal{RC}_{n}\left( x\right) =\frac{x_{n}\sigma _{n}}{\sigma \left(
x\right) }\left( \sum_{j=1}^{n-1}x_{j}\rho _{n,j}\sigma _{j}+\left(
x_{n}\sigma _{n}+x_{n+1}\sigma _{n+1}\right) \right)
\end{equation*}%
and:%
\begin{equation*}
\mathcal{RC}_{n+1}\left( x\right) =\frac{x_{n+1}\sigma _{n+1}}{\sigma \left(
x\right) }\left( \sum_{j=1}^{n-1}x_{j}\rho _{n,j}\sigma _{j}+\left(
x_{n}\sigma _{n}+x_{n+1}\sigma _{n+1}\right) \right)
\end{equation*}%
because $\rho _{n,j}=\rho _{n+1,j}$.

\item We have:%
\begin{equation*}
\mathcal{RC}_{i}\left( y\right) =\frac{y_{i}\sigma _{i}^{\prime }}{\sigma
\left( y\right) }\left( \sum_{j=1}^{n-1}y_{j}\rho _{i,j}^{\prime }\sigma
_{j}^{\prime }+y_{n}\rho _{i,n}^{\prime }\sigma _{n}^{\prime }\right)
\end{equation*}

\item By construction, we have:%
\begin{eqnarray*}
\sigma \left( x\right)  &=&\sum_{i=1}^{n-1}\mathcal{RC}_{i}\left( x\right) +%
\mathcal{RC}_{n}\left( x\right) +\mathcal{RC}_{n+1}\left( x\right)  \\
&=&\sum_{i=1}^{n-1}\mathcal{RC}_{i}\left( y\right) +\mathcal{RC}_{n}\left(
y\right)  \\
&=&\sigma \left( y\right)
\end{eqnarray*}%
For $i<n$, $\mathcal{RC}_{i}\left( y\right) =\mathcal{RC}_{i}\left( x\right)
$ implies then:%
\begin{equation}
\left\{
\begin{array}{l}
y_{i}\sigma _{i}^{\prime }=x_{i}\sigma _{i} \\
y_{j}\rho _{i,j}^{\prime }\sigma _{j}^{\prime }=x_{j}\rho _{i,j}\sigma _{j}
\\
y_{n}\rho _{i,n}^{\prime }\sigma _{n}^{\prime }=\rho _{i,n}\left(
x_{n}\sigma _{n}+x_{n+1}\sigma _{n+1}\right)
\end{array}%
\right.   \label{eq:app2-duplication2}
\end{equation}%
If the previous constraints are verified and if we assume that $\rho
_{i,n}^{\prime }=\rho _{i,n}$, we deduce that the restriction $\mathcal{RC}%
_{n}\left( y\right) =\mathcal{RC}_{n}\left( x\right) +\mathcal{RC}%
_{n+1}\left( x\right) $ is satisfied too:%
\begin{eqnarray*}
S &=&\mathcal{RC}_{n}\left( x\right) +\mathcal{RC}_{n+1}\left( x\right)  \\
&=&\frac{x_{n}\sigma _{n}}{\sigma \left( x\right) }\left(
\sum_{j=1}^{n-1}x_{j}\rho _{n,j}\sigma _{j}+\left( x_{n}\sigma
_{n}+x_{n+1}\sigma _{n+1}\right) \right) + \\
&&\frac{x_{n+1}\sigma _{n+1}}{\sigma \left( x\right) }\left(
\sum_{j=1}^{n-1}x_{j}\rho _{n,j}\sigma _{j}+\left( x_{n}\sigma
_{n}+x_{n+1}\sigma _{n+1}\right) \right)  \\
&=&\left( \frac{x_{n}\sigma _{n}}{\sigma \left( x\right) }+\frac{%
x_{n+1}\sigma _{n+1}}{\sigma \left( x\right) }\right) \left(
\sum_{j=1}^{n-1}y_{j}\rho _{n,j}^{\prime }\sigma _{j}^{\prime }+\frac{\rho
_{i,n}^{\prime }}{\rho _{i,n}}y_{n}\sigma _{n}^{\prime }\right)  \\
&=&\frac{y_{n}\sigma _{n}^{\prime }}{\sigma \left( y\right) }\left(
\sum_{j=1}^{n-1}y_{j}\rho _{n,j}^{\prime }\sigma _{j}^{\prime }+y_{n}\sigma
_{n}^{\prime }\right)  \\
&=&\mathcal{RC}_{n}\left( y\right)
\end{eqnarray*}
A solution of the system (\ref{eq:app2-duplication2}) is:%
\begin{equation*}
\left\{
\begin{array}{l}
y_{i}=x_{i} \\
\sigma _{i}^{\prime }=\sigma _{i} \\
\rho _{i,j}^{\prime }=\rho _{i,j} \\
y_{n}\sigma _{n}^{\prime }=\left( x_{n}\sigma _{n}+x_{n+1}\sigma
_{n+1}\right)
\end{array}%
\right.
\end{equation*}%
In fact, the asset universe and the portfolio are the same for $i<n$. The
only change concerns the $n$-th asset.

\item It suffices to choose an arbitrary value for $\sigma _{n}^{\prime }$
and we have:%
\begin{equation*}
y_{n}=\frac{x_{n}\sigma _{n}+x_{n+1}\sigma _{n+1}}{\sigma _{n}^{\prime }}
\end{equation*}%
It implies that there are infinite solutions. If we set $y_{n}=x_{n}+x_{n+1}$%
, we obtain:%
\begin{equation*}
\sigma _{n}^{\prime }=\frac{x_{n}}{x_{n}+x_{n+1}}\sigma _{n}+\frac{x_{n+1}}{%
x_{n}+x_{n+1}}\sigma _{n+1}
\end{equation*}%
In this case, the volatility $\sigma _{n}^{\prime }$ of the $n$-th asset is
a weighted average of the volatilities of the last two assets. If $\sigma
_{n}^{\prime }=\sigma _{n}+\sigma _{n+1}$, we obtain:%
\begin{equation*}
y_{n}=x_{n}\frac{\sigma _{n}}{\sigma _{n}+\sigma _{n+1}}+x_{n+1}\frac{\sigma
_{n+1}}{\sigma _{n}+\sigma _{n+1}}
\end{equation*}%
The weight $y_{n}$ of the $n$-th asset is a weighted average of the weights
of the last two assets. From a financial point of view, we prefer the first
solution $y_{n}=x_{n}+x_{n+1}$, because the exposures of the portfolio $y$
are coherent with the exposures of the portfolio $x$.

\item We obtain the following results (expressed in \%) for portfolio $x$:%
\begin{equation*}
\begin{tabular}{|c|cccc|}
\hline
Asset & $x_{i}$  & $\mathcal{MR}_{i}$ & $\mathcal{RC}_{i}$ & $\mathcal{RC}^{\star}_{i}$ \\ \hline
$1$   & $20.00$ & ${\bP}8.82$ & ${\bP}1.76$ & $11.60$ \\
$2$   & $30.00$ &     $14.85$ & ${\bP}4.46$ & $29.28$ \\
$3$   & $10.00$ &     $18.40$ & ${\bP}1.84$ & $12.09$ \\
$4$   & $10.00$ &     $23.86$ & ${\bP}2.39$ & $15.68$ \\
$5$   & $30.00$ &     $15.90$ & ${\bP}4.77$ & $31.35$ \\
\hdashline
$\sigma\left(x\right) $ &  &  & $15.22$ &  \\
\hline
\end{tabular}
\end{equation*}
If we consider the first solution, we have:
\begin{equation*}
\begin{tabular}{|c|ccccc|}
\hline
Asset & $y_{i}$  & $\sigma_i^{\prime}$ & $\mathcal{MR}_{i}$ & $\mathcal{RC}_{i}$ & $\mathcal{RC}^{\star}_{i}$ \\ \hline
$1{\bp}$      & $20.00$ & $15.00$ & ${\bP}8.82$ & ${\bP}1.76$ & $11.60$ \\
$2{\bp}$      & $30.00$ & $20.00$ &     $14.85$ & ${\bP}4.46$ & $29.28$ \\
$3{\bp}$      & $10.00$ & $25.00$ &     $18.40$ & ${\bP}1.84$ & $12.09$ \\
$4^{\prime}$  & $40.00$ & $22.50$ &     $17.89$ & ${\bP}7.16$ & $47.03$ \\ \hdashline
$\sigma\left(y\right) $ &  &  & $15.22$ & &  \\
\hline
\end{tabular}
\end{equation*}
For the second solution, we obtain:
\begin{equation*}
\begin{tabular}{|c|ccccc|}
\hline
Asset & $y_{i}$  & $\sigma_i^{\prime}$ & $\mathcal{MR}_{i}$ & $\mathcal{RC}_{i}$ & $\mathcal{RC}^{\star}_{i}$ \\ \hline
$1{\bp}$      & $20.00$ & $15.00$ & ${\bP}8.82$ & ${\bP}1.76$ & $11.60$ \\
$2{\bp}$      & $30.00$ & $20.00$ &     $14.85$ & ${\bP}4.46$ & $29.28$ \\
$3{\bp}$      & $10.00$ & $25.00$ &     $18.40$ & ${\bP}1.84$ & $12.09$ \\
$4^{\prime}$  & $18.00$ & $50.00$ &     $39.76$ & ${\bP}7.16$ & $47.03$ \\ \hdashline
$\sigma\left(y\right) $ &  &  & $15.22$ &  & \\
\hline
\end{tabular}
\end{equation*}
Finally, if we impose $y_4 = 80\%$, we have:
\begin{equation*}
\begin{tabular}{|c|ccccc|}
\hline
Asset & $y_{i}$  & $\sigma_i^{\prime}$ & $\mathcal{MR}_{i}$ & $\mathcal{RC}_{i}$ & $\mathcal{RC}^{\star}_{i}$ \\ \hline
$1{\bp}$      & $20.00$ & $15.00$ & ${\bP}8.82$ & ${\bP}1.76$ & $11.60$ \\
$2{\bp}$      & $30.00$ & $20.00$ &     $14.85$ & ${\bP}4.46$ & $29.28$ \\
$3{\bp}$      & $10.00$ & $25.00$ &     $18.40$ & ${\bP}1.84$ & $12.09$ \\
$4^{\prime}$  & $80.00$ & $11.25$ & ${\bP}8.95$ & ${\bP}7.16$ & $47.03$ \\ \hdashline
$\sigma\left(y\right) $ &  &  & $15.22$ &  & \\
\hline
\end{tabular}
\end{equation*}
\item The previous analysis shows that considering a meta-bond by country is equivalent to consider the complete universe of individual bonds.
\end{enumerate}

\item To find the RB portfolios $y$, we proceed in two steps. First, we
calculate the normalized portfolio $\tilde{y}$ such that the weights are
equal to 1. For that, we optimize the objective function (TR-RPB,
page 102):%
\begin{eqnarray*}
\tilde{y} &=&\arg \min \sum_{i=1}^{n}\sum_{j=1}^{n}\left( \frac{\tilde{y}%
_{i}\cdot \left( \Sigma \tilde{y}\right) _{i}}{b_{i}}-\frac{\tilde{y}%
_{j}\cdot \left( \Sigma \tilde{y}\right) _{j}}{b_{j}}\right) ^{2} \\
\text{} &\text{u.c.}&\mathbf{1}^{\top }\tilde{y}=1\text{ and }\tilde{y}\geq
\mathbf{0}
\end{eqnarray*}%
Then, we deduce the portfolio $y$ in the following way:%
\begin{equation*}
y=\left( \sum x_{i}\right) \cdot \tilde{y}
\end{equation*}

\begin{enumerate}
\item We obtain the following results:%
\begin{equation*}
\begin{tabular}{|c|cccc|}
\hline
Bond & $y^{\left(1\right)}_{i}$  & $\mathcal{MR}_{i}$ & $\mathcal{RC}_{i}$ & $\mathcal{RC}^{\star}_{i}$ \\ \hline
$1$  & ${\bP}9.95$ & $0.023$ & $0.229$ & $20.0\%$ \\
$2$  &     $17.62$ & $0.013$ & $0.229$ & $20.0\%$ \\
$3$  & ${\bP}5.31$ & $0.065$ & $0.344$ & $30.0\%$ \\
$4$  & ${\bP}4.12$ & $0.083$ & $0.344$ & $30.0\%$ \\ \hdashline
$\mathcal{R}\left(y^{\left(1\right)}\right) $ &  &  & $1.145$ &  \\ \hline
\end{tabular}
\end{equation*}

\item We obtain the following results:%
\begin{equation*}
\begin{tabular}{|c|cccc|}
\hline
Bond & $y^{\left(2\right)}_{i}$  & $\mathcal{MR}_{i}$ & $\mathcal{RC}_{i}$ & $\mathcal{RC}^{\star}_{i}$ \\ \hline
$1$  &     $10.15$ & $0.026$ & $0.268$ & $20.0\%$ \\
$2$  &     $22.37$ & $0.012$ & $0.268$ & $20.0\%$ \\
$3$  & ${\bP}6.11$ & $0.066$ & $0.401$ & $30.0\%$ \\
$4$  & ${\bP}3.37$ & $0.119$ & $0.401$ & $30.0\%$ \\ \hdashline
$\mathcal{R}\left(y^{\left(2\right)}\right) $ &  &  & $1.338$ &  \\ \hline
\end{tabular}
\end{equation*}

\item We obtain the following results:%
\begin{equation*}
\begin{tabular}{|c|cccc|}
\hline Bond & $y^{\left(1+2\right)}_{i}$  & $\mathcal{MR}_{i}$ &
$\mathcal{RC}_{i}$ & $\mathcal{RC}^{\star}_{i}$ \\ \hline
$1$ (\#1) & ${\bP}9.95$ & $0.023$ & $0.229$ & ${\bP}9.2\%$ \\
$2$ (\#1) &     $17.62$ & $0.013$ & $0.229$ & ${\bP}9.2\%$ \\
$3$ (\#1) & ${\bP}5.31$ & $0.065$ & $0.344$ &     $13.8\%$ \\
$4$ (\#1) & ${\bP}4.12$ & $0.083$ & $0.344$ &     $13.8\%$ \\
$1$ (\#2) &     $10.15$ & $0.026$ & $0.268$ &     $10.8\%$ \\
$2$ (\#2) &     $22.37$ & $0.012$ & $0.268$ &     $10.8\%$ \\
$3$ (\#2) & ${\bP}6.11$ & $0.066$ & $0.401$ &     $16.2\%$ \\
$4$ (\#2) & ${\bP}3.37$ & $0.119$ & $0.401$ &     $16.2\%$ \\ \hdashline
$\mathcal{R}\left(y^{\left(1+2\right)}\right) $ &  &  & $2.483$ &  \\
\hline
\end{tabular}
\end{equation*}
If we aggregate the exposures by country, we have:
\begin{equation*}
\begin{tabular}{|c|ccc|}
\hline
Country & $y^{\left(1+2\right)}_{j}$  & $\mathcal{RC}_{j}$ & $\mathcal{RC}^{\star}_{j}$ \\ \hline
France  &     $20.10$ & $0.497$ & $20.0\%$ \\
Germany &     $39.99$ & $0.497$ & $20.0\%$ \\
Italy   &     $11.42$ & $0.745$ & $30.0\%$ \\
Spain   & ${\bP}7.49$ & $0.745$ & $30.0\%$ \\
\hline
\end{tabular}
\end{equation*}%

\item The optimization program becomes:%
\begin{eqnarray*}
\tilde{y} &=&\arg \min \sum_{i=1}^{4}\sum_{j=1}^{4}\left( \frac{\tilde{b}_{i}%
}{b_{i}}-\frac{\tilde{b}_{j}}{b_{j}}\right) ^{2} \\
&\text{u.c.}&\left\{
\begin{array}{l}
\tilde{b}_{i}=\tilde{y}_{i}\cdot \left( \Sigma \tilde{y}\right) _{i}+\tilde{y%
}_{i+4}\cdot \left( \Sigma \tilde{y}\right) _{i+4} \\
\mathbf{1}^{\top }\tilde{y}=1 \\
\tilde{y}\geq \mathbf{0}%
\end{array}%
\right.
\end{eqnarray*}
In fact, we have specified four constraints on the risk contributions (one by
country), but we have eight unknown variables. It is obvious that there are
several solutions. The problem with the meta-bonds is that their
characteristics depend on the weights of the portfolio. For instance, in
Question 1(c), we have specified the duration of the meta-bond $i$ as follows:%
\begin{equation}
D_{i}^{\left( 3\right) }=\frac{x_{i}^{\left( 1\right) }D_{i}^{\left(
1\right) }+x_{i}^{\left( 2\right) }D_{i}^{\left( 2\right) }}{x_{i}^{\left(
1\right) }+x_{i}^{\left( 2\right) }}  \label{eq:app2-duplication3}
\end{equation}%
If we assume that:%
\begin{equation}
D_{i}^{\left( 3\right) }=\frac{\tilde{y}_{i}^{\left( 1\right) }D_{i}^{\left(
1\right) }+\tilde{y}_{i}^{\left( 2\right) }D_{i}^{\left( 2\right) }}{\tilde{y%
}_{i}^{\left( 1\right) }+\tilde{y}_{i}^{\left( 2\right) }}
\label{eq:app2-duplication4}
\end{equation}%
we see that we face an endogeneity problem because the duration of the
meta-bond depends on the solution of the RB optimization problem.
Nevertheless, the analysis conducted in Question 2(d) helps us to propose a
practical solution. The idea is to specify the meta-bonds using Equation (%
\ref{eq:app2-duplication3}) and not Equation (\ref{eq:app2-duplication4}).
It means that we maintain the same proportion of individual bonds in the
portfolio $\tilde{y}$ than previously. In this case, we obtain the following
results:%
\begin{equation*}
\begin{tabular}{|c|cccc|}
\hline
Country & $y^{\left(4\right)}_{j}$  & $\mathcal{MR}_{j}$ & $\mathcal{RC}_{j}$ & $\mathcal{RC}^{\star}_{j}$ \\ \hline
France  &     $20.51$ & $0.025$ & $0.502$ & $20.0\%$ \\
Germany &     $39.93$ & $0.013$ & $0.502$ & $20.0\%$ \\
Italy   &     $11.54$ & $0.065$ & $0.754$ & $30.0\%$ \\
Spain   & ${\bP}7.02$ & $0.107$ & $0.754$ & $30.0\%$ \\ \hdashline
$\mathcal{R}\left(y^{\left(4\right)}\right) $ &  &  & $2.512$ &  \\ \hline
\end{tabular}
\end{equation*}
To obtain the allocation in terms of bonds, we define the portfolio $%
y^{\left( 5\right) }$ as follows:%
\begin{equation*}
y^{\left( 5\right) }=\left[
\begin{array}{c}
w\circ y^{\left( 4\right) } \\
\left( 1-w\right) \circ y^{\left( 4\right) }%
\end{array}%
\right]
\end{equation*}%
where $w$ is the vector such that:%
\begin{equation*}
w_{i}=\frac{x_{i}^{\left( 1\right) }}{x_{i}^{\left( 1\right) }+x_{i}^{\left(
2\right) }}
\end{equation*}%
This allocation principle is derived from the specification of the
meta-bonds. Finally, we obtain:%
\begin{equation*}
\begin{tabular}{|c|cccc|}
\hline Bond & $y^{\left(5\right)}_{i}$  & $\mathcal{MR}_{i}$ &
$\mathcal{RC}_{i}$ & $\mathcal{RC}^{\star}_{i}$ \\ \hline
$1$ (\#1) &     $11.39$ & $0.023$ & $0.262$ &     $10.4\%$ \\
$2$ (\#1) &     $23.96$ & $0.013$ & $0.311$ &     $12.4\%$ \\
$3$ (\#1) & ${\bP}4.62$ & $0.065$ & $0.299$ &     $11.9\%$ \\
$4$ (\#1) & ${\bP}2.34$ & $0.083$ & $0.195$ & ${\bP}7.8\%$ \\
$1$ (\#2) & ${\bP}9.11$ & $0.026$ & $0.240$ & ${\bP}9.6\%$ \\
$2$ (\#2) &     $15.97$ & $0.012$ & $0.191$ & ${\bP}7.6\%$ \\
$3$ (\#2) & ${\bP}6.93$ & $0.066$ & $0.455$ &     $18.1\%$ \\
$4$ (\#2) & ${\bP}4.68$ & $0.119$ & $0.559$ &     $22.2\%$ \\ \hdashline
$\mathcal{R}\left(y^{\left(5\right)}\right) $ &  &  & $2.512$ &  \\
\hline
\end{tabular}
\end{equation*}
This solution is different from the previous merged portfolio. We have then found at least two solutions.
By the property of the Euler allocation, it implies that every combination
of these two solutions is also another solution. It means that there are infinite solutions
to this problem.

\end{enumerate}
\end{enumerate}

\section{Risk contributions of long-short portfolios}

\begin{enumerate}
\item
\begin{enumerate}
\item We have:%
\begin{equation*}
\sigma \left( x\right) =\sqrt{x_{1}^{2}\sigma _{1}^{2}+2x_{1}x_{2}\rho
\sigma _{1}\sigma _{2}+x_{2}^{2}\sigma _{2}^{2}}
\end{equation*}%
It follows that:%
\begin{equation*}
\frac{\partial \,\sigma \left( x\right) }{\partial \,x_{1}}=\frac{%
x_{1}\sigma _{1}^{2}+x_{2}\rho \sigma _{1}\sigma _{2}}{\sigma \left(
x\right) }
\end{equation*}%
Finally, we get:%
\begin{equation*}
\mathcal{RC}_{1}=\frac{x_{1}^{2}\sigma _{1}^{2}+x_{1}x_{2}\rho \sigma
_{1}\sigma _{2}}{\sigma \left( x\right) }
\end{equation*}%
and:%
\begin{equation*}
\mathcal{RC}_{2}=\frac{x_{2}^{2}\sigma _{2}^{2}+x_{1}x_{2}\rho \sigma
_{1}\sigma _{2}}{\sigma \left( x\right) }
\end{equation*}

\item The inequality $\mathcal{RC}_{2}\leq 0$ implies that:%
\begin{eqnarray*}
&&\frac{x_{2}^{2}\sigma _{2}^{2}+x_{1}x_{2}\rho \sigma _{1}\sigma _{2}}{%
\sigma \left( x\right) }\leq 0 \\
&\Leftrightarrow &x_{2}^{2}\sigma _{2}^{2}+x_{1}x_{2}\rho \sigma _{1}\sigma
_{2}\leq 0 \\
&\Leftrightarrow &x_{2}\left( x_{2}\sigma _{2}^{2}+x_{1}\rho \sigma
_{1}\sigma _{2}\right) \leq 0 \\
&\Leftrightarrow &\left\{
\begin{array}{ccc}
x_{2}\leq 0 & \text{and} & x_{2}\geq -x_{1}\rho \sigma _{1}/\sigma _{2} \\
x_{2}\geq 0 & \text{and} & x_{2}\leq -x_{1}\rho \sigma _{1}/\sigma _{2}%
\end{array}%
\right.
\end{eqnarray*}%
We distinguish two cases. If $x_{1}\rho <0$, we get $x_{2}\in \left[
0,-x_{1}\rho \sigma _{1}/\sigma _{2}\right] $. If $x_{1}\rho >0$, the
solution set becomes $\left[ -x_{1}\rho \sigma _{1}/\sigma _{2},0\right] $.
We notice that if the risk contribution is negative for a negative (resp.
positive) value of $x_{2}$, it cannot be negative for a positive (resp.
negative) value of $x_{2}$.

\item The relationship between $x_{2}$ and $\mathcal{RC}_{2}$ is reported in
Figure \ref{fig:app2-3-7-1}.

\begin{figure}[tbph]
\centering
\includegraphics[width = \figurewidth, height = \figureheight]{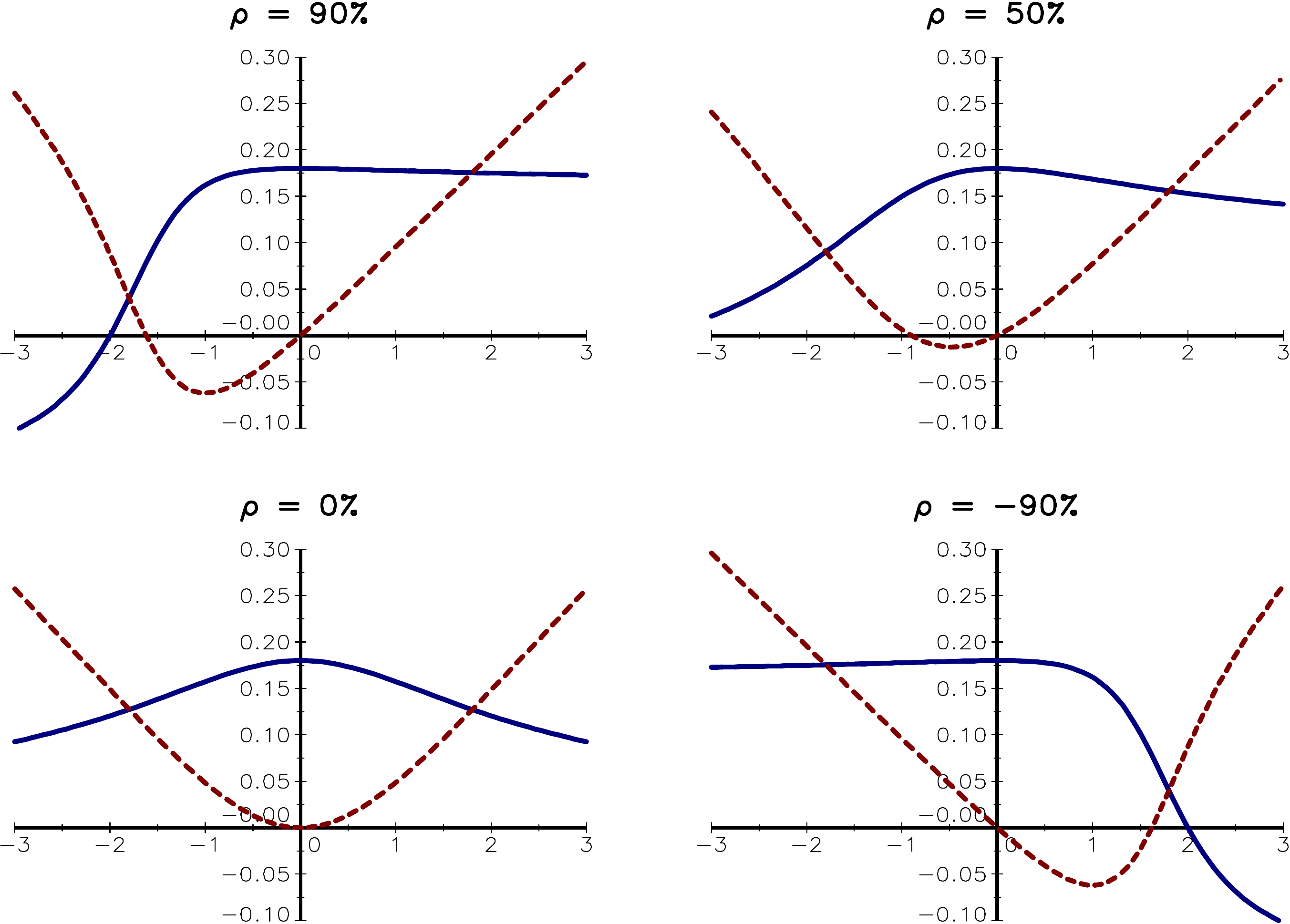}
\caption{Risk contribution $\mathcal{RC}_{2}$}
\label{fig:app2-3-7-1}
\end{figure}

\end{enumerate}

\item

\begin{enumerate}
\item We have:%
\begin{equation*}
\begin{tabular}{|c|rrrr|}
\hline
Asset & \multicolumn{1}{c}{$x_{i}$} & \multicolumn{1}{c}{$\mathcal{MR}_{i}$} &
\multicolumn{1}{c}{$\mathcal{RC}_{i}$} & \multicolumn{1}{c|}{$\mathcal{RC}_{i}^{\star }$} \\
\hline
$1$                          & $ 100.00\%$ & $  4.73\%$ & $ 4.73\%$ & $10.66\%$  \\
$2$                          & $-100.00\%$ & $ -3.94\%$ & $ 3.94\%$ & $ 8.88\%$  \\
$3$                          & $ 100.00\%$ & $ 12.50\%$ & $12.50\%$ & $28.17\%$  \\
$4$                          & $-100.00\%$ & $-20.50\%$ & $20.50\%$ & $46.19\%$  \\
$5$                          & $ 100.00\%$ & $  1.01\%$ & $ 1.01\%$ & $ 2.28\%$  \\
$6$                          & $-100.00\%$ & $ -1.69\%$ & $ 1.69\%$ & $ 3.81\%$  \\ \hdashline
$\sigma\left( x\right)$      &           &              & $44.38\%$ &            \\
\hline
\end{tabular}%
\end{equation*}

\item The covariance matrix of the long-short assets is $A\Sigma A^{\top }$\
with:%
\begin{equation*}
A=\left(
\begin{array}{rrrrrr}
1 & -1 & 0 & 0 & 0 & 0 \\
0 & 0 & 1 & -1 & 0 & 0 \\
0 & 0 & 0 & 0 & 1 & -1%
\end{array}%
\right)
\end{equation*}%
We deduce that the volatilities of the long-short assets are $20.62\%$, $%
38.67\%$ and $10.95\%$ whereas the correlation matrix is:%
\begin{equation*}
C=\left(
\begin{array}{rrr}
100.00\% &  &  \\
-4.39\% & 100.00\% &  \\
-2.21\% & 1.18\% & 100\%%
\end{array}%
\right)
\end{equation*}%
Finally, the risk decomposition is:%
\begin{equation*}
\begin{tabular}{|c|rrrr|}
\hline
L/S asset & \multicolumn{1}{c}{$y_{i}$} & \multicolumn{1}{c}{$\mathcal{MR}_{i}$} &
\multicolumn{1}{c}{$\mathcal{RC}_{i}$} & \multicolumn{1}{c|}{$\mathcal{RC}_{i}^{\star }$} \\
\hline
$1$                          & $100.00\%$ & $ 8.67\%$ & $ 8.67\%$ & $19.54\%$  \\
$2$                          & $100.00\%$ & $33.01\%$ & $33.01\%$ & $74.37\%$  \\
$3$                          & $100.00\%$ & $ 2.70\%$ & $ 2.70\%$ & $ 6.09\%$  \\ \hdashline
$\sigma\left( y\right)$      &           &            & $44.38\%$ &            \\
\hline
\end{tabular}%
\end{equation*}

\item The ERC portfolio is:%
\begin{equation*}
\begin{tabular}{|c|rrrr|}
\hline
L/S asset & \multicolumn{1}{c}{$y_{i}$} & \multicolumn{1}{c}{$\mathcal{MR}_{i}$} &
\multicolumn{1}{c}{$\mathcal{RC}_{i}$} & \multicolumn{1}{c|}{$\mathcal{RC}_{i}^{\star }$} \\
\hline
$1$                          & $29.81\%$ & $11.51\%$ & $ 3.43\%$ & $33.33\%$  \\
$2$                          & $15.63\%$ & $21.95\%$ & $ 3.43\%$ & $33.33\%$  \\
$3$                          & $54.56\%$ & $ 6.29\%$ & $ 3.43\%$ & $33.33\%$  \\ \hdashline
$\sigma\left( y\right)$      &           &           & $10.29\%$ &            \\
\hline
\end{tabular}%
\end{equation*}
If we don't take into account the correlations, we obtain a similar
portfolio:%
\begin{equation*}
\begin{tabular}{|c|rrrr|}
\hline
L/S asset & \multicolumn{1}{c}{$y_{i}$} & \multicolumn{1}{c}{$\mathcal{MR}_{i}$} &
\multicolumn{1}{c}{$\mathcal{RC}_{i}$} & \multicolumn{1}{c|}{$\mathcal{RC}_{i}^{\star }$} \\
\hline
$1$                          & $29.28\%$ & $11.90\%$ & $ 3.49\%$ & $33.33\%$  \\
$2$                          & $15.61\%$ & $22.32\%$ & $ 3.49\%$ & $33.33\%$  \\
$3$                          & $55.11\%$ & $ 6.32\%$ & $ 3.49\%$ & $33.33\%$  \\ \hdashline
$\sigma\left( y\right)$      &           &           & $10.46\%$ &            \\
\hline
\end{tabular}%
\end{equation*}

\item The risk allocation with respect to the six assets is:%
\begin{equation*}
\begin{tabular}{|c|rrrr|}
\hline
Asset & \multicolumn{1}{c}{$x_{i}$} & \multicolumn{1}{c}{$\mathcal{MR}_{i}$} &
\multicolumn{1}{c}{$\mathcal{RC}_{i}$} & \multicolumn{1}{c|}{$\mathcal{RC}_{i}^{\star }$} \\
\hline
$1$                          & $ 29.81\%$ & $  5.32\%$ & $ 1.59\%$ & $15.41\%$  \\
$2$                          & $-29.81\%$ & $ -6.19\%$ & $ 1.84\%$ & $17.92\%$  \\
$3$                          & $ 15.63\%$ & $  7.29\%$ & $ 1.14\%$ & $11.07\%$  \\
$4$                          & $-15.63\%$ & $-14.66\%$ & $ 2.29\%$ & $22.26\%$  \\
$5$                          & $ 54.56\%$ & $  2.88\%$ & $ 1.57\%$ & $15.28\%$  \\
$6$                          & $-54.56\%$ & $ -3.41\%$ & $ 1.86\%$ & $18.06\%$  \\ \hdashline
$\sigma\left( x\right)$      &           &             & $10.29\%$ &            \\
\hline
\end{tabular}%
\end{equation*}
We notice that we do not have $\mathcal{RC}_{1}=\mathcal{RC}_{2}$, $\mathcal{%
RC}_{3}=\mathcal{RC}_{4}$ and $\mathcal{RC}_{5}=\mathcal{RC}_{6}$.

\item A first route is to perform an optimization by tacking into account
the weight constraints $x_{1}x_{2}<0$, $x_{3}x_{4}<0$, $x_{5}x_{6}<0$.
Nevertheless, if there is a solution $x$ to this problem, the portfolio
$y=\alpha x$ is also a solution (TR-RPB, page 256). This is why we need to
impose that:%
\begin{equation*}
\mathcal{RC}_{1}=c
\end{equation*}%
with $c$ a positive scalar. For example, if $c=4\%$, we obtain the following
portfolio:%
\begin{equation*}
\begin{tabular}{|c|rrrr|}
\hline
Asset & \multicolumn{1}{c}{$x_{i}$} & \multicolumn{1}{c}{$\mathcal{MR}_{i}$} &
\multicolumn{1}{c}{$\mathcal{RC}_{i}$} & \multicolumn{1}{c|}{$\mathcal{RC}_{i}^{\star }$} \\
\hline
$1$                          & $  64.77\%$ & $  5.91\%$ & $ 3.82\%$ & $17.09\%$  \\
$2$                          & $ -64.77\%$ & $ -5.66\%$ & $ 3.67\%$ & $16.39\%$  \\
$3$                          & $  38.90\%$ & $  9.48\%$ & $ 3.69\%$ & $16.47\%$  \\
$4$                          & $ -30.42\%$ & $-12.36\%$ & $ 3.76\%$ & $16.80\%$  \\
$5$                          & $ 115.20\%$ & $  2.88\%$ & $ 3.32\%$ & $14.84\%$  \\
$6$                          & $-123.68\%$ & $ -3.33\%$ & $ 4.12\%$ & $18.41\%$  \\ \hdashline
$\sigma\left( x\right)$      &           &              & $22.38\%$ &            \\
\hline
\end{tabular}%
\end{equation*}
We notice that this portfolio does not match the constraints. In
fact, this optimization problem is tricky from a numerical point of view. A
second route consists in using the following parametrization $y=Ax$ with:%
\begin{equation*}
A=\limfunc{diag}\left( 1,-1,1,-1,1,-1\right)
\end{equation*}%
We have then transformed the assets and the new covariance matrix is $A\Sigma
A^{\top }$. We can then compute the ERC portfolio $y$ and deduce the
long-short portfolio $x=A^{-1}y$. In this case, we obtain a long-short
portfolio that matches all the constraints:%
\begin{equation*}
\begin{tabular}{|c|rrrr|}
\hline
Asset & \multicolumn{1}{c}{$x_{i}$} & \multicolumn{1}{c}{$\mathcal{MR}_{i}$} &
\multicolumn{1}{c}{$\mathcal{RC}_{i}$} & \multicolumn{1}{c|}{$\mathcal{RC}_{i}^{\star }$} \\
\hline
$1$                          & $ 14.29\%$ & $  5.95\%$ & $0.85\%$ & $16.67\%$  \\
$2$                          & $-15.03\%$ & $ -5.66\%$ & $0.85\%$ & $16.67\%$  \\
$3$                          & $  8.94\%$ & $  9.51\%$ & $0.85\%$ & $16.67\%$  \\
$4$                          & $ -6.96\%$ & $-12.23\%$ & $0.85\%$ & $16.67\%$  \\
$5$                          & $ 27.68\%$ & $  3.07\%$ & $0.85\%$ & $16.67\%$  \\
$6$                          & $-27.10\%$ & $ -3.14\%$ & $0.85\%$ & $16.67\%$  \\ \hdashline
$\sigma\left( x\right)$      &           &             & $5.11\%$ &            \\
\hline
\end{tabular}%
\end{equation*}
Note that solution $x$ is not unique and every portfolio of the form $%
y = \alpha x$ is also a solution.
\end{enumerate}
\end{enumerate}

\section{Risk parity funds}

\begin{enumerate}
\item
\begin{enumerate}
\item The RP portfolio is defined as follows:%
\begin{equation*}
x_{i}=\frac{\sigma _{i}^{-1}}{\sum_{j=1}^{n}\sigma _{j}^{-1}}
\end{equation*}%
We obtain the following results:%
\begin{equation*}
\begin{tabular}{|c|rrrrrr|}
\hline
Date                    & \multicolumn{1}{c}{$1999$} & \multicolumn{1}{c}{$2002$} &
                          \multicolumn{1}{c}{$2005$} & \multicolumn{1}{c}{$2007$} &
                          \multicolumn{1}{c}{$2008$} & \multicolumn{1}{c|}{$2010$} \\
\hline
S                       & $23.89\%$ & $18.75\%$ & $38.35\%$ & $23.57\%$ & $18.07\%$ & $22.63\%$ \\
B                       & $52.81\%$ & $52.71\%$ & $43.60\%$ & $55.45\%$ & $61.35\%$ & $55.02\%$ \\
C                       & $23.29\%$ & $28.54\%$ & $18.05\%$ & $20.98\%$ & $20.58\%$ & $22.36\%$ \\ \hdashline
$\sigma\left( x\right)$ & $ 4.83\%$ & $ 6.08\%$ & $ 6.26\%$ & $ 5.51\%$ & $11.64\%$ & $ 8.38\%$ \\
\hline
\end{tabular}%
\end{equation*}

\item In the ERC portfolio, the risk contributions are equal for all the
assets:%
\begin{equation*}
\mathcal{RC}_{i}=\mathcal{RC}_{j}
\end{equation*}%
with:%
\begin{equation}
\mathcal{RC}_{i}=\frac{x_{i}\cdot \left( \Sigma x\right) _{i}}{\sqrt{x^{\top
}\Sigma x}}  \label{eq:app2-3-9-1}
\end{equation}%
We obtain the following results:%
\begin{equation*}
\begin{tabular}{|c|rrrrrr|}
\hline
Date                    & \multicolumn{1}{c}{$1999$} & \multicolumn{1}{c}{$2002$} &
                          \multicolumn{1}{c}{$2005$} & \multicolumn{1}{c}{$2007$} &
                          \multicolumn{1}{c}{$2008$} & \multicolumn{1}{c|}{$2010$} \\
\hline
S                       & $23.66\%$ & $18.18\%$ & $37.85\%$ & $23.28\%$ & $17.06\%$ & $20.33\%$ \\
B                       & $53.12\%$ & $58.64\%$ & $43.18\%$ & $59.93\%$ & $66.39\%$ & $59.61\%$ \\
C                       & $23.22\%$ & $23.18\%$ & $18.97\%$ & $16.79\%$ & $16.54\%$ & $20.07\%$ \\ \hdashline
$\sigma\left( x\right)$ & $ 4.82\%$ & $ 5.70\%$ & $ 6.32\%$ & $ 5.16\%$ & $10.77\%$ & $ 7.96\%$ \\
\hline
\end{tabular}%
\end{equation*}

\item We notice that $\sigma \left( x_{\mathrm{erc}}\right) \leq \sigma
\left( x_{\mathrm{rp}}\right) $ except for the year 2005. This date
corresponds to positive correlations between assets. Moreover, the
correlation between stocks and bonds is the highest. Starting from the RP
portfolio, it is then possible to approach the ERC portfolio by reducing the
weights of stocks and bonds and increasing the weight of commodities. At the
end, we find an ERC portfolio that has a slightly higher volatility.

\item The volatility of the RP portfolio is:%
\begin{eqnarray*}
\sigma \left( x\right)  &=&\frac{1}{\sum_{j=1}^{n}\sigma _{j}^{-1}}\sqrt{%
\left( \sigma ^{-1}\right) ^{\top }\Sigma \sigma ^{-1}} \\
&=&\frac{1}{\sum_{j=1}^{n}\sigma _{j}^{-1}}\sqrt{\sum_{i=1}^{n}\sum_{j=1}^{n}%
\frac{1}{\sigma _{i}\sigma _{j}}\rho _{i,j}\sigma _{i}\sigma _{j}} \\
&=&\frac{1}{\sum_{j=1}^{n}\sigma _{j}^{-1}}\sqrt{n+2\sum_{i>j}\rho _{i,j}} \\
&=&\frac{1}{\sum_{j=1}^{n}\sigma _{j}^{-1}}\sqrt{n\left( 1+\left( n-1\right)
\bar{\rho}\right) }
\end{eqnarray*}%
where $\bar{\rho}$ is the average correlation between asset returns. For the
marginal risk, we obtain:%
\begin{eqnarray*}
\mathcal{MR}_{i} &=&\frac{\left( \Sigma \sigma ^{-1}\right) _{i}}{\sigma
\left( x\right) \sum_{j=1}^{n}\sigma _{j}^{-1}} \\
&=&\frac{1}{\sqrt{n\left( 1+\left( n-1\right) \bar{\rho}\right) }}%
\sum_{j=1}^{n}\rho _{i,j}\sigma _{i}\sigma _{j}\frac{1}{\sigma _{j}} \\
&=&\frac{\sigma _{i}}{\sqrt{n\left( 1+\left( n-1\right) \bar{\rho}\right) }}%
\sum_{j=1}^{n}\rho _{i,j} \\
&=&\frac{\sigma _{i}\bar{\rho}_{i}\sqrt{n}}{\sqrt{1+\left( n-1\right) \bar{%
\rho}}}
\end{eqnarray*}%
where $\bar{\rho}_{i}$ is the average correlation of asset $i$ with the
other assets (including itself). The expression of the risk contribution is
then:%
\begin{eqnarray*}
\mathcal{RC}_{i} &=&\frac{\sigma _{i}^{-1}}{\sum_{j=1}^{n}\sigma _{j}^{-1}}%
\frac{\sigma _{i}\bar{\rho}_{i}\sqrt{n}}{\sqrt{1+\left( n-1\right) \bar{\rho}%
}} \\
&=&\frac{\bar{\rho}_{i}\sqrt{n}}{\sqrt{1+\left( n-1\right) \bar{\rho}}%
\sum_{j=1}^{n}\sigma _{j}^{-1}}
\end{eqnarray*}%
We deduce that the normalized risk contribution is:%
\begin{eqnarray*}
\mathcal{RC}_{i}^{\star } &=&\frac{\bar{\rho}_{i}\sqrt{n}}{\sigma \left(
x\right) \sqrt{1+\left( n-1\right) \bar{\rho}}\sum_{j=1}^{n}\sigma _{j}^{-1}}
\\
&=&\frac{\bar{\rho}_{i}}{1+\left( n-1\right) \bar{\rho}}
\end{eqnarray*}

\item We obtain the following normalized risk contributions:%
\begin{equation*}
\begin{tabular}{|c|rrrrrr|}
\hline
Date                    & \multicolumn{1}{c}{$1999$} & \multicolumn{1}{c}{$2002$} &
                          \multicolumn{1}{c}{$2005$} & \multicolumn{1}{c}{$2007$} &
                          \multicolumn{1}{c}{$2008$} & \multicolumn{1}{c|}{$2010$} \\
\hline
S                       & $33.87\%$ & $34.96\%$ & $34.52\%$ & $32.56\%$ & $34.45\%$ & $36.64\%$ \\
B                       & $32.73\%$ & $20.34\%$ & $34.35\%$ & $24.88\%$ & $24.42\%$ & $26.70\%$ \\
C                       & $33.40\%$ & $44.69\%$ & $31.14\%$ & $42.57\%$ & $41.13\%$ & $36.67\%$ \\
\hline
\end{tabular}%
\end{equation*}
We notice that the risk contributions are not exactly equal for all the assets. Generally, the risk contribution
of bonds is lower than the risk contribution of equities, which is itself lower than the risk contribution
of commodities.
\end{enumerate}

\item
\begin{enumerate}
\item We obtain the following RB portfolios:
\begin{equation*}
\hspace*{-1cm}
\begin{tabular}{|c|r:rrrrrr|}
\hline
Date                    & \multicolumn{1}{c:}{$b_i$}  &
                          \multicolumn{1}{c}{$1999$} & \multicolumn{1}{c}{$2002$} &
                          \multicolumn{1}{c}{$2005$} & \multicolumn{1}{c}{$2007$} &
                          \multicolumn{1}{c}{$2008$} & \multicolumn{1}{c|}{$2010$} \\
\hline
S                       & $45\%$ & $26.83\%$ & $22.14\%$ & $42.83\%$ & $27.20\%$ & $20.63\%$ & $25.92\%$ \\
B                       & $45\%$ & $59.78\%$ & $66.10\%$ & $48.77\%$ & $66.15\%$ & $73.35\%$ & $67.03\%$ \\
C                       & $10\%$ & $13.39\%$ & $11.76\%$ & $ 8.40\%$ & $ 6.65\%$ & $ 6.02\%$ & $ 7.05\%$ \\ \hdashline
S                       & $70\%$ & $40.39\%$ & $29.32\%$ & $65.53\%$ & $39.37\%$ & $33.47\%$ & $46.26\%$ \\
B                       & $10\%$ & $37.63\%$ & $51.48\%$ & $19.55\%$ & $47.18\%$ & $52.89\%$ & $37.76\%$ \\
C                       & $20\%$ & $21.98\%$ & $19.20\%$ & $14.93\%$ & $13.45\%$ & $13.64\%$ & $15.98\%$ \\ \hdashline
S                       & $20\%$ & $17.55\%$ & $16.02\%$ & $25.20\%$ & $18.78\%$ & $12.94\%$ & $13.87\%$ \\
B                       & $70\%$ & $69.67\%$ & $71.70\%$ & $66.18\%$ & $74.33\%$ & $80.81\%$ & $78.58\%$ \\
C                       & $10\%$ & $12.78\%$ & $12.28\%$ & $ 8.62\%$ & $ 6.89\%$ & $ 6.24\%$ & $ 7.55\%$ \\ \hdashline
S                       & $25\%$ & $21.69\%$ & $15.76\%$ & $34.47\%$ & $20.55\%$ & $14.59\%$ & $16.65\%$ \\
B                       & $25\%$ & $48.99\%$ & $54.03\%$ & $39.38\%$ & $55.44\%$ & $61.18\%$ & $53.98\%$ \\
C                       & $50\%$ & $29.33\%$ & $30.21\%$ & $26.15\%$ & $24.01\%$ & $24.22\%$ & $29.37\%$ \\
\hline
\end{tabular}%
\end{equation*}

\item To compute the implied risk premium $\tilde{\pi}_{i}$, we use the following
formula (TR-RPB, page 274):%
\begin{eqnarray*}
\tilde{\pi}_{i} &=&\func{SR}\left( x\mid r\right) \cdot \mathcal{MR}_{i} \\
&=&\func{SR}\left( x\mid r\right) \cdot \frac{\left( \Sigma x\right) _{i}}{%
\sigma \left( x\right) }
\end{eqnarray*}%
where $\func{SR}\left( x\mid r\right) $ is the Sharpe ratio of the
portfolio. We obtain the following results:%
\begin{equation*}
\begin{tabular}{|c|r:rrrrrr|}
\hline
Date                    & \multicolumn{1}{c:}{$b_i$}  &
                          \multicolumn{1}{c}{$1999$} & \multicolumn{1}{c}{$2002$} &
                          \multicolumn{1}{c}{$2005$} & \multicolumn{1}{c}{$2007$} &
                          \multicolumn{1}{c}{$2008$} & \multicolumn{1}{c|}{$2010$} \\
\hline
S                       & $45\%$ & $3.19\%$ & $4.60\%$ & $2.49\%$ & $3.15\%$ & $ 8.64\%$ & $5.20\%$ \\
B                       & $45\%$ & $1.43\%$ & $1.54\%$ & $2.19\%$ & $1.29\%$ & $ 2.43\%$ & $2.01\%$ \\
C                       & $10\%$ & $1.42\%$ & $1.92\%$ & $2.82\%$ & $2.86\%$ & $ 6.58\%$ & $4.24\%$ \\ \hdashline
S                       & $70\%$ & $4.05\%$ & $6.45\%$ & $2.86\%$ & $4.31\%$ & $11.56\%$ & $6.32\%$ \\
B                       & $10\%$ & $0.62\%$ & $0.52\%$ & $1.37\%$ & $0.51\%$ & $ 1.04\%$ & $1.11\%$ \\
C                       & $20\%$ & $2.13\%$ & $2.81\%$ & $3.59\%$ & $3.61\%$ & $ 8.11\%$ & $5.23\%$ \\ \hdashline
S                       & $20\%$ & $2.06\%$ & $2.68\%$ & $1.91\%$ & $1.93\%$ & $ 5.61\%$ & $3.91\%$ \\
B                       & $70\%$ & $1.82\%$ & $2.10\%$ & $2.54\%$ & $1.71\%$ & $ 3.14\%$ & $2.42\%$ \\
C                       & $10\%$ & $1.42\%$ & $1.75\%$ & $2.79\%$ & $2.64\%$ & $ 5.82\%$ & $3.60\%$ \\ \hdashline
S                       & $25\%$ & $2.33\%$ & $3.78\%$ & $1.98\%$ & $2.74\%$ & $ 8.06\%$ & $5.13\%$ \\
B                       & $25\%$ & $1.03\%$ & $1.10\%$ & $1.74\%$ & $1.02\%$ & $ 1.92\%$ & $1.58\%$ \\
C                       & $50\%$ & $3.45\%$ & $3.95\%$ & $5.23\%$ & $4.69\%$ & $ 9.71\%$ & $5.82\%$ \\
\hline
\end{tabular}%
\end{equation*}

\item We have:%
\begin{equation*}
x_{i}\tilde{\pi}_{i}=\func{SR}\left( x\mid r\right) \cdot \mathcal{RC}_{i}
\end{equation*}%
We deduce that:%
\begin{equation*}
\tilde{\pi}_{i}\propto \frac{b_{i}}{x_{i}}
\end{equation*}%
$x_{i}$ is generally an increasing function of $b_{i}$. As a
consequence, the relationship between the risk budgets $b_{i}$ and the risk
premiums $\tilde{\pi}_{i}$ is not necessarily increasing. However, we notice that
the bigger the risk budget, the higher the risk premium. This is easily explained.
If an investor allocates more risk budget to one asset class
than another investor, he thinks that the risk premium of this asset class is
higher than the other investor. However, we must be careful. This
interpretation is valid if we compare two sets of risk budgets. It is false
if we compare the risk budgets among themselves. For instance, if we
consider the third parameter set, the risk budget of bonds is $70\%$ whereas
the risk budget of stocks is $20\%$. It does not mean that the risk premium
of bonds is higher than the risk premium of equities. In fact, we observe
the contrary. If we would like to compare risk budgets among themselves, the
right measure is the implied Sharpe ratio, which is equal to:%
\begin{eqnarray*}
\func{SR}_{i} &=&\frac{\tilde{\pi}_{i}}{\sigma _{i}} \\
&=&\func{SR}\left( x\mid r\right) \cdot \frac{\mathcal{MR}_{i}}{\sigma _{i}}
\end{eqnarray*}%
For instance, if we consider the most diversified portfolio, the marginal
risk is proportional to the volatility and we retrieve the result that
Sharpe ratios are equal if the MDP is optimal.
\end{enumerate}
\end{enumerate}

\section{Frazzini-Pedersen model}

\begin{enumerate}
\item
\begin{enumerate}
\item The Lagrange function of the optimization problem is:%
\begin{eqnarray*}
\mathcal{L}\left( x;\lambda \right)  &=&x_{j}^{\top }\mathbb{E}_{t}\left[
P_{t+1}+D_{t+1}-\left( 1+r\right) P_{t}\right] -\frac{\phi _{j}}{2}%
x_{j}^{\top }\Sigma x_{j}- \\
&&\lambda _{j}\left( m_{j}\left( x_{j}^{\top }P_{t}\right) -W_{j}\right)
\end{eqnarray*}
where $\lambda _{j}$ is the Lagrange multiplier associated with the
constraint $m_{j}\left( x_{j}^{\top }P_{t}\right) \leq W_{j}$. The solution $%
x_{j}$ then verifies the first-order condition:%
\begin{equation*}
\partial _{x}\,\mathcal{L}\left( x;\lambda \right) =\mathbb{E}_{t}\left[
P_{t+1}+D_{t+1}-\left( 1+r\right) P_{t}\right] -\phi _{j}\Sigma
x_{j}-\lambda _{j}m_{j}P_{t}=0
\end{equation*}%
We deduce that:%
\begin{equation*}
x_{j}=\frac{1}{\phi _{j}}\Sigma ^{-1}\left( \mathbb{E}_{t}\left[
P_{t+1}+D_{t+1}\right] -\left( 1+r+\lambda _{j}m_{j}\right)
P_{t}\right)
\end{equation*}

\item At the equilibrium, we have:%
\begin{eqnarray*}
\bar{x} &=&\sum_{j=1}^{m}\frac{1}{\phi _{j}}\Sigma ^{-1}\left( \mathbb{E}_{t}%
\left[ P_{t+1}+D_{t+1}\right] -\left( 1+r+\lambda _{j}m_{j}\right)
P_{t}\right)  \\
&=&\sum_{j=1}^{m}\frac{1}{\phi _{j}}\Sigma ^{-1}\mathbb{E}_{t}\left[
P_{t+1}+D_{t+1}\right] -\sum_{j=1}^{m}\frac{1}{\phi _{j}}\left( 1+r+\lambda
_{j}m_{j}\right) \Sigma ^{-1}P_{t} \\
&=&\left( \sum_{j=1}^{m}\phi _{j}^{-1}\right) \Sigma ^{-1}\mathbb{E}_{t}%
\left[ P_{t+1}+D_{t+1}\right] - \\
&&\left( \sum_{j=1}^{m}\phi _{j}^{-1}\right) \sum_{j=1}^{m}\frac{1}{\left(
\sum_{k=1}^{m}\phi _{k}^{-1}\right) \phi _{j}}\left( 1+r+\lambda
_{j}m_{j}\right) P_{t}
\end{eqnarray*}
Frazzini and Pedersen (2010) introduce the notations:%
\begin{equation*}
\phi =\frac{1}{\left( \sum_{j=1}^{m}\phi _{j}^{-1}\right) }
\end{equation*}%
and:%
\begin{eqnarray*}
\psi &=&\sum_{j=1}^{m}\frac{1}{\left( \sum_{k=1}^{m}\phi
_{k}^{-1}\right)
\phi _{j}}\lambda _{j}m_{j} \\
&=&\sum_{j=1}^{m}\phi \phi _{j}^{-1}\lambda _{j}m_{j}
\end{eqnarray*}%
We finally obtain:%
\begin{equation*}
\bar{x}=\frac{1}{\phi }\Sigma ^{-1}\left( \mathbb{E}_{t}\left[
P_{t+1}+D_{t+1}\right] -\left( 1+r+\psi \right) P_{t}\right)
\end{equation*}%
because:%
\begin{eqnarray*}
\sum_{j=1}^{m}\frac{1}{\left( \sum_{k=1}^{m}\phi _{k}^{-1}\right)
\phi _{j}}
&=&\sum_{j=1}^{m}\frac{\phi }{\phi _{j}} \\
&=&\phi \left( \sum_{j=1}^{m}\phi _{j}^{-1}\right) \\
&=&1
\end{eqnarray*}

\item The equilibrium prices are then given by:%
\begin{equation*}
P_{t}=\frac{\mathbb{E}_{t}\left[ P_{t+1}+D_{t+1}\right] -\phi \Sigma \bar{x}%
}{1+r+\psi }
\end{equation*}%
It follows that:%
\begin{equation*}
P_{i,t}=\frac{\mathbb{E}_{t}\left[ P_{i,t+1}+D_{i,t+1}\right] -\phi
\left( \Sigma \bar{x}\right) _{i}}{1+r+\psi }
\end{equation*}

\item Following Frazzini and Pedersen (2010), the asset return $R_{i,t+1}$
satisfies the following equation:%
\begin{eqnarray*}
\mathbb{E}_{t}\left[ R_{i,t+1}\right] &=&\frac{\mathbb{E}_{t}\left[
P_{i,t+1}+D_{i,t+1}\right] }{P_{i,t}}-1 \\
&=&r+\psi +\frac{\phi }{P_{i,t}}\left( \Sigma \bar{x}\right) _{i}
\end{eqnarray*}%
We know that:%
\begin{eqnarray*}
\left( \Sigma \bar{x}\right) _{i} &=&\func{cov}\left( P_{i,t+1}+D_{i,t+1},%
\bar{x}^{\intercal }\left( P_{t+1}+D_{t+1}\right) \right) \\
&=&P_{i,t}\cdot \func{cov}\left( R_{i,t+1},R_{t+1}\left(
\bar{x}\right) \right) \cdot \left( \bar{x}^{\intercal }P_{t}\right)
\end{eqnarray*}%
Let $\bar{w}_{i}=\left( \bar{x}_{i}P_{i,t}\right) /\left(
\bar{x}^{\intercal }P_{t}\right) $ be the weight of asset $i$ in the
market portfolio. It
follows that:%
\begin{eqnarray*}
\mathbb{E}_{t}\left[ R_{t+1}\left( \bar{x}\right) \right] &=&\sum_{j=1}^{m}%
\bar{w}_{j}\mathbb{E}_{t}\left[ R_{i,t+1}\right] \\
&=&r+\psi +\phi \left( \bar{x}^{\intercal }P_{t}\right) \sigma
^{2}\left( \bar{x}\right)
\end{eqnarray*}%
We deduce that:%
\begin{eqnarray*}
\mathbb{E}_{t}\left[ R_{i,t+1}\right] &=&r+\psi +\phi \beta
_{i}\sigma
^{2}\left( \bar{x}\right) \left( \bar{x}^{\intercal }P_{t}\right) \\
&=&r+\psi +\beta _{i}\left( \mathbb{E}_{t}\left[ R_{t+1}\left( \bar{x}%
\right) \right] -r-\psi \right) \\
&=&r+\psi \left( 1-\beta _{i}\right) +\beta _{i}\left(
\mathbb{E}_{t}\left[ R_{t+1}\left( \bar{x}\right) \right] -r\right)
\end{eqnarray*}%
We finally obtain that:%
\begin{equation*}
\mathbb{E}_{t}\left[ R_{i,t+1}\right] -r=\alpha _{i}+\beta
_{i}\left( \mathbb{E}_{t}\left[ R_{t+1}\left( \bar{x}\right) \right]
-r\right)
\end{equation*}%
where $\alpha _{i}=\psi \left( 1-\beta _{i}\right) $.

\item In the CAPM, the traditional relationship between the risk premium and
the beta of asset $i$ is:%
\begin{equation*}
\mathbb{E}_{t}\left[ R_{i,t+1}\right] -r=\beta _{i}\left( \mathbb{E}_{t}%
\left[ R_{t+1}\left( \bar{x}\right) \right] -r\right)
\end{equation*}%
If we compare this equation with the expression obtained by Frazzini
and Pedersen (2010), we notice the presence of a new term $\alpha
_{i}$, which
is Jensen's alpha. Moreover, $\alpha _{i}$ is a decreasing function of $%
\beta _{i}$.
\end{enumerate}

\begin{enumerate}
\item The optimal value of $\phi $ is (TR-RPB, page 14):%
\begin{equation*}
\phi =\mathbf{1}^{\top }\Sigma ^{-1}\left( \mu -r\mathbf{1}\right)
\end{equation*}%
The tangency portfolio is then:%
\begin{equation*}
x^{\star }=\frac{1}{\phi }\Sigma ^{-1}\left( \mu -r\mathbf{1}\right)
\end{equation*}%
The beta of asset $i$ is defined as follows (TR-RPB, page 17):%
\begin{equation*}
\beta _{i}=\frac{\mathbf{e}_{i}\Sigma x^{\star }}{x^{\star ^{\top
}}\Sigma x^{\star }}
\end{equation*}%
We can also compute the beta component of the risk premium:%
\begin{equation*}
\pi \left( \mathbf{e}_{i}\mid x^{\star }\right) =\beta _{i}\left(
\mu \left( x^{\star }\right) -r\right)
\end{equation*}%
In our case, we obtain $\phi =2.1473$. The composition of the portfolio is $%
\left( 47.50\%,19.83\%,27.37\%,5.30\%\right) $. We deduce that the
expected return of the portfolio is $\mu \left( x^{\star }\right)
=6.07\%$. Finally,
we obtain the following results:%
\begin{equation*}
\begin{tabular}{|c|cccc|}
\hline
Asset & $x_{i}^{\star }$ & $\pi _{i}$ & $\beta _{i}$ & $\pi \left( \mathbf{e}_{i}\mid x^{\star }\right)$ \\ \hline
$1$ &     $47.50\%$ & $3.00\%$ & $0.737$ & $3.00\%$ \\
$2$ &     $19.83\%$ & $4.00\%$ & $0.982$ & $4.00\%$ \\
$3$ &     $27.37\%$ & $6.00\%$ & $1.473$ & $6.00\%$ \\
$4$ & ${\bP}5.30\%$ & $4.00\%$ & $0.982$ & $4.00\%$ \\
\hline
\end{tabular}%
\end{equation*}
We verify that:
\begin{equation*}
\pi_i = \mu_i - r = \pi \left( \mathbf{e}_{i}\mid x^{\star }\right) =\beta _{i}\left(
\mu \left( x^{\star }\right) -r\right)
\end{equation*}%

\item We obtain the following portfolio weights%
\footnote{For the second investor, the risky assets only represent $50\%$ of his wealth.}:%
\begin{equation*}
\begin{tabular}{|c|ccc|}
\hline
Asset & $x_{i,1}$ & $x_{i,2}$ & $\bar{x}_{i}$ \\ \hline
$1$ &     $47.50\%$ &     $15.82\%$ &      $42.21\%$ \\
$2$ &     $19.83\%$ & ${\bP}3.72\%$ &      $15.70\%$ \\
$3$ &     $27.37\%$ &     $27.09\%$ &      $36.31\%$ \\
$4$ & ${\bP}5.30\%$ & ${\bP}3.37\%$ &  ${\bP}5.78\%$ \\
\hline
\end{tabular}%
\end{equation*}
The corresponding Lagrange coefficients are $\lambda _{1}=0.9314\%$ and $%
\lambda _{1}=1.7178\%$. The expected return $\mu \left(
\bar{x}\right) $ of
the market portfolio is $6.30\%$. Finally, we obtain the following results:%
\begin{equation*}
\begin{tabular}{|c|cccc|c|}
\hline
Asset & $\pi _{i}$ & $\alpha _{i}$ & $\beta _{i}$ & $\pi \left( \mathbf{e}%
_{i}\mid \bar{x}\right) $ & $\alpha _{i}+\beta _{i}\left( \mu \left(
\bar{x}\right) -r\right) $ \\ \hline
$1$ & $3.00\%$ & ${\bN}0.32\%$ & $0.62$ & $2.68\%$ & $3.00\%$ \\
$2$ & $4.00\%$ & ${\bN}0.07\%$ & $0.91$ & $3.93\%$ & $4.00\%$ \\
$3$ & $6.00\%$ &     $-0.41\%$ & $1.49$ & $6.41\%$ & $6.00\%$ \\
$4$ & $4.00\%$ & ${\bN}0.07\%$ & $0.91$ & $3.93\%$ & $4.00\%$ \\
\hline
\end{tabular}%
\end{equation*}

\item The second investor has a cash constraint and invests only $50\%$
of his wealth in risky assets. His portfolio is then highly exposed to the third asset.
This implies that the market portfolio is overweighted in the third asset with respect
to the tangency portfolio. This asset has then a negative alpha. At the opposite,
the first asset has a positive alpha, because its beta is low and it is underweighted in the market portfolio.
\end{enumerate}
\end{enumerate}

\section{Dynamic risk budgeting portfolios}

\begin{enumerate}
\item
\begin{enumerate}
\item The optimization problem is:%
\begin{eqnarray*}
x^{\star } &=&\arg \max x^{\top }\left( \mu -r\mathbf{1}\right) -\frac{\phi
}{2}x^{\top }\Sigma x \\
&\text{u.c.}&\mathbf{1}^{\top }x=1
\end{eqnarray*}%
The first-order condition is:%
\begin{equation*}
\mu -r\mathbf{1}-\phi \Sigma x=0
\end{equation*}%
We have:%
\begin{equation*}
x=\frac{1}{\phi }\Sigma ^{-1}\left( \mu -r\mathbf{1}\right)
\end{equation*}%
Finally, we obtain:%
\begin{equation*}
x^{\star }=c \cdot \Sigma ^{-1}\left( \mu -r\mathbf{1}\right)
\end{equation*}%
with:%
\begin{equation*}
c=\frac{1}{\mathbf{1}^{\top }\Sigma ^{-1}\left( \mu -r\mathbf{1}\right) }
\end{equation*}%
When the correlations are equal to zero, the optimal weights are
proportional to the risk premium $\pi _{i}=\mu _{i}-r$ and inversely
proportional to the variance $\sigma _{i}^{2}$ of asset returns:%
\begin{equation*}
x_{i}^{\star }=c \cdot \frac{\left( \mu _{i}-r\right) }{\sigma _{i}^{2}}
\end{equation*}

\item Let $\sigma \left( x\right) =\sqrt{x^{\top }\Sigma x}$ be the
volatility of the portfolio. We have:%
\begin{eqnarray*}
\mathcal{RC}_{i} &=&x_{i}\cdot \frac{\partial \,\sigma \left( x\right) }{%
\partial \,x_{i}} \\
&=&\frac{x_{i}\cdot \left( \Sigma x\right) _{i}}{\sigma \left( x\right) } \\
&=&\frac{x_{i}\left( \sum_{j=1}^{n}x_{j}\rho _{i,j}\sigma _{i}\sigma
_{j}\right) }{\sigma \left( x\right) } \\
&=&\frac{x_{i}^{2}\sigma _{i}^{2}+x_{i}\sigma _{i}\left( \sum_{j\neq
i}x_{j}\rho _{i,j}\sigma _{j}\right) }{\sigma \left( x\right) }
\end{eqnarray*}

\item When the correlations are equal to zero, we obtain:%
\begin{equation*}
\mathcal{RC}_{i}=\frac{x_{i}^{2}\sigma _{i}^{2}}{\sigma \left( x\right) }
\end{equation*}%
The RB portfolio satisfies:%
\begin{equation*}
\frac{\mathcal{RC}_{i}}{b_{i}}=\frac{\mathcal{RC}_{j}}{b_{j}}
\end{equation*}%
or:%
\begin{equation*}
\frac{x_{i}^{2}\sigma _{i}^{2}}{b_{i}}=\frac{x_{j}^{2}\sigma _{j}^{2}}{b_{j}}
\end{equation*}%
We deduce that:%
\begin{equation*}
x_{i}\propto \frac{\sqrt{b_{i}}}{\sigma _{i}}
\end{equation*}%
The RB portfolio is the tangency portfolio when the risk budgets are
proportional to the square of the Sharpe ratios:%
\begin{eqnarray*}
b_{i} &\propto &\left( \frac{\mu _{i}-r}{\sigma _{i}}\right) ^{2} \\
&=&\frac{\pi _{i}^{2}}{\sigma _{i}^{2}}
\end{eqnarray*}
\end{enumerate}

\item

\begin{enumerate}
\item If $\alpha =\beta =\gamma =\delta =0$, we get:%
\begin{equation*}
b_{i}\left( t\right) =b_{i}\left( \infty \right) =\frac{\pi _{i}^{2}\left(
\infty \right) }{\sigma _{i}^{2}\left( \infty \right) }
\end{equation*}%
The risk budgets at time $t$ are equal to the long-run risk budgets.
However, it does not mean that the allocation is static, because it will
depend on the covariance matrix. If $\alpha =\beta =\gamma =\delta =2$, we
get:%
\begin{equation*}
b_{i}\left( t\right) =\frac{\pi _{i}^{2}\left( t\right) }{\sigma
_{i}^{2}\left( t\right) }
\end{equation*}%
These risk budgets correspond to those given by the tangency portfolio when
the correlations are equal to zero.

\item If the correlations are equal to zero, we have:%
\begin{eqnarray*}
x_{i}\left( t\right) =\frac{\pi _{i}\left( \infty \right) }{\sigma
_{i}^{2}\left( \infty \right) } &\Leftrightarrow &b_{i}\left( t\right) =%
\frac{\pi _{i}^{2}\left( \infty \right) \sigma _{i}^{2}\left( t\right) }{%
\sigma _{i}^{4}\left( \infty \right) } \\
&\Leftrightarrow &b_{i}\left( t\right) =b_{i}\left( \infty \right) \frac{%
\sigma _{i}^{2}\left( t\right) }{\sigma _{i}^{2}\left( \infty \right) }
\end{eqnarray*}%
It implies that $\alpha =\beta =0$ and $\gamma =\delta =-2$.

\item The long-run tangency portfolio is $x_{1}^{\star }\left( \infty
\right) =38.96\%$, $x_{2}^{\star }\left( \infty \right) =25.97\%$, $%
x_{3}^{\star }\left( \infty \right) =19.48\%$ and $x_{4}^{\star }\left(
\infty \right) =15.58\%$. The risk budgets $b_{i}\left( \infty \right) $ are
all equal to $25\%$ meaning that the long-run tangency portfolio is the ERC
portfolio. The relationship between the parameters $\theta $, the risk
budgets $b_{i}\left( t\right) $ and the RB weights $x_{i}\left( t\right) $
is reported in Figure \ref{fig:app2-3-10-1}. The allocation at time $t$ differs from the
long-run allocation when the parameter $\theta $ increases. $\theta $ may
then be viewed as a parameter that controls the relative weight of the
tactical asset allocation with respect to the strategic asset (or long-run)
allocation.

\begin{figure}[tbph]
\centering
\includegraphics[width = \figurewidth, height = \figureheight]{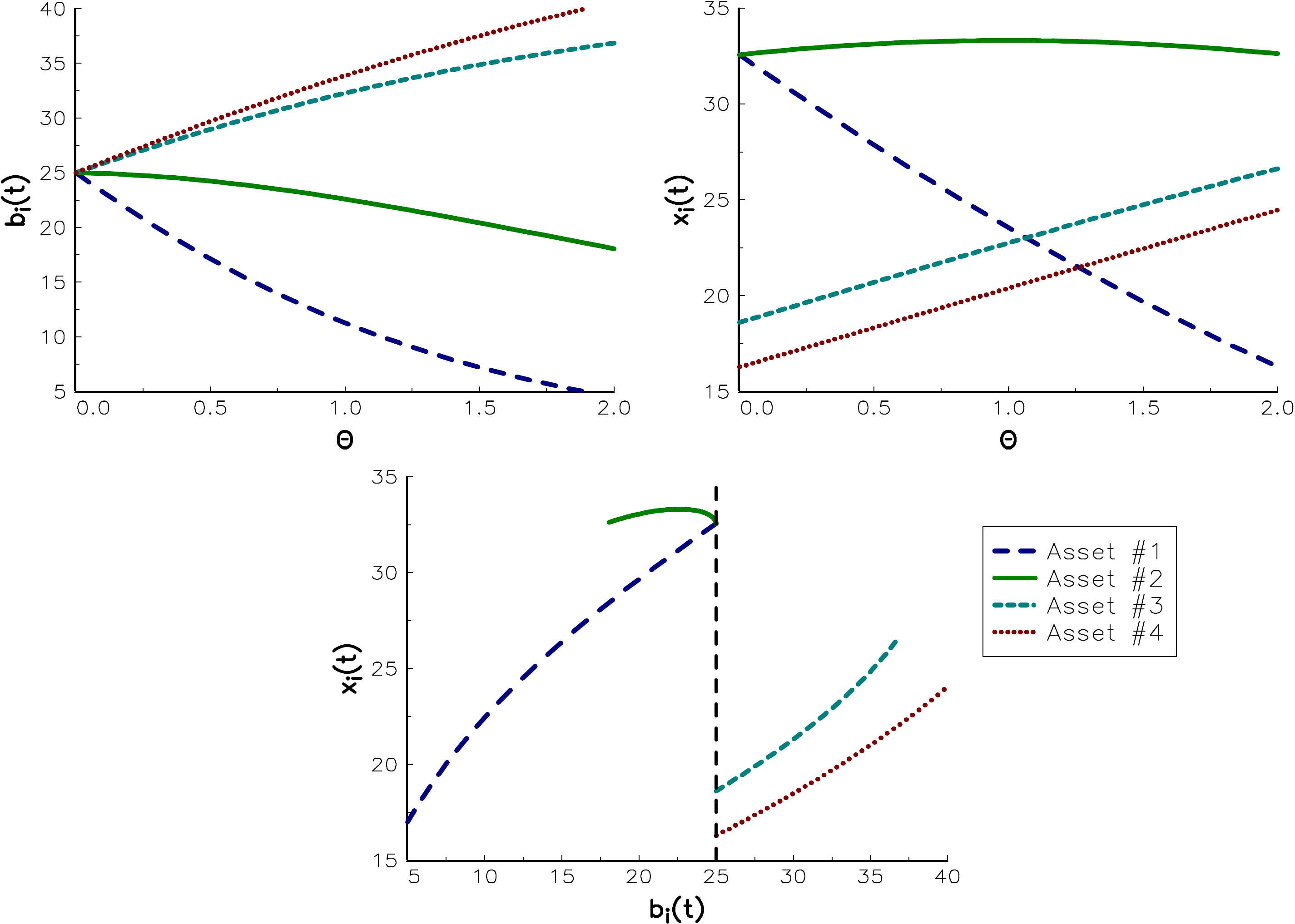}
\caption{Relationship between $\theta $, $b_{i}\left( t\right) $ and $x_{i}\left( t\right) $}
\label{fig:app2-3-10-1}
\end{figure}
\end{enumerate}

\item

\begin{enumerate}
\item The backtests are reported in Figure \ref{fig:app2-3-10-2}. We notice that the risk
budgets change from one period to another period in the case of the dynamic
risk parity strategy. However, the weights are not so far from those
obtained with the ERC strategy. We also observe that the performance is better
for the dynamic risk parity strategy whereas it has the same volatility than
the ERC strategy.

\begin{figure}[tbph]
\centering
\includegraphics[width = \figurewidth, height = \figureheight]{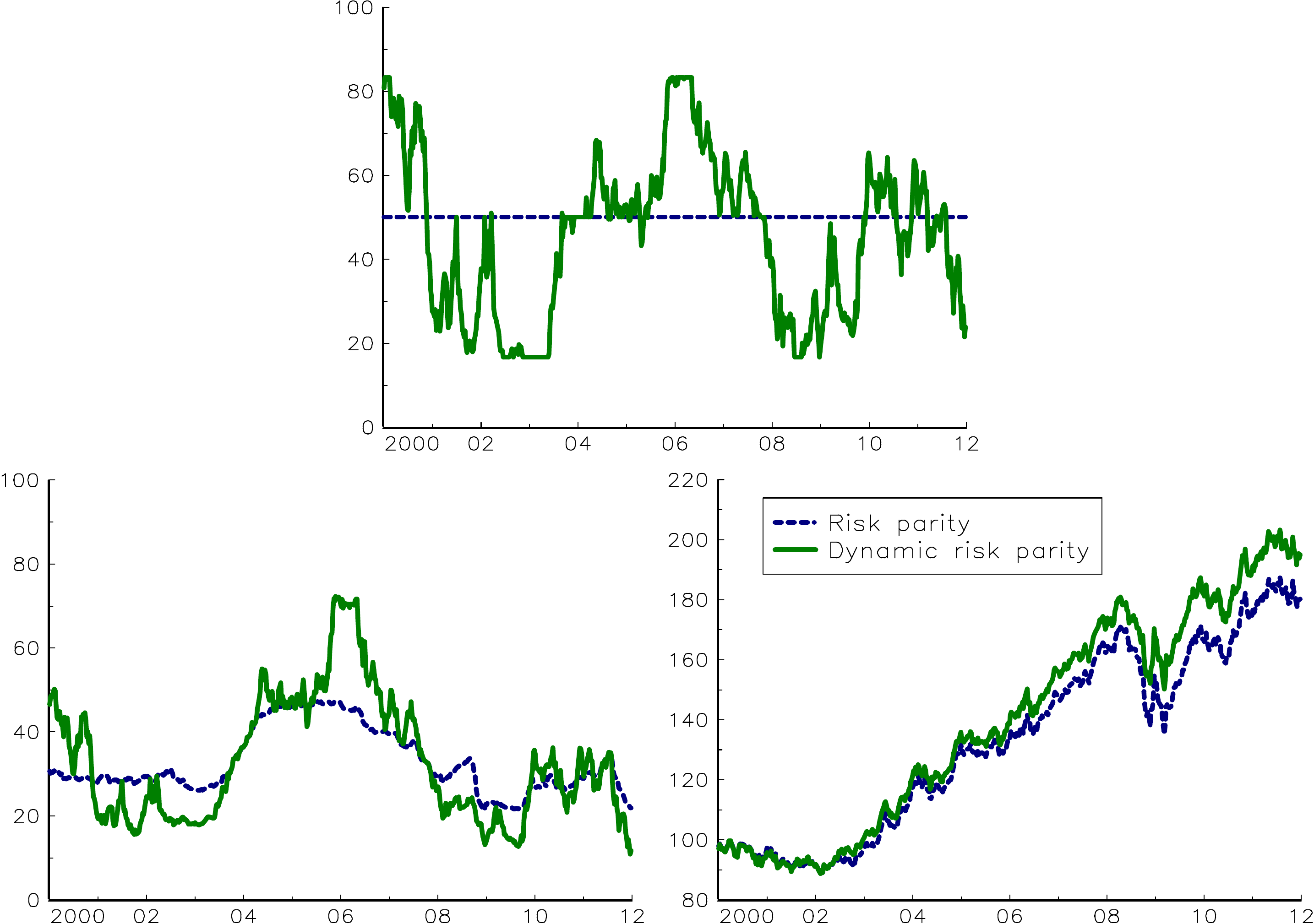}
\caption{Simulation of the dynamic risk parity strategy}
\label{fig:app2-3-10-2}
\end{figure}

\item We have reported the evolution of equity weights in Figure \ref{fig:app2-3-10-3}. We
notice that the tangency portfolio produces a higher turnover. Moreover, it
is generally invested in only one asset class, either stocks or bonds.

\begin{figure}[tbph]
\centering
\includegraphics[width = \figurewidth, height = \figureheight]{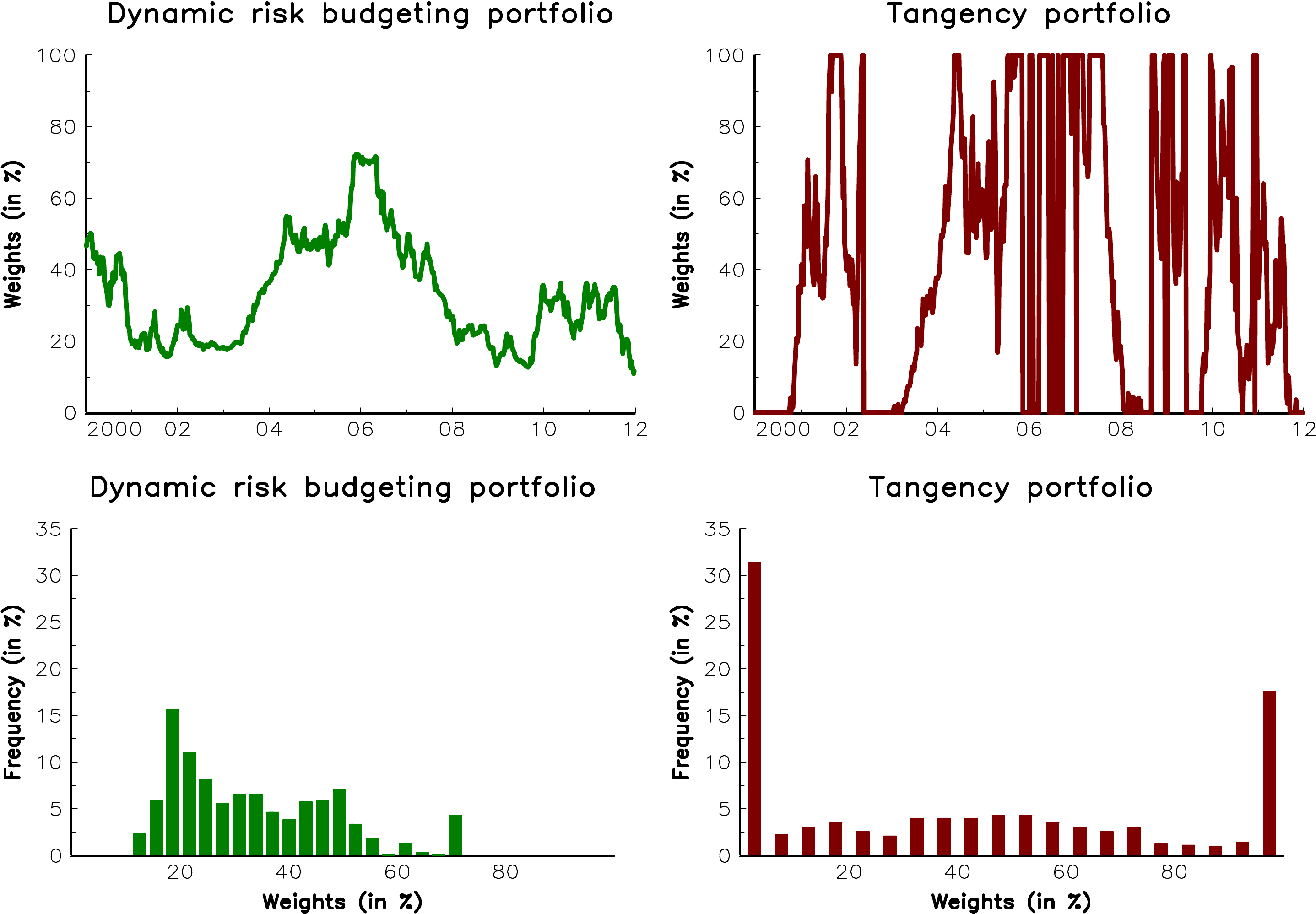}
\caption{Comparison of the weights}
\label{fig:app2-3-10-3}
\end{figure}

\item Results are given in Figure \ref{fig:app2-3-10-4}.

\begin{figure}[tbph]
\centering
\includegraphics[width = \figurewidth, height = \figureheight]{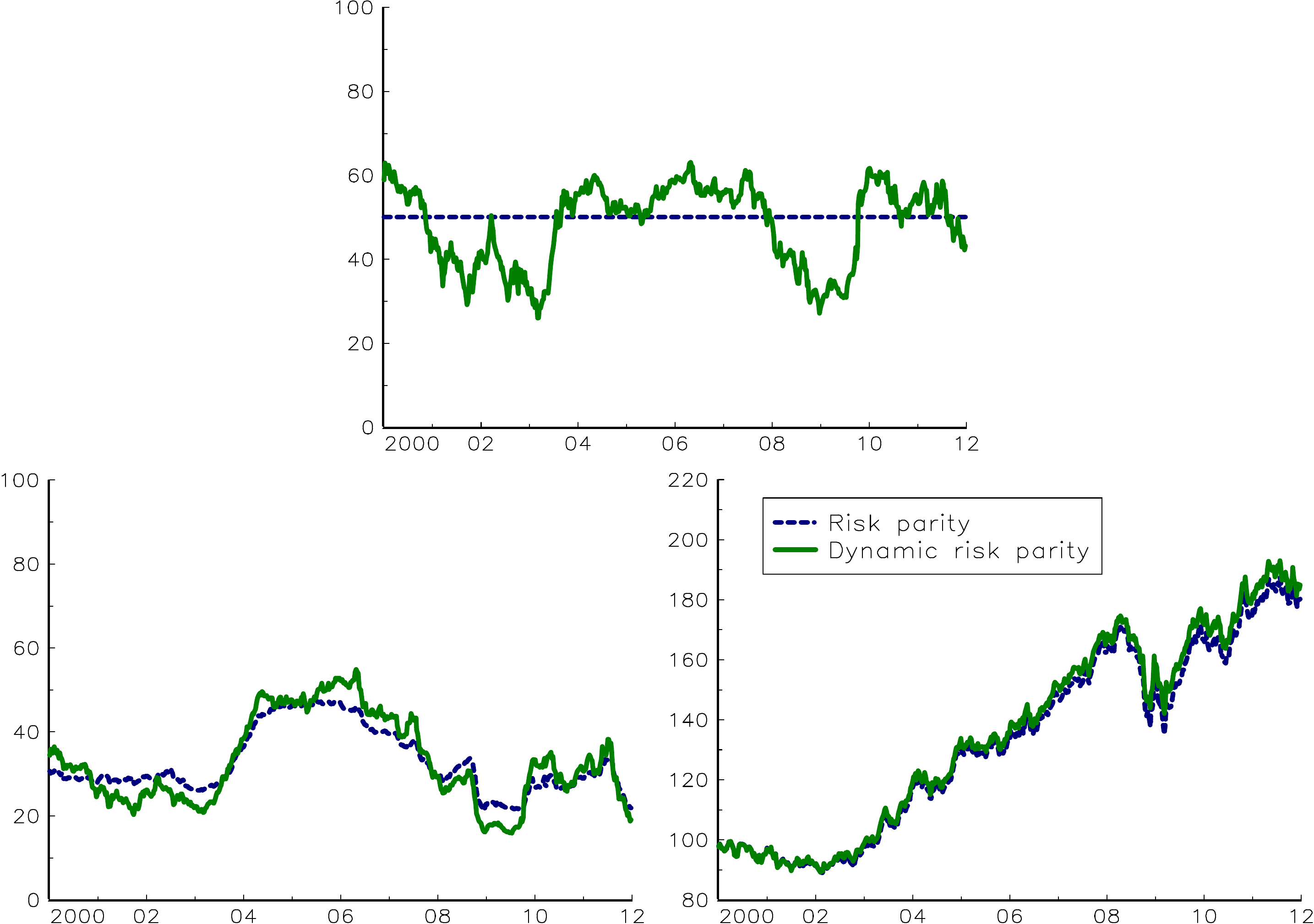}
\caption{Simulation of the second dynamic risk parity strategy}
\label{fig:app2-3-10-4}
\end{figure}

\item We notice that the first simulation is based on the Sharpe ratio
whereas the second simulation considers the risk premium. Nevertheless, the
risk premium of stocks is not homogeneous to the risk premium of bonds. This
is why it is better to use the Sharpe ratio.
\end{enumerate}
\end{enumerate}

\end{document}